\documentclass[twocolumn,secnumarabic,amssymb, aps,prx,showpacs]{revtex4-2}
\usepackage{array}[=2016-10-06]
\usepackage{makecell}
\usepackage[justification=raggedright,singlelinecheck=false]{caption}
\usepackage{tabularx}

\usepackage{xcolor}
\usepackage{float}
\usepackage{comment}
\usepackage{amsmath,amsthm,amssymb}

\usepackage{graphicx}
\usepackage[abs]{overpic}
\usepackage[labelfont=bf]{caption} 
\usepackage{subcaption} 
\usepackage{tikz}

\newcommand{\KeyholeMoundIcon}{%
\begin{tikzpicture}[x=1mm,y=1mm,baseline=-0.5ex]
  \draw[line width=0.35pt, fill=white]
    (0,0.2) -- (6.0,1.0) -- (6.0,3.0) -- (0,3.8) -- cycle;
  \draw[line width=0.35pt, fill=white]
    (8.8,2.0) circle (2.8);
\end{tikzpicture}%
}

\newcommand{\CircularMoundIcon}{%
\begin{tikzpicture}[x=1mm,y=1mm,baseline=-0.5ex]
  \draw[line width=0.35pt, fill=white] (3,2) circle (2.8);
\end{tikzpicture}%
}

\newcommand{\SquareFrontedMoundIcon}{%
\begin{tikzpicture}[x=1mm,y=1mm,baseline=-0.5ex]
  \draw[line width=0.35pt, fill=white]
    (0,0.2) -- (6.0,1.0) -- (6.0,3.0) -- (0,3.8) -- cycle;
  \draw[line width=0.35pt, fill=white]
    (5.9,0.0) rectangle (9.9,4.0);
\end{tikzpicture}%
}

\newcommand{\FirmIcon}{%
\begin{tikzpicture}[x=1mm,y=1mm,baseline=-0.5ex]
  \draw[line width=0.35pt, fill=white] (0,0) rectangle (2.6,6.0);
  \draw[line width=0.35pt, fill=white] (3.2,0) rectangle (5.4,4.2);
\end{tikzpicture}%
}

\newcommand{\PanelLabel}[1]{%
  \color{black}\Large\sffamily\bfseries (#1)%
}

\newcommand{\ShapeLabel}[2]{%
\begin{tikzpicture}[baseline=-0.5ex]
  \node[
    fill=white,
    fill opacity=0.85,
    text opacity=1,
    inner sep=1.5pt,
    anchor=west
  ] {\scriptsize\sffamily #1\hspace{2pt}#2};
\end{tikzpicture}%
}

\makeatletter
\let\oldappendix\appendix

\renewcommand{\appendix}{%
  \oldappendix
  
  \gdef\p@subsection{\thesection.}%
  \gdef\p@subsubsection{\thesection.\thesubsection.}%
}
\makeatother
\usepackage[colorlinks=true, allcolors=blue]{hyperref}

\begin{document}
\title{
Zipf's law before the monetary economy and written administration: volume distribution of kofun, ancient Japanese burial mounds
}
\author{Hayafumi Watanabe$^{1,2}$}\email[E-mail: ]{hayafumi.watanabe@gmail.com}

\affiliation{Department of Economics, Seijo University, Setagaya-ku, Tokyo 157-8511, Japan}
\affiliation{The Institute of Statistical Mathematics, Tachikawa-shi, Tokyo 190-8562,Japan}


\begin{abstract}
We analyze the volume distribution of kofun, large mounded tombs constructed in the Japanese archipelago mainly from the mid-third through the seventh century CE. We ask whether the distribution of collectively mobilized resources in an ancient society without a monetary economy or written administration could follow Zipf's law, as observed in modern firm sales. Here, Zipf's law refers to a power-law distribution with an exponent close to one. Using a nationwide database, we estimate kofun volumes as a proxy for labor and resources mobilized by the constructing groups, and analyze the volume distributions archipelago-wide and by region, period, and mound type. We find that the volume distribution of keyhole-shaped kofun exhibits a Zipf-like upper tail with a cumulative power-law exponent close to unity, while its central part is close to log-normal---a shape that resembles the distribution of modern firm sales. Moreover, many regional and temporal differences appear primarily as scale differences: after median normalization, most distributions collapse onto a common curve and remain approximately Zipf-like. However, some exceptional groups exist---such as the politically central Kinki region---that show heavier-than-Zipf tails (cumulative exponents below unity), indicating stronger concentration among the largest kofun. To interpret these regularities, we introduce a Kesten-type stochastic growth model with stopping and reorganization. The model provides a unified account in which the log-normal-like body, Zipf-like tail, and regional/temporal variations arise from a common growth process. Together, these findings raise the possibility that collectively mobilized resources exhibited a Zipf-like structure already in the Kofun period.
\end{abstract}

\maketitle
\section{Introduction}
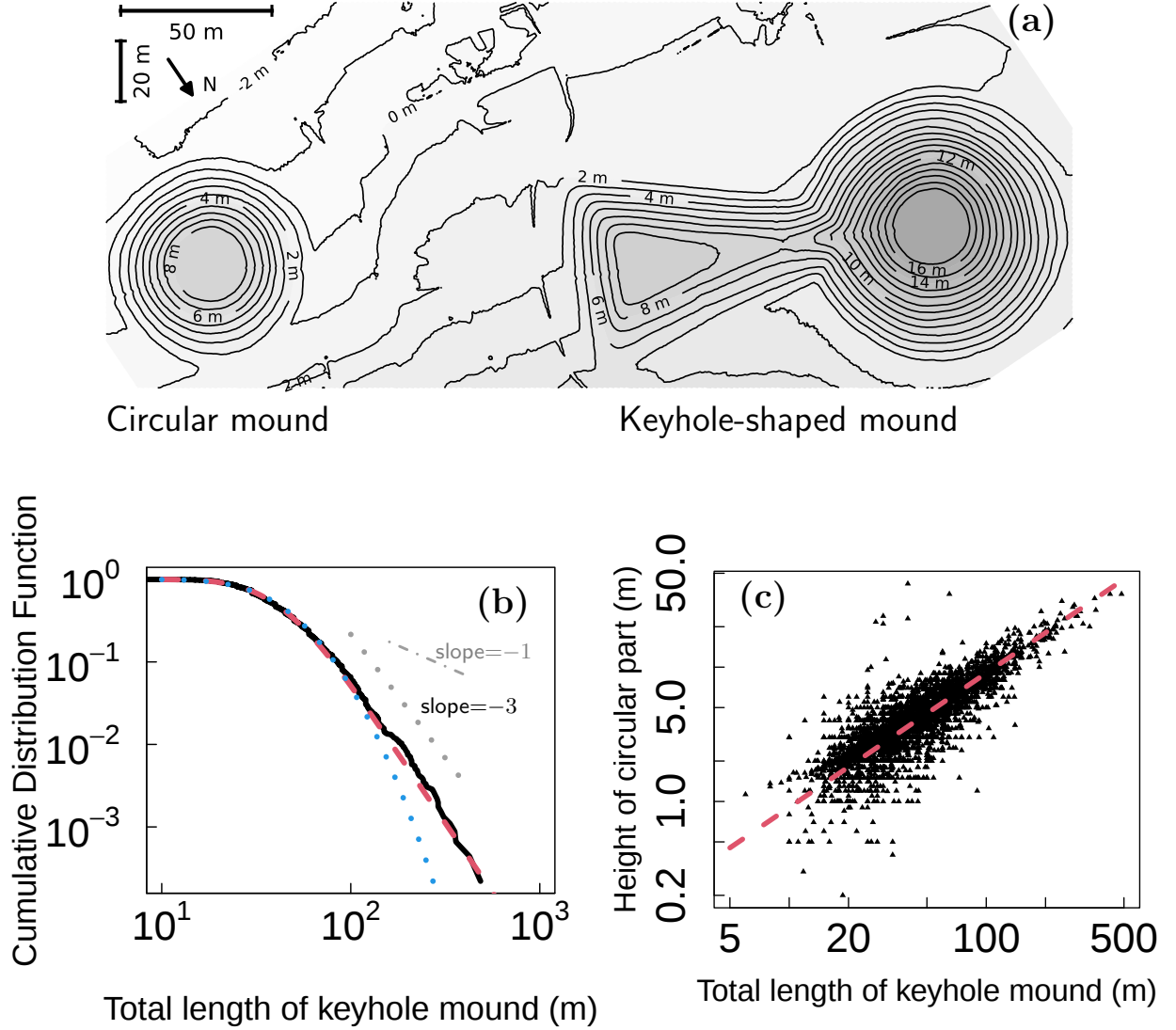
\begin{figure*}[!t]
    \begin{overpic}[width=15.7cm]{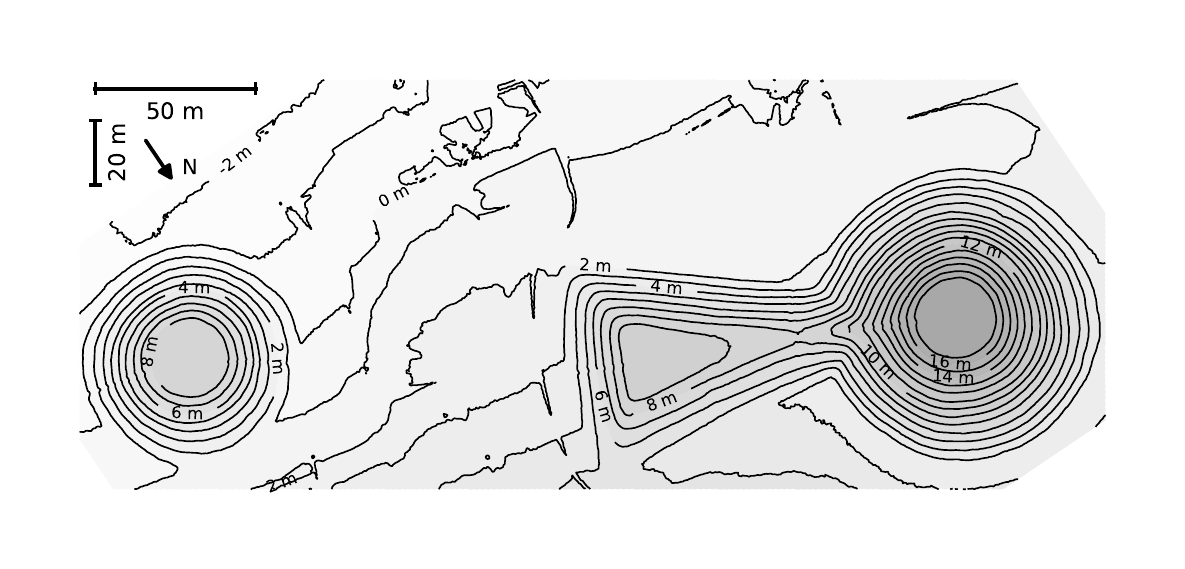}
         \put(390,173){\color{black}\Large\bfseries (a)}
        \put(235,13){\color{black}\Large \sffamily Keyhole-shaped mound}
          \put(30,13){\color{black}\Large \sffamily Circular mound}
        
    \end{overpic}
      \begin{overpic}[width=8.7cm]{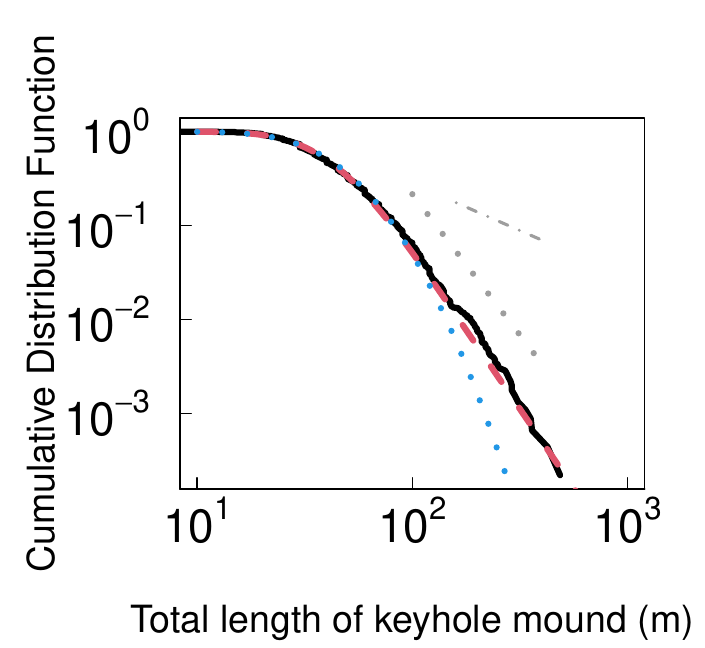}
        \put(195,168){\color{black}\Large\bfseries (b)}
         \put(179,150){\color{black!50}\sffamily slope=$-1$}
         \put(173,128){\color{black}\sffamily slope=$-3$}
    \end{overpic} 
        \begin{overpic}[width=8.1cm]{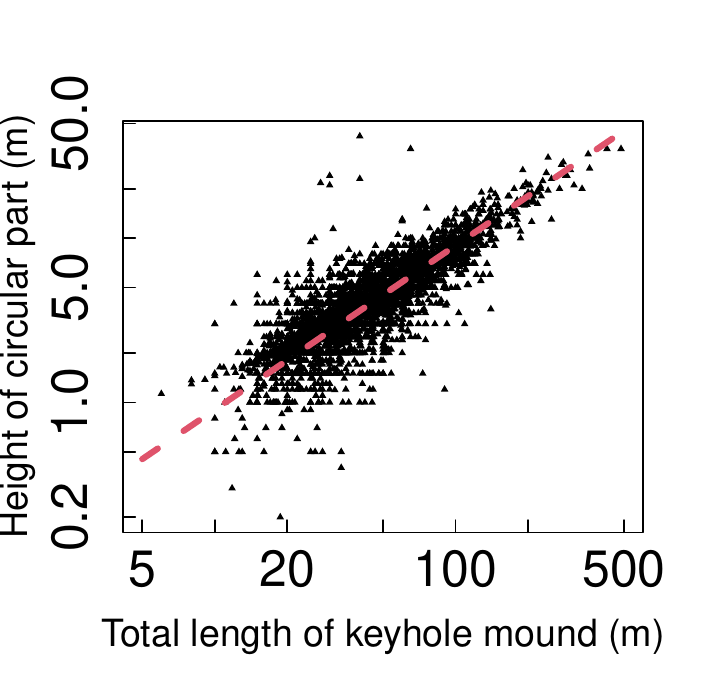}
        \put(50,170){\color{black}\Large\bfseries (c)}
    \end{overpic} 
    %
\caption{
Shapes and size distributions of kofun.
(a) Examples of kofun. A round kofun is shown on the left, and an example of a large keyhole-shaped kofun is shown on the right. The kofun on the right is a massive earthen structure, with a mound length exceeding 100 m and a rear circular part whose height exceeds 16 m. The figure was produced from the Yamanashi Prefecture point-cloud data \cite{YamanashiPrefecture2024PointCloud}. 
Fig.~\ref{app_fig_kofun_image} shows a photograph of a restored keyhole-shaped kofun and a model illustrating its construction process.
(b) Complementary cumulative distribution of mound length for keyhole-shaped kofun. The black line shows the empirical data, the red dashed line shows the one-sided dPlN distribution (Eq.~\eqref{app_eq_dPlN}; $\mu=3.42,\sigma=0.435,\alpha=3.36$), and the blue dotted line shows the lognormal distribution ($\mu=3.72,\sigma=0.537$). The gray guide lines indicate $x^{-1}$ and $x^{-3}$. The empirical tail is close to a power law with exponent 3.
(c) Relation between mound length and the height of the rear circular part for keyhole-shaped kofun. The red dashed line indicates $y=0.09x$. Mound length and the height of the rear circular part are approximately proportional.
}
\label{fig_shape}
\end{figure*}

At first glance, twenty-first-century firms and kofun appear to have little in common. Kofun are large mounded tombs constructed across the Japanese archipelago mainly from the mid-third century through the seventh century CE (Fig.~\ref{fig_shape}(a), Fig. \ref{app_fig_kofun_image}) \cite{Matsugi2025Kofun}. Yet, the annual sales distributions of modern firms and kofun volume distributions share a common statistical regularity: both follow a Zipf-like distribution, in which a small number of top-ranked units account for a disproportionately large share of the total (Fig. \ref{app_fig_firm_kofun}). Formally, the upper cumulative distribution follows a power-law distribution with an exponent of approximately 1,  
\begin{equation}
P(X \geq x) \propto x^{-\alpha}, \qquad \alpha \approx 1.
\label{eq:zipf}
\end{equation}
Such concentration patterns have so far been studied mainly in socioeconomic data recorded under institutional infrastructures such as written administration, monetary economies, taxation systems, and corporate accounting \cite{DiGiovanni2013ZipfsWorld,Axtell2001ZipfFirmSizes,Fujiwara2004ParetoZipfGibrat,Mizuno2002StatisticalLaws,Cristelli2012MoreThanPowerLaw,Arvanitidis2016ZipfMilitaryExpenditures}. Here, we show that a comparable Zipf-like pattern can also be observed in a society in which written administration and a monetary economy had not yet become fully established. \par
%
Inequality in ancient societies has been widely studied using archaeological proxies such as house size and grave size, yet much less is known about the distributional form of such inequality. In particular, it remains unclear whether resources mobilized or accumulated by social groups can show a concentration pattern comparable to that of modern collective units, or whether such patterns instead require the institutional infrastructures discussed above. This question is fundamental for understanding the distributional structure of competition and inequality in human societies, but direct quantitative records comparable to income, sales, tax records, or GDP are rarely preserved for societies that predate such institutions.\par

A key distinction underlying this question is between inequality among individuals or households and resource distributions among collective units such as political groups, firms, and states. The upper tails of individual income and wealth distributions often show heavy-tailed power-law behavior. However, their estimated exponents vary across income types, countries, and time periods \cite{Souma2001UniversalStructure,Aoyama2000ParetoLawIncomeDebt,DeVries2022CapitalLaborPareto}.
By contrast, collective-unit resource distributions such as firm sales and cross-country distributions of GDP and military expenditures more consistently exhibit Zipf-like behavior---that is, an upper cumulative exponent close to 1---than individual income and wealth distributions.
\par

For premodern wealth and inequality, some studies have also examined power-law, or Pareto-type, distributions using evidence such as house sizes, grave sizes, landholding records, and property registers \cite{AbulMagd2002,Danon2022PompeiiSenatorialWealth,Danon2025ReconstructingWealthDistributions,YuEtAl2019GraveSizesInequality,NohKim2026SouthKoreaInequality,StrawinskaZankoEtAl2018MayaInequality,BrownEtAl2012PoorMayapan}. However, many of these studies focus mainly on individual- or household-level distributions, or on distributions confined to a single site, city, or region. Premodern cases in which resource distributions among collective units show Zipf-like concentration over broad spatial scales are very limited. One of the few examples is the distribution of the number of serf families owned by medieval Hungarian nobles \cite{HegyiEtAl2007}, but this case also comes from a society with written records and administrative institutions. Whether collective-unit resource distributions can exhibit Zipf-like concentration over such broad spatial scales in societies where writing and money were not yet fully established therefore remains unknown. A summary of related studies is provided in the SI Appendix~\ref{app_sec_old}. \par

This study addresses this gap using kofun volume distributions from Japan's Kofun period, a phase of state formation in the Japanese archipelago generally dated from the mid- to late third century through the seventh century CE. 
During this period, regional chiefdoms were gradually reorganized into a broader political order against the background of expanding agricultural societies, interregional exchange, and the circulation of metal tools and prestige goods. Written administrative records and a monetary economy had not yet become central institutions of society, particularly during the fourth to sixth centuries CE, the core period of this study. At the same time, many kofun constructed under the authority of ruling elites and influential local groups, including keyhole-shaped kofun (Fig.~\ref{fig_shape}(a)), were large-scale construction projects that could not have been achieved through individual labor alone (Fig. \ref{app_fig_kofun_image}). The volumes of these mounds can therefore be interpreted not as direct measures of the personal wealth of the deceased, but as material traces of the labor, resources, and organizational coordination mobilized by the groups responsible for their construction \cite{Obayashi1985Nintoku}. \par

Several previous studies have provided important insights into kofun size distributions. Ozawa reported that the volume distribution of 23 early keyhole-shaped kofun shows power-law behavior \cite{Ozawa2012MathematicalHistory}. 
In addition, Okubo reported an archipelago-wide analysis of keyhole-shaped kofun mound lengths, identifying power-law behavior in the upper range and a log-normal body in the bulk of the distribution \cite{Okubo2019RankSizeKofun}.
These studies suggest that kofun size distributions may contain both a power-law upper tail and a log-normal body. However, the existing evidence remains limited in scope. The volume-based analysis was limited to this early subset, whereas the archipelago-wide analysis used mound length rather than volume. Therefore, it has not yet been systematically tested to what extent kofun volume itself exhibits a Zipf-like upper tail across the Japanese archipelago and by region, period, and mound shape. \par

\begin{figure*}[t]
\centering
\setlength{\tabcolsep}{2pt}
\renewcommand{\arraystretch}{0.3}
\begin{tabular}{cc}
\begin{overpic}[percent,width=7.0cm]{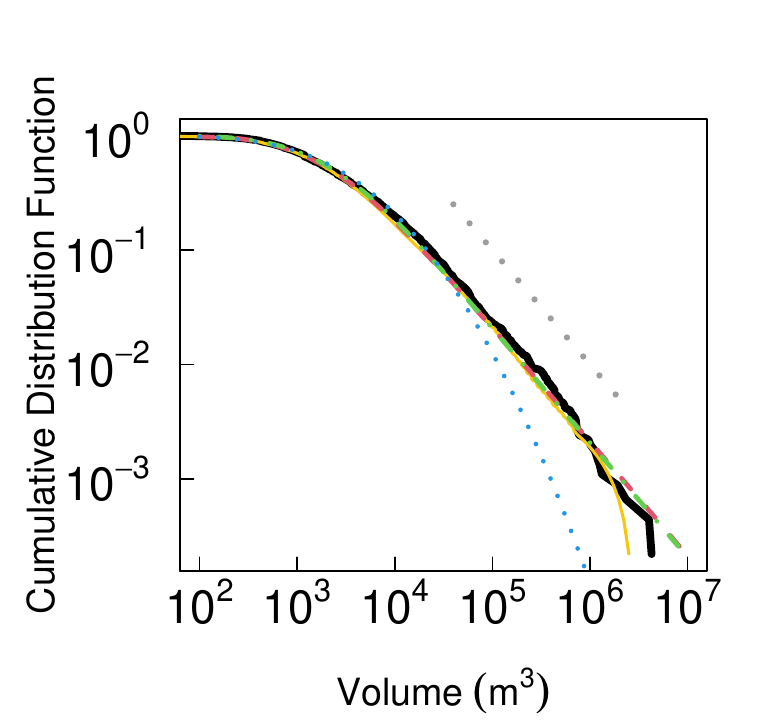}
  \put(26,65){\PanelLabel{a}}
  \put(28,30){\ShapeLabel{\KeyholeMoundIcon}}
  \put(28,23){Keyhole-shaped mound}
  \put(65,64){\color{black!75}\large\sffamily slope=$-1$}
\end{overpic}
&
\begin{overpic}[percent,width=7.0cm]{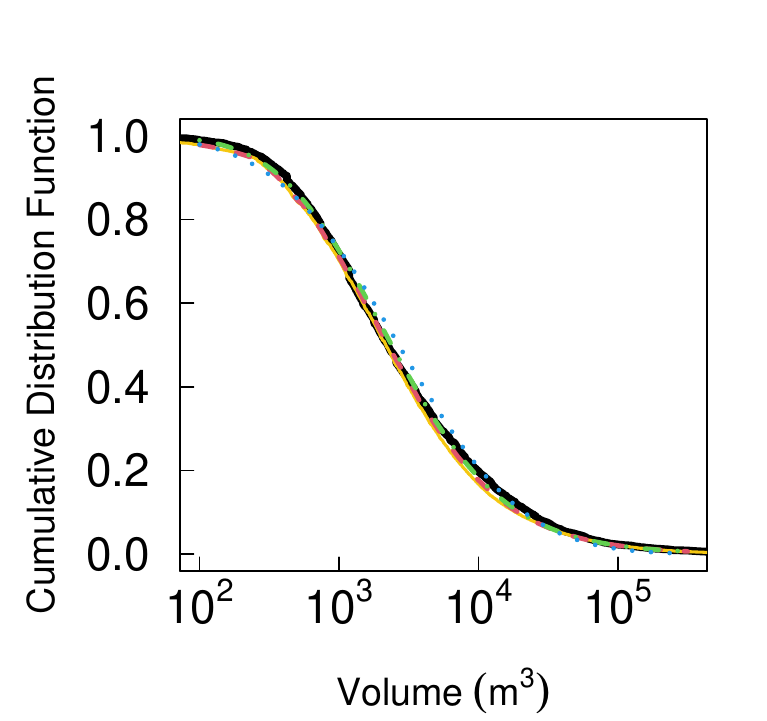}
  \put(26,65){\PanelLabel{b}}
  \put(56,70){\large Semi-log plot}
  \put(27,30){\ShapeLabel{\KeyholeMoundIcon}}
  \put(26,23){Keyhole-shaped mound}
  
\end{overpic}
\\
\begin{overpic}[percent,width=7.0cm]{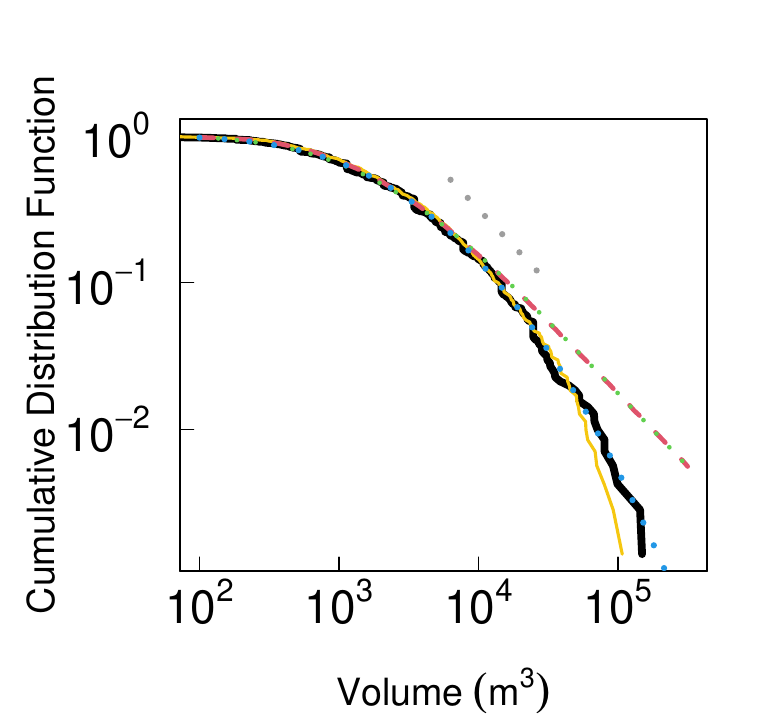}
  \put(26,65){\PanelLabel{c}}
  \put(28,30){\ShapeLabel{\CircularMoundIcon}}
  \put(28,23){Circular mound}
  \put(68,64){\color{black!75}\large\sffamily slope=$-1$}
\end{overpic}
&
\begin{overpic}[percent,width=7.0cm]{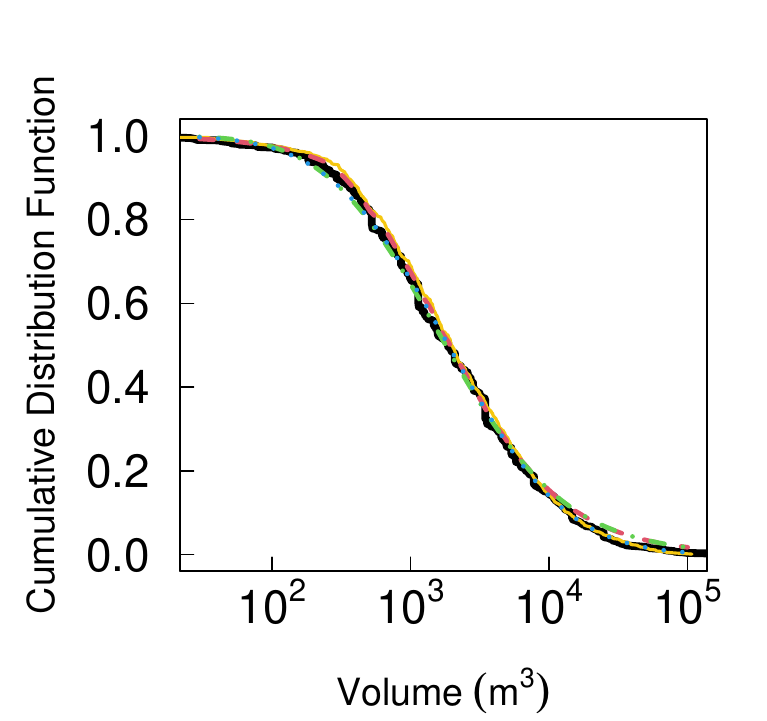}
  \put(26,65){\PanelLabel{d}}
  \put(56,70){\large Semi-log plot}
  \put(27,30){\ShapeLabel{\CircularMoundIcon}}
   \put(26,23){Circular mound}
\end{overpic}
\end{tabular}
\caption{
Volume distributions of kofun.
The black solid lines show the empirical data, the red dashed lines show the theoretical distributions generated by the Kesten process (Eq.~\eqref{eq:kesten}; $A_0$, $b_0$, $\alpha$), the green dash-dotted lines show the one-sided dPlN distributions (Eq.~\eqref{app_eq_dPlN}; $\mu_1$, $\sigma_1$, $\alpha$), the blue dotted lines show the lognormal distributions ($\mu_2$, $\sigma_2$), and the thin yellow solid lines show simulated samples generated from the Kesten process. The gray dashed lines are guide lines with slope $-1$. The left column shows log--log plots, and the right column shows semi-log plots.
(a),(b) Keyhole-shaped kofun: $\alpha=1$, $A_0=225$, $b_0=1.13$, $\mu_1=6.88$, $\sigma_1=1.23$, $\mu_2=7.89$, $\sigma_2=1.62$.
(c),(d) Round kofun: $\alpha=1$, $A_0=263$, $b_0=1.16$, $\mu_1=6.58$, $\sigma_1=1.34$, $\mu_2=7.54$, $\sigma_2=1.55$.
The figure shows that, for all three types, the central part of the distribution is close to a lognormal distribution, whereas the tail exhibits a power-law-like form with an exponent close to 1. In particular, for the keyhole-shaped kofun in (a), the theoretical distribution generated by the Kesten process agrees well with the empirical data in both the central part and the tail. In contrast, for the round kofun in (c), the tail is somewhat rounded and the overall shape is closer to a lognormal distribution. However, the yellow line in (c) shows that, in finite-sample simulations, data generated from the Kesten process can also appear close to a lognormal distribution.
}
\label{fig_volume}
\end{figure*}

In this study, we estimate kofun volumes from a country-wide database that includes mound length, height, and shape, and analyze their distributions across the Japanese archipelago (Fig.~\ref{fig_map}(a)).  We show that the volume distribution of keyhole-shaped kofun has both a Zipf-like tail, with a cumulative exponent close to 1, and a log-normal body (Fig.~\ref{fig_volume}(a), (b)). We also show that many regional and temporal differences approximately collapse onto a common distributional form after normalization by the median(Fig.~\ref{fig_area_cdf}). At the same time, the distributions for several periods in the politically central Kinki region and for some periods in the Sanyo region, which includes areas that were semi-central during the Kofun period, exhibit thick tails that clearly deviate from the typical distribution(Fig.~\ref{fig_map}(b)-(e)). These features can be modeled in a unified way by a simple stochastic process in which the scale of the group behind the construction of each kofun (hereafter referred to as a local group) grows multiplicatively, while growth stops or the group is reorganized at random times (Sec. \ref{sec_model}). Mathematically, this model is a special case of a Kesten process. \par 

These results show that, even in societies where institutional infrastructures based on writing and money had not yet been fully established, the distribution of politico-economic resources can exhibit a statistical structure similar to that of modern firm sales distributions. We further situate kofun volume distributions within the same framework as two later cases: distributions of kokudaka---a rice-yield-based measure of the productive capacity of political units in early modern Japan---and modern firm sales distributions (Secs. \ref{sec_nowfirm} and \ref{sec_kokudaka}). Through this comparison, we discuss the possibility that Zipf-like concentration is not a product specific to particular institutions, but can arise from general processes of growth, competition, persistence, and reorganization in human groups.\par

\section{Empirical analysis of kofun size distributions}
\label{sec_results}
\subsection{Kofun shapes and basic size distributions}
\label{sec_shape}
Fig. ~\ref{fig_shape}(a) shows representative examples of the kofun analyzed in this study: a round kofun and a keyhole-shaped kofun. Although kofun shapes vary by mound type, the basic structure within each type is broadly shared despite stylistic variation across periods and regions. This study focuses on the three major mound types introduced in Sec.~\ref{sec_method_data}: keyhole-shaped kofun ($n=4545$), round kofun ($n=705$), and square-fronted, square-rear kofun ($n=455$; hereafter, the square-fronted type). Schematic icons for each mound type are shown in Figs.~\ref{fig_volume} and \ref{app_fig_volume}.\par
As the contour lines in Fig.~\ref{fig_shape}(a) indicate, kofun are not flat archaeological features but three-dimensional earthen mounds. The Kai Ch\={o}shizuka Kofun, shown on the right of the figure, is a large keyhole-shaped mound constructed around the fourth century, with a total mound length of 169\,m and a rear circular part 15\,m in height; it ranks among the largest kofun nationally. Across all kofun, mound height is approximately proportional to mound length (Fig.~\ref{fig_shape}(c)); round kofun show a similar relationship (Fig.~\ref{app_fig_shape_en}(c)).\par
Fig. ~\ref{fig_shape}(b) shows the distribution of mound length $L$ for keyhole-shaped kofun. On a log--log plot, the tail closely resembles a power-law distribution with an exponent of approximately 3. That is, mound length itself does not follow Zipf's law, but shows a power-law tail with a different exponent. Similar behavior is observed for the height and diameter of the rear circular part and for the width of the front rectangular part (SI Appendix~\ref{app_fig_shape_xy}).
\subsection{Distribution of kofun volumes}
As three-dimensional earthen mounds, kofun volume is a natural measure of their size. Moreover, the principal cost of construction, driven by the amount of earth moved, is roughly proportional to volume \cite{Obayashi1985Nintoku}. We therefore treat kofun volume as a proxy for the resources or mobilizing capacity invested in construction; this assumption is discussed further in Sec.~\ref{app_sec_model_volume}. Volumes were estimated using the method described in Sec.~\ref{sec_method_volume} and SI Appendix~\ref{app_sec_method_volume}.
\subsubsection{Keyhole-shaped kofun}
\paragraph{Zipf's law in the right tail}
Figure~\ref{fig_volume}(a) shows the distribution of estimated volume $V$ for keyhole-shaped kofun on a log--log plot.
The tail follows a power-law distribution with an exponent of 1, corresponding to Zipf's law (Clauset-method estimate: $\alpha = 1.00$; see Sec.~\ref{sec_method_powerlaw} for the estimation method). Under the proxy interpretation of kofun volume, this result suggests that the distribution of resources or mobilizing capacity may have followed a Zipf-like form even before writing and monetary exchange became central organizing institutions.

\paragraph{Log-normal body and approximation of the whole distribution}
In the semi-log plot of Fig.~\ref{fig_volume}(b), the central and lower-size range of the distribution closely resembles a log-normal distribution (blue dotted line). Together with the Zipf-like tail described above, this indicates that the volume distribution combines a log-normal body with a power-law tail.

To describe this overall shape, we use a one-sided double Pareto-lognormal (dPlN) distribution, a distributional form that combines a log-normal body with a Pareto-type tail. The green dash-dotted line shows the one-sided dPlN fit, which approximates the observed distribution well. The dPlN and one-sided dPlN distributions are described in SI Appendix~\ref{app_sec_dPlN}. The red dashed line shows the distribution generated by the theoretical model introduced in Sec.~\ref{sec_model}, which gives a similar approximation to the observed distribution.

Distributions with a log-normal body and a Zipf-like power-law tail have also been observed in modern firm sales distributions \cite{Ishikawa2008FirmSizeDisplacement}, as well as in early modern \textit{kokudaka} distributions and Japanese prefectural GDP distributions analyzed in SI Appendix~\ref{a_sec_kokudaka}. Here, \textit{kokudaka} refers to a rice-yield-based measure of the economic scale of early modern Japanese domains. 

\paragraph{Relation between mound length and volume}
\label{sec_scaling}
The tail exponent near 1 for volume is consistent with the exponent near 3 for mound length.
Because mound height is approximately proportional to mound length, kofun volume can be approximated as $V \propto L^3$.
A simple change of variables in the upper cumulative distribution shows that a cumulative power-law exponent $\alpha_L \approx 3$ for mound length corresponds to $\alpha_V = \alpha_L/3 \approx 1$ for volume; the derivation is given in SI Appendix~\ref{app_sec_powerlaw}.
\subsubsection{Round kofun}
Results for round kofun are shown in Figs.~\ref{fig_volume}(c,d). In the semi-log plot, the central portion of the distribution again resembles a log-normal distribution (blue dotted line). The overall distribution is also well approximated by a single log-normal distribution across its full range (Kolmogorov--Smirnov goodness-of-fit test, $p=0.52$), and unlike keyhole-shaped kofun, round kofun show no strong evidence of a distinct power-law tail. \par

Given the small sample size, however, it is difficult to distinguish a purely log-normal distribution from one with a log-normal body and a power-law tail. As a diagnostic restricted to the tail alone, the Clauset-method estimate gives $\alpha = 0.99$ for round kofun—consistent with the exponent found for keyhole-shaped kofun. 
As a reference, we also compare the round-kofun distribution with the theoretical model introduced in Sec.~\ref{sec_model}, with parameters chosen so that the asymptotic tail exponent equals 1. A finite-sample simulation under these conditions (yellow solid line) produces a distribution close to a log-normal form. The corresponding theoretical distribution (red dashed line) shows some deviation from the observed round-kofun distribution, but this deviation is not statistically significant (goodness-of-fit test, $p=0.41$). We return to this issue of tail shape in Sec.~\ref{sec_teisi}, where it is discussed in relation to finite-sample effects and stopping processes.  Results for the square-fronted type are provided in SI Appendix~\ref{app_sec_square}.

\begin{figure*}[t]
    \centering
    \begin{overpic}[width=5.8cm]{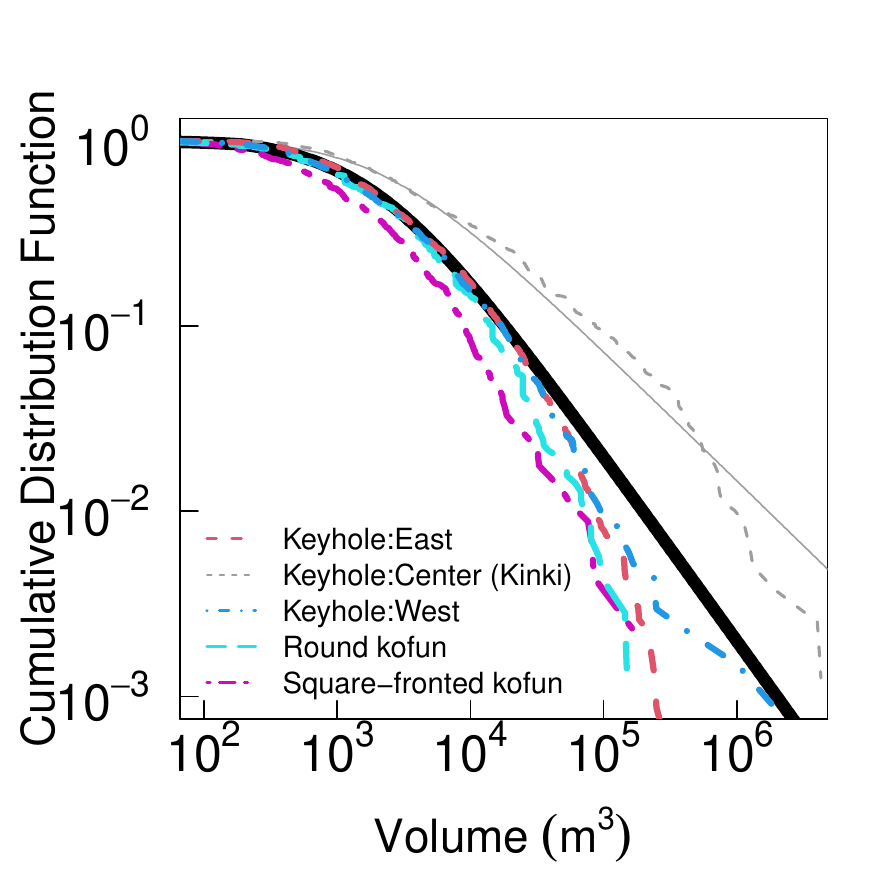}
         \put(10,150){\textbullet \color{black} Grouped by Region / mound type }
        \put(120,120){\color{black}\Large\bfseries (a)}
         \put(35,80){\color{black}\bfseries Original}
         \put(50,70){\color{black}\bfseries volume}
    \end{overpic}
      \begin{overpic}[width=5.8cm]{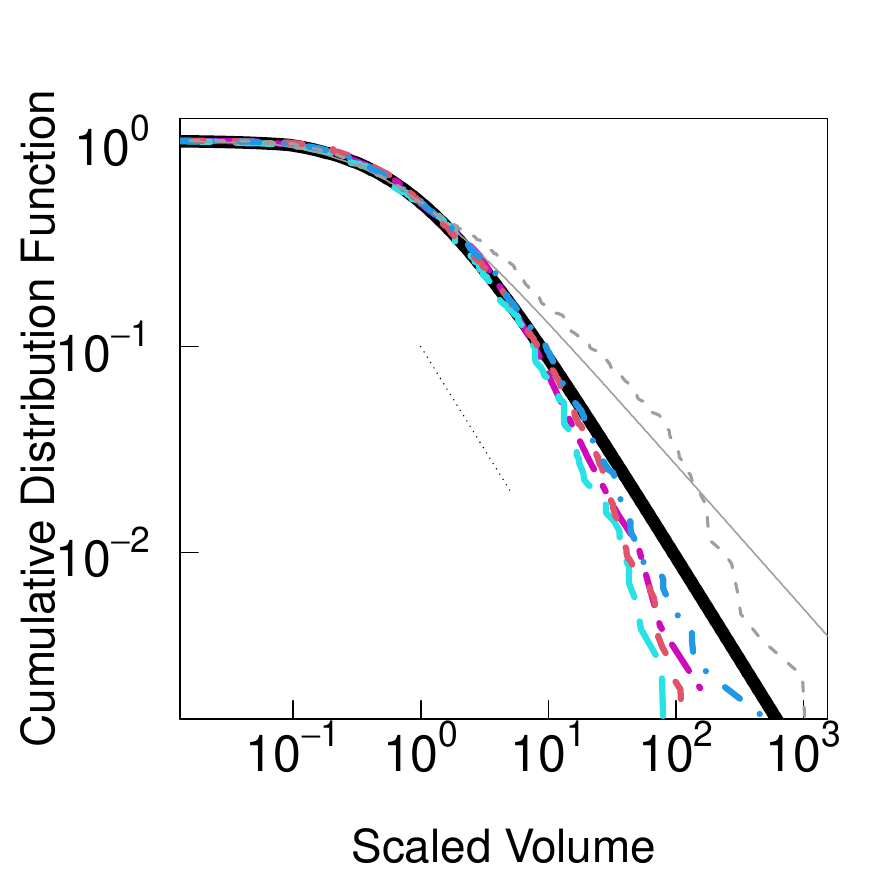}
        \put(120,120){\color{black}\Large\bfseries (b)}
        \put(38,40){\color{black}\bfseries Scaled, log-log}
    \end{overpic} 
    \begin{overpic}[width=5.8cm]{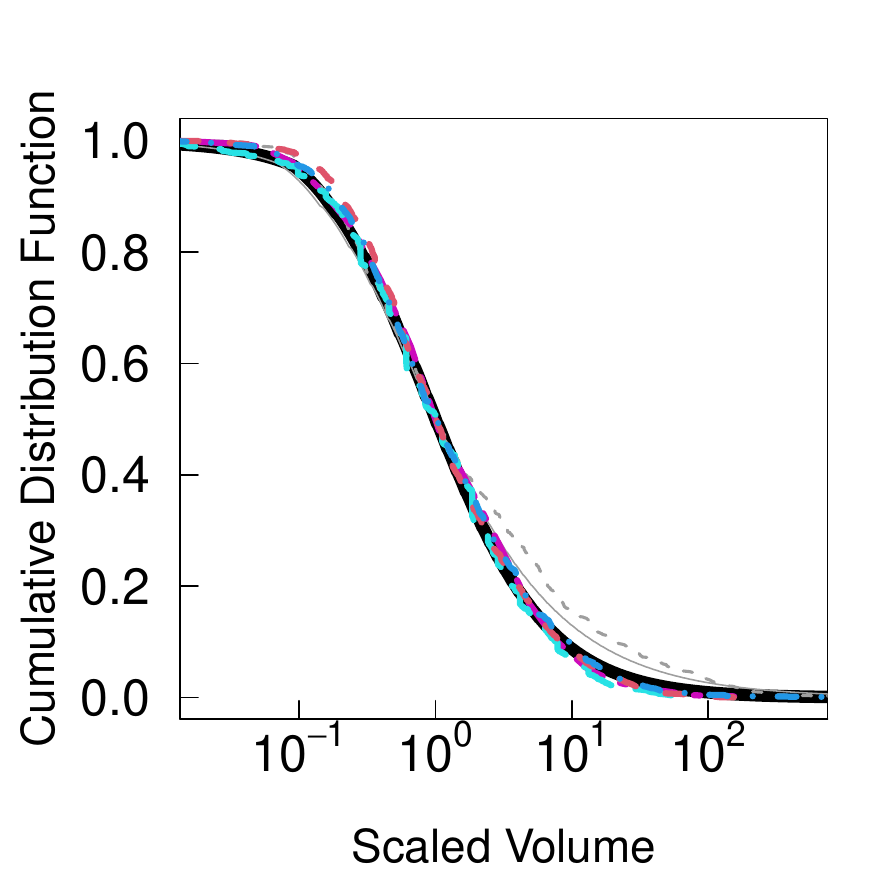}
        \put(120,124){\color{black}\Large\bfseries (c)}
        \put(75,107){\color{black}\bfseries Scaled, semi-log}
    \end{overpic} 
       \begin{overpic}[width=5.8cm]{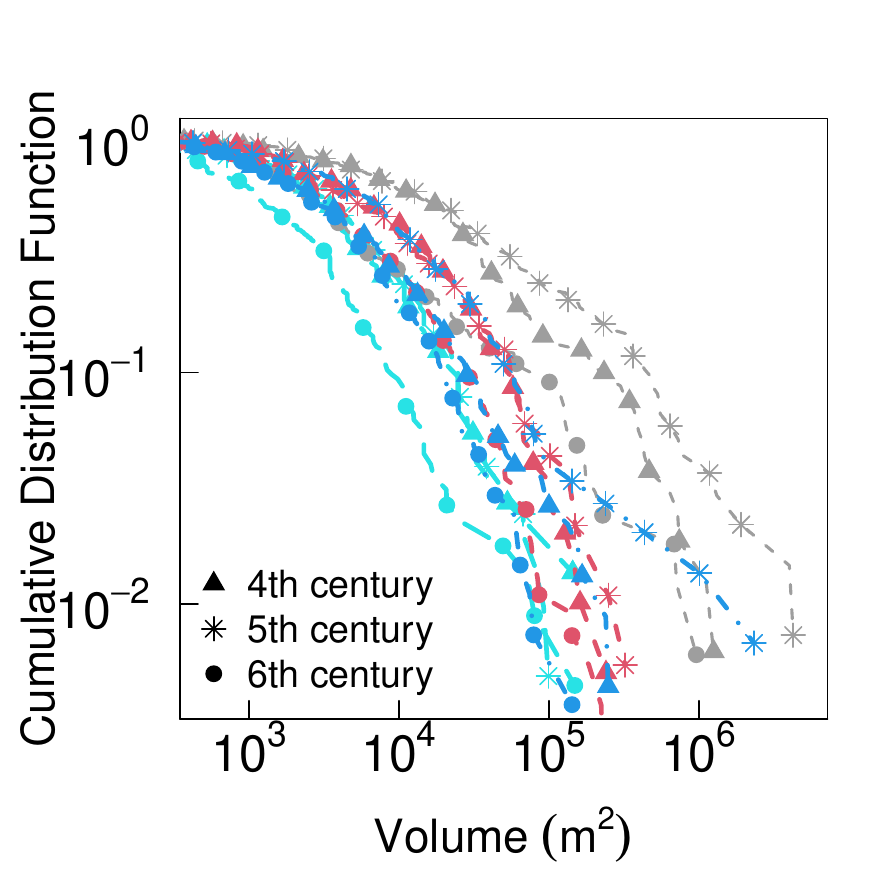}
         \put(10,150){\textbullet \color{black} Grouped by Region / mound type $\times$ century }
        \put(120,120){\color{black}\Large\bfseries (d)}
        \put(35,72){\color{black}\bfseries Original}
        \put(39,62){\color{black}\bfseries volume}
       
    \end{overpic}
        \begin{overpic}[width=5.8cm]{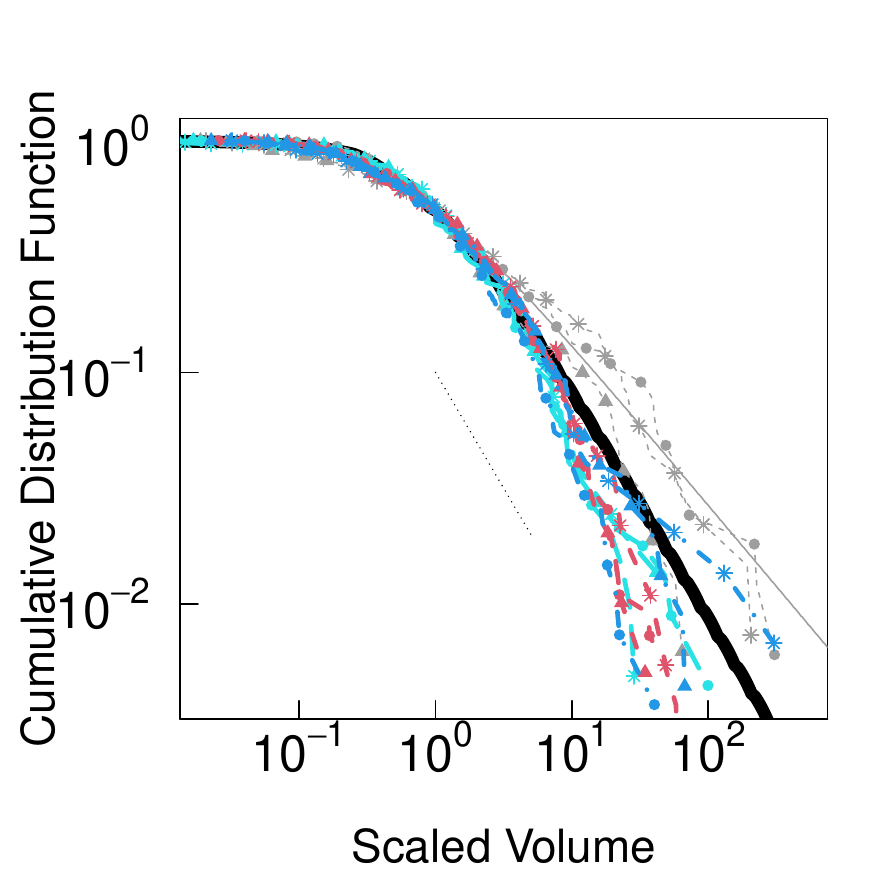}
        \put(120,120){\color{black}\Large\bfseries (e)}
         \put(38,40){\color{black}\bfseries Scaled, log-log}
    \end{overpic} 
      \begin{overpic}[width=5.8cm]{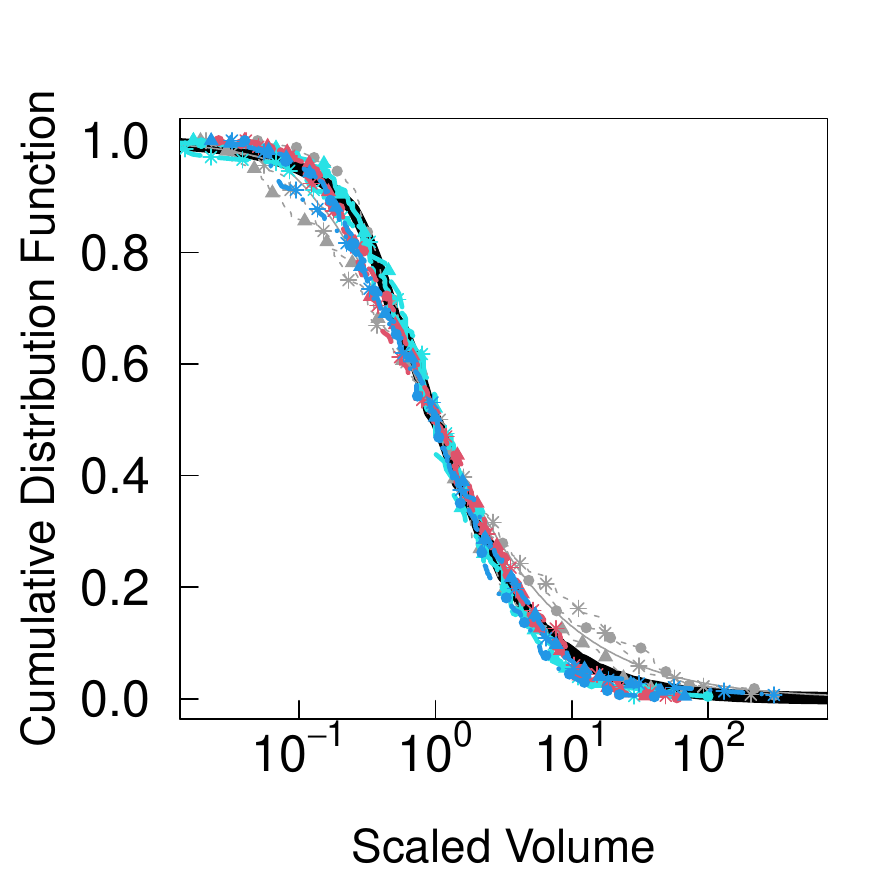}
        \put(120,124){\color{black}\Large\bfseries (f)}
        \put(75,107){\color{black}\bfseries Scaled, semi-log}
    \end{overpic}
\caption{
Complementary cumulative distributions of kofun volume by region, mound type, and period.
The regional classification follows Fig.~\ref{fig_map}. The red dashed lines indicate eastern Japan, the blue dash-dotted lines western Japan, the thin gray lines the political center region (Kinki), the cyan long-dashed lines round kofun, and the pink long-dash-dotted lines square-fronted kofun. The thick black solid line shows the theoretical distribution of the Kesten process (Eq.~\eqref{eq:kesten}; $\alpha=1.0$, $b_0=1.1$, $A_0=190$), and the thin gray dashed line shows a reference theoretical distribution with a heavier tail ($\alpha=0.7$, $b_0=1.1$, $A_0=298$).
(a)--(c) show the distributions by region and mound type. Panel (a) shows the original volume distributions, (b) shows the log--log plot after scaling by the median. The black dotted line indicates a reference slope proportional to $1/x$.  (c) shows the corresponding semi-log plot. Except for the political center region, the distributions overlap well after median scaling. They share a similar shape, with a central part close to a lognormal distribution and a tail close to a power law with an exponent of about 1. In contrast, the political center region has a heavier tail, close to a power law with an exponent of about 0.7.
(d)--(f) show the distributions further divided by period, in addition to region and mound type. Triangles, asterisks, and circles denote the fourth, fifth, and sixth centuries, respectively. The line colors and line types are the same as in (a)--(c). Panel (d) shows the original volume distributions, (e) shows the log--log plot after scaling by the median, and (f) shows the corresponding semi-log plot. Panel (d) shows that, for many regions and mound types, the fifth-century distributions are shifted toward larger volumes. Panels (e) and (f) show that, except for the political center region, the distributions largely overlap after scaling. This indicates that many of the differences among periods and regions appear mainly as differences in the median, rather than as differences in distributional shape.
}
\label{fig_area_cdf}
\end{figure*}
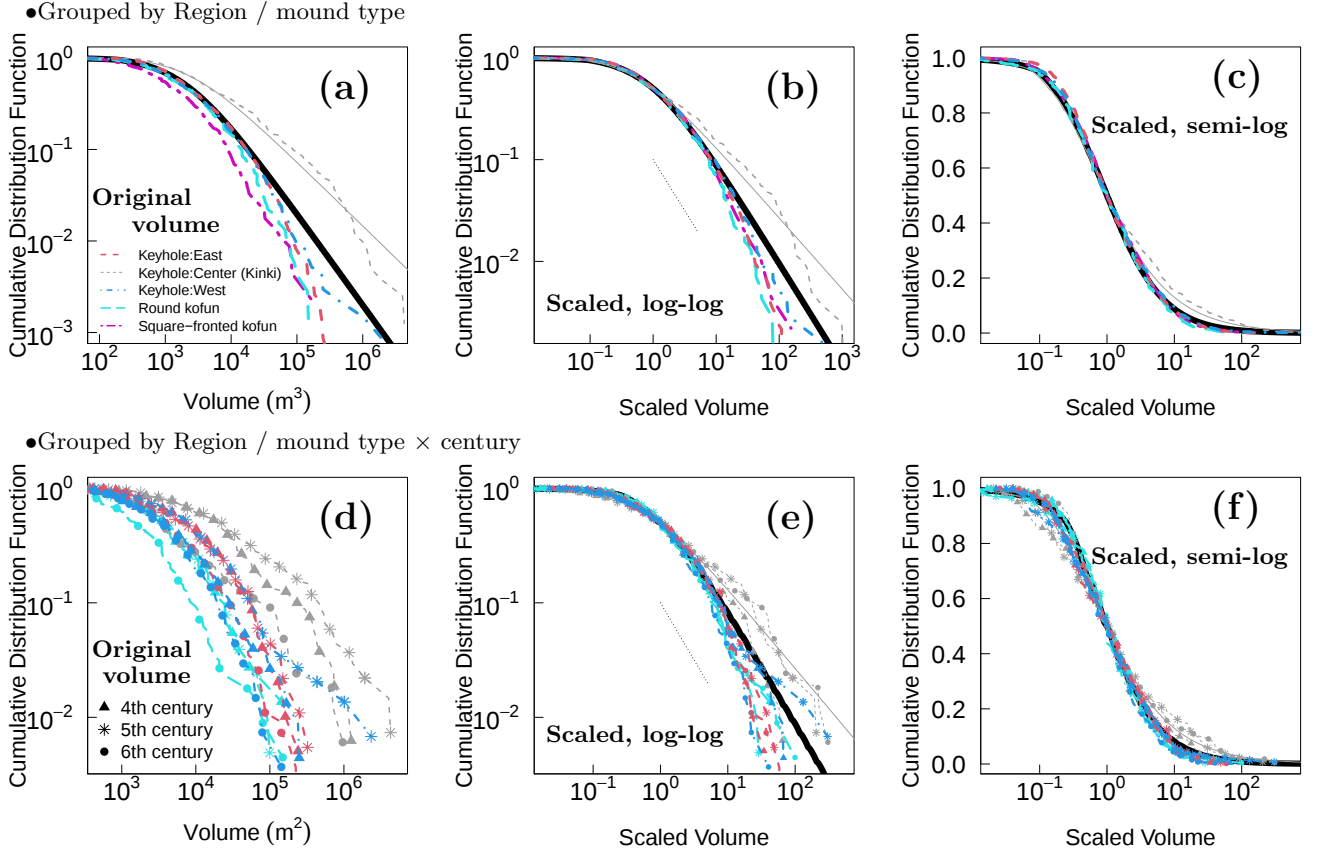
\subsection{Regional and temporal scaling of volume distributions}
\subsubsection{Scale shifts and consistent distributional shapes across regions and periods}
\begin{figure*}[!htbp]
    \centering
    \begin{overpic}[width=13cm]{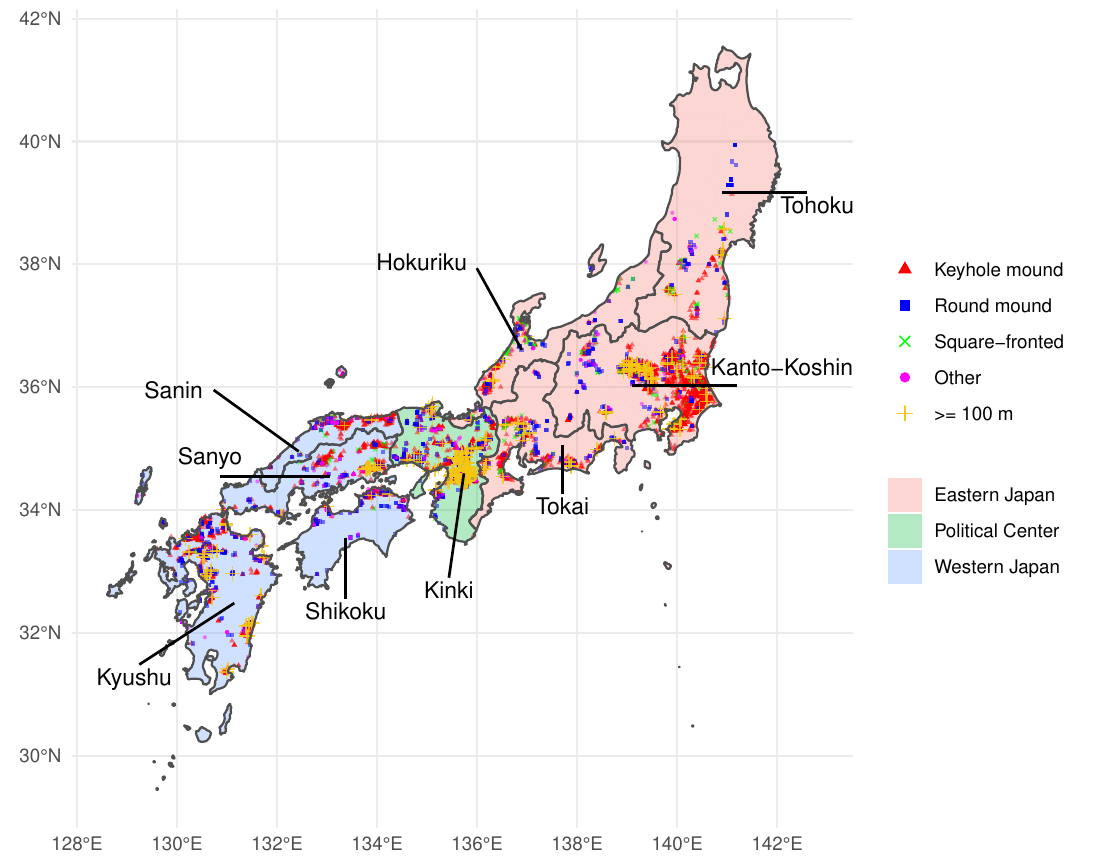}
        \put(40,266){\color{black}\Large\bfseries (a)}
    \end{overpic}
      \begin{overpic}[width=10.1cm]{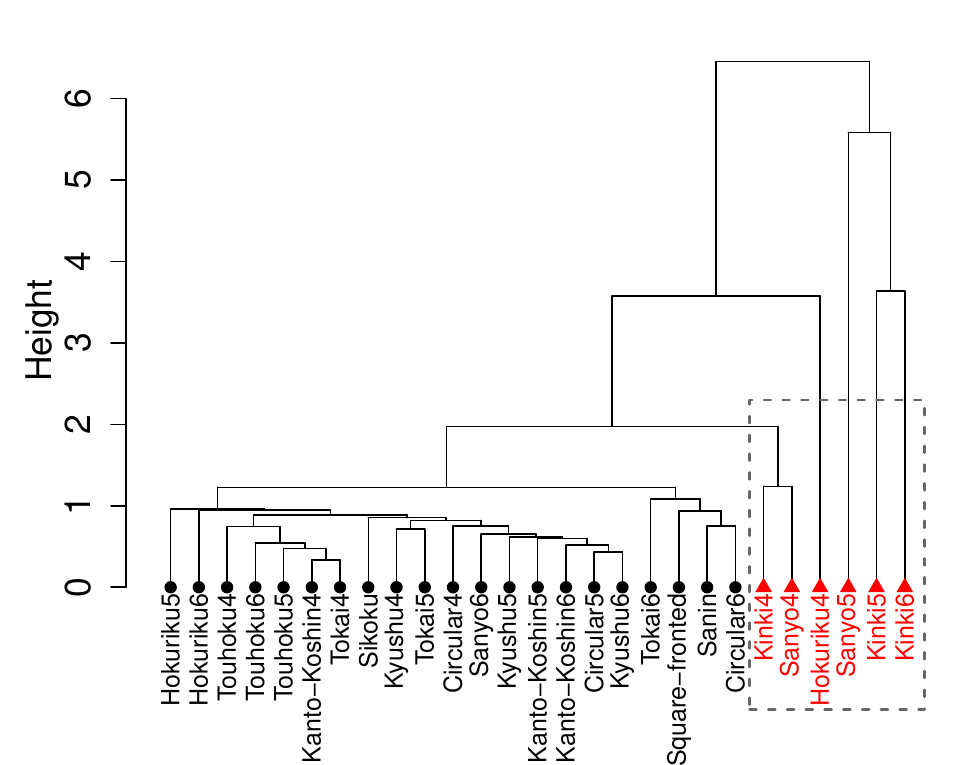}
        \put(45,192){\color{black}\Large\bfseries (b)}
    \end{overpic} 
    \begin{overpic}[width=7.6cm]{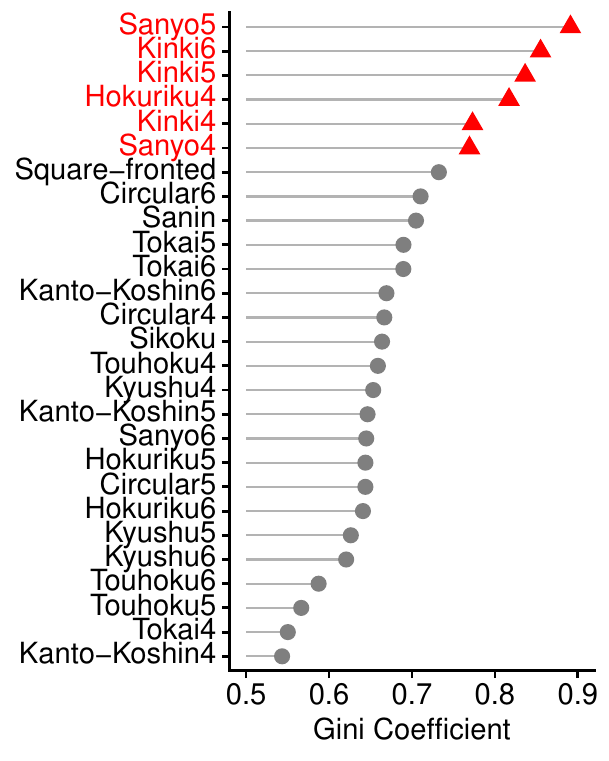}
        \put(200,228){\color{black}\Large\bfseries (c)}
         \put(134,35){\includegraphics[width=4.2cm]{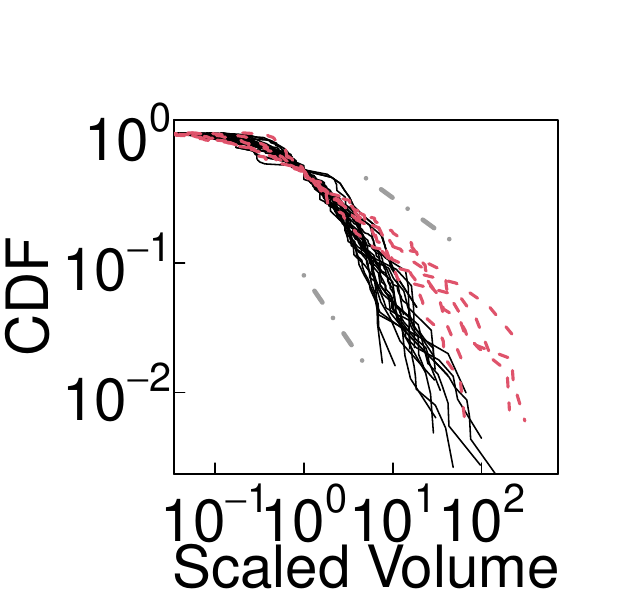}}
          \put(220,110){\color{black}\Large\bfseries (d)}
      \end{overpic} 
\caption{
Regional classification of kofun distributions and comparison of volume distributions by region and period.
(a) Regional classification used in this study and the spatial distribution of kofun. The background colors indicate eastern Japan, western Japan, and the political center region (Kinki), and the points show the locations of kofun. Red triangles indicate keyhole-shaped kofun, blue squares round kofun, green crosses square-fronted kofun, pink dots other mound types, and yellow crosses large kofun with mound lengths of 100 m or more.
(b) Hierarchical clustering based on the volume distributions by region, mound type, and period. 
Labels such as Sanyo4 and Hokuriku5 indicate the region and century (e.g., Sanyo region in the 4th century CE). The distributions are broadly divided into a typical group and exceptional groups with heavier tails, shown in red.  For details of the method, see SI Appendix \ref{app_sec_distribution_clustering}. 
(c) Gini coefficients of each distribution. Red triangles indicate the exceptional groups, and gray circles indicate the typical groups. The exceptional groups have higher Gini coefficients, indicating stronger concentration in large kofun.
(d) Complementary cumulative distributions of median-scaled volume. The black solid line shows the typical group, and the red dashed line shows the exceptional groups. The exceptional groups have heavier tails, consistent with the higher Gini coefficients shown in (c).
The exceptional groups include many distributions from the political center region, Kinki, as well as from Sanyo, which includes Kibi, often regarded as a secondary central region.
}
\label{fig_map}
\end{figure*}

The upper panels of Fig.~\ref{fig_area_cdf} compare the volume distributions of keyhole-shaped kofun across three broad regions: eastern Japan, the central region (Kinki), and western Japan (regional boundaries shown in Fig.~\ref{fig_map}(a)). For comparison, the volume distributions of round kofun and the square-fronted type are also shown.\par
Fig.~\ref{fig_area_cdf}(a) shows the cumulative distributions of volume $V$ itself. Figs.~\ref{fig_area_cdf}(b) and (c) show the same distributions after scaling by the median volume of each group, $V'=V/\mathrm{median}(V)$, plotted on log-log and semi-log axes, respectively. After this scaling, most distributions approximately collapse onto a common curve; the one clear exception is the central region, Kinki (light gray in the figure). This indicates that, for most regions and mound types other than Kinki, differences in the raw distributions mainly reflect differences in scale rather than in distributional shape. The black solid line shows the theoretical model curve described in Sec.~\ref{sec_model}, with a right-tail power-law exponent of 1; the gray solid line shows the corresponding curve for an exponent of 0.7; the remaining curves show empirical distributions. \par
The lower panels extend this comparison by dividing keyhole-shaped kofun in eastern Japan, the central region (Kinki), and western Japan, as well as round kofun, into three periods: the fourth, fifth, and sixth centuries. In the raw-volume distributions of Fig.~\ref{fig_area_cdf}(d), curves for the same region or mound type (same color and line style) shift left or right depending on the century. Fifth-century distributions, marked with an asterisk, are in most cases shifted furthest toward larger volumes. This is consistent with previous archaeological findings that the fifth century was a peak period in the construction of large kofun, and it matches the century-by-century changes in natural-log median reported in Table~\ref{tab_table1}.\par

Once volume is rescaled by the median, however, this shift largely disappears. As shown in Figs.~\ref{fig_area_cdf}(e) and (f), most distributions, again with the exception of the central region (Kinki), approximately collapse onto a common curve. In other words, the typical scale of kofun volume changed substantially across regions and periods, but the shape of the distribution after removing scale remained relatively stable. This stability is also reflected in Table~\ref{tab_table1}: the natural-log interquartile range clusters around 2 for most regions, periods, and mound types, with the central region again a notable exception.\par
\subsubsection{Clustering by detailed regional divisions and inequality in kofun volumes}
Fig.~\ref{fig_map} compares distributional shapes based on more detailed regional divisions. Fig.~\ref{fig_map}(a) is a scatter plot showing the locations of the kofun analyzed in this study on a map of Japan, together with the regional names used in the following analyses. The kofun analyzed here are not concentrated in a few areas but are distributed across a wide part of the Japanese archipelago.\par

Fig.~\ref{fig_map}(b) shows the results of hierarchical clustering, based on the method described in Sec.~\ref{app_sec_distribution_clustering}. To compare differences in distributional shape across regions and periods, we first normalized volume by the median within each group, calculated distances between the resulting distributions, and then performed hierarchical clustering on the distance matrix. This clustering therefore reflects differences in distributional shape after rescaling, rather than differences in the typical scale of each region-period group. Cutting the dendrogram at height 1.5 divides the groups into a typical group (black circles) and an exceptional group (red triangles). The exceptional group consists of Kinki, the fourth- and fifth-century Sanyo groups, and the fourth-century Hokuriku group. \par

Fig.~\ref{fig_map}(c) shows the Gini coefficient for each region-period group. The exceptional group (red triangles) has consistently higher Gini coefficients, indicating greater inequality in kofun volume than in the typical group. This confirms that the region-period groups identified as exceptional in the clustering differ not merely in typical scale, but in distributional shape itself—particularly in the degree of concentration among upper-ranked kofun. \par

Fig.~\ref{fig_map}(d) compares the size distributions of the typical and exceptional groups after normalizing volume by the median. The typical group closely follows the reference line with a right-tail power-law exponent of 1, whereas the exceptional group has a heavier tail, closer to the reference line with an exponent of 0.5—consistent with the higher Gini coefficients above.\par

Tables~\ref{app_tab_table1} and \ref{app_tab_table2} summarize the power-law exponent $\alpha$ and Gini coefficient for each region-period group. Because sample sizes become small once the data are divided by detailed region and period, these estimates should be treated as reference values rather than precise measurements.\par

These exceptional regions may share a common feature: political centrality within the Kofun period. Kinki was the political center of the Yamato polity, so it is plausible that it shows a distributional shape different from that of typical regions. Sanyo, which includes Kibi, can be regarded as a semi-central region: Kibi was one of the major regions outside Kinki where kofun of the largest size class were constructed, consistent with a distribution in which upper-ranked kofun are relatively large. Later historical traditions add a further, if tentative, point of consistency: the so-called Kibi rebellion, said to have occurred around the late fifth century and to have been suppressed by the central Yamato polity. Although the historicity of this tradition must be treated with caution, it is not inconsistent with our finding that the sixth-century Sanyo group shifts closer to the typical group. 
Among the outlier groups, the political centrality of the Hokuriku region in the 4th century is not as clear as that of the Kinki and Sanyo regions. The distribution shown in Fig.~\ref{app_fig_sanyo}(e) suggests that this region may have been classified as an outlier group due to the influence of three large, anomalous kofun.
The difference between these power-law exponents is interpreted in terms of the model in Sec.~\ref{sec_model_power}; its possible social interpretation is discussed in Sec.~\ref{sec_nowfirm}, through comparison with industry-level sales distributions of contemporary Japanese firms.\par

\begin{table*}[!t]
\begin{tabular}{llrrrrrrlrrrr}
\hline
Region
& Century
& $n$
& $\alpha$
& 95\% CI
& logMedi.
& logIQR
& AIC
& Gini
& Ln p-val.
& Kes(1) p-val.
& Kes($\alpha$) p-val. \\
\hline
\multicolumn{12}{l}{\textbf{Keyhole-shaped mounds}} \\
\hline
All    & All & 4545 & 1.000 & [0.865, 1.100] & 7.711 & 2.113 & Dp(1) & 0.854 & $1.32{\times}10^{-7}$ & $1.33{\times}10^{-5}$ & $1.08{\times}10^{-5}$ \\
All    & 4   &  597 & 1.007 & [0.858, 1.222] & 8.843 & 2.443 & Kes($\alpha$), Kes(1) & 0.282 & 0.274 & 0.170 & 0.0692 \\
All    & 5   &  486 & 0.797 & [0.633, 1.009] & 9.040 & 2.479 & Kes($\alpha$) & 0.862 & 0.143 & 0.0945 & 0.175 \\
All    & 6   & 1005 & 1.174 & [0.840, 1.460] & 8.177 & 2.033 & Dp($\alpha$), Dp(1) & 0.755 & 0.0998 & 0.0205 & 0.393 \\

East   & All & 2306 & 1.130 & [0.798, 2.244] & 7.631 & 1.993 & Dp($\alpha$), Dp(1) & 0.732 & $1.94{\times}10^{-4}$ & 0.00147 & $3.83{\times}10^{-4}$ \\
East   & 4   &  197 & 1.144 & [0.718, 3.390] & 8.037 & 2.352 & Ln & 0.654 & 0.402 & 0.274 & 0.371 \\
East   & 5   &  183 & 0.830 & [0.519, 2.434] & 7.667 & 2.284 & Kes(1) & 0.705 & 0.489 & 0.513 & 0.124 \\
East   & 6   &  546 & 1.241 & [0.756, 2.470] & 8.256 & 2.075 & Ln & 0.682 & 0.643 & 0.113 & 0.567 \\

Center & All &  794 & 0.701 & [0.453, 1.253] & 8.319 & 2.683 & Kes($\alpha$) & 0.878 & $7.10{\times}10^{-5}$ & $1.44{\times}10^{-5}$ & 0.0197 \\
Center & 4   &  160 & 0.803 & [0.625, 1.065] & 9.879 & 2.102 & Ln, Dp(1), Kes($\alpha$)  & 0.773 & 0.674 & 0.497  & 0.640 \\
Center & 5   &  136 & 0.562 & [0.420, 1.901] & 8.351 & 2.814 & Ln, Dp(1), Kes($\alpha$) & 0.836 & 0.586 & 0.0670 & 0.866 \\
Center & 6   &  165 & 0.571 & [0.468, 0.745] & 6.725 & 2.150 & Dp($\alpha$) & 0.855 & 0.0144 & 0.0874 & 0.0147 \\

West   & All & 1349 & 1.223 & [0.897, 1.668] & 7.491 & 2.134 & Kes($\alpha$) & 0.813 & 0.0299 & 0.386 & 0.0389 \\
West   & 4   &  226 & 1.125 & [0.663, 1.778] & 8.217 & 2.227 & Kes(1) & 0.735 & 0.783 & 0.746 & 0.737 \\
West   & 5   &  147 & 0.799 & [0.625, 1.864] & 8.950 & 2.255 & Kes(1) & 0.838 & 0.932 & 0.820 & 0.618 \\
West   & 6   &  271 & 1.058 & [0.863, 2.501] & 8.173 & 1.900 & Ln & 0.653 & 0.873 & 0.115 & 0.140 \\
\hline
\multicolumn{12}{l}{\textbf{Circular mounds}} \\
\hline
All    & All & 705 & 0.990 & [0.876, 1.910] & 7.532 & 2.025 & Kes(1) & 0.708 & 0.531 & 0.408 & 0.0656 \\
All    & 4   &  73 & 0.925 & [0.650, 2.041] & 8.150 & 1.599 & Kes(1) & 0.667 & 0.949 & 0.409 & 0.302 \\
All    & 5   & 204 & 0.897 & [0.719, 2.496] & 8.150 & 1.906 & Kes(1) & 0.644 & 0.359 & 0.310 & 0.153 \\
All    & 6   & 224 & 1.171 & [0.908, 1.953] & 7.310 & 1.932 & Kes($\alpha$), Kes(1) & 0.710 & 0.422 & 0.0990 & 0.639 \\

East   & All & 257 & 0.897 & [0.719, 2.496] & 7.562 & 2.092 & Kes($\alpha$), Kes(1) & 0.733 & 0.686 & 0.340 & 0.729 \\
Center & All & 171 & 0.871 & [0.648, 2.032] & 7.879 & 1.964 & Kes(1) & 0.661 & 0.581 & 0.406 & 0.112 \\
West   & All & 255 & 1.193 & [0.825, 1.889] & 7.254 & 1.976 & Kes($\alpha$) & 0.480 & 0.934 & 0.393 & 0.389 \\

\hline
\multicolumn{12}{l}{\textbf{Square-fronted mounds}} \\
\hline
All    & All & 455 & 1.396 & [0.823, 1.819] & 7.022 & 2.093 & Kes($\alpha$) & 0.732 & 0.821 & 0.264 & 0.366 \\
\hline
\end{tabular}

\begin{flushleft}
\footnotesize
Notes:
Region indicates the spatial grouping: All = all Japan or period (including unknown period), East = eastern Japan,
Center = capital region, and West = western Japan. 
For circular mounds, East, Center, and West indicate the corresponding circular-mound subsets.
Square-fronted mounds indicate square-fronted and square-backed mounds.
$\alpha$ is the estimated upper-tail exponent.
The 95\% CI reports the bootstrap interval of $\alpha$.
logMedi. and logIQR denote the median and interquartile range of log mound size, respectively.
AIC indicates the model selected by AIC:
Ln = lognormal; Dp($\alpha$) = dPlN with the estimated exponent;
Dp(1) = dPlN with $\alpha=1$;
Kes($\alpha$) = Kesten process with the estimated exponent;
Kes(1) = Kesten process with $\alpha=1$.
The last three columns give Anderson--Darling goodness-of-fit p-values
for the lognormal model, the Kesten process with $\alpha=1$, and the
Kesten process with the estimated exponent, respectively.
Detailed regional results are provided in SI Appendix Table~\ref{app_tab_table1}, \ref{app_tab_table2}.
\end{flushleft}
\caption{Statistics for kofun volume distributions by mound type, region, and period. The table reports tail exponents, log-scale summary statistics, AIC-selected models, Gini coefficients, and goodness-of-fit p-values.}
\label{tab_table1}
\end{table*}
\section{Modeling kofun volume distributions with a Kesten process}
\label{sec_model}
\subsection{Model formulation}
Following the principle of parsimony, we describe the distribution of kofun volumes using a minimal discrete-time model based on a multiplicative random growth process, the Kesten process \cite{Kesten1973RandomDifference,SornetteCont1997Multiplicative,TakayasuEtAl1997RandomAmplification}. Let $x(t)$ denote the politico-economic scale of a chief or local group at time $t$. 
At each step, $x(t)$ either grows multiplicatively by a factor $b_0$ and receives an additive resource $A_0$ (with probability $p_0$), or the group is reset---reflecting extinction or reorganization---and replaced by a new entrant whose size is proportional to $A_0$ (with probability $1-p_0$):
\begin{equation}
x(t+1) =
\begin{cases}
b_0 \, x(t) + A_0 & \text{with probability } p_0, \\
A_0 \cdot x_{\mathrm{new}} & \text{with probability } 1-p_0,
\end{cases}
\label{eq:kesten}
\end{equation}
where $b_0>1$ is the growth factor and $x_{\mathrm{new}}\sim\mathcal{U}(0,2)$ represents the relative scale of a new entrant. Kofun volume is then given by $y(t)=Q\cdot x(t)$, where $Q$ is the fraction of economic scale allocated to kofun construction (SI Appendix~\ref{app_sec_model_volume}). 
This process provides a minimal mechanism that generates a rounded, log-normal-like body together with a Pareto-type upper tail. Further details are provided in the SI Appendix, including the relation of this process to the dPlN distribution \cite{ReedJorgensen2004dPlN} and to a related multiplicative-growth model of inequality \cite{Jones2018}, as well as its dependence on the distribution of $x_{\mathrm{new}}$ (SI Appendix~\ref{app_sec_model_dpln} and \ref{app_sec_additive}).

\subsection{Size distribution of the model}
\label{subsec:kesten_analysis}
Eq.~\eqref{eq:kesten} is a special case of the Kesten process $x(t+1)=b(t)x(t)+f(t)$, whose upper-tail cumulative distribution follows a power law, $\Pr(X>x)\propto x^{-\alpha}$, whenever $\langle b(t)^\alpha\rangle=1$ \cite{SornetteCont1997Multiplicative,TakayasuEtAl1997RandomAmplification}. For the binary multiplier used here, this condition gives
\begin{equation}
b_0 = p_0^{-1/\alpha},
\label{eq:b0_alpha}
\end{equation}
so that $\alpha=1$ (Zipf's law) corresponds to $b_0=1/p_0$ (full derivation in SI Appendix~\ref{app_sec_model_power}).

This Zipf condition, $p_0 b_0=1$, has an intuitive zero-sum-like interpretation: for example, when $p_0=1/2$ and $b_0=2$, half of the groups are reset on average, and surviving groups double their relative share by absorbing the shares of the reset groups. Because the additive term $A_0$ is also present, the process is not strictly zero-sum but is better described as a quasi-zero-sum competitive process, in which externally supplied resources are added and then distributed through competition. The multiplicative term may be understood as capturing share competition through conflict, political reorganization, or differential access to scarce resources such as iron; the additive term may capture gradual resource accumulation such as annual agricultural surplus.\par

The model's log-normal-like body arises because, for groups with short lifetimes $T$, the additive term $A_0$ dominates over multiplicative amplification. Intuitively, when $b_0\simeq 1$, as in the fit to keyhole-shaped kofun ($b_0=1.13$; Fig.~\ref{fig_volume}(a)), scale grows approximately linearly at each step, $x(t+1) \approx x(t)+A_0$, so that after $T$ steps $x_T \approx A_0(T+x_{\mathrm{new}})$.  Because the lifetime $T$ follows a geometric distribution, which is approximately exponential in continuous time, the distribution of $\log x_T$ takes a rounded unimodal shape, giving the body of the distribution its approximately log-normal-like appearance (SI Appendix~\ref{app_sec_kesten_lnorm}). \par

Finally, we give the full stationary density of the model. As derived in SI Appendix~\ref{app_sec_model_decomposition}, the stationary distribution of kofun volume can be written as
\begin{equation}
y_T = QA_0 \cdot J_T ,
\label{eq:QA0J}
\end{equation}
where $J_T$ is a dimensionless random variable with probability density function
\begin{equation}
f_J(j)
=
\sum_{k=0}^{\infty}
(1-p_0)\,p_0^k\,
\frac{1}{u_k-\ell_k}\,
\mathbf{1}_{\{\ell_k \le j \le u_k\}},
\label{eq:fJ}
\end{equation}
where $\ell_k=(b_0^k-1)/(b_0-1)$, $u_k=\ell_k+2b_0^k$, and $\mathbf{1}_{\{A\}}$ equals 1 if $A$ holds and 0 otherwise—that is, a mixture of uniform distributions over intervals $[\ell_k,u_k]$, weighted by the geometric probability of surviving $k$ growth steps (derivation in SI Appendix~\ref{app_sec_model_prob}).  The red dashed line in Fig.~\ref{fig_volume}(a) shows this theoretical distribution. \par
\subsection{Model-based description of changes in kofun volume distributions}
\label{subsec:model_distributional_change}
This decomposition separates kofun volume into a scale component $QA_0$ and a dimensionless competitive outcome $J_T$ given in Eq.~\eqref{eq:QA0J}. The factor $QA_0$ represents the resource scale $A_0$ available in a given region and period, multiplied by the fraction $Q$ allocated to kofun construction. In contrast, $J_T$ captures the outcome of growth, reset, and re-entry, and determines the distributional shape once this scale factor is removed. \par

\subsubsection{Distributional shifts and historical changes in scale}
\label{subsec:scaling_relation}
This framework naturally explains empirical variations across space and time. As shown by the variations in log-median volume (Fig.~\ref{fig_area_cdf}(d), Table~\ref{tab_table1}), the distribution of kofun volume shifts left or right depending on the region and period. 
Under this interpretation, these regional and temporal shifts correspond, to a first approximation, to variations in the scale component $QA_0$.  
Specifically, for keyhole-shaped kofun, the volume distribution shifts toward larger values in many regions from the fourth to the fifth century. 
In the model, this is described as an increase in $QA_0$, which is historically consistent with the fifth century being the peak of keyhole-shaped kofun culture, when many exceptionally large mounds were constructed. Further discussion of how observed changes in $QA_0$ may be decomposed into changes in $A_0$ and $Q$, including the contrasting behavior of keyhole-shaped and round kofun between the fourth and fifth centuries, is provided in SI Appendix~\ref{app_sec_QA}. \par
In the sixth century, both keyhole-shaped and round kofun shift toward smaller volumes, represented in the model as a decline in $QA_0$. This is consistent with a decline in the relative investment fraction $Q$ for kofun, reflecting historical changes in mortuary ritual, the acceptance of Buddhism, and political integration under the emerging central authority in the Japanese archipelago. \par

\subsubsection{Distributional shape and competitive structure}
\label{sec_model_power}
Whereas $QA_0$ governs shifts in scale, the empirical distributions rescaled by the median retain nearly the same shape across many typical region-period groups. This relative stability of shape suggests that the competitive process represented by $J_T$ was broadly similar across these typical groups.

The power-law exponent $\alpha$ admits a model-based interpretation within this competitive framework. For typical kofun groups outside Kinki and Sanyo, the estimated exponents are broadly consistent with $\alpha\approx1$ (black symbols, Figs.~\ref{fig_map}(a,d)). Through Eq.~\eqref{eq:b0_alpha}, this corresponds to $b_0=1/p_0$, corresponding to a quasi-zero-sum competitive process. Such a process naturally generates the common distributional shape discussed above: a rounded, log-normal-like body that asymptotically gives way to a power-law tail with $\alpha\approx1$. \par

By contrast, non-typical groups such as Kinki and Sanyo show $\alpha<1$ (red symbols), indicating a heavier tail in which large kofun are relatively more prominent. Within the model, $\alpha<1$ corresponds to a lower effective reset probability $1-p_0$. That is, established groups were more likely to persist than under the quasi-zero-sum competitive process corresponding to Zipf's law. We return to this point in Sec.~\ref{sec_nowfirm}, relating these exceptional groups to political centrality and semi-centrality, and to the distributional differences between competitive and regulated industries in the modern economy.
\par
\section{Summary and discussion}
\subsection{Summary of the main findings}
Zipf's law—the tendency for indicators of collective resource mobilization or economic scale to follow a power-law distribution with an exponent close to 1—has been documented mainly in societies with written administration and monetary economies. Here, we show that the volume distribution of kofun, large mounded tombs constructed in ancient Japan before these institutions were fully established, likewise follows Zipf's law. Using kofun volume as a proxy for the resource mobilization capacity of chiefly groups, we find that this distribution—not merely the presence of inequality—closely resembles that of modern firm sales, even in the absence of such institutions. Specifically, for keyhole-shaped kofun, this similarity extends across the entire distribution, with both the upper tail following a power law with a cumulative exponent close to 1 and the body approximating a log-normal distribution (Fig.~\ref{fig_volume}). \par

Beyond this overall pattern, regional and temporal comparisons reveal that most differences are differences in scale rather than in distributional form (Fig.~\ref{fig_area_cdf}). When raw volume is used, the distributions shift left or right depending on region and century. However, when volume is instead normalized by the median, the distributions for most regions, periods, and mound types collapse approximately onto a common curve with a tail exponent of $\alpha \approx 1$, except for a few exceptional groups. These exceptions primarily comprise Kinki, the political center of the period, and Sanyo during the fourth and fifth centuries---a region that includes Kibi, often regarded as a semi-central region (Fig.~\ref{fig_map}). These exceptional groups exhibit a power-law exponent $\alpha$ smaller than 1, that is, a heavier tail, indicating a stronger concentration of resources among top-ranked kofun than in the typical group. \par

These features are naturally explained within a single framework by a Kesten-type process of growth, stopping, and reorganization (Sec. \ref{sec_model}). Under this model, entities that stop growing after a short duration mainly driven by additive growth form the log-normal-like central part of the distribution, while the small number that continue growing over a long period accumulate the effects of multiplicative growth and generate the power-law tail. Shifts in the distributions across regions and periods can be interpreted as changes in a scale factor reflecting the resources available and the fraction allocated to kofun construction. The stability of the distributional shape after median normalization, by contrast, suggests that the underlying growth, stopping, and reorganization processes were broadly shared across regions and periods.
In particular, the $\alpha \approx 1$ tail of the typical group corresponds to a quasi-zero-sum competitive process, in which resources are largely redistributed through competition among chiefly groups. The heavier tails observed for the exceptional groups, by contrast, indicate a departure from this process, which we interpret as suggesting that existing groups were more likely to persist over longer periods. \par

Taken together, assuming the archaeological understanding that kofun size reflects the politico-economic resources and mobilizing capacity of chiefly groups, these results can be regarded as quantifying the uneven distribution of such resources and capacity through the shape of a statistical distribution. This commonality of Zipf's law in distributional shape, moreover, provides a basis for quantitatively comparing kofun volume distributions with distributions of collective economic scale from other periods, such as the sales of modern firms and early-modern kokudaka, a rice-yield-based measure of domain economic scale (Sec. \ref{sec_discussion}, Fig. \ref{app_fig_firm_kofun}).\par

\subsection{Discussion}
\label{sec_discussion}
\subsubsection{Modern Japanese firm-size distributions and interpretation of kofun distributional shape}
\label{sec_nowfirm}
Our findings also invite comparison with modern Japanese firm-size distributions. Studies of the sales and profit distributions of modern Japanese firms have reported tails with power-law exponents close to one, scale shifts of the entire distribution over time, and log-normal-like bodies \cite{Mizuno2002StatisticalLaws,Ishikawa2008FirmSizeDisplacement,Watanabe2013AllometricScalings}—features that parallel our findings for kofun volume distributions. Although kofun volume is a material proxy rather than a monetary or documentary record, the two are comparable in overall distributional shape (Fig. \ref{app_fig_firm_kofun}). \par

A further parallel emerges at the industry level. Many sectors, such as construction, manufacturing, and wholesale and retail trade, show sales distributions with power-law exponents close to one, whereas more regulated industries—electricity, gas, water utilities, and finance—show substantially smaller exponents, in some cases closer to $\alpha \approx 0.5$, and correspondingly heavier tails (SI Appendix Figs.~\ref{app_fig_nowfirm_1}--\ref{app_fig_nowfirm_3}). Within the Kesten-type framework of Sec.~\ref{sec_model}, industries with Zipf-like exponents are broadly consistent with a quasi-zero-sum competitive process, while smaller-exponent industries correspond to cases in which existing firms are more likely to persist than under this baseline—consistent with these industries typically being subject to strong entry regulation and institutional protection. \par

This analogy suggests one interpretation for the typical and exceptional kofun groups: the typical group's Zipf-like tail parallels the exponents observed in unregulated industries, while the exceptional groups—Kinki, the political center, and Sanyo, including the semi-central Kibi region—show heavier tails resembling those of regulated industries. Sanyo's temporal trajectory is notable: exceptional in the fourth and fifth centuries, it shifts toward the typical group in the sixth. Within the Kesten-type framework of Sec.~\ref{sec_model}, this shift corresponds to a weakening of the conditions favoring incumbent groups in Kibi, moving the region toward the more general quasi-zero-sum process — a change loosely reminiscent of deregulation. It also resonates with the traditional chronology of the Kibi rebellion, in which Kibi's powerful groups were said to be suppressed by the central Yamato polity in the late fifth century. However, both the narrative's historicity and the direct evidence for such institutional conditions remain uncertain, so this connection should be treated as a tentative analogy. \par

\subsubsection{Timing of share formation inferred from comparison with kokudaka distributions}
\label{sec_kokudaka}
A Zipf-like distribution does not necessarily mean that competition over shares was still ongoing during the period of kofun construction. For simplicity, our model assumes that competition over shares and kofun construction proceeded during the same period; in reality, however, the two may be temporally separated, with shares first forming through competition and then remaining largely fixed as the distribution shifts with economic growth. A similar pattern appears in two other Japanese distributions of collective economic scale: kokudaka, a measure of domain economic scale in early modern Japan, and postwar prefectural GDP. Like kofun volume, both are approximated by a distribution combining a log-normal body with a Zipf-like tail. In the case of kokudaka, under the Tokugawa order established after the territorial competition of the Sengoku period, military competition among domains was suppressed, and relative shares remained comparatively stable even as the distribution shifted rightward with economic growth (SI Appendix~\ref{a_sec_kokudaka}; Fig. \ref{a_fig_kokudaka_gdp}). \par

Kofun volume distributions may likewise partly inherit a power-share structure that took shape during the Yayoi period before the Kofun period. The Yayoi period was a time when power structures formed as wet-rice agriculture spread and politico-economic inequality grew, and Chinese historical records suggest that near the end of the Yayoi period, a conflict reminiscent of the Sengoku period — often referred to as the Wakoku War — took place.  In terms of our model and the analogy with Edo-period kokudaka, one possibility is that the shape of the distribution ($J_T$ in Eq.~\ref{eq:QA0J}) partly reflects a power-share structure inherited from the Yayoi period and the Wakoku War, and that shifts in the distribution across the Kofun period were driven mainly by overall scale change ($QA_0$ in Eq.~\ref{eq:QA0J}), with relative shares remaining largely fixed. \par

However, distributional shape alone cannot determine whether this share structure was established before the Kofun period, reorganized within it, or both. That said, the fact that Sanyo, including Kibi, shifted from the exceptional group in the fourth and fifth centuries to the typical group by the sixth may suggest that kofun volume distributions reflect not only an inherited share structure but also political reorganization within the Kofun period itself. \par

\subsubsection{Log-normal distributions and stopping processes}
\label{sec_teisi}
We have described kofun volume distributions in a unified way as combining a log-normal-like central part with a power-law tail. Some distributions, however, particularly those of round kofun, are also well described by a simple log-normal distribution alone (Fig.~\ref{fig_volume}(b), Table~\ref{tab_table1}). This does not contradict our interpretation. Finite samples generated by the Kesten process can show substantial sample-to-sample fluctuation, and when the sample size is small, the power-law tail may not be clearly represented within the observed range (SI Appendix~\ref{app_sec_teisi_yuragi}; Fig.~\ref{app_fig_teisi_yuragi}). \par

There is also a more substantive interpretation of the rounded, log-normal-like tails observed in some distributions, especially round kofun. In the main model of this study, we used a Kesten process with a constant growth rate and a constant stopping probability, corresponding to an exponentially distributed lifetime. If instead the stopping probability increases with growth duration, as in Weibull-type or Gompertz-type stopping processes, larger entities become increasingly likely to be suppressed. As a result, the tail becomes thinner and approaches a more rounded, log-normal-like shape (SI Appendix~\ref{app_sec_teisi_dist}; Fig.~\ref{app_fig_teisi_main}). \par

This perspective offers a unified way to organize differences among kofun volume distributions in terms of the persistence, or stopping process, of local groups. Within this framework, politically central or semi-central regions such as Kinki and Sanyo show heavy tails, consistent with a low stopping probability relative to growth and high persistence among existing groups. The standard type found in most regions has $\alpha\approx1$, corresponding to a quasi-zero-sum process in which growth and stopping are balanced. Round kofun show the strongest rounding of the tail, which may be consistent with a stopping probability that increases with growth duration; that is, groups may have become more likely to be suppressed as their growth duration increased. \par

This last pattern may also relate to the possibility that the builders of round kofun included many groups of lower political and ritual status than the builders of keyhole-shaped kofun. In that case, the scale of round kofun may have been kept from becoming extremely large by constraints tied to social position and available resources. These interpretations, however, are inferred from distributional shape alone and should be examined further against archaeological evidence. \par

\subsubsection{Dimensionality of proxy variables and power-law exponents}
Power-law exponents can differ without necessarily reflecting different underlying processes. One useful perspective is the dimensionality of the proxy variable: different measures of the same system are often related by a nonlinear scaling relation. A classic example is the allometric relation in biology, $\text{energy use} \propto \text{body mass}^{3/4}$ \cite{West1997AllometricScaling}. A similar relation has been reported in archaeological work on Pompeian houses, $\text{room number} \propto \text{area}^{0.6}$ \cite{Hanson2024ScalingPompeii}. \par

This perspective also applies to modern Japanese firms \cite{Watanabe2013AllometricScalings}. Sales $s$ follow a Zipf-like distribution with an exponent close to one, whereas employee count $l$ has an exponent of about $1.3$. This difference does not reflect a different underlying process; rather, it arises because the two indicators are related by a nonlinear scaling relation, $s \propto l^{1.3}$, which transforms one exponent into the other (SI Appendix~\ref{a_sec_firm_dimension}). In other words, the same underlying entity can exhibit different exponents depending on which aspect of its size is measured. \par
The same logic applies to the relation we established in Sec.~\ref{sec_scaling} between mound length and volume: a mound-length exponent of about three corresponds to a volume exponent of about one. This relation follows mathematically from $V \propto L^3$, since mound height is roughly proportional to mound length (Sec.~\ref{sec_scaling}). As with firm sales and employee counts, the difference between the two exponents need not reflect different underlying processes, but simply the different dimensionality of the two measures. As with sales, volume is thus the more natural proxy for resource mobilization capacity, being more directly tied to construction cost, while mound length is better understood as a derived indicator. \par
This perspective also suggests two readings of the smaller exponents observed in Kinki and Sanyo. One possibility is that existing groups there were more likely to persist, reflecting a distinct competitive structure. Another is that the relation between resources and volume was itself nonlinear in these regions: if exceptionally large kofun in political centers had few precedents to scale against, small differences in underlying resources or status could have been amplified when expressed as volume. Data are currently too limited to distinguish these possibilities directly; we adopt the former, more parsimonious hypothesis as our main interpretation. \par 
\paragraph{Volume distribution and the size-imitation hypothesis}
The dimensionality argument also bears on an alternative explanation for the Zipf-like tail: imitation, in which builders adjust new mound lengths or heights relative to earlier ones, rather than resource competition. In SI Appendix~\ref{app_sec_imitation}, we show that this imitation hypothesis, taken at face value, predicts a mound-length exponent closer to one—inconsistent with the observed exponent of about three. Reconciling the hypothesis with the data would require assuming that builders imitated volume rather than the more directly perceptible mound length. By contrast, the resource-competition interpretation requires no such assumption and naturally predicts the observed Zipf-like tail. \par
\paragraph{Proxy variables for wealth and scaling dimensions} 
More generally, when wealth cannot be directly quantified through sales, income, or tax records, researchers often rely on material proxies such as house floor area, grave goods, and tomb size. Comparisons using the same proxy variable are relatively straightforward to interpret; Kohler et al.'s cross-cultural comparison of Gini coefficients based on house floor area is one such example \cite{Kohler2025EconomicInequality}. However, cross-comparisons between different proxy variables are more difficult. \par
In this study, we position the proxy variable for mobilization capacity as the main size variable, using it as a foothold for cross-comparisons between different proxy indicators such as kofun volume and firm sales. Taking this main variable for mobilization capacity---which is expected to follow a Zipf-like distribution---as a reference, and considering and correcting for dimensional differences among proxy variables that scale nonlinearly with it, may offer a clue for comparing inequality across different indicators (SI Appendix~\ref{app_sec_dimension}).
At the same time, examining how such proxy indicators correspond to wealth and mobilization capacity remains, of course, an important direction for future research \cite{Nortoft2022GraveWealth}. 
\section{Materials and methods}
\label{sec_method}
\subsection{Dataset}
\label{sec_method_data}
We used the National Kofun Database compiled by the Center for Ancient Studies and Sacred Sites, Nara Women's University \cite{NaraWomensUniversityKofunDatabase} (March 31, 2025 version; 6779 kofun in total), which records mound shape, location, chronological period, and physical measurements. We focus on the three mound types with sufficient sample size: keyhole-shaped, round, and square-fronted kofun. \par
Because approximately 160,000 kofun and horizontal tombs have been identified across Japan \cite{AgencyCulturalAffairs2021BuriedSites}, the database is not exhaustive. Coverage is high for keyhole-shaped and square-fronted kofun—4809 and 469 entries, respectively, broadly consistent with nationwide estimates of about 4800 \cite{Izuta2016KeyholeDistribution} and 500 \cite{Wada2009OtherWorldKofun}. By contrast, the database includes mainly major, often historically designated, round mounds and should not be treated as a complete population of round kofun nationwide. Kofun missing the physical measurements needed for volume estimation, or with unknown chronological period, were excluded from the corresponding analyses; sample sizes are given in Table~\ref{app_tab:mound_counts}, \ref{tab_table1}, \ref{app_tab_table1} and \ref{app_tab_table2}.

\subsection{Missing-value imputation and estimation of mound volume}
\label{sec_method_volume}
Mound volume was estimated from each kofun's principal measurements using simple geometric approximations: rectangular prisms and cylinders for the component parts of each mound type. Full formulas are given in SI Appendix~\ref{app_sec_method_volume}. Missing measurements were imputed by PCA-based imputation (\texttt{imputePCA}, R package \texttt{missMDA} \cite{Josse2016MissMDA}, one principal component), exploiting the strong common-scale correlation among mound measurements (SI Appendix~\ref{app_sec_shape}); kofun missing all principal measurements were excluded rather than imputed. These approximations are not intended to reconstruct precise mound geometry but are sufficient for comparing relative size and distributional shape.

\subsection{Estimation of power-law exponents}
\label{sec_method_powerlaw}
Power-law exponents were estimated using the standard method of Clauset et al. \cite{Clauset2009PowerLaw}, implemented in the R package \texttt{poweRlaw} \cite{Gillespie2015Powerlaw}, which jointly estimates the lower cutoff $x_{\min}$ and the maximum-likelihood exponent for $x\ge x_{\min}$. We report upper cumulative exponents, $\alpha=\zeta-1$, where $\zeta$ is the probability-density exponent returned by \texttt{poweRlaw}. Point estimates and 95\% confidence intervals were obtained by bootstrap resampling (1000 replicates; full procedure in SI Appendix~\ref{app_sec_method_powerlaw}).

%
%
\begin{acknowledgments}
We thank Aya Minoura (Research Assistant, School of Arts and Letters, Meiji University) for helpful discussions on the Kofun period and for checking our terminology usage.
\end{acknowledgments}
%
\bibliography{model14_pre}

\begin{thebibliography}{45}%
\makeatletter
\providecommand \@ifxundefined [1]{%
 \@ifx{#1\undefined}
}%
\providecommand \@ifnum [1]{%
 \ifnum #1\expandafter \@firstoftwo
 \else \expandafter \@secondoftwo
 \fi
}%
\providecommand \@ifx [1]{%
 \ifx #1\expandafter \@firstoftwo
 \else \expandafter \@secondoftwo
 \fi
}%
\providecommand \natexlab [1]{#1}%
\providecommand \enquote  [1]{``#1''}%
\providecommand \bibnamefont  [1]{#1}%
\providecommand \bibfnamefont [1]{#1}%
\providecommand \citenamefont [1]{#1}%
\providecommand \href@noop [0]{\@secondoftwo}%
\providecommand \href [0]{\begingroup \@sanitize@url \@href}%
\providecommand \@href[1]{\@@startlink{#1}\@@href}%
\providecommand \@@href[1]{\endgroup#1\@@endlink}%
\providecommand \@sanitize@url [0]{\catcode `\\12\catcode `\$12\catcode
  `\&12\catcode `\#12\catcode `\^12\catcode `\_12\catcode `\%12\relax}%
\providecommand \@@startlink[1]{}%
\providecommand \@@endlink[0]{}%
\providecommand \url  [0]{\begingroup\@sanitize@url \@url }%
\providecommand \@url [1]{\endgroup\@href {#1}{\urlprefix }}%
\providecommand \urlprefix  [0]{URL }%
\providecommand \Eprint [0]{\href }%
\providecommand \doibase [0]{https://doi.org/}%
\providecommand \selectlanguage [0]{\@gobble}%
\providecommand \bibinfo  [0]{\@secondoftwo}%
\providecommand \bibfield  [0]{\@secondoftwo}%
\providecommand \translation [1]{[#1]}%
\providecommand \BibitemOpen [0]{}%
\providecommand \bibitemStop [0]{}%
\providecommand \bibitemNoStop [0]{.\EOS\space}%
\providecommand \EOS [0]{\spacefactor3000\relax}%
\providecommand \BibitemShut  [1]{\csname bibitem#1\endcsname}%
\let\auto@bib@innerbib\@empty
\bibitem [{\citenamefont {Matsugi}(2025)}]{Matsugi2025Kofun}%
  \BibitemOpen
  \bibfield  {author} {\bibinfo {author} {\bibfnamefont {T.}~\bibnamefont
  {Matsugi}},\ }\href@noop {} {\emph {\bibinfo {title} {Kofun jidai no rekishi
  [The History of the Kofun Period]}}},\ \bibinfo {series} {Kodansha Gendai
  Shinsho}\ No.\ \bibinfo {number} {2792}\ (\bibinfo  {publisher} {Kodansha},\
  \bibinfo {address} {Tokyo, Japan},\ \bibinfo {year} {2025})\ \bibinfo {note}
  {in Japanese}\BibitemShut {NoStop}%
\bibitem [{\citenamefont {di~Giovanni}\ and\ \citenamefont
  {Levchenko}(2013)}]{DiGiovanni2013ZipfsWorld}%
  \BibitemOpen
  \bibfield  {author} {\bibinfo {author} {\bibfnamefont {J.}~\bibnamefont
  {di~Giovanni}}\ and\ \bibinfo {author} {\bibfnamefont {A.~A.}\ \bibnamefont
  {Levchenko}},\ }\bibfield  {title} {\bibinfo {title} {Firm entry, trade, and
  welfare in zipf's world},\ }\href
  {https://doi.org/10.1016/j.jinteco.2012.08.002} {\bibfield  {journal}
  {\bibinfo  {journal} {Journal of International Economics}\ }\textbf {\bibinfo
  {volume} {89}},\ \bibinfo {pages} {283} (\bibinfo {year} {2013})}\BibitemShut
  {NoStop}%
\bibitem [{\citenamefont {Axtell}(2001)}]{Axtell2001ZipfFirmSizes}%
  \BibitemOpen
  \bibfield  {author} {\bibinfo {author} {\bibfnamefont {R.~L.}\ \bibnamefont
  {Axtell}},\ }\bibfield  {title} {\bibinfo {title} {Zipf distribution of u.s.
  firm sizes},\ }\href {https://doi.org/10.1126/science.1062081} {\bibfield
  {journal} {\bibinfo  {journal} {Science}\ }\textbf {\bibinfo {volume}
  {293}},\ \bibinfo {pages} {1818} (\bibinfo {year} {2001})}\BibitemShut
  {NoStop}%
\bibitem [{\citenamefont {Fujiwara}\ \emph {et~al.}(2004)\citenamefont
  {Fujiwara}, \citenamefont {Di~Guilmi}, \citenamefont {Aoyama}, \citenamefont
  {Gallegati},\ and\ \citenamefont {Souma}}]{Fujiwara2004ParetoZipfGibrat}%
  \BibitemOpen
  \bibfield  {author} {\bibinfo {author} {\bibfnamefont {Y.}~\bibnamefont
  {Fujiwara}}, \bibinfo {author} {\bibfnamefont {C.}~\bibnamefont {Di~Guilmi}},
  \bibinfo {author} {\bibfnamefont {H.}~\bibnamefont {Aoyama}}, \bibinfo
  {author} {\bibfnamefont {M.}~\bibnamefont {Gallegati}},\ and\ \bibinfo
  {author} {\bibfnamefont {W.}~\bibnamefont {Souma}},\ }\bibfield  {title}
  {\bibinfo {title} {Do pareto--zipf and gibrat laws hold true? an analysis
  with european firms},\ }\href {https://doi.org/10.1016/j.physa.2003.12.015}
  {\bibfield  {journal} {\bibinfo  {journal} {Physica A: Statistical Mechanics
  and its Applications}\ }\textbf {\bibinfo {volume} {335}},\ \bibinfo {pages}
  {197} (\bibinfo {year} {2004})}\BibitemShut {NoStop}%
\bibitem [{\citenamefont {Mizuno}\ \emph {et~al.}(2002)\citenamefont {Mizuno},
  \citenamefont {Katori}, \citenamefont {Takayasu},\ and\ \citenamefont
  {Takayasu}}]{Mizuno2002StatisticalLaws}%
  \BibitemOpen
  \bibfield  {author} {\bibinfo {author} {\bibfnamefont {T.}~\bibnamefont
  {Mizuno}}, \bibinfo {author} {\bibfnamefont {M.}~\bibnamefont {Katori}},
  \bibinfo {author} {\bibfnamefont {H.}~\bibnamefont {Takayasu}},\ and\
  \bibinfo {author} {\bibfnamefont {M.}~\bibnamefont {Takayasu}},\ }\bibfield
  {title} {\bibinfo {title} {Statistical laws in the income of japanese
  companies},\ }in\ \href {https://doi.org/10.1007/978-4-431-66993-7_35} {\emph
  {\bibinfo {booktitle} {Empirical Science of Financial Fluctuations}}},\
  \bibinfo {editor} {edited by\ \bibinfo {editor} {\bibfnamefont
  {H.}~\bibnamefont {Takayasu}}}\ (\bibinfo  {publisher} {Springer},\ \bibinfo
  {address} {Tokyo},\ \bibinfo {year} {2002})\ pp.\ \bibinfo {pages}
  {321--330}\BibitemShut {NoStop}%
\bibitem [{\citenamefont {Cristelli}\ \emph {et~al.}(2012)\citenamefont
  {Cristelli}, \citenamefont {Batty},\ and\ \citenamefont
  {Pietronero}}]{Cristelli2012MoreThanPowerLaw}%
  \BibitemOpen
  \bibfield  {author} {\bibinfo {author} {\bibfnamefont {M.}~\bibnamefont
  {Cristelli}}, \bibinfo {author} {\bibfnamefont {M.}~\bibnamefont {Batty}},\
  and\ \bibinfo {author} {\bibfnamefont {L.}~\bibnamefont {Pietronero}},\
  }\bibfield  {title} {\bibinfo {title} {There is more than a power law in
  zipf},\ }\href {https://doi.org/10.1038/srep00812} {\bibfield  {journal}
  {\bibinfo  {journal} {Scientific Reports}\ }\textbf {\bibinfo {volume} {2}},\
  \bibinfo {pages} {812} (\bibinfo {year} {2012})}\BibitemShut {NoStop}%
\bibitem [{\citenamefont {Arvanitidis}\ and\ \citenamefont
  {Kollias}(2016)}]{Arvanitidis2016ZipfMilitaryExpenditures}%
  \BibitemOpen
  \bibfield  {author} {\bibinfo {author} {\bibfnamefont {P.}~\bibnamefont
  {Arvanitidis}}\ and\ \bibinfo {author} {\bibfnamefont {C.}~\bibnamefont
  {Kollias}},\ }\bibfield  {title} {\bibinfo {title} {Zipf's law and world
  military expenditures},\ }\href {https://doi.org/10.1515/peps-2015-0016}
  {\bibfield  {journal} {\bibinfo  {journal} {Peace Economics, Peace Science
  and Public Policy}\ }\textbf {\bibinfo {volume} {22}},\ \bibinfo {pages} {41}
  (\bibinfo {year} {2016})}\BibitemShut {NoStop}%
\bibitem [{\citenamefont {Souma}(2001)}]{Souma2001UniversalStructure}%
  \BibitemOpen
  \bibfield  {author} {\bibinfo {author} {\bibfnamefont {W.}~\bibnamefont
  {Souma}},\ }\bibfield  {title} {\bibinfo {title} {Universal structure of the
  personal income distribution},\ }\href
  {https://doi.org/10.1142/S0218348X01000816} {\bibfield  {journal} {\bibinfo
  {journal} {Fractals}\ }\textbf {\bibinfo {volume} {9}},\ \bibinfo {pages}
  {463} (\bibinfo {year} {2001})}\BibitemShut {NoStop}%
\bibitem [{\citenamefont {Aoyama}\ \emph {et~al.}(2000)\citenamefont {Aoyama},
  \citenamefont {Souma}, \citenamefont {Nagahara}, \citenamefont {Okazaki},
  \citenamefont {Takayasu},\ and\ \citenamefont
  {Takayasu}}]{Aoyama2000ParetoLawIncomeDebt}%
  \BibitemOpen
  \bibfield  {author} {\bibinfo {author} {\bibfnamefont {H.}~\bibnamefont
  {Aoyama}}, \bibinfo {author} {\bibfnamefont {W.}~\bibnamefont {Souma}},
  \bibinfo {author} {\bibfnamefont {Y.}~\bibnamefont {Nagahara}}, \bibinfo
  {author} {\bibfnamefont {M.~P.}\ \bibnamefont {Okazaki}}, \bibinfo {author}
  {\bibfnamefont {H.}~\bibnamefont {Takayasu}},\ and\ \bibinfo {author}
  {\bibfnamefont {M.}~\bibnamefont {Takayasu}},\ }\bibfield  {title} {\bibinfo
  {title} {Pareto's law for income of individuals and debt of bankrupt
  companies},\ }\href {https://doi.org/10.1142/S0218348X0000038X} {\bibfield
  {journal} {\bibinfo  {journal} {Fractals}\ }\textbf {\bibinfo {volume} {8}},\
  \bibinfo {pages} {293} (\bibinfo {year} {2000})}\BibitemShut {NoStop}%
\bibitem [{\citenamefont {de~Vries}\ and\ \citenamefont
  {Toda}(2022)}]{DeVries2022CapitalLaborPareto}%
  \BibitemOpen
  \bibfield  {author} {\bibinfo {author} {\bibfnamefont {T.}~\bibnamefont
  {de~Vries}}\ and\ \bibinfo {author} {\bibfnamefont {A.~A.}\ \bibnamefont
  {Toda}},\ }\bibfield  {title} {\bibinfo {title} {Capital and labor income
  pareto exponents across time and space},\ }\href
  {https://doi.org/10.1111/roiw.12556} {\bibfield  {journal} {\bibinfo
  {journal} {Review of Income and Wealth}\ }\textbf {\bibinfo {volume} {68}},\
  \bibinfo {pages} {1058} (\bibinfo {year} {2022})}\BibitemShut {NoStop}%
\bibitem [{\citenamefont {Abul-Magd}(2002)}]{AbulMagd2002}%
  \BibitemOpen
  \bibfield  {author} {\bibinfo {author} {\bibfnamefont {A.~Y.}\ \bibnamefont
  {Abul-Magd}},\ }\bibfield  {title} {\bibinfo {title} {Wealth distribution in
  an ancient egyptian society},\ }\href
  {https://doi.org/10.1103/PhysRevE.66.057104} {\bibfield  {journal} {\bibinfo
  {journal} {Physical Review E}\ }\textbf {\bibinfo {volume} {66}},\ \bibinfo
  {pages} {057104} (\bibinfo {year} {2002})}\BibitemShut {NoStop}%
\bibitem [{\citenamefont {Danon}(2022)}]{Danon2022PompeiiSenatorialWealth}%
  \BibitemOpen
  \bibfield  {author} {\bibinfo {author} {\bibfnamefont {B.}~\bibnamefont
  {Danon}},\ }\bibfield  {title} {\bibinfo {title} {Senators and senatorial
  wealth at pompeii: Reconstructing the local wealth distribution},\ }in\ \href
  {https://doi.org/10.1017/9781009121873.004} {\emph {\bibinfo {booktitle} {The
  Uncertain Past: Probability in Ancient History}}},\ \bibinfo {editor} {edited
  by\ \bibinfo {editor} {\bibfnamefont {M.}~\bibnamefont {Lavan}}, \bibinfo
  {editor} {\bibfnamefont {D.}~\bibnamefont {Jew}},\ and\ \bibinfo {editor}
  {\bibfnamefont {B.}~\bibnamefont {Danon}}}\ (\bibinfo  {publisher} {Cambridge
  University Press},\ \bibinfo {year} {2022})\ pp.\ \bibinfo {pages}
  {93--134}\BibitemShut {NoStop}%
\bibitem [{\citenamefont
  {Danon}(2025)}]{Danon2025ReconstructingWealthDistributions}%
  \BibitemOpen
  \bibfield  {author} {\bibinfo {author} {\bibfnamefont {B.}~\bibnamefont
  {Danon}},\ }\bibfield  {title} {\bibinfo {title} {Reconstructing historical
  wealth distributions},\ }in\ \href
  {https://doi.org/10.1017/9781009496940.004} {\emph {\bibinfo {booktitle}
  {Wealth, Office and Rank in Roman Italy}}}\ (\bibinfo  {publisher} {Cambridge
  University Press},\ \bibinfo {year} {2025})\ pp.\ \bibinfo {pages}
  {40--61}\BibitemShut {NoStop}%
\bibitem [{\citenamefont {Yu}\ \emph {et~al.}(2019)\citenamefont {Yu},
  \citenamefont {Chen},\ and\ \citenamefont
  {Fang}}]{YuEtAl2019GraveSizesInequality}%
  \BibitemOpen
  \bibfield  {author} {\bibinfo {author} {\bibfnamefont {S.-Y.}\ \bibnamefont
  {Yu}}, \bibinfo {author} {\bibfnamefont {X.-X.}\ \bibnamefont {Chen}},\ and\
  \bibinfo {author} {\bibfnamefont {H.}~\bibnamefont {Fang}},\ }\bibfield
  {title} {\bibinfo {title} {Inferring inequality in prehistoric societies from
  grave sizes: a methodological framework},\ }\href
  {https://doi.org/10.1007/s12520-019-00845-0} {\bibfield  {journal} {\bibinfo
  {journal} {Archaeological and Anthropological Sciences}\ }\textbf {\bibinfo
  {volume} {11}},\ \bibinfo {pages} {4947} (\bibinfo {year}
  {2019})}\BibitemShut {NoStop}%
\bibitem [{\citenamefont {Noh}\ and\ \citenamefont
  {Kim}(2026)}]{NohKim2026SouthKoreaInequality}%
  \BibitemOpen
  \bibfield  {author} {\bibinfo {author} {\bibfnamefont {Y.}~\bibnamefont
  {Noh}}\ and\ \bibinfo {author} {\bibfnamefont {S.}~\bibnamefont {Kim}},\
  }\bibfield  {title} {\bibinfo {title} {Quantitative assessment of social
  stratification and inequality in the late bronze age and early iron age south
  korea},\ }\bibfield  {journal} {\bibinfo  {journal} {Archaeometry}\ }\href
  {https://doi.org/10.1111/arcm.70161} {10.1111/arcm.70161} (\bibinfo {year}
  {2026}),\ \bibinfo {note} {early View}\BibitemShut {NoStop}%
\bibitem [{\citenamefont {Strawinska-Zanko}\ \emph {et~al.}(2018)\citenamefont
  {Strawinska-Zanko}, \citenamefont {Liebovitch}, \citenamefont {Watson},\ and\
  \citenamefont {Brown}}]{StrawinskaZankoEtAl2018MayaInequality}%
  \BibitemOpen
  \bibfield  {author} {\bibinfo {author} {\bibfnamefont {U.}~\bibnamefont
  {Strawinska-Zanko}}, \bibinfo {author} {\bibfnamefont {L.~S.}\ \bibnamefont
  {Liebovitch}}, \bibinfo {author} {\bibfnamefont {A.~A.}\ \bibnamefont
  {Watson}},\ and\ \bibinfo {author} {\bibfnamefont {C.~T.}\ \bibnamefont
  {Brown}},\ }\bibfield  {title} {\bibinfo {title} {Capital in the first
  century: The evolution of inequality in ancient maya society},\ }in\ \href
  {https://doi.org/10.1007/978-3-319-76765-9_9} {\emph {\bibinfo {booktitle}
  {Mathematical Modeling of Social Relationships: What Mathematics Can Tell Us
  about People}}},\ \bibinfo {series and number} {Computational Social
  Sciences},\ \bibinfo {editor} {edited by\ \bibinfo {editor} {\bibfnamefont
  {U.}~\bibnamefont {Strawinska-Zanko}}\ and\ \bibinfo {editor} {\bibfnamefont
  {L.~S.}\ \bibnamefont {Liebovitch}}}\ (\bibinfo  {publisher} {Springer},\
  \bibinfo {address} {Cham},\ \bibinfo {year} {2018})\ pp.\ \bibinfo {pages}
  {161--192}\BibitemShut {NoStop}%
\bibitem [{\citenamefont {Brown}\ \emph {et~al.}(2012)\citenamefont {Brown},
  \citenamefont {Watson}, \citenamefont {Gravlin-Beman},\ and\ \citenamefont
  {Liebovitch}}]{BrownEtAl2012PoorMayapan}%
  \BibitemOpen
  \bibfield  {author} {\bibinfo {author} {\bibfnamefont {C.~T.}\ \bibnamefont
  {Brown}}, \bibinfo {author} {\bibfnamefont {A.~A.}\ \bibnamefont {Watson}},
  \bibinfo {author} {\bibfnamefont {A.}~\bibnamefont {Gravlin-Beman}},\ and\
  \bibinfo {author} {\bibfnamefont {L.~S.}\ \bibnamefont {Liebovitch}},\
  }\bibfield  {title} {\bibinfo {title} {Poor mayapan},\ }in\ \href@noop {}
  {\emph {\bibinfo {booktitle} {The Ancient Maya of Mexico: Reinterpreting the
  Past of the Northern Maya Lowlands}}},\ \bibinfo {series and number}
  {Approaches to Anthropological Archaeology},\ \bibinfo {editor} {edited by\
  \bibinfo {editor} {\bibfnamefont {G.~E.}\ \bibnamefont {Braswell}}}\
  (\bibinfo  {publisher} {Equinox Publishing},\ \bibinfo {year} {2012})\ pp.\
  \bibinfo {pages} {306--324}\BibitemShut {NoStop}%
\bibitem [{\citenamefont {Hegyi}\ \emph {et~al.}(2007)\citenamefont {Hegyi},
  \citenamefont {N{\'e}da},\ and\ \citenamefont {Santos}}]{HegyiEtAl2007}%
  \BibitemOpen
  \bibfield  {author} {\bibinfo {author} {\bibfnamefont {G.}~\bibnamefont
  {Hegyi}}, \bibinfo {author} {\bibfnamefont {Z.}~\bibnamefont {N{\'e}da}},\
  and\ \bibinfo {author} {\bibfnamefont {M.~A.}\ \bibnamefont {Santos}},\
  }\bibfield  {title} {\bibinfo {title} {Wealth distribution and pareto's law
  in the hungarian medieval society},\ }\href
  {https://doi.org/10.1016/j.physa.2007.02.094} {\bibfield  {journal} {\bibinfo
   {journal} {Physica A: Statistical Mechanics and its Applications}\ }\textbf
  {\bibinfo {volume} {380}},\ \bibinfo {pages} {271} (\bibinfo {year}
  {2007})}\BibitemShut {NoStop}%
\bibitem [{\citenamefont {{Obayashi Corporation Project
  Team}}(1985)}]{Obayashi1985Nintoku}%
  \BibitemOpen
  \bibfield  {author} {\bibinfo {author} {\bibnamefont {{Obayashi Corporation
  Project Team}}},\ }\bibfield  {title} {{\bibinfo
  {title} {Gendai gijutsu to kodai gijutsu no hikaku ni yoru nintoku tenno-ryo
  no kensetsu [construction of the emperor nintoku tomb by comparing modern and
  ancient technologies]}},\ }\href@noop {} {\bibfield  {journal} {\bibinfo
  {journal} {Kikan Obayashi}\ ,\ \bibinfo {pages} {2}} (\bibinfo {year}
  {1985})}\BibitemShut {NoStop}%
\bibitem [{\citenamefont {Ozawa}(2012)}]{Ozawa2012MathematicalHistory}%
  \BibitemOpen
  \bibfield  {author} {\bibinfo {author} {\bibfnamefont {K.}~\bibnamefont
  {Ozawa}},\ }\bibfield  {title} {\bibinfo {title} {Mathematical understanding
  of history},\ }in\ \href@noop {} {\emph {\bibinfo {booktitle} {Proceedings of
  the 18th Public Symposium ``Humanities and Databases''}}}\ (\bibinfo {year}
  {2012})\ pp.\ \bibinfo {pages} {81--91},\ \bibinfo {note} {in
  Japanese}\BibitemShut {NoStop}%
\bibitem [{\citenamefont {Okubo}(2019)}]{Okubo2019RankSizeKofun}%
  \BibitemOpen
  \bibfield  {author} {\bibinfo {author} {\bibfnamefont {K.}~\bibnamefont
  {Okubo}},\ }\bibfield  {title} {\bibinfo {title} {Rank-size rule in large
  keyhole-shaped tumulus},\ }in\ \href
  {https://doi.org/10.11316/jpsgaiyo.74.2.0_2380} {\emph {\bibinfo {booktitle}
  {Meeting Abstracts of the Physical Society of Japan, 2019 Autumn Meeting}}},\
  Vol.~\bibinfo {volume} {74}\ (\bibinfo {year} {2019})\ p.\ \bibinfo {pages}
  {2380},\ \bibinfo {note} {11pK26-5, in Japanese}\BibitemShut {NoStop}%
\bibitem [{\citenamefont {{Yamanashi
  Prefecture}}(2024)}]{YamanashiPrefecture2024PointCloud}%
  \BibitemOpen
  \bibfield  {author} {\bibinfo {author} {\bibnamefont {{Yamanashi
  Prefecture}}},\ }\href@noop {} {\bibinfo {title} {{Yamanashi-ken ten-gun
  d{\=e}ta [Yamanashi Prefecture Point Cloud Data]}}},\ \bibinfo {howpublished}
  {\url{https://www.geospatial.jp/ckan/dataset/yamanashi-pointcloud-2024}}
  (\bibinfo {year} {2024}),\ \bibinfo {note} {g-Spatial Information Center.
  Accessed: 2026-06-18. In Japanese}\BibitemShut {NoStop}%
\bibitem [{\citenamefont {Ishikawa}(2008)}]{Ishikawa2008FirmSizeDisplacement}%
  \BibitemOpen
  \bibfield  {author} {\bibinfo {author} {\bibfnamefont {A.}~\bibnamefont
  {Ishikawa}},\ }\href@noop {} {\emph {\bibinfo {title} {Power-Law and
  Log-Normal Distributions in Firm Size Displacement Data}}},\ \bibinfo {type}
  {Economics Discussion Papers}\ \bibinfo {number} {2008-45}\ (\bibinfo
  {institution} {Kiel Institute for the World Economy},\ \bibinfo {year}
  {2008})\BibitemShut {NoStop}%
\bibitem [{\citenamefont {Kesten}(1973)}]{Kesten1973RandomDifference}%
  \BibitemOpen
  \bibfield  {author} {\bibinfo {author} {\bibfnamefont {H.}~\bibnamefont
  {Kesten}},\ }\bibfield  {title} {\bibinfo {title} {Random difference
  equations and renewal theory for products of random matrices},\ }\href
  {https://doi.org/10.1007/BF02392040} {\bibfield  {journal} {\bibinfo
  {journal} {Acta Mathematica}\ }\textbf {\bibinfo {volume} {131}},\ \bibinfo
  {pages} {207} (\bibinfo {year} {1973})}\BibitemShut {NoStop}%
\bibitem [{\citenamefont {Sornette}\ and\ \citenamefont
  {Cont}(1997)}]{SornetteCont1997Multiplicative}%
  \BibitemOpen
  \bibfield  {author} {\bibinfo {author} {\bibfnamefont {D.}~\bibnamefont
  {Sornette}}\ and\ \bibinfo {author} {\bibfnamefont {R.}~\bibnamefont
  {Cont}},\ }\bibfield  {title} {\bibinfo {title} {Convergent multiplicative
  processes repelled from zero: power laws and truncated power laws},\ }\href
  {https://doi.org/10.1051/jp1:1997169} {\bibfield  {journal} {\bibinfo
  {journal} {Journal de Physique I}\ }\textbf {\bibinfo {volume} {7}},\
  \bibinfo {pages} {431} (\bibinfo {year} {1997})}\BibitemShut {NoStop}%
\bibitem [{\citenamefont {Takayasu}\ \emph {et~al.}(1997)\citenamefont
  {Takayasu}, \citenamefont {Sato},\ and\ \citenamefont
  {Takayasu}}]{TakayasuEtAl1997RandomAmplification}%
  \BibitemOpen
  \bibfield  {author} {\bibinfo {author} {\bibfnamefont {H.}~\bibnamefont
  {Takayasu}}, \bibinfo {author} {\bibfnamefont {A.-H.}\ \bibnamefont {Sato}},\
  and\ \bibinfo {author} {\bibfnamefont {M.}~\bibnamefont {Takayasu}},\
  }\bibfield  {title} {\bibinfo {title} {Stable infinite variance fluctuations
  in randomly amplified langevin systems},\ }\href
  {https://doi.org/10.1103/PhysRevLett.79.966} {\bibfield  {journal} {\bibinfo
  {journal} {Physical Review Letters}\ }\textbf {\bibinfo {volume} {79}},\
  \bibinfo {pages} {966} (\bibinfo {year} {1997})}\BibitemShut {NoStop}%
\bibitem [{\citenamefont {Reed}\ and\ \citenamefont
  {Jorgensen}(2004)}]{ReedJorgensen2004dPlN}%
  \BibitemOpen
  \bibfield  {author} {\bibinfo {author} {\bibfnamefont {W.~J.}\ \bibnamefont
  {Reed}}\ and\ \bibinfo {author} {\bibfnamefont {M.}~\bibnamefont
  {Jorgensen}},\ }\bibfield  {title} {\bibinfo {title} {The double
  pareto-lognormal distribution---a new parametric model for size
  distributions},\ }\href {https://doi.org/10.1081/STA-120037438} {\bibfield
  {journal} {\bibinfo  {journal} {Communications in Statistics - Theory and
  Methods}\ }\textbf {\bibinfo {volume} {33}},\ \bibinfo {pages} {1733}
  (\bibinfo {year} {2004})}\BibitemShut {NoStop}%
\bibitem [{\citenamefont {Jones}\ and\ \citenamefont {Kim}(2018)}]{Jones2018}%
  \BibitemOpen
  \bibfield  {author} {\bibinfo {author} {\bibfnamefont {C.~I.}\ \bibnamefont
  {Jones}}\ and\ \bibinfo {author} {\bibfnamefont {J.}~\bibnamefont {Kim}},\
  }\bibfield  {title} {\bibinfo {title} {A schumpeterian model of top income
  inequality},\ }\href {https://doi.org/10.1086/699190} {\bibfield  {journal}
  {\bibinfo  {journal} {Journal of Political Economy}\ }\textbf {\bibinfo
  {volume} {126}},\ \bibinfo {pages} {1785} (\bibinfo {year}
  {2018})}\BibitemShut {NoStop}%
\bibitem [{\citenamefont {Watanabe}\ \emph {et~al.}(2013)\citenamefont
  {Watanabe}, \citenamefont {Takayasu},\ and\ \citenamefont
  {Takayasu}}]{Watanabe2013AllometricScalings}%
  \BibitemOpen
  \bibfield  {author} {\bibinfo {author} {\bibfnamefont {H.}~\bibnamefont
  {Watanabe}}, \bibinfo {author} {\bibfnamefont {H.}~\bibnamefont {Takayasu}},\
  and\ \bibinfo {author} {\bibfnamefont {M.}~\bibnamefont {Takayasu}},\
  }\bibfield  {title} {\bibinfo {title} {Relations between allometric scalings
  and fluctuations in complex systems: The case of japanese firms},\ }\href
  {https://doi.org/10.1016/j.physa.2012.10.020} {\bibfield  {journal} {\bibinfo
   {journal} {Physica A: Statistical Mechanics and its Applications}\ }\textbf
  {\bibinfo {volume} {392}},\ \bibinfo {pages} {741} (\bibinfo {year}
  {2013})}\BibitemShut {NoStop}%
\bibitem [{\citenamefont {West}\ \emph {et~al.}(1997)\citenamefont {West},
  \citenamefont {Brown},\ and\ \citenamefont
  {Enquist}}]{West1997AllometricScaling}%
  \BibitemOpen
  \bibfield  {author} {\bibinfo {author} {\bibfnamefont {G.~B.}\ \bibnamefont
  {West}}, \bibinfo {author} {\bibfnamefont {J.~H.}\ \bibnamefont {Brown}},\
  and\ \bibinfo {author} {\bibfnamefont {B.~J.}\ \bibnamefont {Enquist}},\
  }\bibfield  {title} {\bibinfo {title} {A general model for the origin of
  allometric scaling laws in biology},\ }\href
  {https://doi.org/10.1126/science.276.5309.122} {\bibfield  {journal}
  {\bibinfo  {journal} {Science}\ }\textbf {\bibinfo {volume} {276}},\ \bibinfo
  {pages} {122} (\bibinfo {year} {1997})}\BibitemShut {NoStop}%
\bibitem [{\citenamefont {Hanson}(2024)}]{Hanson2024ScalingPompeii}%
  \BibitemOpen
  \bibfield  {author} {\bibinfo {author} {\bibfnamefont {J.~W.}\ \bibnamefont
  {Hanson}},\ }\bibfield  {title} {\bibinfo {title} {Scaling in pompeii:
  Preliminary evidence for the occurrence of scaling phenomena within an
  ancient built environment},\ }\href
  {https://doi.org/10.1007/s10816-023-09604-x} {\bibfield  {journal} {\bibinfo
  {journal} {Journal of Archaeological Method and Theory}\ }\textbf {\bibinfo
  {volume} {31}},\ \bibinfo {pages} {448} (\bibinfo {year} {2024})}\BibitemShut
  {NoStop}%
\bibitem [{\citenamefont {Kohler}\ \emph {et~al.}(2025)\citenamefont {Kohler},
  \citenamefont {Bogaard}, \citenamefont {Ortman}, \citenamefont {Crema},
  \citenamefont {Chirikure}, \citenamefont {Cruz}, \citenamefont {Green},
  \citenamefont {Kerig}, \citenamefont {McCoy}, \citenamefont {Munson},
  \citenamefont {Petrie}, \citenamefont {Thompson}, \citenamefont {Birch},
  \citenamefont {Cervantes~Quequezana}, \citenamefont {Feinman}, \citenamefont
  {Fochesato}, \citenamefont {Gronenborn}, \citenamefont {Hamerow},
  \citenamefont {Jin}, \citenamefont {Lawrence}, \citenamefont {Roscoe},
  \citenamefont {Rosenstock}, \citenamefont {Erny}, \citenamefont {Kim},
  \citenamefont {Ohlrau}, \citenamefont {Hanson}, \citenamefont
  {Fargher~Navarro},\ and\ \citenamefont
  {Pailes}}]{Kohler2025EconomicInequality}%
  \BibitemOpen
  \bibfield  {author} {\bibinfo {author} {\bibfnamefont {T.~A.}\ \bibnamefont
  {Kohler}}, \bibinfo {author} {\bibfnamefont {A.}~\bibnamefont {Bogaard}},
  \bibinfo {author} {\bibfnamefont {S.~G.}\ \bibnamefont {Ortman}}, \bibinfo
  {author} {\bibfnamefont {E.~R.}\ \bibnamefont {Crema}}, \bibinfo {author}
  {\bibfnamefont {S.}~\bibnamefont {Chirikure}}, \bibinfo {author}
  {\bibfnamefont {P.}~\bibnamefont {Cruz}}, \bibinfo {author} {\bibfnamefont
  {A.}~\bibnamefont {Green}}, \bibinfo {author} {\bibfnamefont
  {T.}~\bibnamefont {Kerig}}, \bibinfo {author} {\bibfnamefont {M.~D.}\
  \bibnamefont {McCoy}}, \bibinfo {author} {\bibfnamefont {J.}~\bibnamefont
  {Munson}}, \bibinfo {author} {\bibfnamefont {C.}~\bibnamefont {Petrie}},
  \bibinfo {author} {\bibfnamefont {A.~E.}\ \bibnamefont {Thompson}}, \bibinfo
  {author} {\bibfnamefont {J.}~\bibnamefont {Birch}}, \bibinfo {author}
  {\bibfnamefont {G.}~\bibnamefont {Cervantes~Quequezana}}, \bibinfo {author}
  {\bibfnamefont {G.~M.}\ \bibnamefont {Feinman}}, \bibinfo {author}
  {\bibfnamefont {M.}~\bibnamefont {Fochesato}}, \bibinfo {author}
  {\bibfnamefont {D.}~\bibnamefont {Gronenborn}}, \bibinfo {author}
  {\bibfnamefont {H.}~\bibnamefont {Hamerow}}, \bibinfo {author} {\bibfnamefont
  {G.}~\bibnamefont {Jin}}, \bibinfo {author} {\bibfnamefont {D.}~\bibnamefont
  {Lawrence}}, \bibinfo {author} {\bibfnamefont {P.~B.}\ \bibnamefont
  {Roscoe}}, \bibinfo {author} {\bibfnamefont {E.}~\bibnamefont {Rosenstock}},
  \bibinfo {author} {\bibfnamefont {G.~K.}\ \bibnamefont {Erny}}, \bibinfo
  {author} {\bibfnamefont {H.}~\bibnamefont {Kim}}, \bibinfo {author}
  {\bibfnamefont {R.}~\bibnamefont {Ohlrau}}, \bibinfo {author} {\bibfnamefont
  {J.~W.}\ \bibnamefont {Hanson}}, \bibinfo {author} {\bibfnamefont
  {L.}~\bibnamefont {Fargher~Navarro}},\ and\ \bibinfo {author} {\bibfnamefont
  {M.}~\bibnamefont {Pailes}},\ }\bibfield  {title} {\bibinfo {title} {Economic
  inequality is fueled by population scale, land-limited production, and
  settlement hierarchies across the archaeological record},\ }\href
  {https://doi.org/10.1073/pnas.2400691122} {\bibfield  {journal} {\bibinfo
  {journal} {Proceedings of the National Academy of Sciences of the United
  States of America}\ }\textbf {\bibinfo {volume} {122}},\ \bibinfo {pages}
  {e2400691122} (\bibinfo {year} {2025})}\BibitemShut {NoStop}%
\bibitem [{\citenamefont {N{\o}rtoft}(2022)}]{Nortoft2022GraveWealth}%
  \BibitemOpen
  \bibfield  {author} {\bibinfo {author} {\bibfnamefont {M.}~\bibnamefont
  {N{\o}rtoft}},\ }\bibfield  {title} {\bibinfo {title} {A new framework for
  quantifying prehistoric grave wealth},\ }\href
  {https://doi.org/10.5334/jcaa.86} {\bibfield  {journal} {\bibinfo  {journal}
  {Journal of Computer Applications in Archaeology}\ }\textbf {\bibinfo
  {volume} {5}},\ \bibinfo {pages} {123} (\bibinfo {year} {2022})}\BibitemShut
  {NoStop}%
\bibitem [{\citenamefont {{Center for Ancient Studies and Sacred Sites, Nara
  Women's University}}(2019)}]{NaraWomensUniversityKofunDatabase}%
  \BibitemOpen
  \bibfield  {author} {\bibinfo {author} {\bibnamefont {{Center for Ancient
  Studies and Sacred Sites, Nara Women's University}}},\ }\href@noop {}
  {\bibinfo {title} {{Zenkoku Kofun Database}}},\ \bibinfo {howpublished}
  {\url{https://zenkoku-kofun.nara-hgis.jp/}} (\bibinfo {year} {2019}),\
  \bibinfo {note} {accessed: 2026-06-18}\BibitemShut {NoStop}%
\bibitem [{\citenamefont {{Agency for Cultural Affairs, Government of
  Japan}}(2021)}]{AgencyCulturalAffairs2021BuriedSites}%
  \BibitemOpen
  \bibfield  {author} {\bibinfo {author} {\bibnamefont {{Agency for Cultural
  Affairs, Government of Japan}}},\ }\href@noop {} {\bibinfo {title} {Reiwa 3
  nendo shuchi no maiz{\=o} bunkazai h{\=o}z{\=o}chi-s{\=u} [number of known
  buried cultural property sites, fy2021]}},\ \bibinfo {howpublished}
  {\url{https://www.bunka.go.jp/seisaku/bunkazai/shokai/pdf/93717701_02.pdf}}
  (\bibinfo {year} {2021}),\ \bibinfo {note} {accessed: 2026-06-18. In
  Japanese}\BibitemShut {NoStop}%
\bibitem [{\citenamefont {Izuta}(2016)}]{Izuta2016KeyholeDistribution}%
  \BibitemOpen
  \bibfield  {author} {\bibinfo {author} {\bibfnamefont {K.}~\bibnamefont
  {Izuta}},\ }\bibfield  {title} {\bibinfo {title} {Zenp{\=o}-k{\=o}enfun no
  chiikisei: bunpuronteki apur{\=o}chi kara [regional characteristics of
  keyhole-shaped tombs: A distributional approach]},\ }in\ \href@noop {} {\emph
  {\bibinfo {booktitle} {2015-nen taikai tokubetsu kenky{\=u} happy{\=o}:
  h{\=o}koku, t{\=o}ron no y{\=o}shi oyobi zach{\=o} no shoken}}},\
  Vol.~\bibinfo {volume} {68}\ (\bibinfo {year} {2016})\ pp.\ \bibinfo {pages}
  {102--107},\ \bibinfo {note} {japanese}\BibitemShut {NoStop}%
\bibitem [{\citenamefont {Wada}(2009)}]{Wada2009OtherWorldKofun}%
  \BibitemOpen
  \bibfield  {author} {\bibinfo {author} {\bibfnamefont {S.}~\bibnamefont
  {Wada}},\ }\bibfield  {title} {{\bibinfo {title}
  {Kofun no takaikan [the concept of the other world in kofun]}},\ }\href
  {https://doi.org/10.15024/00001712} {\bibfield  {journal} {\bibinfo
  {journal} {Bulletin of the National Museum of Japanese History}\ }\textbf
  {\bibinfo {volume} {152}},\ \bibinfo {pages} {247} (\bibinfo {year}
  {2009})}\BibitemShut {NoStop}%
\bibitem [{\citenamefont {Josse}\ and\ \citenamefont
  {Husson}(2016)}]{Josse2016MissMDA}%
  \BibitemOpen
  \bibfield  {author} {\bibinfo {author} {\bibfnamefont {J.}~\bibnamefont
  {Josse}}\ and\ \bibinfo {author} {\bibfnamefont {F.}~\bibnamefont {Husson}},\
  }\bibfield  {title} {\bibinfo {title} {{missMDA}: A package for handling
  missing values in multivariate data analysis},\ }\href
  {https://doi.org/10.18637/jss.v070.i01} {\bibfield  {journal} {\bibinfo
  {journal} {Journal of Statistical Software}\ }\textbf {\bibinfo {volume}
  {70}},\ \bibinfo {pages} {1} (\bibinfo {year} {2016})}\BibitemShut {NoStop}%
\bibitem [{\citenamefont {Clauset}\ \emph {et~al.}(2009)\citenamefont
  {Clauset}, \citenamefont {Shalizi},\ and\ \citenamefont
  {Newman}}]{Clauset2009PowerLaw}%
  \BibitemOpen
  \bibfield  {author} {\bibinfo {author} {\bibfnamefont {A.}~\bibnamefont
  {Clauset}}, \bibinfo {author} {\bibfnamefont {C.~R.}\ \bibnamefont
  {Shalizi}},\ and\ \bibinfo {author} {\bibfnamefont {M.~E.~J.}\ \bibnamefont
  {Newman}},\ }\bibfield  {title} {\bibinfo {title} {Power-law distributions in
  empirical data},\ }\href {https://doi.org/10.1137/070710111} {\bibfield
  {journal} {\bibinfo  {journal} {SIAM Review}\ }\textbf {\bibinfo {volume}
  {51}},\ \bibinfo {pages} {661} (\bibinfo {year} {2009})}\BibitemShut
  {NoStop}%
\bibitem [{\citenamefont {Gillespie}(2015)}]{Gillespie2015Powerlaw}%
  \BibitemOpen
  \bibfield  {author} {\bibinfo {author} {\bibfnamefont {C.~S.}\ \bibnamefont
  {Gillespie}},\ }\bibfield  {title} {\bibinfo {title} {Fitting heavy tailed
  distributions: The {poweRlaw} package},\ }\href
  {https://doi.org/10.18637/jss.v064.i02} {\bibfield  {journal} {\bibinfo
  {journal} {Journal of Statistical Software}\ }\textbf {\bibinfo {volume}
  {64}},\ \bibinfo {pages} {1} (\bibinfo {year} {2015})}\BibitemShut {NoStop}%
\bibitem [{\citenamefont {Watanabe}\ \emph {et~al.}(2012)\citenamefont
  {Watanabe}, \citenamefont {Takayasu},\ and\ \citenamefont
  {Takayasu}}]{Watanabe2012BiasedDiffusion}%
  \BibitemOpen
  \bibfield  {author} {\bibinfo {author} {\bibfnamefont {H.}~\bibnamefont
  {Watanabe}}, \bibinfo {author} {\bibfnamefont {H.}~\bibnamefont {Takayasu}},\
  and\ \bibinfo {author} {\bibfnamefont {M.}~\bibnamefont {Takayasu}},\
  }\bibfield  {title} {\bibinfo {title} {Biased diffusion on the japanese
  inter-firm trading network: estimation of sales from the network structure},\
  }\href {https://doi.org/10.1088/1367-2630/14/4/043034} {\bibfield  {journal}
  {\bibinfo  {journal} {New Journal of Physics}\ }\textbf {\bibinfo {volume}
  {14}},\ \bibinfo {pages} {043034} (\bibinfo {year} {2012})}\BibitemShut
  {NoStop}%
\bibitem [{\citenamefont {Watanabe}(2013)}]{Watanabe2013DoctoralThesis}%
  \BibitemOpen
  \bibfield  {author} {\bibinfo {author} {\bibfnamefont {H.}~\bibnamefont
  {Watanabe}},\ }\emph {\bibinfo {title} {Analysis and Modeling of Fluctuations
  in Corporate Credit Research Data Using Statistical Physics Methods}},\ \href
  {https://researchmap.jp/7941192/misc/54220182?lang=en} {\bibinfo {type}
  {Ph.d. dissertation}},\ \bibinfo  {school} {Tokyo Institute of Technology},
  \bibinfo {address} {Tokyo, Japan} (\bibinfo {year} {2013}),\ \bibinfo {note}
  {original title in Japanese: Kigy{\=o} shin'y{\=o} ch{\=o}sa d{\=e}ta ni
  mirareru yuragi no t{\=o}kei butsurigakuteki na shuh{\=o} ni yoru kaiseki to
  moderuka. Dissertation no. Ko 9049}\BibitemShut {NoStop}%
\bibitem [{\citenamefont {Watanabe}(2014)}]{Watanabe2014MeanField}%
  \BibitemOpen
  \bibfield  {author} {\bibinfo {author} {\bibfnamefont {H.}~\bibnamefont
  {Watanabe}},\ }\bibfield  {title} {\bibinfo {title} {Mean field approximation
  for biased diffusion on japanese inter-firm trading network},\ }\href
  {https://doi.org/10.1371/journal.pone.0091704} {\bibfield  {journal}
  {\bibinfo  {journal} {PLOS ONE}\ }\textbf {\bibinfo {volume} {9}},\ \bibinfo
  {pages} {e91704} (\bibinfo {year} {2014})}\BibitemShut {NoStop}%
\bibitem [{\citenamefont {Satake}(2002)}]{Satake2002Jitsudaka}%
  \BibitemOpen
  \bibfield  {author} {\bibinfo {author} {\bibfnamefont {K.}~\bibnamefont
  {Satake}},\ }\href@noop {} {\bibinfo {title} {{Edo daimyo jitsudaka 60}}},\
  \bibinfo {howpublished}
  {\url{https://homepage-nifty.com/ksatake/libinde.html}} (\bibinfo {year}
  {2002}),\ \bibinfo {note} {online table, revised and updated July 27, 2002;
  accessed June 21, 2026}\BibitemShut {NoStop}%
\bibitem [{\citenamefont {{Economic and Social Research Institute, Cabinet
  Office, Government of Japan}}()}]{ESRI_PrefecturalAccounts}%
  \BibitemOpen
  \bibfield  {author} {\bibinfo {author} {\bibnamefont {{Economic and Social
  Research Institute, Cabinet Office, Government of Japan}}},\ }\href@noop {}
  {\bibinfo {title} {Prefectural accounts: Statistical tables}},\ \bibinfo
  {howpublished}
  {\url{https://www.esri.cao.go.jp/jp/sna/data/data_list/kenmin/files/files_kenmin.html}},\
  \bibinfo {note} {accessed June 21, 2026}\BibitemShut {NoStop}%
\end{thebibliography}%

\clearpage
\appendix

\makeatletter
\@addtoreset{equation}{section}
\@addtoreset{figure}{section}
\@addtoreset{table}{section}
\makeatother

\renewcommand{\theequation}{\thesection\arabic{equation}}
\renewcommand{\thefigure}{\thesection\arabic{figure}}
\renewcommand{\thetable}{\thesection\arabic{table}}
\section{Images of Kofun and Kofun Construction}
\begin{figure*}[hbt]
    \centering
    \begin{overpic}[width=8.8cm]{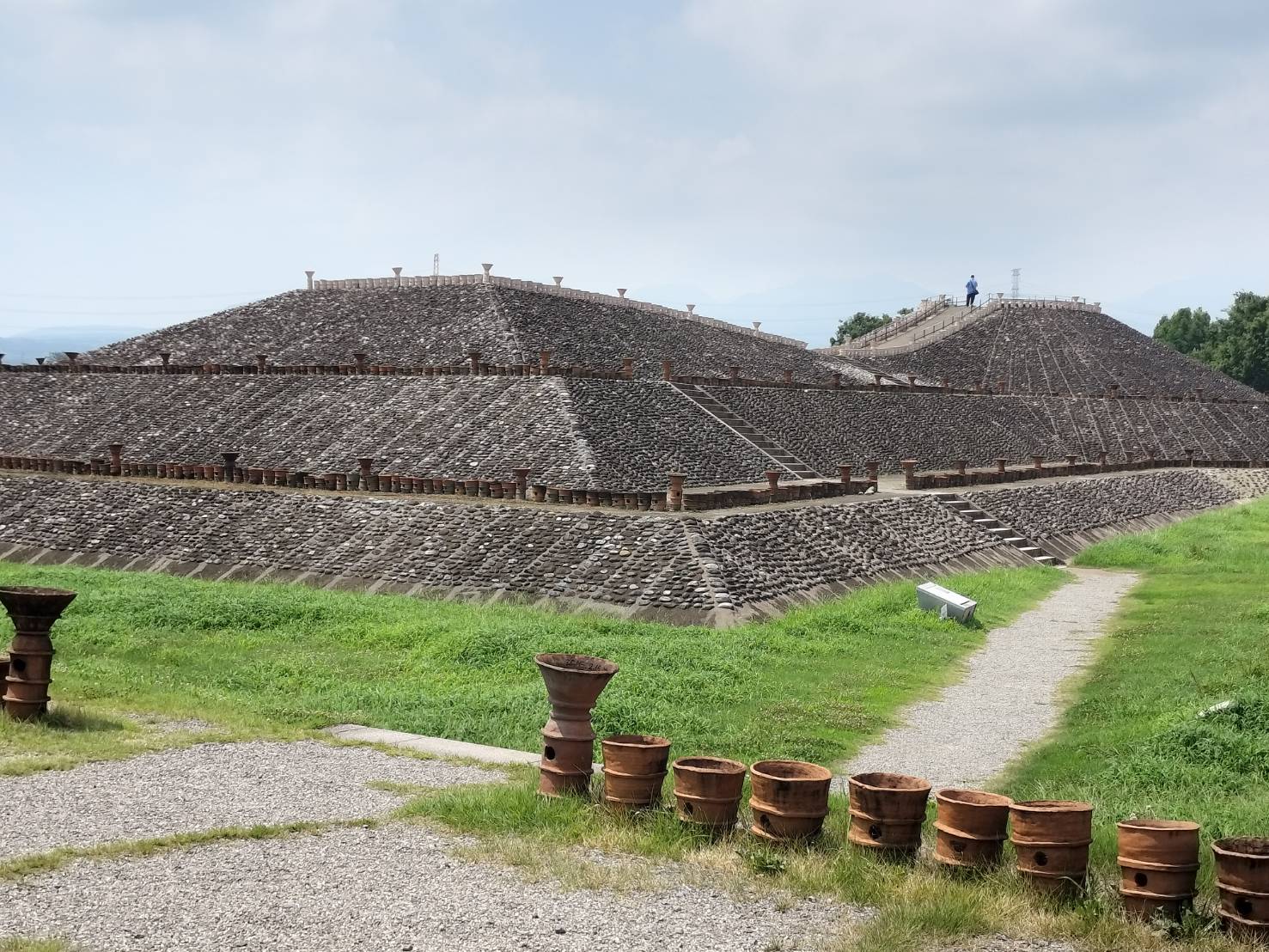}
      \put(10,193){\color{black}\Large\bfseries (a)}
    \end{overpic}
  \begin{overpic}[width=8.8cm]{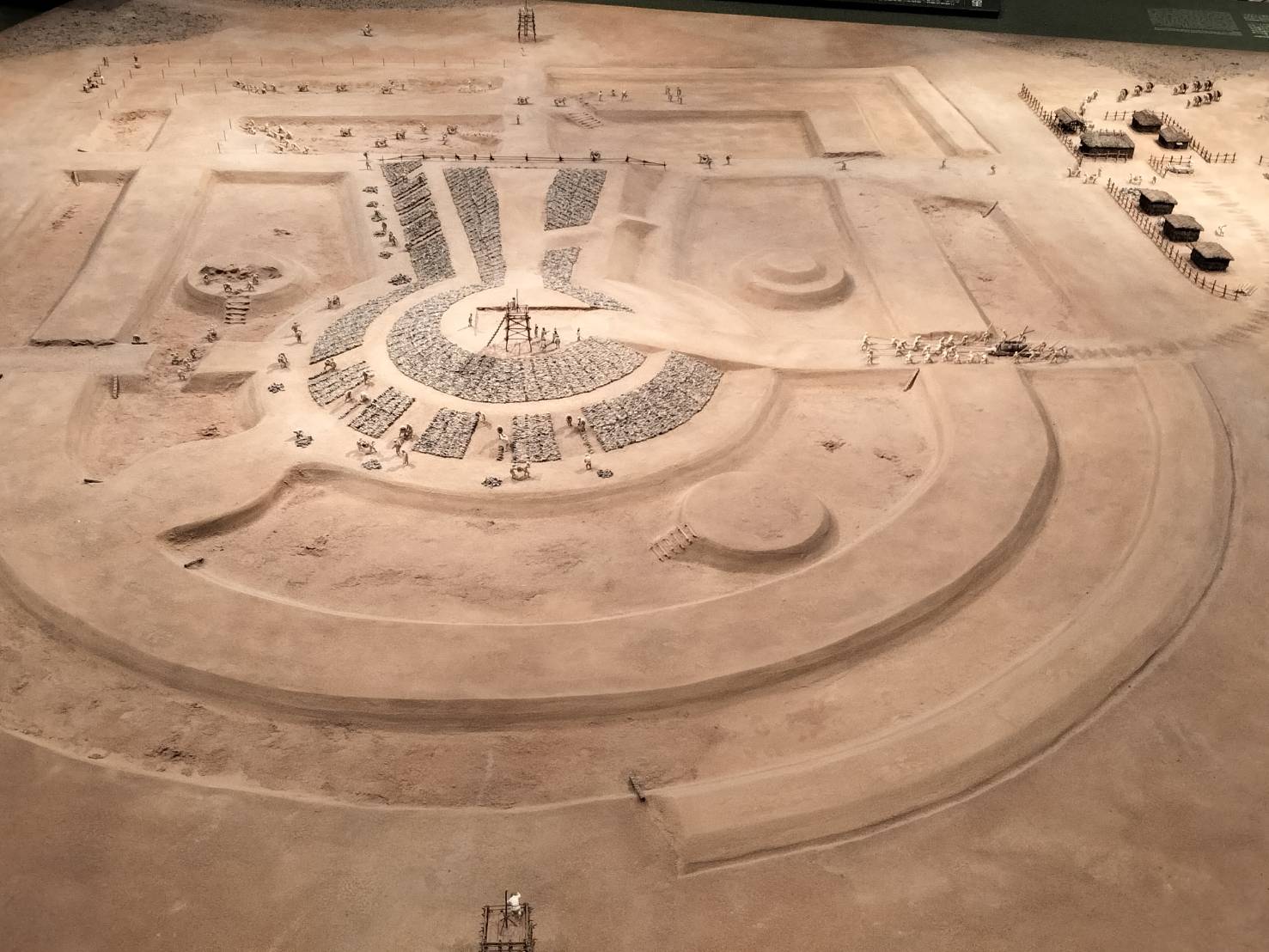}
   \put(10,193){\color{black}\Large\bfseries (b)}
  \end{overpic} 
   \begin{overpic}[width=8.8cm]{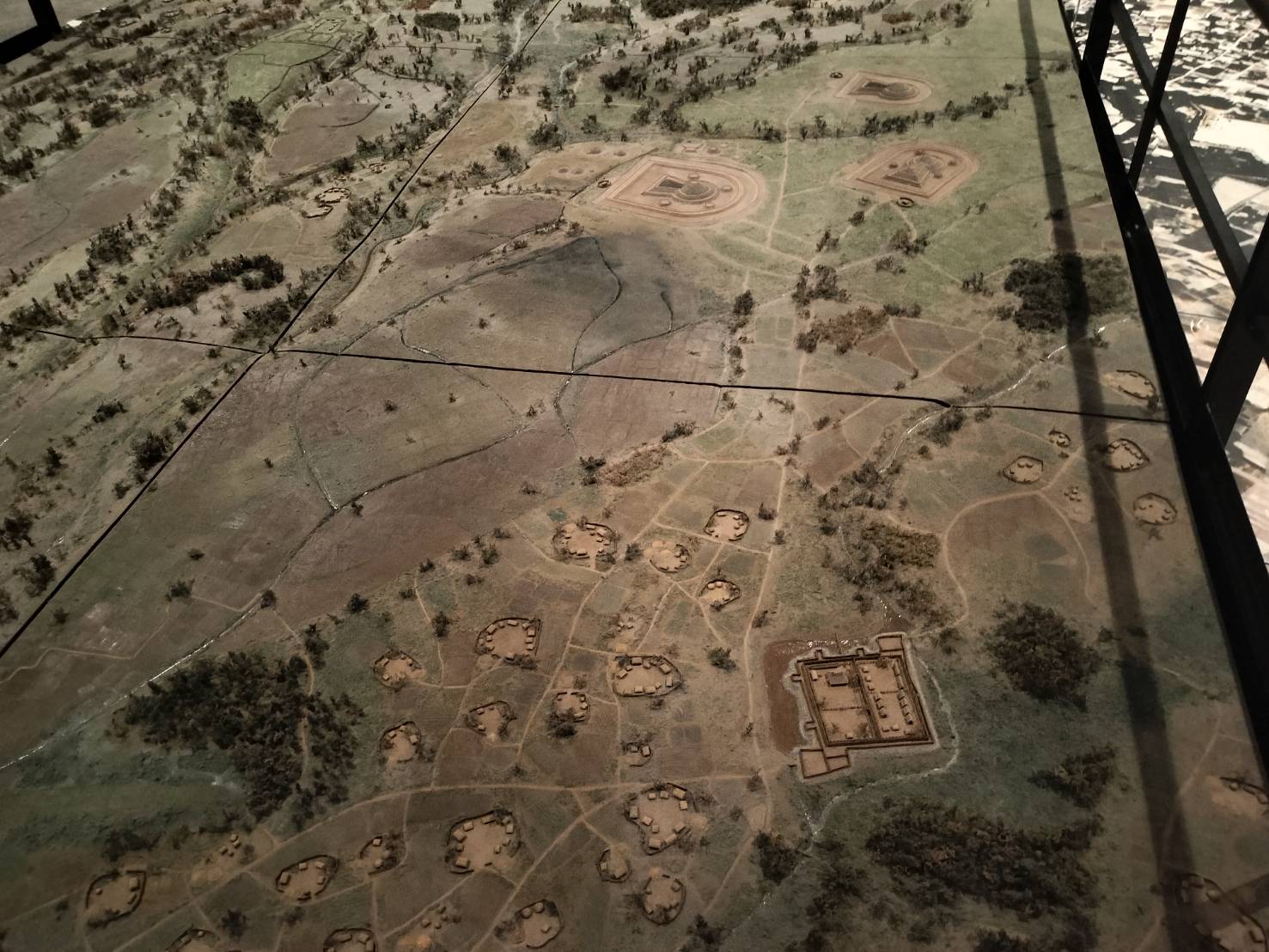}
    \put(-20,175){\color{black}\Large\bfseries (c)}
  \end{overpic}     
\caption{
Keyhole-shaped kofun and kofun construction. Hodota Hachiman-zuka Kofun.
(a) The restored mound, showing the earthen construction covered with fukiishi (facing stones).
(b) Model of kofun construction, by Kamitsukenosato Museum.
(c) Model of a 5th-century village, by Kamitsukenosato Museum. Three keyhole-shaped kofun are visible in the background. The large rectangular, fenced enclosure in the foreground is the chieftain's residence; the smaller round enclosures are residents' compounds.
(b) and (c) are used with permission from the Takasaki City Board of Education.
}
\label{app_fig_kofun_image}
\end{figure*}

To provide a visual overview of kofun and their construction, Fig.~\ref{app_fig_kofun_image} presents a restored keyhole-shaped kofun, 
a model of the construction of a keyhole-shaped kofun, and a model of the 5th-century village responsible for the kofun's construction.
\clearpage

\section{Additional results of the kofun data analysis}
\subsection{Kofun shape statistics and scaling relations}
\label{app_sec_kofun}
\subsubsection{Empirical statistics of kofun shapes}
\label{app_sec_shape}
In this section, we describe in more detail the statistical properties of kofun shapes discussed in Sec.~\ref{sec_shape}.\par

\begin{figure*}[t]
    \begin{center}
    \begin{overpic}[width=5.8cm]{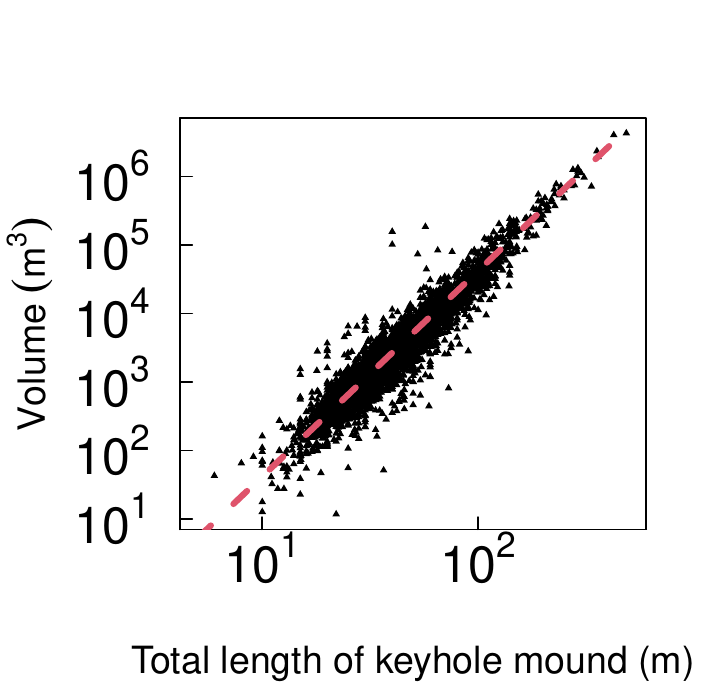}
         \put(50,118){\color{black}\Large\bfseries (a)}
         \put(55,100){\color{black}\sffamily slope=$3$}
    \end{overpic}
      \begin{overpic}[width=5.8cm]{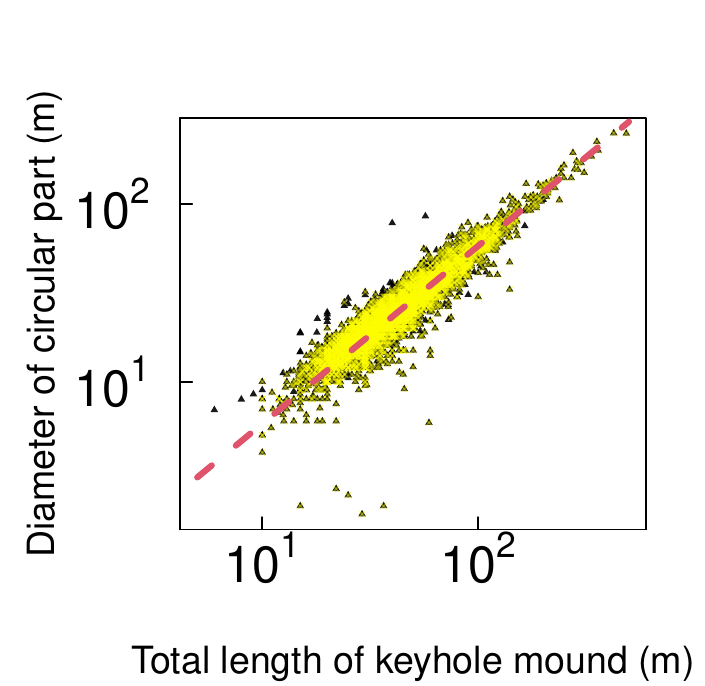}
        \put(50,118){\color{black}\Large\bfseries (b)}
    \end{overpic} 
        \begin{overpic}[width=5.8cm]{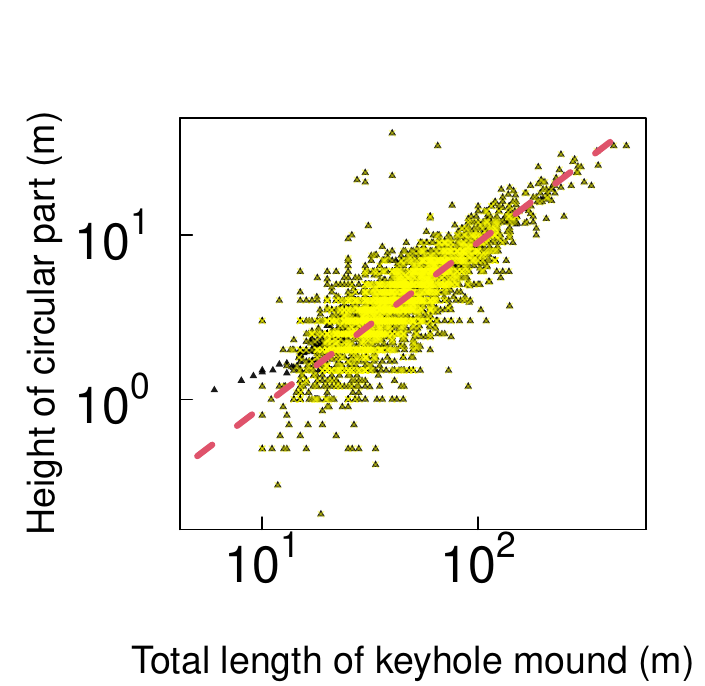}
        \put(50,118){\color{black}\Large\bfseries (c)}
    \end{overpic}
     \begin{overpic}[width=5.8cm]{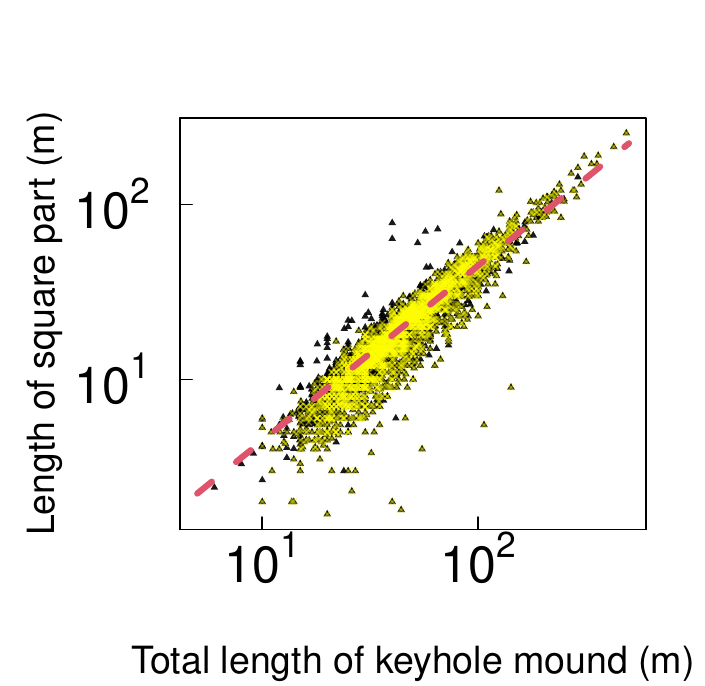}
         \put(50,118){\color{black}\Large\bfseries (d)}
    \end{overpic}
      \begin{overpic}[width=5.8cm]{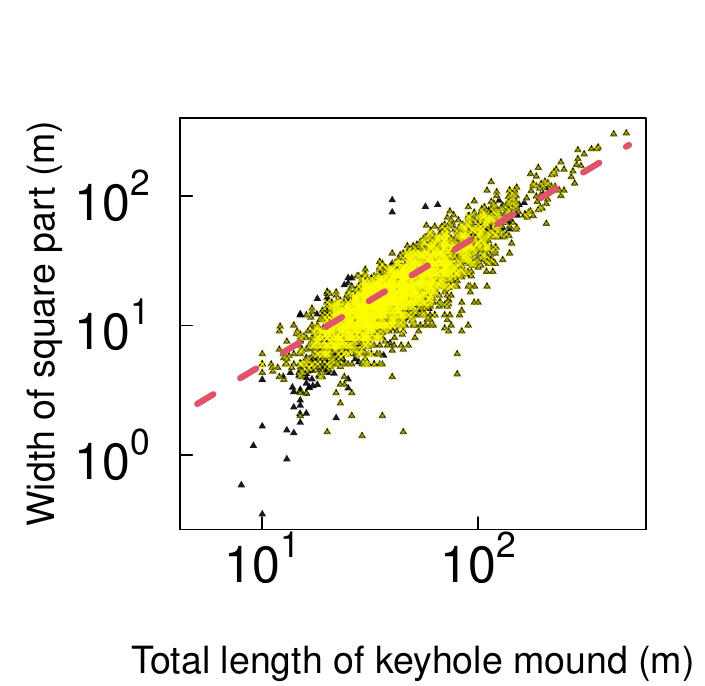}
        \put(50,118){\color{black}\Large\bfseries (e)}
    \end{overpic} 
        \begin{overpic}[width=5.8cm]{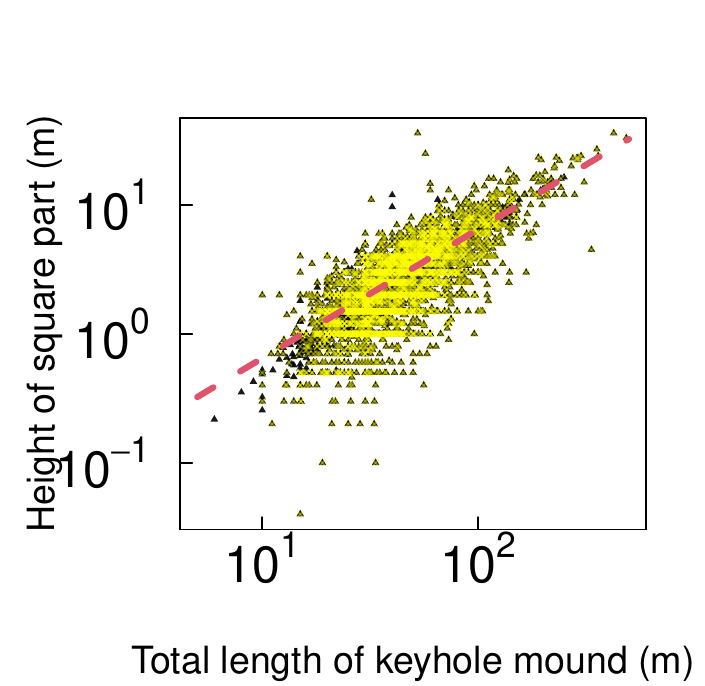}
        \put(50,118){\color{black}\Large\bfseries (f)}
    \end{overpic} 
 \end{center} 
   \caption{
Relationship between mound length and mound volume or the dimensions of individual mound components for keyhole-shaped kofun. The red dashed lines indicate (a) $y=ax^3$ and (b)--(f) $y=ax$. The fitted coefficients are: (a) mound volume, $a=0.0417$; (b) diameter of the posterior circular part, $a=0.580$; (c) height of the posterior circular part, $a=0.0903$; (d) length of the anterior square part, $a=0.446$; (e) width of the anterior square part, $a=0.494$; and (f) height of the anterior square part, $a=0.0648$. Black points denote the data after PCA-based imputation, whereas yellow points denote the values observed before imputation. The red dashed lines are reference lines showing that volume is proportional to the cube of mound length in (a), and that each component dimension is proportional to mound length in (b)--(f). Mound volume is approximately proportional to the cube of mound length, while the principal dimensions of both the posterior circular part and the anterior square part are nearly proportional to mound length. The same tendency is also observed in the pre-imputation data, indicating that the main mound dimensions are well described by a common size scale. This property supports the validity of missing-value imputation based on the first principal component.
}
\label{app_fig_shape_xy}
\end{figure*}

Fig.~\ref{app_fig_shape_xy} shows the relation between mound length and volume for keyhole-shaped kofun. It also shows the relation between mound length and the dimensions of each component. Panel (a) shows that the volume of keyhole-shaped kofun is approximately proportional to the cube of mound length. This relation is expected when the components of a kofun expand almost geometrically with the total mound length.\par

Panels (b)--(f) show the relation between mound length and each component dimension. These dimensions are (b) the diameter of the rear round part, (c) the height of the rear round part, (d) the length of the square front part, (e) the width of the square  front part, and (f) the height of the square front part. All dimensions increase approximately in proportion to mound length, although there is some scatter. This indicates that the main dimensions of keyhole-shaped kofun can be described well by a common size scale. The yellow points represent the data observed before the missing-value imputation described in Sec.~\ref{sec_method_volume}. Even before imputation, each dimension shows an approximately proportional relation with mound length. This supports the validity of the missing-value imputation based on the first principal component.\par

\begin{figure}[th]
    \begin{center}
    \begin{overpic}[width=8cm]{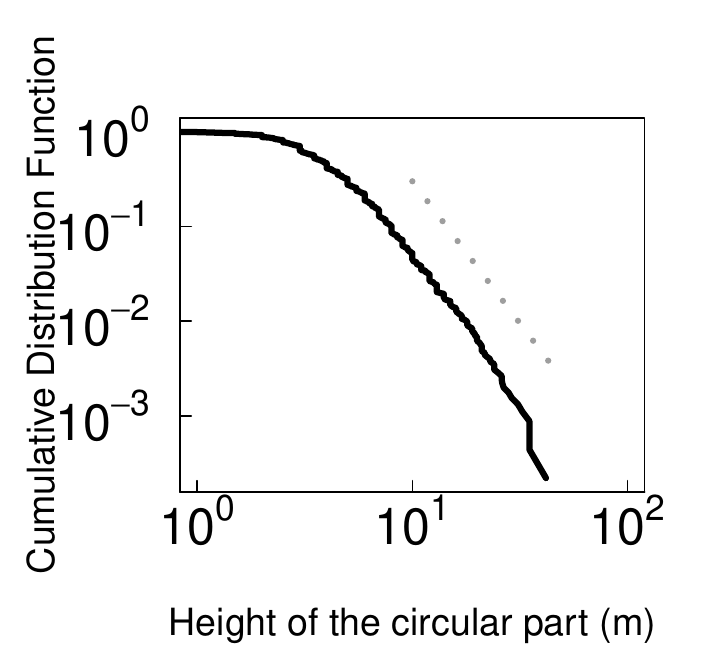}
         \put(70,145){\color{black}\Large\bfseries (a)}
         \put(160,130){\color{black}\sffamily slope=$-3$}
    \end{overpic}
\end{center}
    \caption{
        Upper cumulative distribution of the height of the posterior circular part of keyhole-shaped kofun. The black line shows the empirical data, and the gray dashed line is a reference line representing a power-law distribution with cumulative exponent 3. Similar to the mound-length distribution shown in Fig.~\ref{fig_shape}, the distribution of posterior circular-part height exhibits an approximately power-law-like shape with cumulative exponent 3 in the upper tail. This result is consistent with the nearly proportional relationship between posterior circular-part height and mound length, as confirmed in Fig.~\ref{app_fig_shape_xy}.
}
\label{app_fig_shape_takasa}
\end{figure}

Fig.~\ref{app_fig_shape_takasa} shows the upper cumulative distribution, of the height of the rear round part of keyhole-shaped kofun. This height distribution is also close to a power-law distribution with an upper cumulative exponent of about three. This is similar to the mound-length distribution shown in Fig.~\ref{fig_shape}(b). The agreement of these exponents is consistent with the proportional relation between mound length and rear-part height shown in Fig.~\ref{app_fig_shape_xy}(c). If height is proportional to mound length, the two distributions are expected to have the same power-law exponent.\par

\begin{figure*}[t]
    \begin{center}
    \begin{overpic}[width=8.0cm]{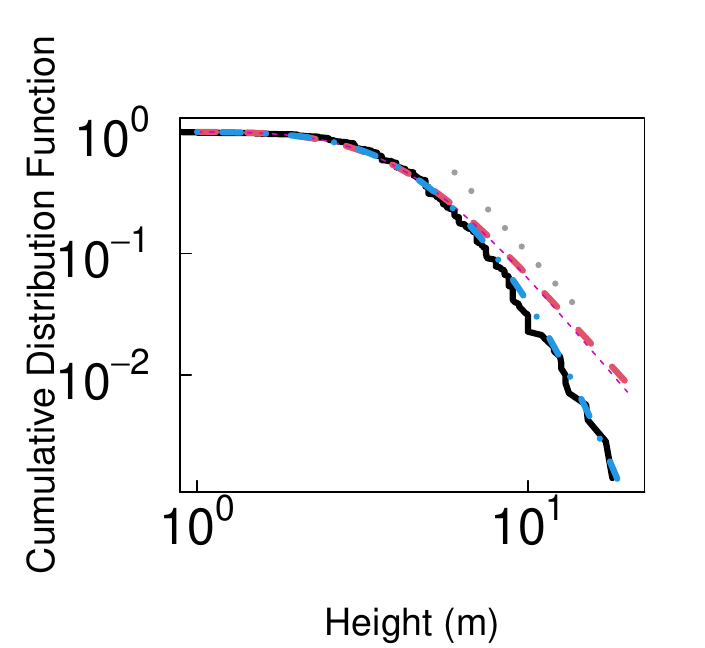}
         \put(60,142){\color{black}\Large\bfseries (a)}
         \put(173,134){\color{black}\sffamily slope=$3$}
    \end{overpic}
      \begin{overpic}[width=8.0cm]{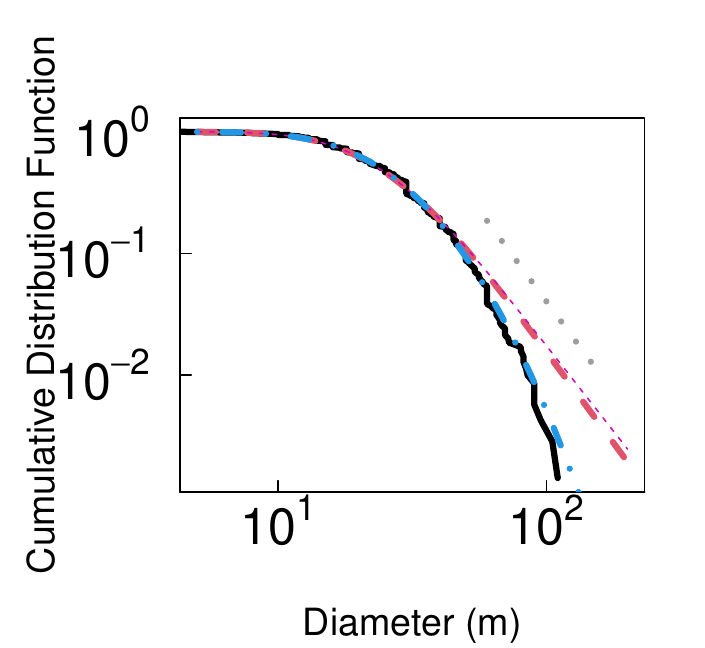}
        \put(60,142){\color{black}\Large\bfseries (b)}
        \put(173,134){\color{black}\sffamily slope=$3$}
    \end{overpic} 
        \begin{overpic}[width=8.0cm]{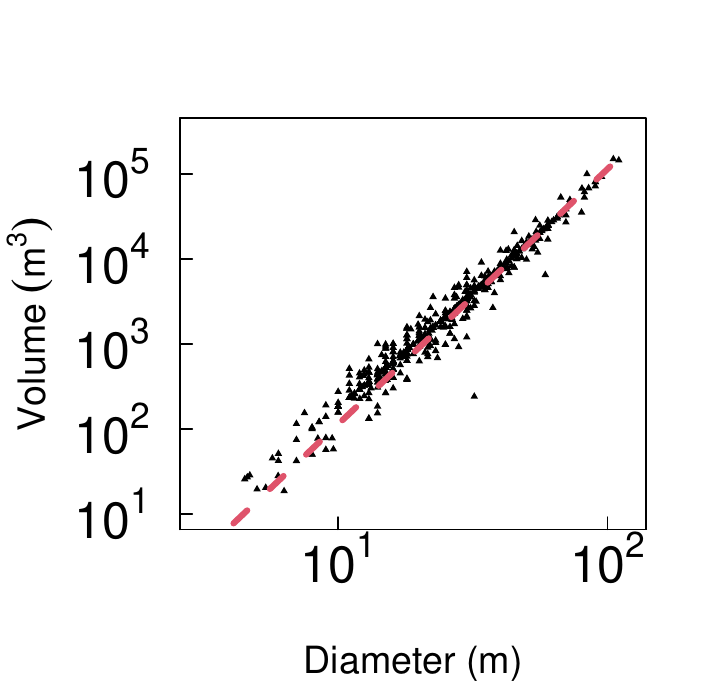}
        \put(65,162){\color{black}\Large\bfseries (c)}
    \end{overpic}
        \begin{overpic}[width=8.0cm]{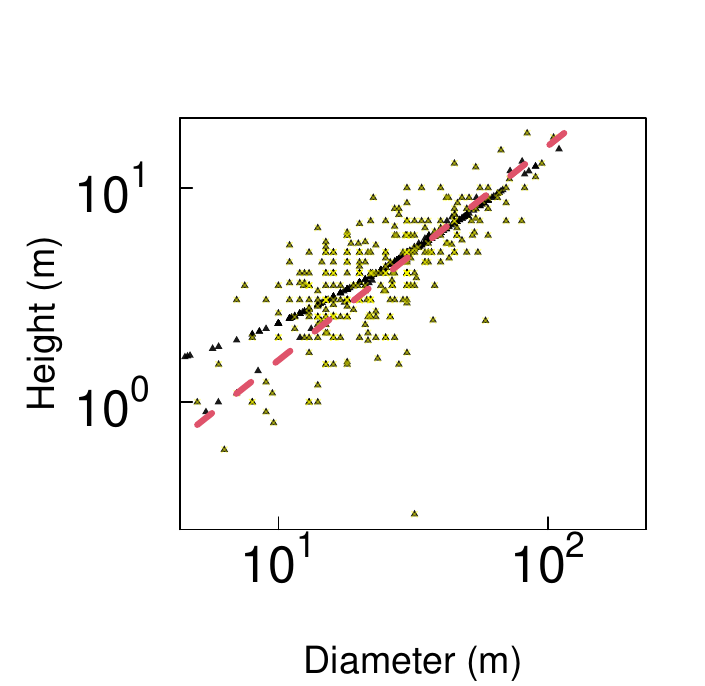}
        \put(65,162){\color{black}\Large\bfseries (d)}
    \end{overpic}
\end{center}
\caption{
    Shape statistics of round burial mounds. Panels show (a) the upper cumulative distribution of height, (b) the upper cumulative distribution of diameter, (c) the relationship between diameter and volume, and (d) the relationship between diameter and height. The black lines and black points represent the imputed data, whereas the yellow points represent the values observed before imputation.
In (a) and (b), the blue dotted lines indicate lognormal distributions, the red dashed lines indicate one-sided dPlN distributions (Eq.~\eqref{app_eq_dPlN}) with the upper-tail cumulative exponent fixed at $\alpha=3$, and the thin pink dashed lines indicate one-sided dPlN distributions with $\alpha$ estimated freely. For the height distribution, the lognormal parameters are $\mu=1.41$ and $\sigma=0.51$; the parameters of the red dashed one-sided dPlN distribution are $\mu=1.10$, $\sigma=0.45$, and $\alpha=3.00$; and those of the thin pink dashed distribution are $\mu=1.12$, $\sigma=0.45$, and $\alpha=3.17$. For the diameter distribution, the lognormal parameters are $\mu=3.18$ and $\sigma=0.55$; the parameters of the red dashed one-sided dPlN distribution are $\mu=2.86$, $\sigma=0.48$, and $\alpha=3.00$; and those of the thin pink dashed distribution are $\mu=2.84$, $\sigma=0.48$, and $\alpha=2.82$. The gray dashed lines are reference lines representing a power-law distribution with cumulative exponent 3.
The distributions of both height and diameter are close to lognormal in the central body and approximately follow a power law with cumulative exponent 3 in the upper tail.
In (c), the red dashed line represents the relation $V=0.114D^3$, indicating that volume is proportional to the cube of diameter. In (d), the red dashed line represents the relation $H=0.157D$, indicating that height is proportional to diameter. Thus, for round burial mounds as well, volume is approximately proportional to the cube of the principal linear dimension, namely diameter, while height is nearly proportional to diameter.
}
\label{app_fig_shape_en}
\end{figure*}

Fig.~\ref{app_fig_shape_en}(a) and (b) show the distributions of diameter and height for round mounds. These distributions are well approximated by log-normal distributions. At the same time, their upper tails show power-law-like shapes, as in keyhole-shaped kofun. Figs.~\ref{app_fig_shape_en}(c) and (d) show the relations of volume and height to diameter for round mounds. As in the case of keyhole-shaped kofun, volume is approximately proportional to the cube of diameter, and height is approximately proportional to diameter. These results indicate that, for round mounds as well, volume can be approximated as a cubic-scale measure of the main linear dimension.\par

\subsubsection{Relationship between power-law exponents for mound length and volume}
\label{app_sec_powerlaw}
The relationship between the power-law exponent of mound length $L$ and that of volume $V$
can be explained from the geometric shape of kofun.
Here, the power-law exponent $\alpha$ refers to the exponent in the upper cumulative distribution,
$\Pr(X>x)\propto x^{-\alpha}$.\par

In general, when there is a relationship $Y \propto X^{\gamma}$,
and the upper cumulative distribution of $X$ follows a power law with exponent $\alpha_X$,
$\Pr(X>x)\propto x^{-\alpha_X}$,
we have
$\Pr(Y>y)=\Pr(X>y^{1/\gamma})\propto y^{-\alpha_X/\gamma}$.
Therefore, the power-law exponent of the upper cumulative distribution of $Y$ is given by
\begin{equation}
  \alpha_Y = \frac{\alpha_X}{\gamma}.
  \label{eq:alpha_transform}
\end{equation}
\par

Because mound height is approximately proportional to mound length,
kofun volume can be approximated as $V \propto L^3$
(the direct relationship between mound length and volume is shown in Figs.~\ref{app_fig_shape_xy}(a) and \ref{app_fig_shape_en}(c)).
Thus, substituting the power-law exponent of mound length, $\alpha_L \approx 3$, gives
\begin{equation}
  \alpha_V = \frac{\alpha_L}{3} \approx 1 .
\end{equation}
This is consistent with the exponent of the tail of the volume distribution observed in the main text and the previous section.

\subsection{Square-fronted and square-backed kofun}
\label{app_sec_square}
\begin{figure*}[t]
\centering
\setlength{\tabcolsep}{2pt}
\renewcommand{\arraystretch}{0.3}
\begin{tabular}{cc}
\begin{overpic}[percent,width=7.0cm]{zenpo_all_loglog.pdf}
  \put(26,65){\PanelLabel{a}}
  \put(28,30){\ShapeLabel{\KeyholeMoundIcon}}
  \put(28,23){Keyhole-shaped mound}
  \put(65,64){\color{black!75}\large\sffamily slope=$-1$}
\end{overpic}
&
\begin{overpic}[percent,width=7.0cm]{zenpo_all_semilog.pdf}
  \put(26,65){\PanelLabel{b}}
  \put(56,70){\large Semi-log plot}
  \put(27,30){\ShapeLabel{\KeyholeMoundIcon}}
  \put(26,23){Keyhole-shaped mound}
  
\end{overpic}
\\
\begin{overpic}[percent,width=7.0cm]{en_all_loglog.pdf}
  \put(26,65){\PanelLabel{c}}
  \put(28,30){\ShapeLabel{\CircularMoundIcon}}
  \put(28,23){Circular mound}
  \put(68,64){\color{black!75}\large\sffamily slope=$-1$}
\end{overpic}
&
\begin{overpic}[percent,width=7.0cm]{en_all_semilog.pdf}
  \put(26,65){\PanelLabel{d}}
  \put(56,70){\large Semi-log plot}
  \put(27,30){\ShapeLabel{\CircularMoundIcon}}
   \put(26,23){Circular mound}
\end{overpic}
\\
\begin{overpic}[percent,width=7.0cm]{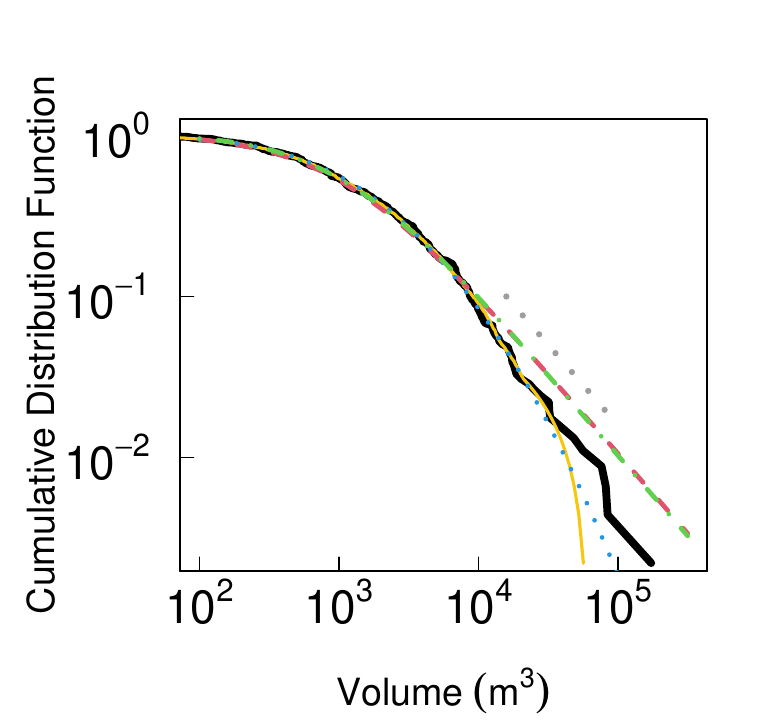}
  \put(26,65){\PanelLabel{e}}
  \put(28,30){\ShapeLabel{\SquareFrontedMoundIcon}}
  \put(28,23){Square-fronted mound}
   \put(68,54){\color{black!75}\large\sffamily slope=$-1$}
\end{overpic}
&
\begin{overpic}[percent,width=7.0cm]{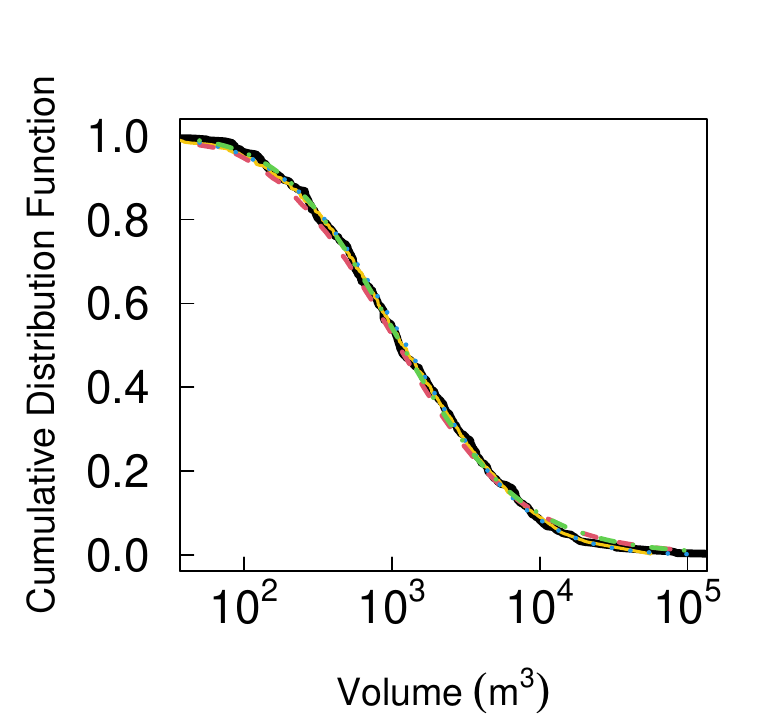}
  \put(26,65){\PanelLabel{f}}
   \put(56,70){\large Semi-log plot}
  \put(27,30){\ShapeLabel{\SquareFrontedMoundIcon}}
  \put(26,23){Square-fronted mound}
\end{overpic}
\end{tabular}
\caption{
Volume distributions for the three major types of kofun.
The black solid lines show the empirical data, the red dashed lines show the theoretical distributions generated by the Kesten process (Eq.~\eqref{eq:kesten}; $A_0$, $b_0$, $\alpha$), the green dash-dotted lines show the one-sided dPlN distributions (Eq.~\eqref{app_eq_dPlN}; $\mu_1$, $\sigma_1$, $\alpha$), the blue dotted lines show the lognormal distributions ($\mu_2$, $\sigma_2$), and the thin yellow solid lines show simulated samples generated from the Kesten process. The gray dashed lines are guide lines with slope $-1$. The left column shows log--log plots, and the right column shows semi-log plots.
(a),(b) Keyhole-shaped kofun: $\alpha=1$, $A_0=225$, $b_0=1.13$, $\mu_1=6.88$, $\sigma_1=1.23$, $\mu_2=7.89$, $\sigma_2=1.62$.
(c),(d) Round kofun: $\alpha=1$, $A_0=263$, $b_0=1.16$, $\mu_1=6.58$, $\sigma_1=1.34$, $\mu_2=7.54$, $\sigma_2=1.55$.
(e),(f) Square-fronted kofun: $\alpha=1$, $A_0=78.7$, $b_0=1.08$, $\mu_1=6.17$, $\sigma_1=1.24$, $\mu_2=7.13$, $\sigma_2=1.51$.
The figure shows that, for all three types, the central part of the distribution is close to a lognormal distribution, whereas the tail exhibits a power-law-like form with an exponent close to 1. In particular, for the keyhole-shaped kofun in (a), the theoretical distribution generated by the Kesten process agrees well with the empirical data in both the central part and the tail. In contrast, for the round kofun in (c), the tail is somewhat rounded and the overall shape is closer to a lognormal distribution. However, the yellow line in (c) shows that, in finite-sample simulations, data generated from the Kesten process can also appear close to a lognormal distribution.
}
\label{app_fig_volume}
\end{figure*}

This section presents the results for square-fronted and square-backed kofun, hereafter referred to simply as the square-fronted type.
We analyzed 455 mounds of this type in the dataset.
A schematic icon of this mound type is shown in Fig.~\ref{app_fig_volume}(e).
\par

Fig.~\ref{app_fig_volume} extends Fig.~\ref{fig_volume} by adding the results for the square-fronted type.
Figs.~\ref{app_fig_volume}(e) and \ref{app_fig_volume}(f) show the size distribution of this type.
In the semi-log plot (Fig.~\ref{app_fig_volume}(f)), the central part of the distribution has a shape close to a log-normal distribution (blue dotted line).
In the log--log plot (Fig.~\ref{app_fig_volume}(e)), the tail shows a power-law-like form, with an exponent not far from 1.
The Clauset method gives an estimated exponent of $\alpha=1.40$ for the square-fronted type.
However, the Kesten-process model with exponent 1 is not rejected by the Anderson--Darling goodness-of-fit test ($p=0.264$; Table~\ref{tab_table1}).
As in the case of round kofun discussed in Sec.~\ref{app_sec_teisi}, this deviation from exponent 1 may reflect the small sample size and the effects of the stopping process.
\par

\subsection{CCDFs for Kinki-type and exceptional groups}
\label{app_sec_kinkicdf}
\begin{figure*}[t]
\centering
\setlength{\tabcolsep}{1pt}
\renewcommand{\arraystretch}{0.1}
\begin{tabular}{ccc}
\begin{overpic}[percent,width=5.5cm]{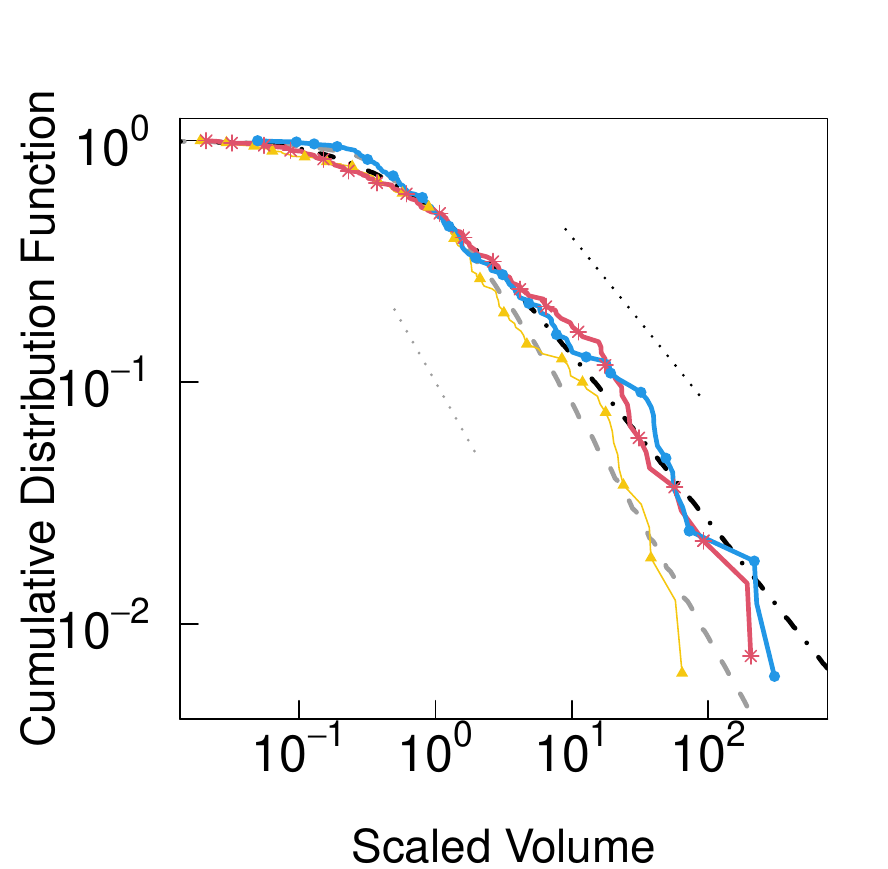}
  \put(26,65){\PanelLabel{a}}
  \put(28,23){Kinki Log-log}
\end{overpic}
&
\begin{overpic}[percent,width=5.5cm]{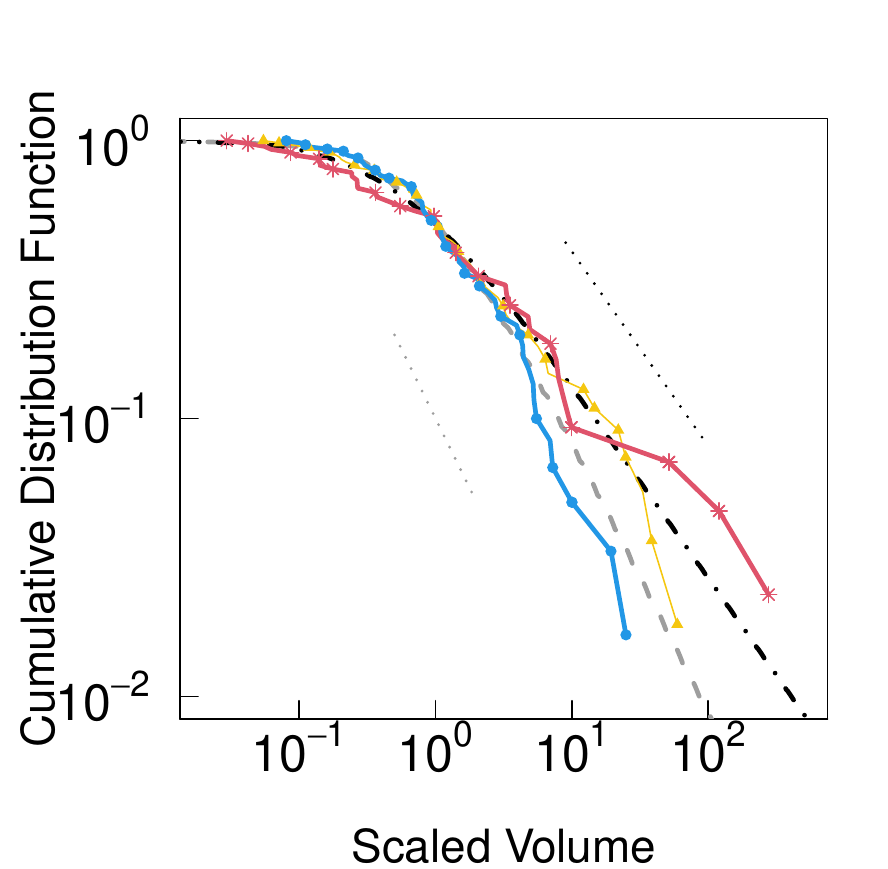}
  \put(26,65){\PanelLabel{c}}
  \put(28,23){Sanyo Log-log}
\end{overpic}
&
\begin{overpic}[percent,width=5.5cm]{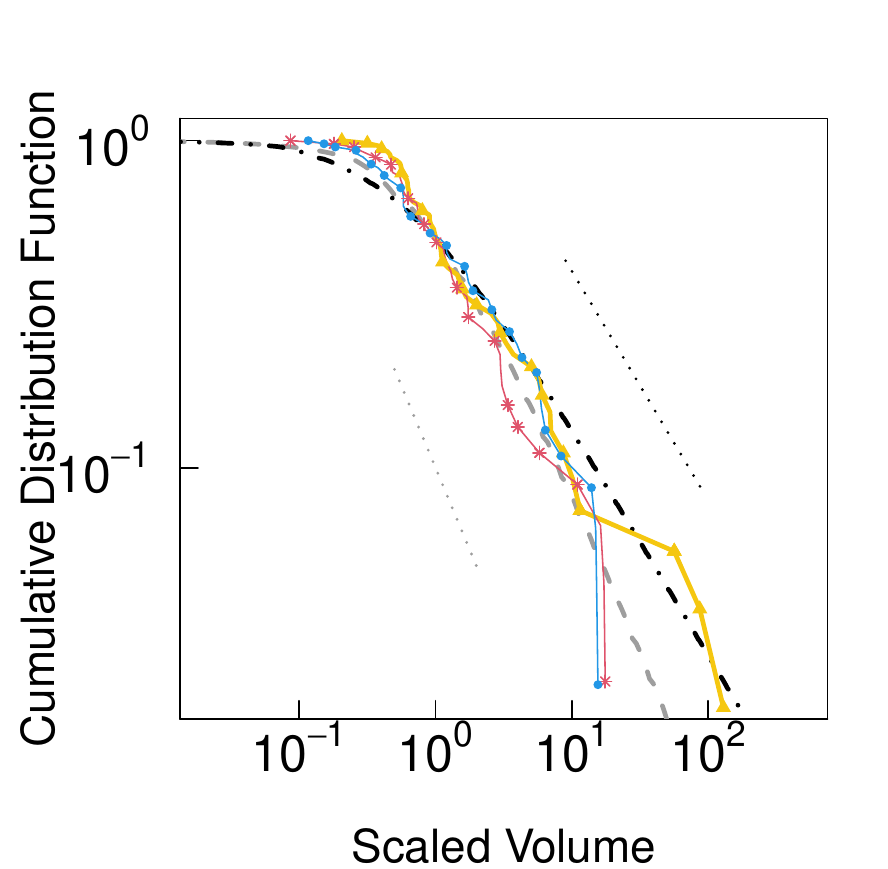}
  \put(26,65){\PanelLabel{e}}
  \put(21,23){Hokuriku Log-log}
\end{overpic}
\\
\begin{overpic}[percent,width=5.5cm]{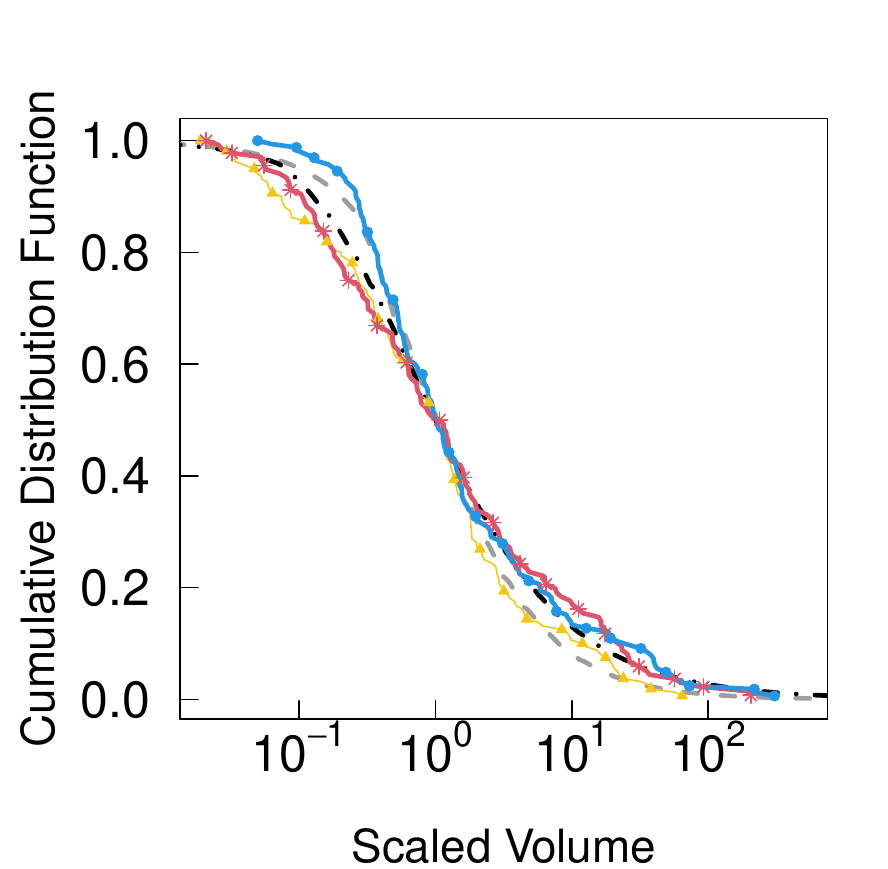}
  \put(26,65){\PanelLabel{b}}
  \put(26,23){Kinki Semi-log}
\end{overpic}
&
\begin{overpic}[percent,width=5.5cm]{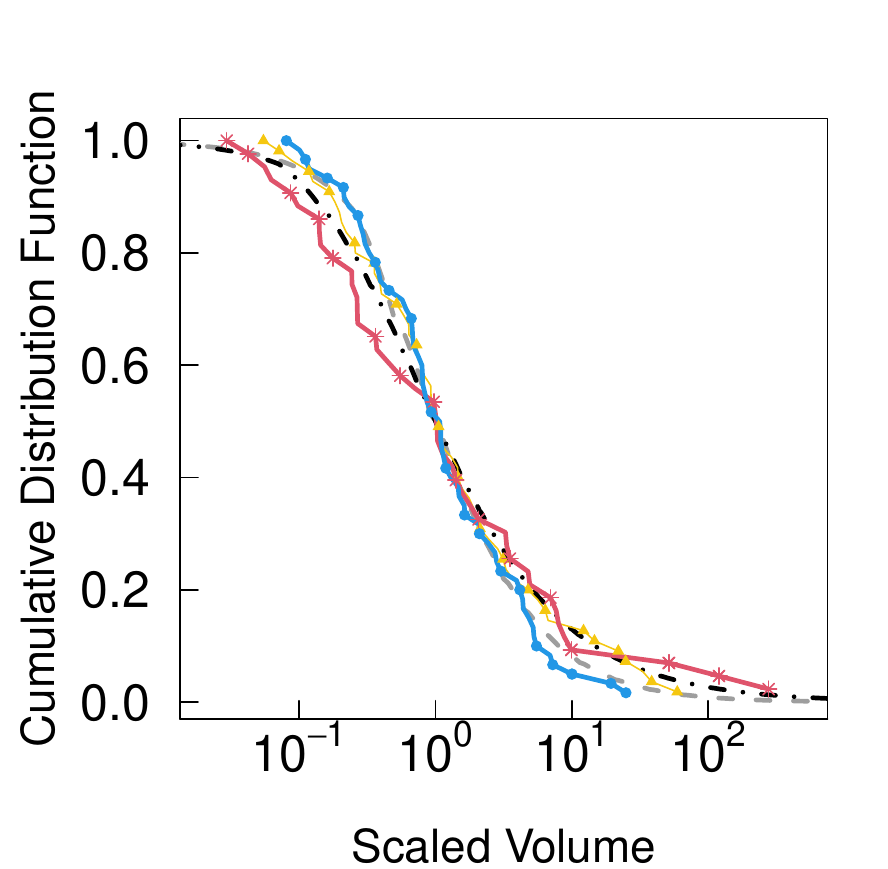}
  \put(26,65){\PanelLabel{d}}
   \put(26,23){Sanyo Log-log}
\end{overpic}
&
\begin{overpic}[percent,width=5.5cm]{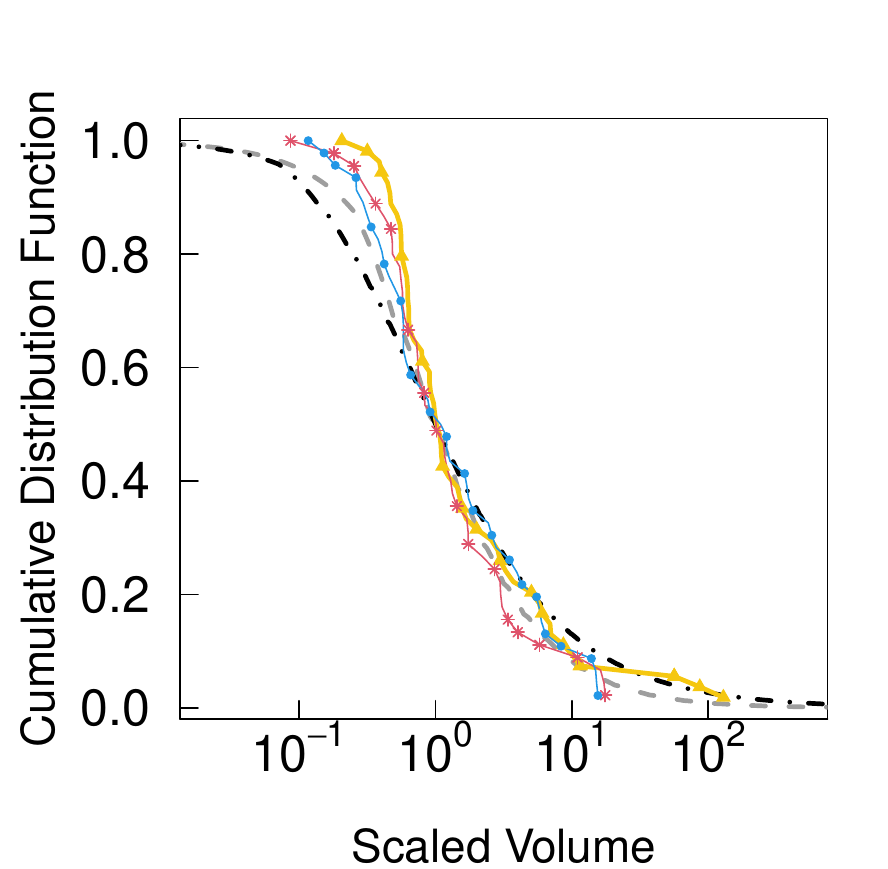}
  \put(26,65){\PanelLabel{f}}
  \put(21,23){Hokuriku Semi-log}
\end{overpic}
\\
&
\begin{overpic}[percent,width=5.5cm]{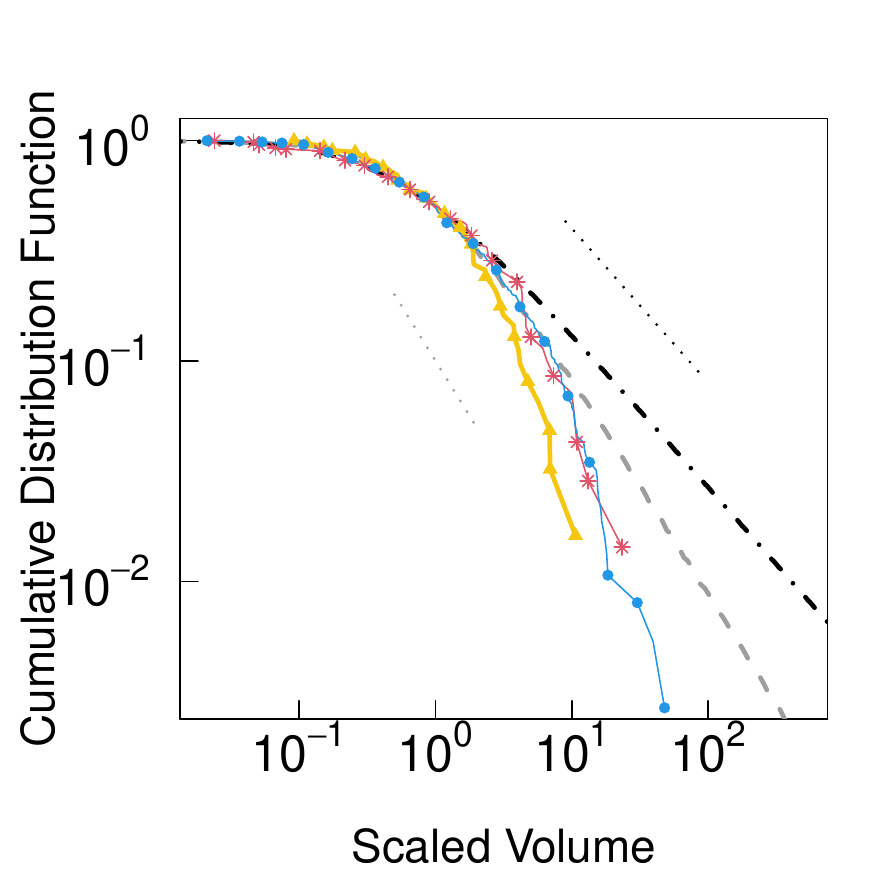}
  \put(26,65){\PanelLabel{g}}
  \put(21,23){Kanto-Koshin Log-log}
\end{overpic}
&
\begin{overpic}[percent,width=5.5cm]{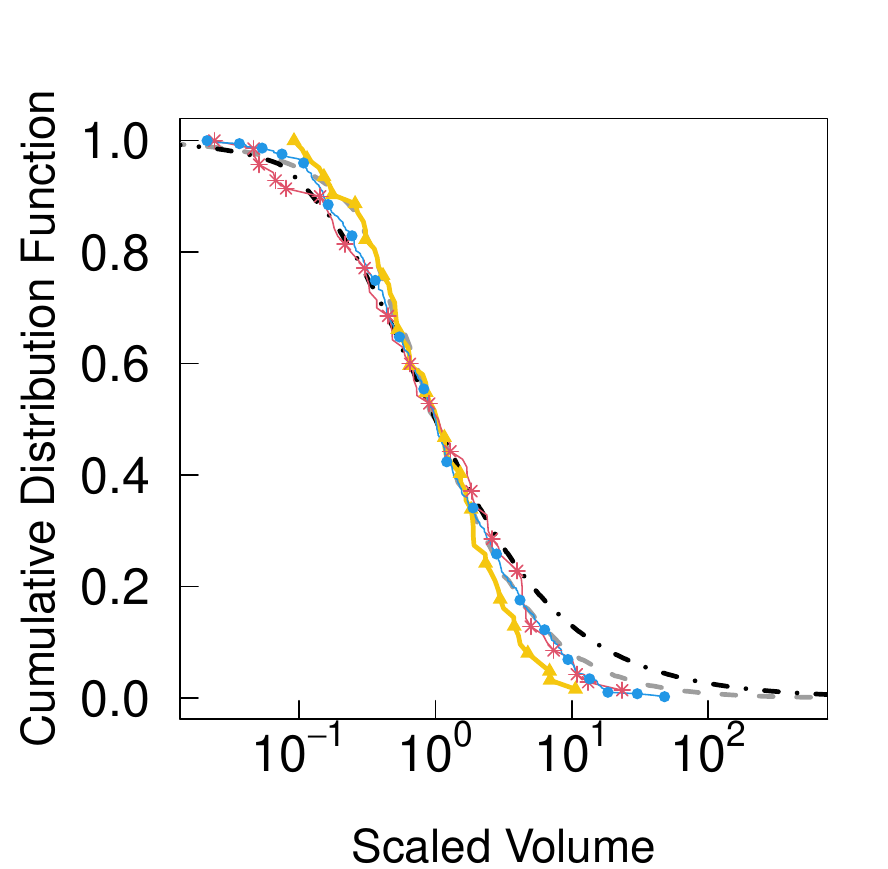}
  \put(26,65){\PanelLabel{h}}
  \put(24,30){Kanto-Koshin}
   \put(24,23){Semi-log} 
\end{overpic}
\end{tabular}
\caption{
Complementary cumulative distributions of kofun volumes by region and century. The left panels show the distributions on log–log axes, and the right panels show the same distributions on semi-logarithmic axes. Yellow triangles, red asterisks, and blue circles indicate kofun constructed in the fourth, fifth, and sixth centuries, respectively. In each group, kofun volumes are normalized by the median. The thick gray dashed line shows the median-normalized theoretical distribution generated by the Kesten process (Eq.~\eqref{eq:kesten}; $\alpha=1.0$, $b_0=1.3$, and $A_0=276$), while the black dash-dotted line shows a median-normalized reference distribution generated with a heavier tail ($\alpha=0.70$, $b_0=1.31$, and $A_0=298$). The gray dotted lines indicate the reference power-law slopes $x^{-0.7}$ and $x^{-1.0}$. 
Panels (a,b), (c,d), (e,f), and (g,h) show the results for Kinki, Sanyo, Hokuriku, and Kanto-Koshin, respectively. The regional classification follows Fig.~\ref{fig_map}. 
Although the limited sample sizes preclude firm conclusions, several suggestive patterns can be observed. In both Kinki and Sanyo, the fifth-century distributions exhibit tails heavier than the Zipf form and close to the $x^{-0.7}$ reference slope. Their fourth-century distributions appear intermediate between the $x^{-1.0}$ and $x^{-0.7}$ slopes. In the sixth century, the Kinki distribution remains close to the $x^{-0.7}$ slope, whereas the Sanyo distribution shifts toward the Zipf-like $x^{-1.0}$ slope. For fourth-century Hokuriku, the apparent heavy tail is strongly influenced by the three largest kofun. By contrast, in Kanto, a typical group, the tail is approximately consistent with the $x^{-1.0}$ slope in all three centuries.
}
\label{app_fig_sanyo}
\end{figure*}

Figure~\ref{app_fig_sanyo} shows the CCDFs of the exceptional groups identified in Fig.~\ref{fig_map}. Specifically, it presents the distributions by century for Kinki, Sanyo, and Hokuriku. For comparison, the figure also includes the CCDFs for Kanto-Koshin, which belongs to the typical group. \par
Although the limited sample sizes preclude firm conclusions, several suggestive patterns can be observed. In both Kinki and Sanyo, the fifth-century distributions have tails heavier than the Zipf form. They are close to the $x^{-0.7}$ reference slope. In the fourth century, the distributions in both regions appear intermediate between the $x^{-1.0}$ and $x^{-0.7}$ slopes. 
In the sixth century, the Kinki distribution remains close to the $x^{-0.7}$ slope. By contrast, the Sanyo distribution shifts toward the Zipf-like $x^{-1.0}$ slope. \par 
For fourth-century Hokuriku, the apparent heavy tail is strongly influenced by the three largest kofun. 
Thus, fourth-century Hokuriku may have been classified as an exceptional group because its limited sample size and a few extreme observations produced an apparently high level of inequality. However, its distribution might be consistent with that of the typical group once those extreme observations are excluded.
In Kanto-Koshin, a typical group, the tail is approximately consistent with the $x^{-1.0}$ slope in all three centuries. \par

\subsection{Detailed regional model-fitting results}
\label{app_sec_table}
More detailed regional model-fitting results are shown in Table~\ref{app_tab_table1} for eastern Japan and Table~\ref{app_tab_table2} for western Japan. These results should be regarded as supplementary, because many of the regional subsets have small sample sizes. The main results, based on the broader classification into eastern Japan, the political center region (Kinki), and western Japan, are given in Table~\ref{tab_table1}.
\subsection{Interpreting the contrasting temporal changes in keyhole-shaped and round kofun through the decomposition of $QA_0$}\label{app_sec_QA}
For keyhole-shaped kofun, the distribution shifts to the right, that is, toward larger volumes, in many regions from the fourth to the fifth century
(the increase in the median values shown in Table~\ref{tab_table1} corresponds to this shift; Fig.~\ref{fig_area_cdf}(d)).
From the viewpoint of the model, this can be described as an increase in the product $QA_0$ given by Eq.~\eqref{eq:QA0J},
that is, the product of the total amount of social resources and the fraction allocated to kofun construction.
By contrast, for round kofun in the same period, the median changes little, suggesting that $QA_0$ remained nearly constant
(Table~\ref{tab_table1}; Fig.~\ref{fig_area_cdf}(d)).\par
If the overall resource scale $A_0$ is assumed to be common to keyhole-shaped and round kofun,
then the near constancy of $QA_0$ for round kofun suggests that it may be possible, in principle,
to separate the effects of $A_0$ and $Q$ on the fifth-century enlargement of keyhole-shaped kofun.
Specifically, if the allocation fraction $Q$ for round kofun is also assumed to have remained roughly
constant, the constancy of $QA_0$ implies that $A_0$ itself did not increase substantially over this
period. Under this assumption, the rightward shift of keyhole-shaped kofun in the fifth century, that
is, their increase in size, cannot be attributed to a rise in $A_0$, or economic growth, and is instead
attributable to an increase in the allocation fraction $Q$ specific to keyhole-shaped kofun.
In other words, the result may reflect an increase in the share of costs devoted to keyhole-shaped
kofun as representations of social authority. 
This interpretation is consistent with the fact that the fifth century was the peak period of
keyhole-shaped kofun culture, when many extremely large kofun were constructed. \par
However, because what is observed is only the product $QA_0$, this interpretation relies on the
assumption that the size of round kofun reflected the resource scale $A_0$ to some extent.
If the size of round kofun was instead kept nearly constant independently of economic scale, for
example by ritual norms, then the fifth-century enlargement of keyhole-shaped kofun cannot be
separated into the effects of a higher allocation fraction $Q$ and a larger resource scale $A_0$.\par

The leftward shift, that is, the shift toward smaller volumes, shared by both keyhole-shaped and
round kofun in the sixth century is consistent with a possible decline in the relative investment
fraction $Q$ for all kofun.
Such a decline may have occurred against the background of changes in mortuary rituals, the
acceptance of Buddhism, the institutionalization of status order, and the progress of political
integration under the Yamato polity, the emerging central authority of the Kofun period.
These changes may have reduced the authoritative and religious importance of kofun.\par

\begin{table*}[htb]
\centering
\caption{
Detailed Statistics for volume distributions of keyhole-shaped kofun in eastern Japan, divided by subregion and period. The main table is Table~\ref{tab_table1}, and the table for western Japan is Table~\ref{app_tab_table2}. 
}

\begin{tabular}{llrrrrrrlrrrr}
\hline
Region
& Century
& $n$
& $\alpha$
& 95\% CI 
& logMedi.
& logIQR
& AIC
& Gini
& Ln p-val.
& Kes(1) p-val.
& Kes($\alpha$) p-val. \\
\hline
Tohoku & All & 106 & 0.876 & [0.466, 1.747] & 7.457 & 2.413 & Kes(1) & 0.749 & 0.330 & 0.564 & 0.656 \\
Tohoku & 4   &  22 & 0.455 & [0.325, 3.856] & 8.826 & 2.913 & Ln, Dp(1), Kes(1;$\alpha$) & 0.659 & 0.906 & 0.626 & 0.715 \\
Tohoku & 5   &  16 & 1.142 & [0.453, 2.764] & 8.197 & 2.066 & Ln, Kes($\alpha$), Kes(1) & 0.567 & 0.979 & 0.947 & 0.887 \\
Tohoku & 6   &  25 & 0.692 & [0.545, 2.658] & 7.071 & 1.595 & Kes(1) & 0.587 & 0.951 & 0.868 & 0.586 \\

Kanto-Koshin & All & 1731 & 1.166 & [0.709, 1.996] & 7.566 & 1.945 & Kes($\alpha$), Dp(1 or $\alpha$) & 0.730 & 0.000484 & 0.00434 & 0.00874 \\
Kanto-Koshin & 4   &   62 & 1.502 & [0.684, 3.861] & 9.531 & 1.656 & Ln & 0.543 & 0.985 & 0.421 & 0.0595 \\
Kanto-Koshin & 5   &   70 & 0.647 & [0.469, 3.861] & 9.532 & 2.246 & Ln & 0.646 & 0.942 & 0.879 & 0.385 \\
Kanto-koshin & 6   &  375 & 0.900 & [0.696, 2.853] & 8.451 & 2.078 & Ln & 0.689 & 0.497 & 0.0774 & 0.146 \\

Tokai & All & 307 & 1.217 & [0.730, 1.729] & 7.995 & 2.055 & Kes(1) & 0.716 & 0.974 & 0.576 & 0.624 \\
Tokai & 4   &  59 & 0.832 & [0.633, 2.411] & 9.220 & 1.747 & Ln & 0.550 & 0.974 & 0.254 & 0.138 \\
Tokai & 5   &  52 & 1.058 & [0.513, 2.391] & 9.069 & 2.288 & Ln, Dp(1), Kes(1) & 0.690 & 0.464 & 0.366 & 0.474 \\
Tokai & 6   & 100 & 0.889 & [0.722, 3.940] & 8.133 & 1.515 & Ln, Dp(1), Kes(1) & 0.690 & 0.926 & 0.166 & 0.0942 \\

Hokuriku & All & 162 & 0.747 & [0.627, 1.409] & 7.633 & 1.626 & Dp(1) & 0.747 & 0.149 & 0.0700 & 0.00691 \\
Hokuriku & 4   &  54 & 0.684 & [0.547, 1.153] & 7.529 & 1.563 & Dp($\alpha$), Dp(1) & 0.817 & 0.0556 & 0.0819 & 0.0356 \\
Hokuriku & 5   &  45 & 0.865 & [0.654, 1.632] & 8.045 & 1.376 & Dp(1) & 0.644 & 0.566 & 0.178 & 0.134 \\
Hokuriku & 6   &  46 & 0.649 & [0.489, 1.628] & 7.615 & 1.956 & Kes(1) & 0.641 & 0.656 & 0.281 & 0.193 \\
\hline
\end{tabular}

\begin{flushleft}
\footnotesize
Notes:
The 95\% CI reports the bootstrap interval of $\alpha$.
logMedi. and logIQR denote the median and interquartile range of log mound size.
AIC indicates the model selected by AIC:
Ln = lognormal; Dp($\alpha$) = dPlN with the estimated exponent;
Dp(1) = dPlN with $\alpha=1$;
Kes($\alpha$) = Kesten process with the estimated exponent;
Kes(1) = Kesten process with $\alpha=1$.
The last three columns give Anderson--Darling goodness-of-fit p-values.
\end{flushleft}
\label{app_tab_table1}
\end{table*}

\begin{table*}[htb]
\centering
\scriptsize
\caption{
Detailed Statistics for volume distributions of keyhole-shaped kofun in western Japan, divided by subregion and period. The main table is Table~\ref{tab_table1}, and the table for eastern Japan is Table~\ref{app_tab_table1}.
}
\begin{tabular}{llrrrrrrlrrrr}
\hline
Region
& Century
& $n$
& $\alpha$
& 95\% CI
& logMedi.
& logIQR
& AIC
& Gini
& Ln p-val.
& Kes(1) p-val.
& Kes($\alpha$) p-val. \\
\hline
Kinki & All & 794 & 0.701 & [0.453, 1.253] & 8.319 & 2.683 & Kes($\alpha$) & 0.878 & $7.10{\times}10^{-5}$ & $1.44{\times}10^{-5}$ & 0.0197 \\
Kinki & 4   & 160 & 0.803 & [0.625, 1.065] & 9.879 & 2.102 & Ln, Dp(1), Kes($\alpha$)  & 0.773 & 0.674 & 0.497  & 0.640 \\
Kinki & 5   & 136 & 0.562 & [0.420, 1.901] & 9.940 & 2.814 & Ln, Dp(1), Kes($\alpha$) & 0.836 & 0.586 & 0.0670 & 0.866 \\
Kinki & 6   & 165 & 0.571 & [0.468, 0.745] & 8.054 & 2.150 & Dp($\alpha$) & 0.855 & 0.0144 & 0.0874 & 0.0147 \\

Chugoku & All & 292 & 0.600 & [0.513, 0.932] & 7.248 & 2.016 & Dp($\alpha$) & 0.911 & 0.00250 & 0.113 & 0.000454 \\
Chugoku & 4   &  55 & 0.668 & [0.488, 1.139] & 8.305 & 2.081 & Kes($\alpha$), Kes(1) & 0.769 & 0.850 & 0.968 & 0.864 \\
Chugoku & 5   &  43 & 0.527 & [0.340, 1.355] & 9.038 & 2.643 & Kes($\alpha$) & 0.891 & 0.877 & 0.445 & 0.985 \\
Chugoku & 6   &  60 & 0.881 & [0.607, 3.127] & 7.468 & 1.951 & Ln, Dp(1), Kes(1) & 0.645 & 0.979 & 0.495 & 0.656 \\

Sanin & All & 361 & 1.026 & [0.852, 1.308] & 6.707 & 1.733 & Kes($\alpha$), Kes(1) & 0.705 & 0.596 & 0.153 & 0.293 \\
Sanin & 4   &   9 & 0.635 & [0.449, 2.960] & 8.695 & 2.251 & Ln, Dp(1) & 0.535 & 0.941 & 0.914 & 0.750 \\
Sanin & 5   &  10 & 0.757 & [0.528, 3.290] & 9.211 & 1.653 & Ln, Dp(1) & 0.568 & 0.993 & 0.899 & 0.772 \\
Sanin & 6   &  40 & 0.679 & [0.448, 6.979] & 7.488 & 2.184 & Ln, Dp(1) & 0.594 & 0.686 & 0.459 & 0.381 \\

Shikoku & All & 118 & 0.863 & [0.658, 1.306] & 7.135 & 1.586 & Dp(1), Kes(1) & 0.757 & 0.505 & 0.300 & 0.128 \\
Shikoku & 4   &  66 & 0.884 & [0.706, 1.772] & 7.134 & 1.354 & Ln, Dp(1) & 0.664 & 0.626 & 0.268 & 0.119 \\
Shikoku & 5   &  14 & 0.768 & [0.372, 2.884] & 8.274 & 2.183 & Kes($\alpha$), Kes(1) & 0.778 & 0.986 & 0.983 & 0.990 \\
Shikoku & 6   &  15 & 0.796 & [0.534, 6.834] & 7.836 & 1.412 & Dp(1) & 0.714 & 0.724 & 0.542 & 0.356 \\

Kyushu & All & 578 & 1.534 & [0.846, 2.220] & 8.276 & 1.926 & Ln, Kes($\alpha$) & 0.668 & 0.667 & 0.0171 & 0.192 \\
Kyushu & 4   &  96 & 1.265 & [0.781, 2.729] & 8.891 & 1.731 & Ln, Kes($\alpha$) & 0.653 & 0.894 & 0.305 & 0.665 \\
Kyushu & 5   &  80 & 0.876 & [0.664, 4.692] & 8.995 & 1.818 & Ln, Kes(1) & 0.626 & 0.852 & 0.555 & 0.380 \\
Kyushu & 6   & 156 & 1.773 & [0.783, 2.885] & 8.454 & 1.889 & Ln, Kes($\alpha$) & 0.621 & 0.955 & 0.160 & 0.851 \\
\hline
\end{tabular}
\begin{flushleft}
\footnotesize
Notes:
The 95\% CI reports the bootstrap interval of $\alpha$.
logMedi. and logIQR denote the median and interquartile range of log mound size.
AIC indicates the model selected by AIC:
Ln = lognormal; Dp($\alpha$) = dPlN with the estimated exponent;
Dp(1) = dPlN with $\alpha=1$;
Kes($\alpha$) = Kesten process with the estimated exponent;
Kes(1) = Kesten process with $\alpha=1$.
The last three columns give Anderson--Darling goodness-of-fit p-values.
\end{flushleft}
\label{app_tab_table2}
\end{table*}
%
%

\clearpage
%
%
%
\section{Comparison with modern Japanese firms}
\label{app_sec_nowfirm}

In this section, we present sales distributions of modern Japanese firms by industry. These distributions are used as comparative material for interpreting kofun volume distributions. \par
\begin{figure*}[t]
\centering
\begin{overpic}[percent,width=7.3cm]{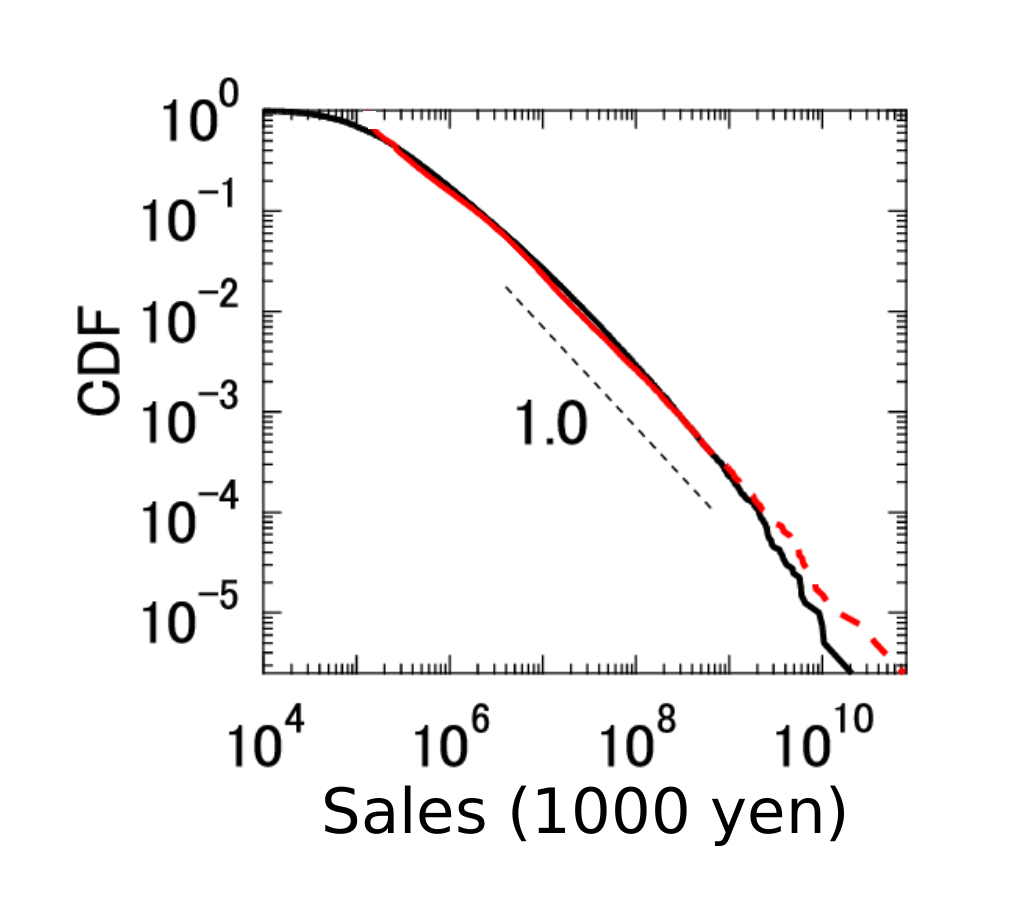}
  \put(29,65){\PanelLabel{a}}
   \put(65,73){Modern}
    \put(65,69){(21st c. CE)}
  \put(30,35){\ShapeLabel{\FirmIcon}}
  \put(31,27){Modern Firm}
\end{overpic}
\begin{overpic}[percent,width=7.0cm]{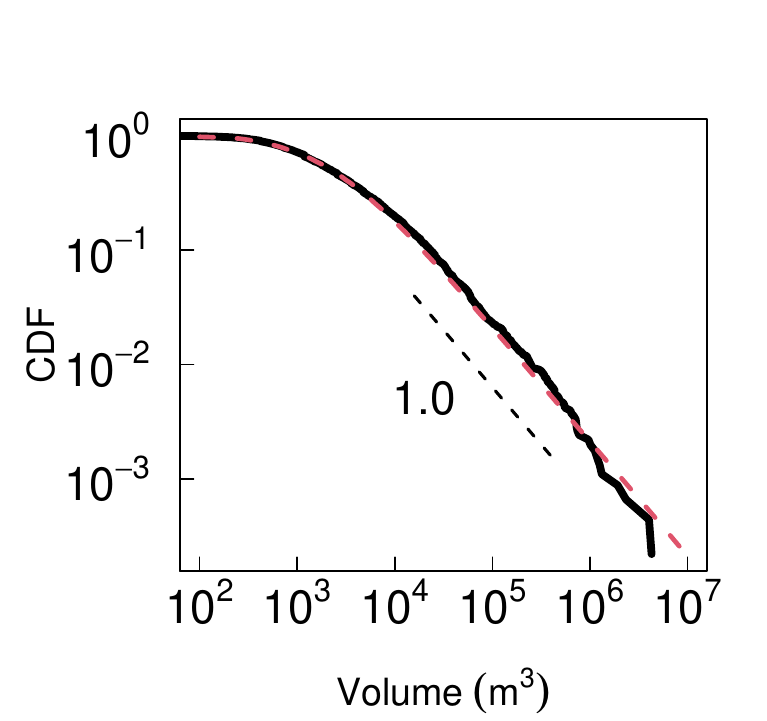}
  \put(26,65){\PanelLabel{b}}
  \put(60,73){Kofun period}
  \put(60,69){(3rd–7th c. CE)}
  \put(28,30){\ShapeLabel{\KeyholeMoundIcon}}
  \put(28,23){Keyhole-shaped mound}
\end{overpic}
\caption{Comparison of the sales distribution of Japanese firms and the volume distribution of kofun.
Panel (a) is adapted from Figure~1 of H.~Watanabe, H.~Takayasu, and M.~Takayasu, ``Biased diffusion on the Japanese inter-firm trading network: estimation of sales from the network structure,'' \textit{New J.\ Phys.} \textbf{14}, 043034 (25 April 2012), \url{https://doi.org/10.1088/1367-2630/14/4/043034}. \copyright\ Deutsche Physikalische Gesellschaft. Reproduced by permission of IOP Publishing. 
 (a) Sales distribution of approximately 700,000 Japanese firms (solid black line: empirical data; red dashed line: theoretical curve).  The red dashed line represents the simulation results from the paper, but it is not particularly relevant to the present study. Accordingly, the left-hand portion has been cropped so that the black solid line (actual data) remains visible.
(b) Volume distribution of keyhole-shaped kofun (solid black line: empirical data; red dashed line: theoretical curve from the Kesten process). Based on the same data as Fig.~\ref{fig_volume}(a) (see that figure for details). In both panels, the thin black dashed guide line indicates a slope of $\propto x^{-1}$. These two quantities—firm sales in the 21st century and kofun volume from the fourth to seventh centuries CE, both related to the scale of mobilized resources—appear visually similar in distributional shape.
}
\label{app_fig_firm_kofun}
\end{figure*}

First, Fig. \ref{app_fig_firm_kofun} compares the sales distribution of modern firms with the volume distribution of keyhole-shaped kofun. The sales distribution of modern Japanese firms is based on sales statistics for approximately 700,000 major Japanese companies compiled by Tokyo Shoko Research \cite{Watanabe2012BiasedDiffusion}.  The same dataset is used in the other figures of present-day firms in this section.
The black solid line in the figure represents the sales distribution of Japanese firms in 2005. This figure shows that the two distributions—corporate sales in the 21st century and kofun volume from the 3rd to 7th centuries, both of which reflect a form of mobilization capacity—closely resemble each other.
\subsection{Power-law exponents and industries}
\label{app_sec_newfirm_cdf}
\begin{figure*}[t]
\begin{center}
    \begin{overpic}[
        width=16cm,
        keepaspectratio,
        trim=40mm 10mm 30mm 100mm,
        clip
    ]{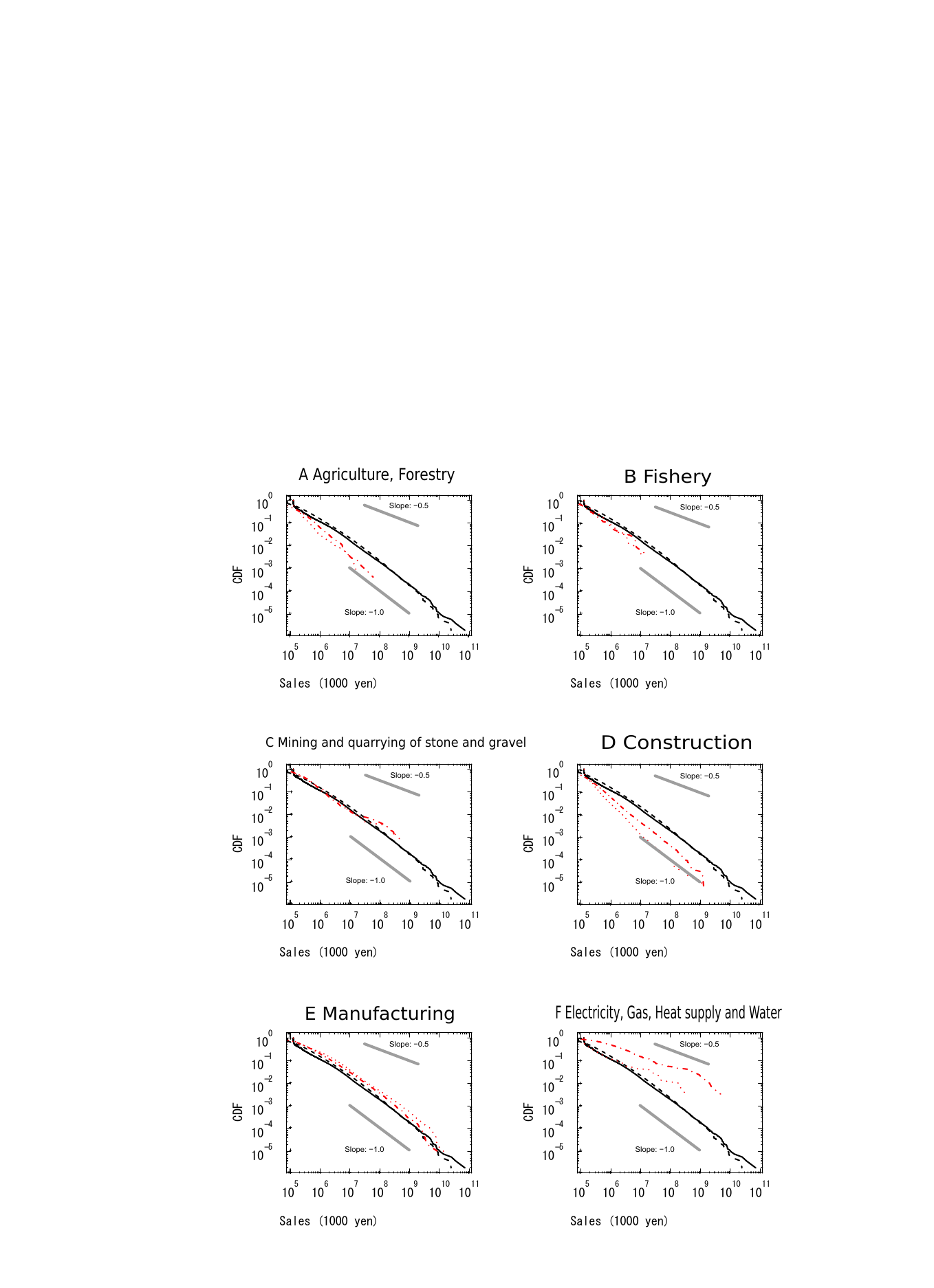}
    \end{overpic}
\end{center}
     \caption{
        Sales distributions of Japanese firms by industry (Part 1 of 3). Each panel shows the upper cumulative distribution of firm sales for a given industry. The industry name is shown at the top of each panel. The black solid line represents the sales distribution for all firms, whereas the red dotted line represents the sales distribution for each industry. The black dashed and red dash-dotted lines are theoretical simulation results reported in previous work \cite{Watanabe2012BiasedDiffusion}; in the present study, we mainly refer to the distributional shapes of the empirical data. The gray reference lines represent power-law distributions proportional to $1/x$ and $1/x^{0.5}$, respectively. In the electricity, gas, heat supply, and water industry, the upper tail is heavier and shows a shape closer to $1/x^{0.5}$. In the other industries, the upper tails are generally close to $1/x$. This result suggests that, in highly regulated or infrastructure-related industries, the concentration of firm sales may become higher than that expected under the standard Zipf law.
}
\label{app_fig_nowfirm_1}
\end{figure*}

\begin{figure*}[t]
    \begin{center}
    \begin{overpic}[
        width=0.93\textwidth,
        height=0.93\textheight,
        keepaspectratio,
        trim=40mm 50mm 50mm 70mm,
        clip
    ]{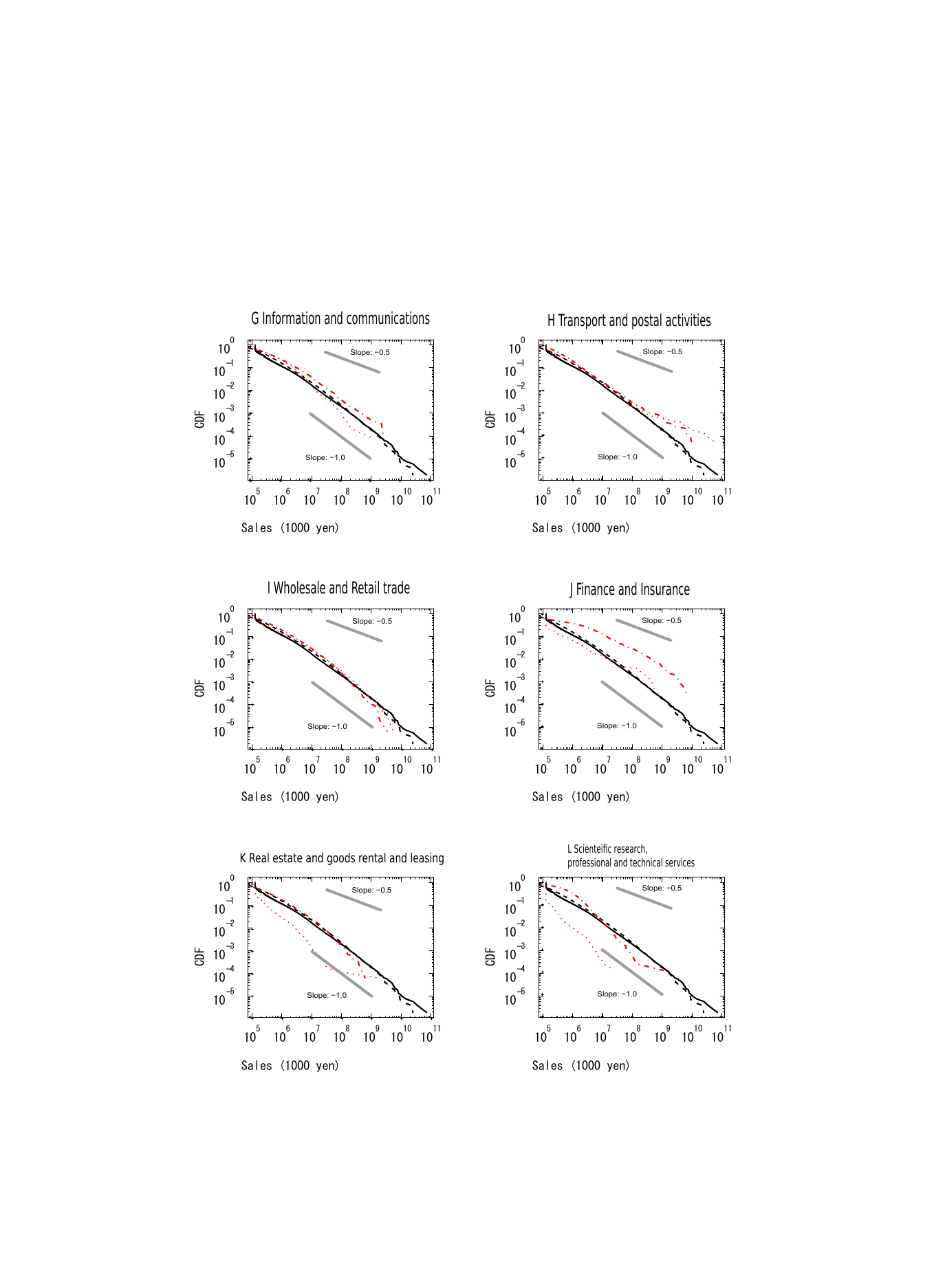}
    \end{overpic}
\end{center}
    \caption{
Sales distributions of Japanese firms by industry (Part 2 of 3). Each panel shows the upper cumulative distribution of firm sales for a given industry. The industry name is shown at the top of each panel. The black solid line represents the sales distribution for all firms, whereas the red dotted line represents the sales distribution for each industry. The black dashed and red dash-dotted lines are theoretical simulation results reported in previous work \cite{Watanabe2012BiasedDiffusion}; in the present study, we mainly refer to the distributional shapes of the empirical data. The gray reference lines represent power-law distributions proportional to $1/x$ and $1/x^{0.5}$, respectively. In the finance and insurance industry, the upper tail is heavier and shows a shape closer to $1/x^{0.5}$. In the other industries, the upper tails are generally close to $1/x$.
}
    \label{app_fig_nowfirm_2}
\end{figure*}

\begin{figure*}[t]
  \begin{center}
    \begin{overpic}[
        width=0.93\textwidth,
        height=0.93\textheight,
        keepaspectratio,
        trim=40mm 50mm 50mm 60mm,
        clip
    ]{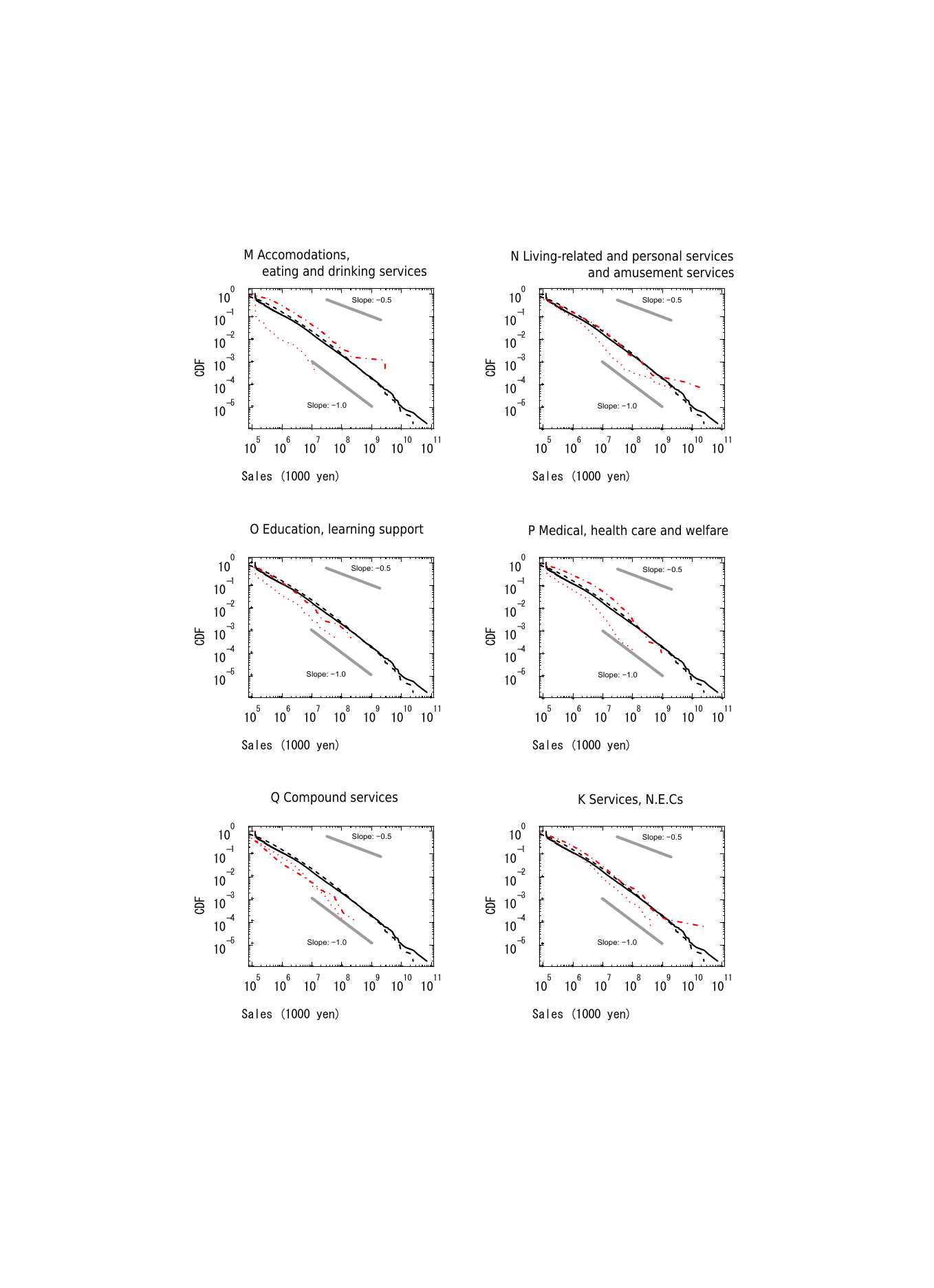}
    \end{overpic} 
\end{center}
    \caption{
        Sales distributions of Japanese firms by industry (Part 3 of 3). Each panel shows the upper cumulative distribution of firm sales for a given industry. The industry name is shown at the top of each panel. The black solid line represents the sales distribution for all firms, whereas the red dotted line represents the sales distribution for each industry. The black dashed and red dash-dotted lines are theoretical simulation results reported in previous work \cite{Watanabe2012BiasedDiffusion}; in the present study, we mainly refer to the distributional shapes of the empirical data. The gray reference lines represent power-law distributions proportional to $1/x$ and $1/x^{0.5}$, respectively. Overall, clear $1/x^{0.5}$-type heavy tails, such as those observed in the electricity, gas, heat supply, and water industry or in the finance and insurance industry, are relatively limited among the industries shown in this figure.
}
\label{app_fig_nowfirm_3}
\end{figure*}

Figs. \ref{app_fig_nowfirm_1}, \ref{app_fig_nowfirm_2} and \ref{app_fig_nowfirm_3} are based on figures in the doctoral thesis of \cite{Watanabe2013DoctoralThesis}. Support lines with power-law exponents $\alpha=1$ and $\alpha=0.5$ for the upper cumulative distribution were added to the original figures. 
In the figure, the black solid line shows the sales distribution of all Japanese firms in 2005. The red dashed line shows the sales distribution of firms belonging to the industry shown in each panel. The other curves show numerical simulation results from the original study. They are not used directly in the present discussion.\par

These figures show that, in many manufacturing, wholesale and retail, and transport industries, the tail of the sales distribution is close to the support line with $\alpha=1$. These distributions are therefore close to Zipf's law. By contrast, in electricity, gas, heat supply, and water, as well as in finance and insurance, the tail is closer to the support line with $\alpha=0.5$ than to that with $\alpha=1$. A similar tendency is also seen in transport, especially among the largest firms.\par

These results suggest that differences in institutional environments across industries can be reflected in the shape of firm sales distributions. In relatively competitive industries, such as manufacturing and wholesale and retail, firm size distributions tend to be close to Zipf's law. By contrast, in industries with stronger entry regulation or institutional constraints, such as electricity, gas, heat supply, water, finance, and insurance, existing firms may be more likely to maintain their size. As a result, the distribution can have an exponent smaller than $\alpha=1$, and hence a heavier tail.\par

This interpretation is consistent with the Kesten-process explanation described in Sec.~\ref{sec_model}. In a competitive environment, existing firms or social groups compete for limited shares. When growth and exit or reorganization are balanced under approximately zero-sum competitive conditions, $\langle b \rangle \sim 1$, a distribution close to Zipf's law can emerge. By contrast, when exit or replacement is less likely relative to growth, existing agents can persist for longer periods. This effect is especially important for agents that have already reached large sizes. As a result, the tail of the distribution can become heavier than Zipf's law and may approach a form with $\alpha=0.5$.\par
This point is analogous to the kofun volume distribution in the Kinki region, which was the political center. Compared with the nationwide distribution, the Kinki distribution deviates from $\alpha=1$ and has a heavier tail. Therefore, industry-specific sales distributions of modern firms provide a comparative example. They show that, in institutionally or politically protected environments, size distributions can systematically deviate from ordinary Zipf's law. \par
\subsubsection{Analogy with kofun}
This analogy provides one bold intuition for understanding the difference between the typical and exceptional groups in the kofun distributions.
In the kofun distributions, the typical group shows Zipf-like tails, and can be interpreted as being close to a quasi-zero-sum competitive process among local groups.
By contrast, the exceptional groups, such as Kinki, the political center, and Sanyo, which includes the semi-political center of Kibi, show smaller exponents and heavier tails.
This may suggest that existing local groups in these regions were placed under more favorable political or institutional conditions than in the quasi-zero-sum baseline. \par

From this perspective, the distributional change in Sanyo, which includes Kibi as a semi-central political region, is particularly suggestive.
Sanyo belongs to the exceptional group in the fourth and fifth centuries, showing a heavier tail, but shifts toward the typical group in the sixth century.
By analogy with modern firm distributions, this change may be understood as a metaphorical analogue of deregulation: a shift away from a ``regulated'' environment favorable to existing firms or groups toward conditions closer to free competition.
In other words, rather than simply reflecting the removal of powerful groups in Kibi, this transition may suggest that political and institutional conditions favorable to existing local groups in Sanyo weakened, and that the distribution moved closer to a more general Zipf-like 
competitive process.\par

A related tradition is the Kibi rebellion, traditionally dated to around the latter half of the fifth century.
The Kibi rebellion refers to a tradition in which powerful groups in Kibi are said to have been suppressed by the central Yamato polity.
The interpretation above can be read as a metaphorical analogue of deregulation.
In this reading, the Yamato polity may have weakened, dismantled, or reorganized special political and institutional conditions that may have favored existing groups in Kibi and Sanyo relative to other regions.
If so, the Kibi rebellion tradition is broadly consistent with the observed change in distributional shape.
However, the historicity of the Kibi rebellion and its specific details are uncertain, and the region-specific data for Sanyo are also limited.
Moreover, because direct evidence for such special political and institutional conditions is lacking, their existence cannot be independently confirmed.
Therefore, this interpretation should be treated as a bold hypothetical analogy.\par


\subsection{Dimensionality of firm-size indicators and power-law exponents in Japanese firms}
\label{a_sec_firm_dimension}
This section supplements the discussion in the main text on the relation between the dimensionality of proxy variables and power-law exponents. For this purpose, we examine the distributions and scaling relations of sales, the number of employees, and the number of trading partners in Japanese firm data. \par 

For modern firms, firm size can be measured by several quantities. These include sales, the number of employees, and degree in a transaction network. All of these variables reflect firm size, but they are not the same quantity. Sales reflect the amount of economic activity allocated and acquired in the market. The number of employees reflects the scale of labor input. The number of trading partners reflects connections in the inter-firm transaction network. Thus, even for the same firms, the observed power-law exponent can change depending on the dimension used to measure size.\par

\begin{figure*}[t]
    \begin{center}
    \begin{overpic}[
        width=7cm
    ]{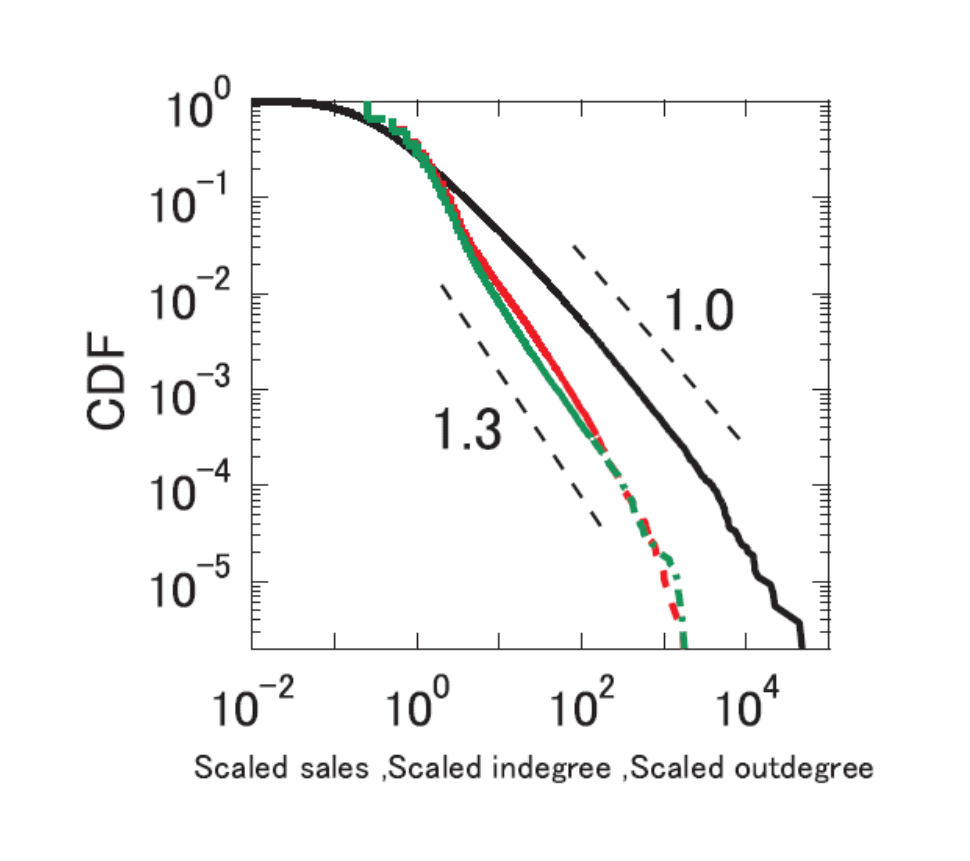}
        \put(60,128){\color{black}\Large\bfseries (a)}
    \end{overpic} 
     \begin{overpic}[
        width=7cm
    ]{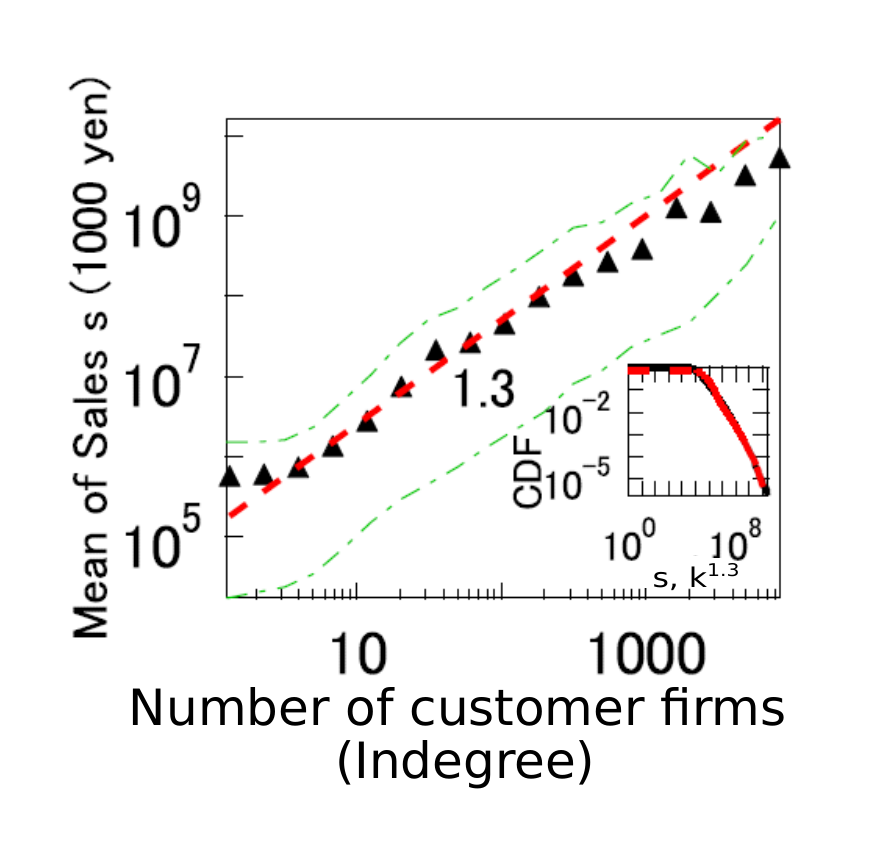}
        \put(60,144){\color{black}\Large\bfseries (b)}
    \end{overpic} 
   \begin{overpic}[
        width=7cm
    ]{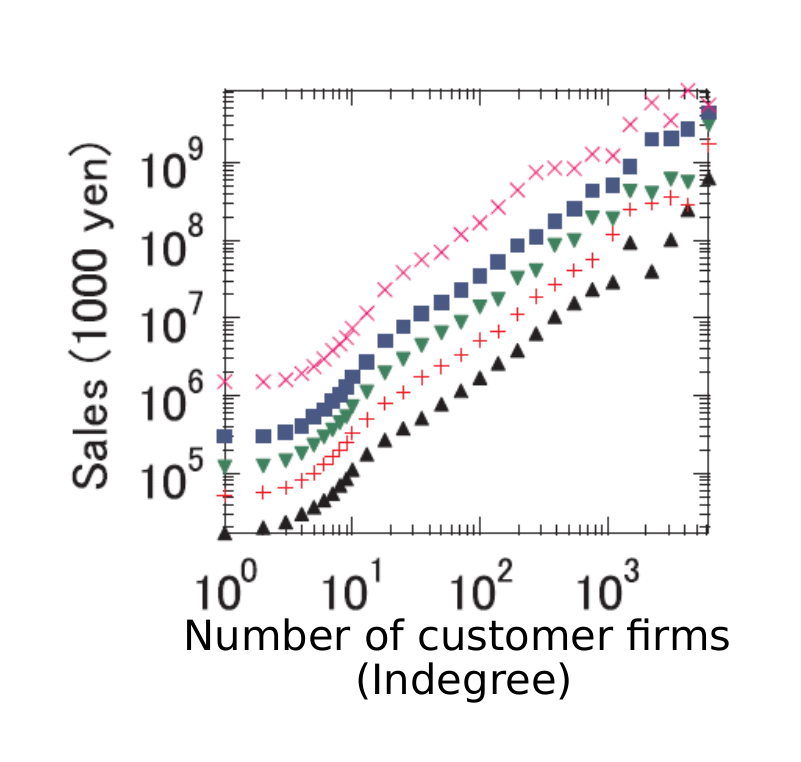}
        \put(60,142){\color{black}\Large\bfseries (c)}
    \end{overpic} 
    \begin{overpic}[
        width=7cm
    ]{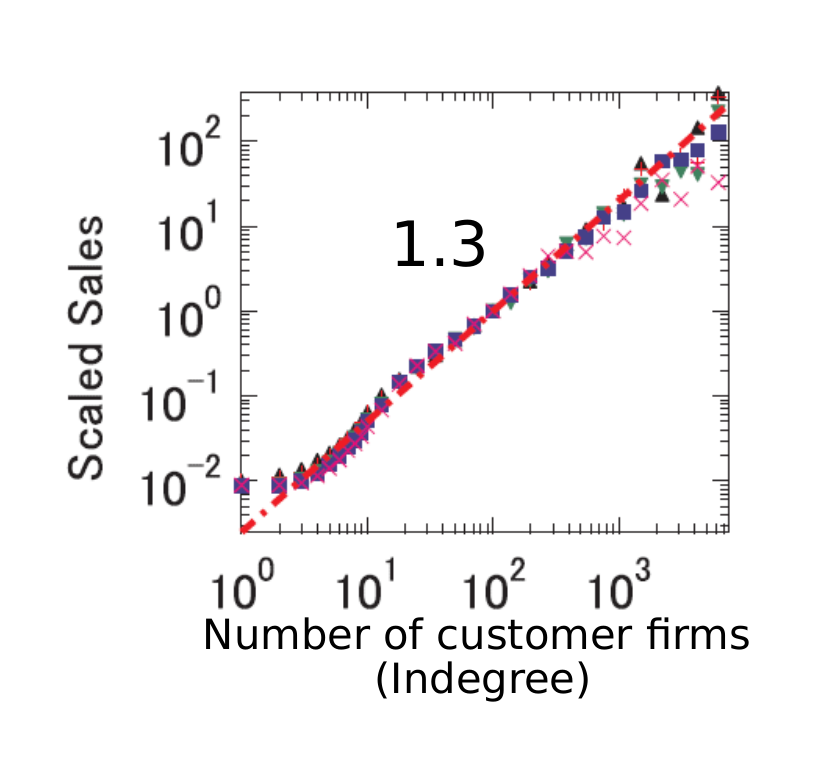}
        \put(63,142){\color{black}\Large\bfseries (d)}
    \end{overpic}
\end{center} 
\caption{
Panels (a) and (b) are adapted from Fig.~1 of H.~Watanabe, H.~Takayasu, and M.~Takayasu, ``Biased diffusion on the Japanese inter-firm trading network: estimation of sales from the network structure,'' \textit{New J.\ Phys.} \textbf{14}, 043034 (25 April 2012), \url{https://doi.org/10.1088/1367-2630/14/4/043034} \cite{Watanabe2012BiasedDiffusion}. \copyright\ Deutsche Physikalische Gesellschaft. Reproduced by permission of IOP Publishing.
Panels (c) and (d) are adapted from Figure~2 of H.~Watanabe, H.~Takayasu, and M.~Takayasu, \textit{Physica A} \textbf{392}(4), 741--756 (2013) \cite{Watanabe2013AllometricScalings}, Copyright (2013), with permission from Elsevier. 
(a) Upper cumulative distributions of sales, in-degree, and out-degree. The black solid line represents sales, the red dashed line represents in-degree, and the green dash-dotted line represents out-degree. Here, in-degree and out-degree are degrees in an inter-firm transaction network defined according to the direction of fund flows: in-degree corresponds to the number of customer firms, whereas out-degree corresponds to the number of supplier firms. Each distribution is normalized by its interquartile range. The upper cumulative exponent of the sales distribution is approximately 1.0, while those of the in-degree and out-degree distributions are approximately 1.3, indicating that the distributional exponents differ between sales and transaction-network degree.
(b) Mean sales conditional on in-degree $k_{\rm in}$. Black triangles represent the conditional mean, the red dotted line is a reference line given by $s=10^{5.1}k_{\rm in}^{1.3}$, and the green dash-dotted lines represent the conditional 5th and 95th percentiles. The conditional mean is approximately proportional to the 1.3 power of in-degree. Thus, sales are not simply proportional to the number of customer firms; rather, firms with more customer firms tend to have larger sales per customer. The inset compares the sales distribution with the distribution obtained by transforming in-degree as $10^{5.1}k_{\rm in}^{1.3}$. The close overlap between the two distributions indicates that the scaling relationship between sales and in-degree is consistent with the transformation of the distributional exponent.
(c) Conditional percentiles of the sales distribution given in-degree. Black triangles, red crosses, green downward triangles, blue squares, and purple crosses represent the 5th, 25th, 50th, 75th, and 95th percentiles, respectively.
(d) Conditional percentiles in (c), normalized so that the value at $k_{\rm in}=100$ is equal to 1. The black dash-dotted line represents a reference line proportional to $k_{\rm in}^{1.3}$. The fact that each percentile increases with approximately the same slope indicates that the relationship between sales and in-degree appears not only in the mean but also as a scaling relationship over the entire conditional distribution.
}
\label{a_fig_firm_indegree}
\end{figure*}

Fig.~\ref{a_fig_firm_indegree} shows the relation between indegree in the transaction network and sales. Here, indegree represents the number of customer firms in an inter-firm transaction network defined by the direction of money flow. The upper cumulative exponent of the sales distribution is about $1.0$ in panel (a). By contrast, the upper cumulative exponents of both indegree and outdegree are about $1.3$. At the same time, the average sales conditional on indegree $k_{\mathrm{in}}$ in panel (b) approximately follow
\begin{equation}
s \propto k_{\mathrm{in}}^{1.3}.
\end{equation}
This relation indicates that sales are not simply proportional to the number of customer firms. Firms with more customers also tend to have larger sales per customer firm. This nonlinearity is thought to arise from factors such as inter-firm transaction networks and differences in capital intensity \cite{Watanabe2012BiasedDiffusion,Watanabe2014MeanField}.   \par

This $1.3$-power scaling is consistent with the difference in distributional exponents. As described by Eq.~\eqref{eq:alpha_transform}, if $Y\propto X^\gamma$, the power-law exponent of the upper cumulative distribution is transformed as $\alpha_Y=\alpha_X/\gamma$. Therefore, if the upper cumulative exponent of indegree is $\alpha_k\simeq 1.3$ and sales follow $s\propto k_{\mathrm{in}}^{1.3}$, the exponent of the sales distribution becomes
\begin{equation}
\alpha_s \simeq \frac{1.3}{1.3} \simeq 1.0 .
\end{equation}
As shown in the inset of Fig.~\ref{a_fig_firm_indegree}(b), the distribution obtained by transforming indegree into $10^{5.1}k_{\mathrm{in}}^{1.3}$ overlaps well with the actual sales distribution. This shows that the Zipf-like exponent of the sales distribution can be understood through nonlinear scaling with the degree of the transaction network.  \par

\begin{figure*}[t]
    \begin{center}
    \begin{overpic}[
        width=5.8cm
    ]{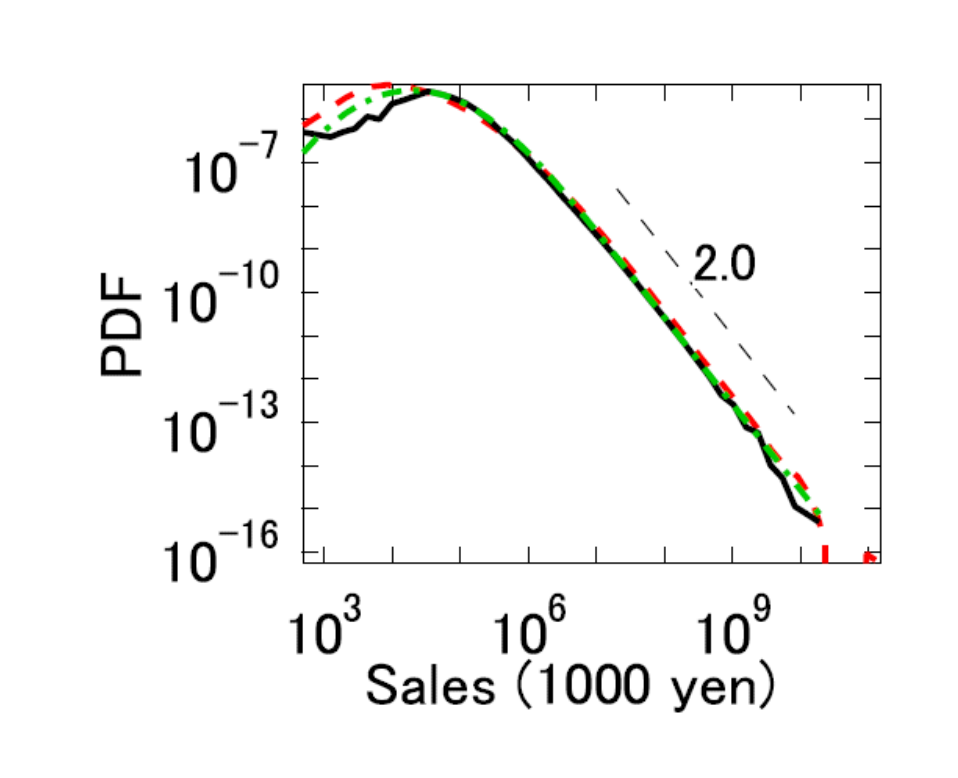}
        \put(55,90){\color{black}\Large\bfseries (a)}
    \end{overpic} 
   \begin{overpic}[
        width=5.8cm
    ]{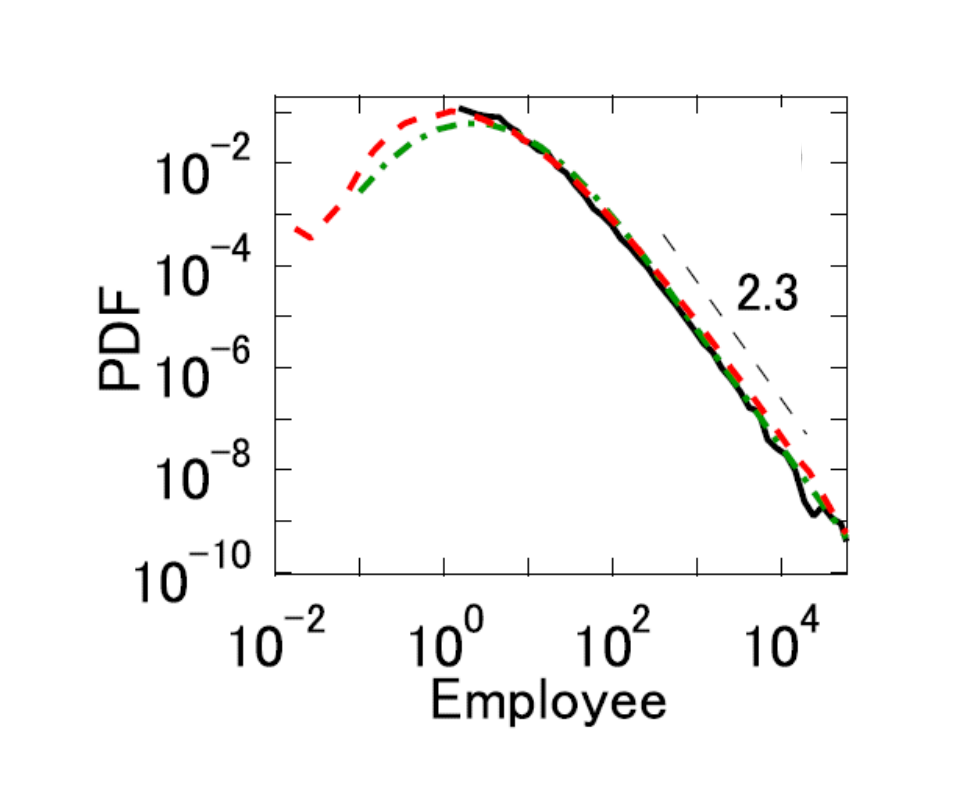}
        \put(122,105){\color{black}\Large\bfseries (b)}
    \end{overpic} 
        \begin{overpic}[
        width=5.8cm
    ]{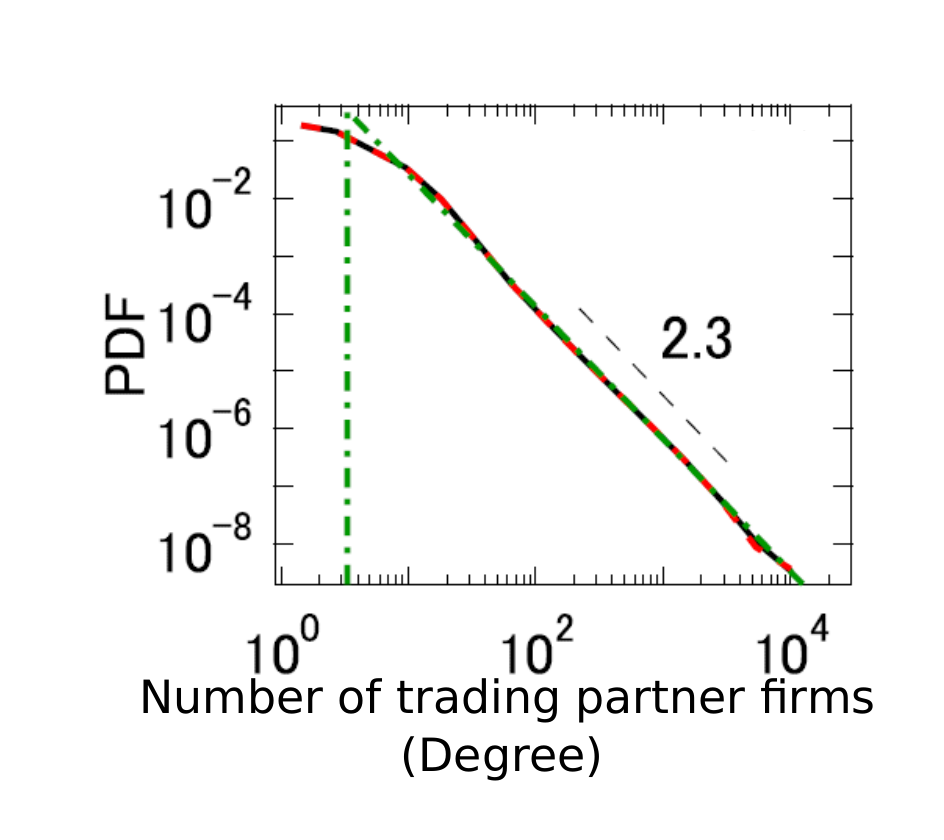}
        \put(122,105){\color{black}\Large\bfseries (c)}
    \end{overpic} 
     \begin{overpic}[
        width=5.8cm
    ]{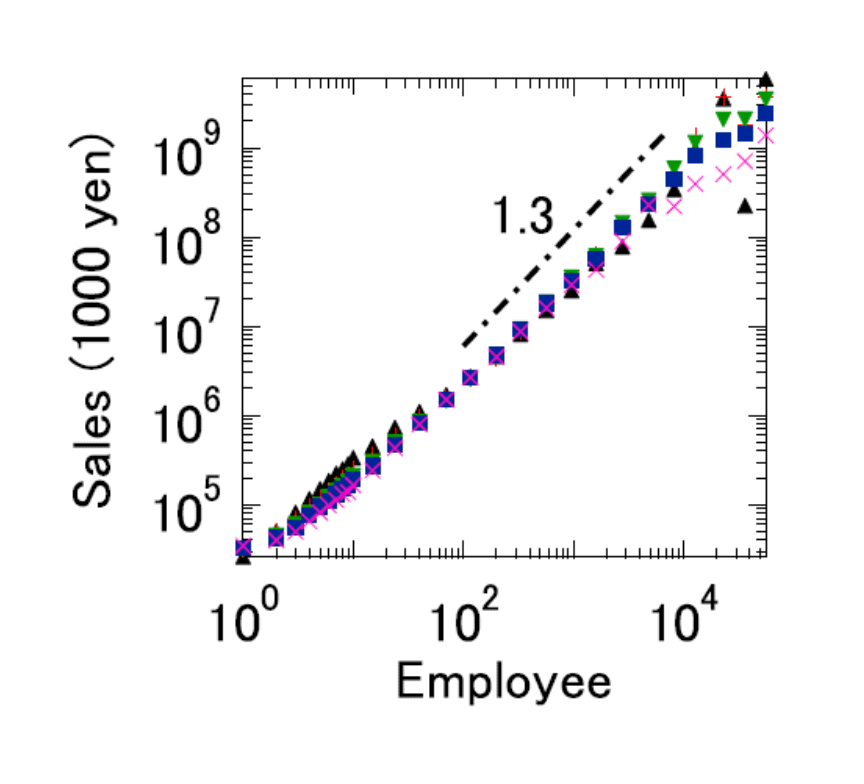}
        \put(50,118){\color{black}\Large\bfseries (d)}
    \end{overpic} 
    \begin{overpic}[
        width=5.8cm
    ]{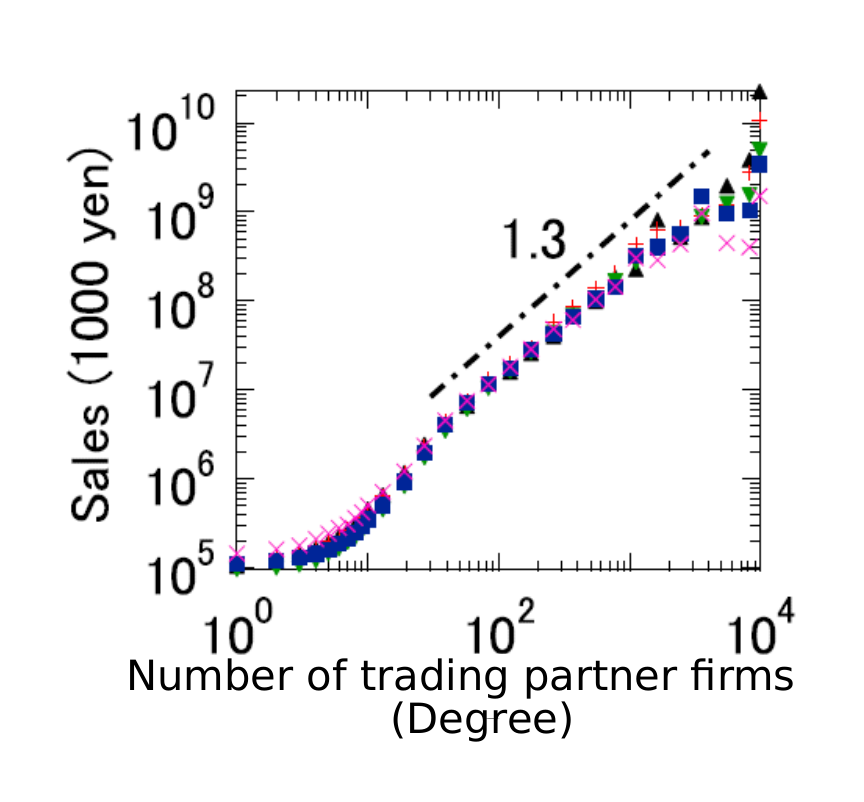}
        \put(50,118){\color{black}\Large\bfseries (e)}
    \end{overpic} 
    \begin{overpic}[
        width=5.8cm
    ]{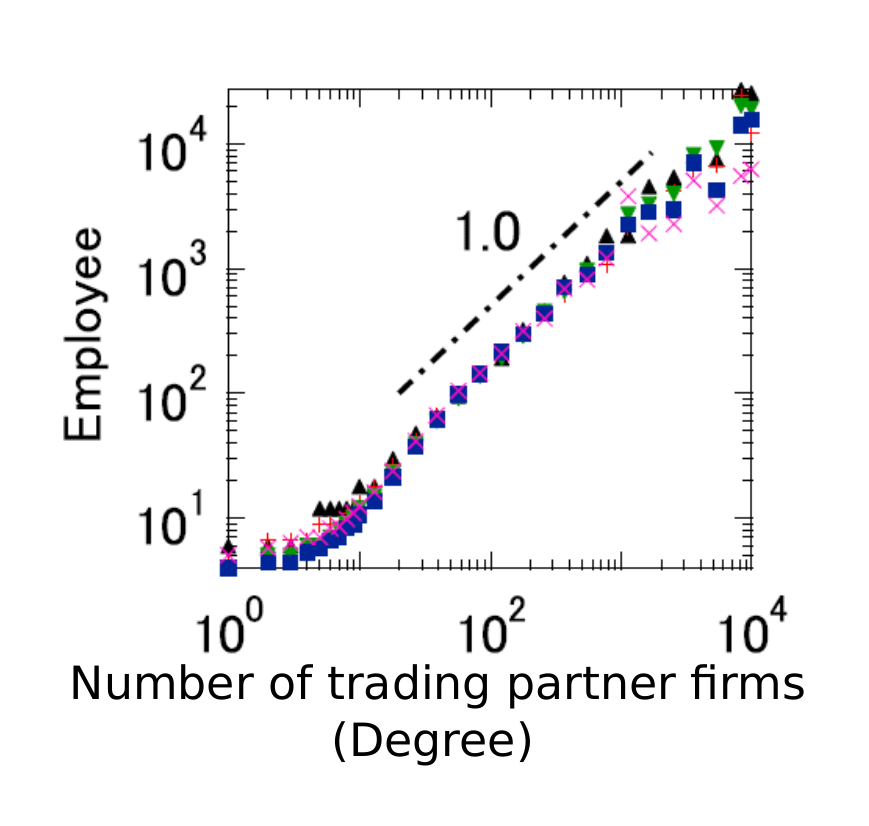}
        \put(50,118){\color{black}\Large\bfseries (f)}
    \end{overpic} 
\end{center}
    \caption{
Distributions and scaling relationships of multiple quantities representing firm size.
Panels (a-c), (d),(e) and (f) are adapted from Figs. ~4,and 3, 2, 1 of H.~Watanabe, H.~Takayasu, and M.~Takayasu, \textit{Physica A} \textbf{392}(4), 741--756 (2013) \cite{Watanabe2013AllometricScalings}, Copyright (2013), with permission from Elsevier. 
(a)--(c) Probability density distributions. The black solid lines represent the empirical data, while the red dashed and green dash-dotted lines represent theoretical curves reported in previous work \cite{Watanabe2013AllometricScalings}. Panels (a), (b), and (c) show sales, number of employees, and number of transaction partners, respectively. Here, the number of transaction partners is the degree in the inter-firm transaction network and corresponds to the sum of in-degree and out-degree.
For sales and number of employees, the green dash-dotted lines show shapes corresponding to one-sided dPlN distributions (Eq.~\eqref{app_eq_dPlN}), indicating that the central body of the distribution is lognormal-like, while the tail is power-law-like. The tail of the sales distribution has a PDF exponent of approximately $2.0$, corresponding to Zipf's law with an upper cumulative exponent of approximately $1.0$. In contrast, the distributions of the number of employees and the number of transaction partners both have PDF exponents of approximately $2.3$, corresponding to upper cumulative exponents of approximately $1.3$. Thus, the number of employees and the number of transaction partners have nearly the same tail exponent, whereas the sales distribution has a heavier tail than either of them.
(d)--(f) Scaling relationships among firm-size indicators. Each point represents a conditional percentile and is normalized so that the curves overlap at a reference point, as in Fig.~\ref{a_fig_firm_indegree}(d). Panels (d), (e), and (f) show the relationships between sales and number of employees, sales and number of transaction partners, and number of employees and number of transaction partners, respectively.
Sales increase approximately as the $1.3$ power of both the number of employees and the number of transaction partners. By contrast, the number of employees and the number of transaction partners are nearly proportional to each other and can be regarded as satisfying $\mathrm{employees}\propto \mathrm{degree}$. This proportional relationship is consistent with the fact that the distributions of both the number of employees and the number of transaction partners have upper cumulative exponents of approximately $1.3$.
The relationships between sales and the number of employees or transaction partners are also consistent with Eq.~\eqref{eq:alpha_transform}. That is, the upper cumulative exponent $\alpha_X\simeq 1.3$ for the number of employees or transaction partners is transformed into the upper cumulative exponent $\alpha_s\simeq 1.0$ for sales through the scaling relation $s\propto X^{1.3}$.
Thus, sales, number of employees, and number of transaction partners can be interpreted as indicators that measure the same underlying firm size from different aspects, and the differences in their distributional exponents can be explained by the scaling relationships among these indicators.
}
\label{a_fig_firm_dimensions}
\end{figure*}

Fig.~\ref{a_fig_firm_dimensions} summarizes the distributions and mutual scaling relations of several firm-size indicators: sales, the number of employees, and the number of trading partners. Here, the number of trading partners is the sum of indegree and outdegree in the inter-firm transaction network. It therefore represents the total number of customer firms and supplier firms. The tail of the sales distribution corresponds to a PDF exponent of about $2.0$, or an upper cumulative exponent of about $1.0$. By contrast, the distributions of the number of employees and the number of trading partners both correspond to a PDF exponent of about $2.3$, or an upper cumulative exponent of about $1.3$.\par

The correspondence among these exponents is also confirmed by the scaling relations shown in the lower panels. Sales increase approximately as the $1.3$ power of both the number of employees and the number of trading partners. By contrast, the number of employees and the number of trading partners are almost proportional to each other. This is consistent with the fact that their distributions have the same upper cumulative exponent. Thus, the number of employees and the number of trading partners are firm-size indicators with similar dimensionality. Sales, on the other hand, can be regarded as an amount of economic activity that is nonlinearly related to them.\par

The distributions of sales and the number of employees also have log-normal-like central bodies and power-law tails. In the terminology of this study, the theoretical curves in the previous study can be understood as shapes corresponding to the one-sided dPlN distribution (Eq.~\eqref{app_eq_dPlN}). This feature is similar to the result of the present study that mound-length and volume distributions of kofun have log-normal-like central bodies and power-law tails. Therefore, the comparison with firm data is suggestive not only for distributional shapes, but also for the fact that power-law exponents can change depending on which proxy variable is used to measure size. The number of employees and the number of trading partners can instead be understood as derived indicators. They represent the components and connections that support that activity.
\par
%

\subsubsection{Analogy with kofun and other archaeological data}
The comparison with modern Japanese firms is also suggestive for interpreting a feature of the kofun data: the mound-length distribution has an exponent of about three, whereas the volume distribution has an exponent of about one. 
It is also useful for considering the dimensionality of proxy variables. The important point is that differences in power-law exponents do not necessarily imply different distribution-generating processes. Even for the same objects, the observed exponent can change depending on which aspect of size is measured.  This is analogous to allometric relations in biology \cite{West1997AllometricScaling}, and related approaches have also been applied to archaeological data such as Pompeian houses \cite{Hanson2024ScalingPompeii}.  In other words, even for the same objects, the observed power-law exponent can change through a transformation of variables, depending on which quantity is used as the measure of size.\par
\paragraph{Dimensional scaling and alternative interpretations of exponent deviations}
The comparison discussed in Sec.~\ref{a_sec_firm_dimension} shows that differences in power-law exponents do not necessarily imply different distribution-generating processes. They can also arise from nonlinear transformations between proxy variables that measure different dimensions of the same objects. 
In the case of kofun, mound length $L$ and volume $V$ are approximately related as $V\propto L^3$. Therefore, even if the upper-tail cumulative exponent of the mound-length distribution is about three, the corresponding volume distribution can appear as a Zipf-like distribution with an exponent of about one. 
In this sense, the relation among sales, employees, and trading partners in Japanese firms provides a modern comparative example for understanding the relation between mound length and volume in kofun. \par

From this perspective, there are two possible interpretations of the deviations from the Zipf-like exponent observed in Kinki and Sanyo. The first is the interpretation mainly adopted in this study. In politically central or semi-central regions, existing groups may have been more likely to persist. These regions may therefore have had a competitive structure different from that of regions with more fluid competition. 

The second interpretation is that the relation between politico-economic resources and kofun volume itself differed nonlinearly among regions. For example, exceptionally large kofun in political centers such as Kinki may have had few comparable precedents. Their size may therefore have reflected extrapolation beyond ordinary experience or symbolic competition for authority. In this case, differences in resources could have been amplified when expressed as differences in kofun volume. \par

At present, however, the data needed to directly test regional differences in the nonlinear relation between politico-economic resources and kofun volume are limited. For this reason, this study adopts the first interpretation as its main working interpretation. That is, it emphasizes regional differences in competitive structure and in the persistence of existing groups.\par

\paragraph{Volume distribution and the size-imitation hypothesis}
\label{app_sec_imitation}
The dimensionality of size measures also allows us to examine the
size-imitation hypothesis in quantitative terms.
Here, the size-imitation hypothesis refers to the interpretation that
the Zipf law in kofun volume does not directly reflect the distribution of
political-economic resources assumed in this study.
Instead, it arises from a process of imitation and competition based on the
sizes of earlier mounds.
For example, if builders referred to earlier mounds and constructed new mounds
that were slightly larger or slightly smaller, this process could be represented
by a neutral multiplicative process,
$x(t+1)=b(t)x(t)$.
If the duration of each construction sequence also has an exponential tail,
the resulting size distribution can have a power-law upper tail.\par

However, this interpretation requires an additional assumption.
If builders referred mainly to visible mound dimensions, such as length and height, then the mound-length distribution itself might be expected to approach Zipf's law. 
If the exponent of the mound-length distribution were $1$, and if volume were
approximately proportional to the cube of mound length, then the exponent of the
volume distribution would be about $1/3$.
In the data, however, the power-law exponent of mound length is approximately
$3$, whereas the volume distribution is close to Zipf law.
Thus, to explain the Zipf law in volume by the size-imitation hypothesis,
one must assume that builders imitated and adjusted not mound length, but volume,
which is less directly perceptible than mound length.\par

By contrast, the interpretation adopted in this study assumes that
political-economic resources, or wealth in a broad sense, were competitively
allocated and acquired.
The scale of these resources was then reflected in kofun volume, which is close
to construction cost.
Under this interpretation, the fact that Zipf law appears in the volume
distribution rather than in the mound-length distribution is naturally consistent
with the model.
It does not require an additional assumption about which size measure was chosen
for imitation.\par

\paragraph{Proxy variables for wealth and scaling dimensions}
\label{app_sec_dimension}
More generally, in periods and societies where wealth or economic scale cannot be
directly quantified from sales, income, assets, or tax records, material indicators
are often used as proxy variables.
Examples include residential floor area, the number or variety of grave goods,
and the size of tombs.
When comparisons are based on the same proxy variable, measures such as the Gini
coefficient and power-law exponents may be interpreted relatively directly.
For example, Kohler et al. used residential floor area as a common proxy variable
for archaeological sites across the world.
They compared Gini coefficients at the household-unit level across regions and
periods \cite{Kohler2025EconomicInequality}.\par

However, when different types of proxy variables are compared, it is necessary to
consider their dimensions and scaling relations.
These variables may scale differently with wealth, resource investment, or
mobilizing capacity.
Indeed, it is not self-evident which dimension is appropriate for wealth,
social status, or mobilizing capacity.
Examining how each proxy variable reflects latent economic scale or social
hierarchy is itself an important research question
\cite{Nortoft2022GraveWealth}.
Otherwise, it is difficult to determine whether differences in observed
inequality measures reflect real differences in inequality, or merely differences
in the dimensions or transformations of the proxy variables.\par

The relation between the main size variable assumed in this study and the
associated quantities that scale with it provides one clue for defining a common
dimension for comparing proxy variables.
Here, quantities such as firm sales and kofun volume are interpreted as being
close to the dimension of political-economic resources in a broad sense.
They represent resources allocated or acquired among competing agents, or cost
measures close to such resources.
Such quantities may also empirically show distributions close to Zipf law.
Therefore, using this main variable as a reference and correcting for dimensional
differences with other associated indicators may provide one approach to comparing
different proxy variables on a common scale.\par


\clearpage

\section{Comparison with regional economic-size distributions: early-modern kokudaka and modern prefectural GDP}
\label{a_sec_kokudaka}

In this section, we present distributions of regional economic size for comparison with temporal changes in kofun size distributions. For early-modern Japan, we use kokudaka by domain in the Edo period. A domain, or han, was a local government or regional political unit in early-modern Japan. Kokudaka was a representative measure of the political and economic scale of each domain. About 300 domains existed in Japan at that time. Kokudaka corresponds to regional productive capacity, or the tax base, expressed in terms of rice yield. In this study, we use omotedaka, the official domain scale in the early seventeenth century, and jitsudaka, which is closer to the actual productive scale around the nineteenth century. For modern Japan, we use nominal prefectural GDP for the 47 prefectures at ten-year intervals from 1960 to 2020.\par

The kokudaka data were taken from an online table compiled by Satake Koji \cite{Satake2002Jitsudaka}. According to Satake's notes, omotedaka was referenced from \textit{Hanshi soran}, while jitsudaka was calculated mainly from the \textit{Kyudaka kyuryo torishirabe-cho}. 
The prefectural GDP data were obtained from the Prefectural Accounts statistical tables published by the Economic and Social Research Institute, Cabinet Office, Government of Japan \cite{ESRI_PrefecturalAccounts}. Because the long-run series covers several statistical standards and benchmark-year revisions, the data were used as a comparative indicator of regional economic scale rather than as a strictly continuous time series. 
\par
\begin{figure*}[t]
    \begin{center}
    \begin{overpic}[
        width=4.3cm
    ]{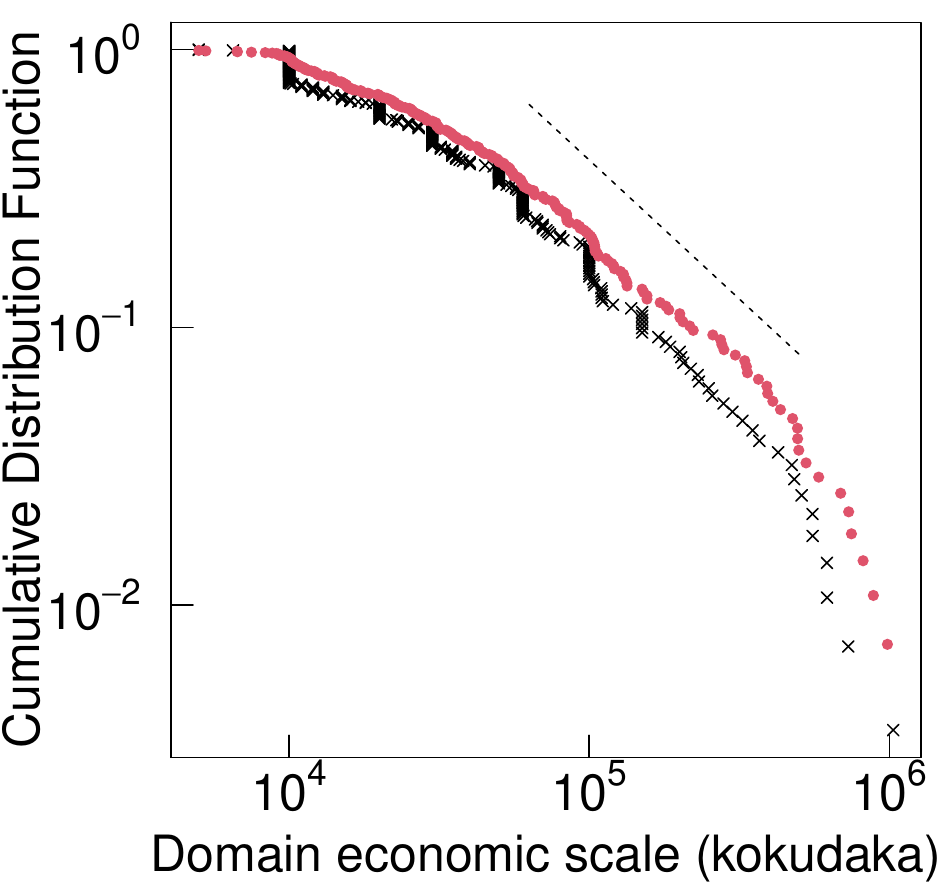}
        \put(30,80){\color{black}\Large\bfseries (a)}
    \end{overpic} 
    \begin{overpic}[
        width=4.3cm
    ]{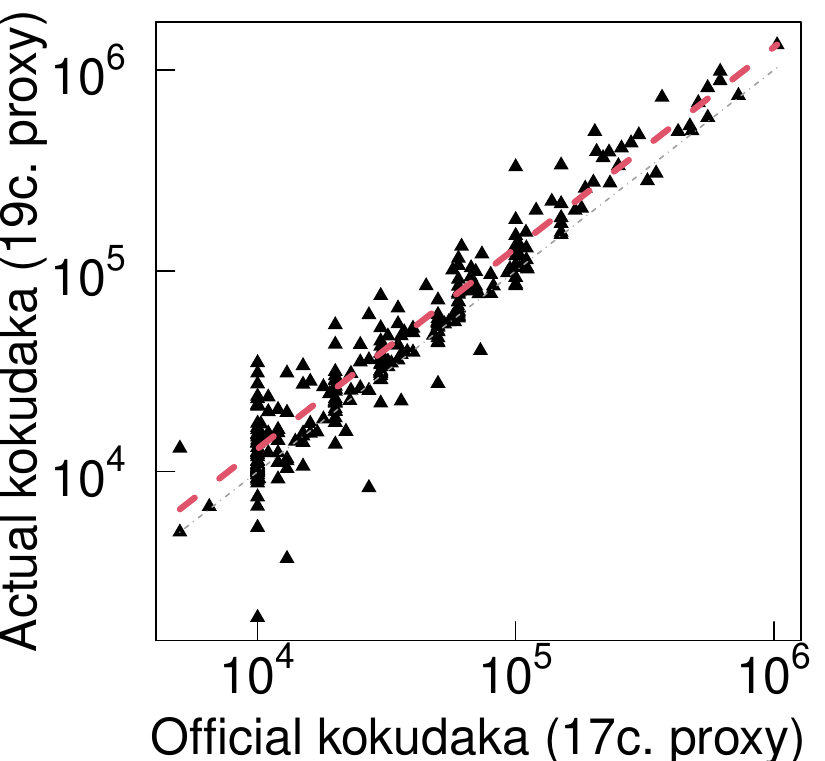}
        \put(30,90){\color{black}\Large\bfseries (b)}
    \end{overpic} 
      \begin{overpic}[
        width=4.3cm,height=4.0cm
    ]{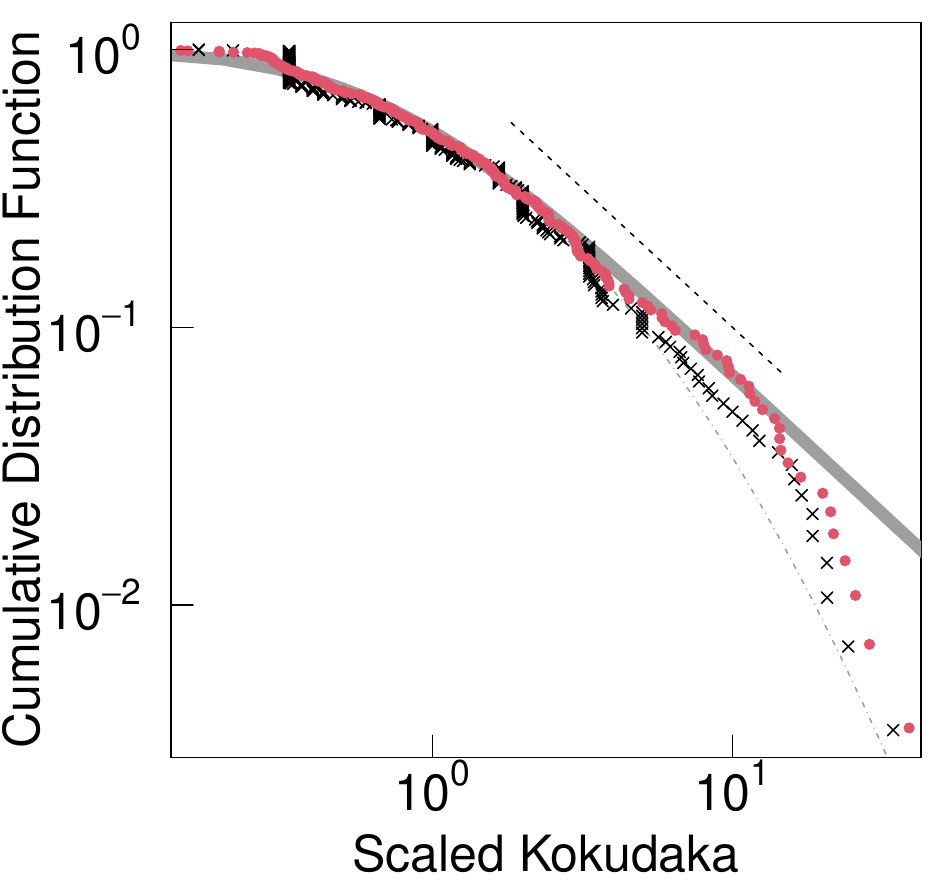}
        \put(30,80){\color{black}\Large\bfseries (c)}
    \end{overpic}
     \begin{overpic}[
        width=4.0cm,height=3.8cm
    ]{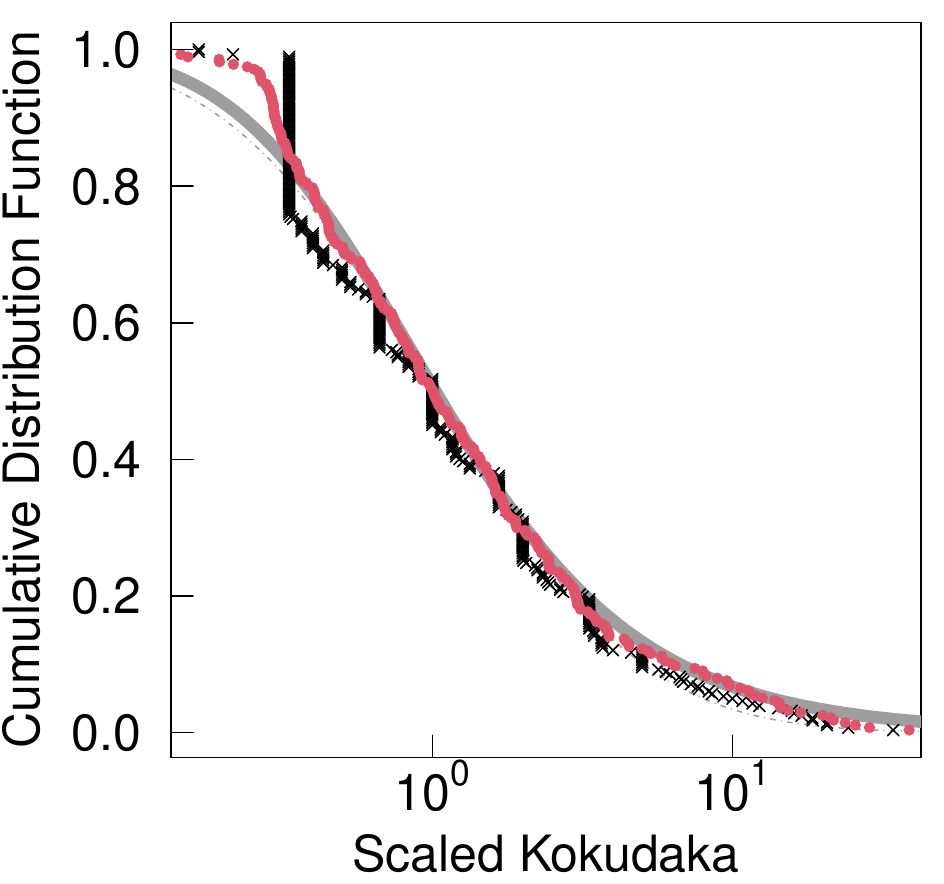}
        \put(80,90){\color{black}\Large\bfseries (d)}
    \end{overpic}
     \begin{overpic}[
        width=4.3cm
    ]{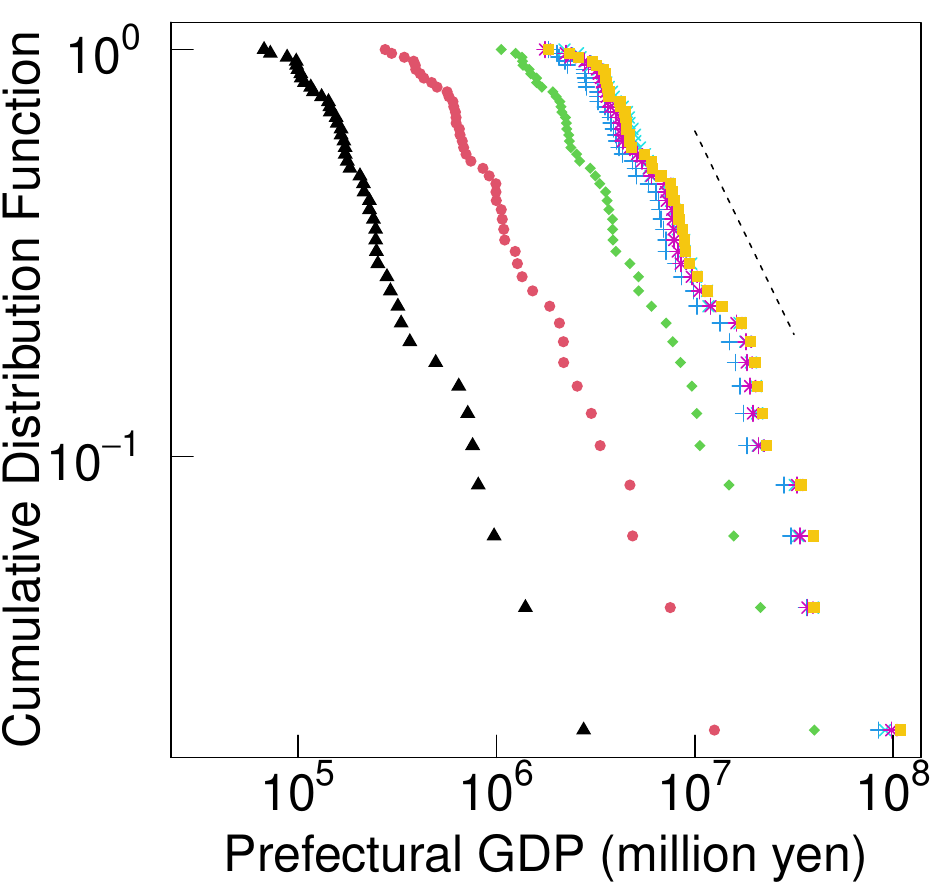}
        \put(25,80){\color{black}\Large\bfseries (e)}
    \end{overpic} 
    \begin{overpic}[
        width=4.3cm
    ]{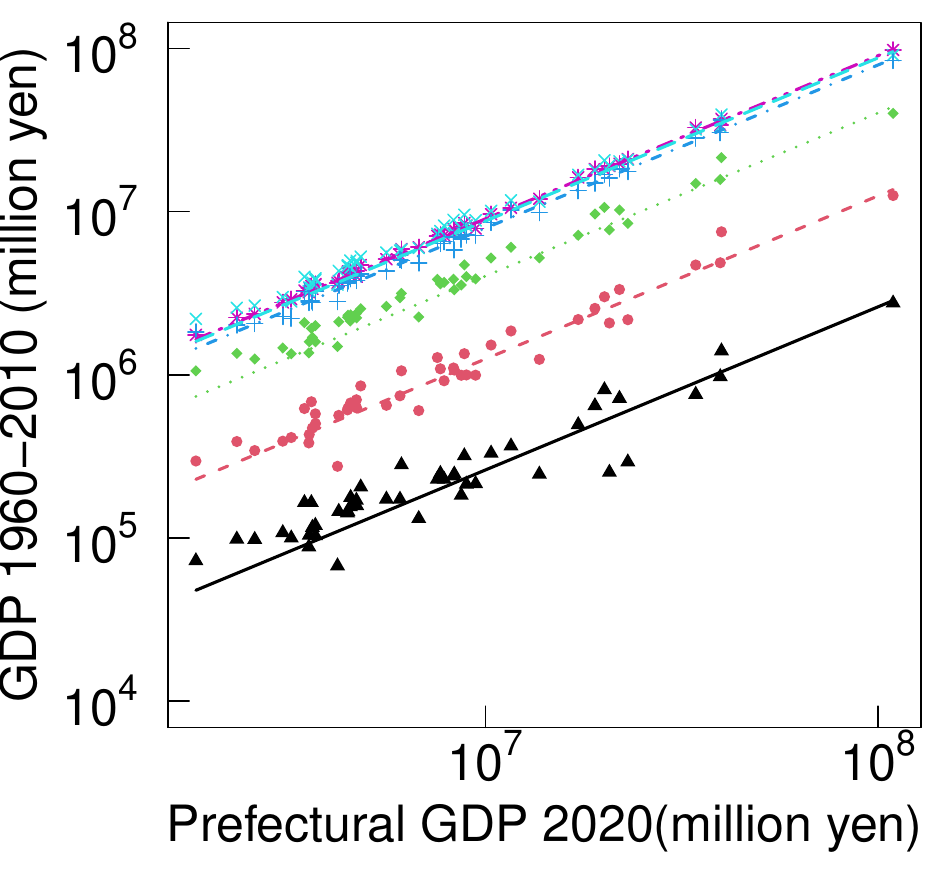}
        \put(30,95){\color{black}\Large\bfseries (f)}
    \end{overpic} 
     \begin{overpic}[
        width=4.3cm,height=4.0cm
    ]{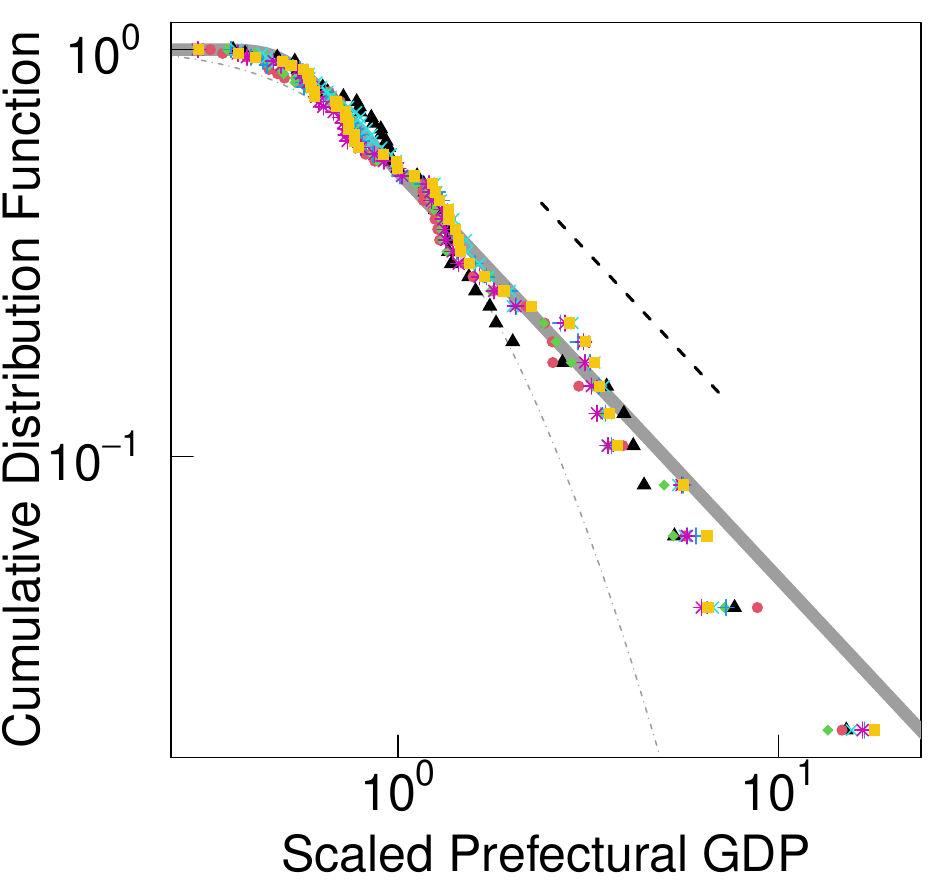}
        \put(95,95){\color{black}\Large\bfseries (g)}
    \end{overpic}  
     \begin{overpic}[
        width=4.3cm,height=4.0cm
    ]{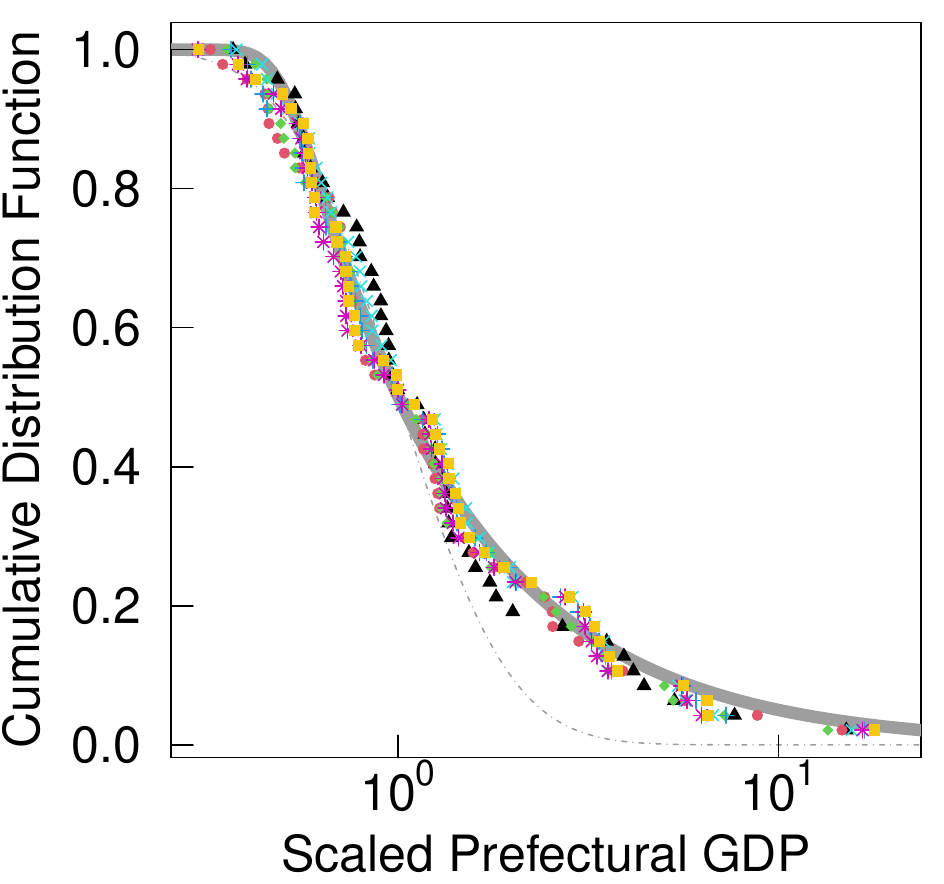}
        \put(95,95){\color{black}\Large\bfseries (h)}
    \end{overpic}  
\end{center}
   \caption{
Distributions of kokudaka in the Edo period and prefectural gross domestic product in modern Japan. 
Kokudaka is a rice-yield-based measure of the productive and economic scale of each domain. 
The upper panels show kokudaka in the Edo period, and the lower panels show nominal gross prefectural domestic product by prefecture in modern Japan.
(a) Upper cumulative distribution of kokudaka by domain. Black crosses represent omotedaka, corresponding to the early seventeenth century, and red circles represent jitsudaka, corresponding to around the nineteenth century.
(b) Relationship between omotedaka and jitsudaka. The red dashed line represents $y=1.3x$, and the thin gray dash-dotted line represents $y=x$. For many domains, jitsudaka is approximately 1.3 times omotedaka.
(c)(d) Upper cumulative distributions of kokudaka normalized by the median. Panel (c) uses log--log axes, and panel (d) uses a semi-logarithmic representation.
(e) Upper cumulative distributions of nominal gross prefectural domestic product by prefecture from 1960 to 2020. Black triangles, red circles, green diamonds, blue crosses, cyan crosses, pink asterisks, and yellow squares represent 1960, 1970, 1980, 1990, 2000, 2010, and 2020, respectively.
(f) Relationship between gross prefectural domestic product in 2020 and that in each previous year. Points of each color correspond to the same years as in (e). The straight lines represent proportional relationships to the 2020 values, with proportionality coefficients of $0.903$, $0.876$, $0.792$, $0.403$, $0.125$, and $0.0261$, respectively, from 2010 back to 1960.
(g)(h) Upper cumulative distributions of gross prefectural domestic product normalized by the median. Panel (g) uses log--log axes, and panel (h) uses a semi-logarithmic representation.
The black dashed lines in (a), (c), (e), and (g) are reference lines proportional to $1/x$. The thick gray lines in (c), (d), (g), and (h) represent one-sided dPlN distributions with upper cumulative exponent $\alpha=1$ (Eq.~\eqref{app_eq_dPlN}; kokudaka: $\mu_1=-0.847$, $\sigma_1=0.950$; prefectural GDP: $\mu_1=-0.70$, $\sigma_1=0.15$). The thin gray dash-dotted lines represent lognormal distributions whose central parts are matched by the log-median and log-interquartile deviation (kokudaka: $\mu_2=0$, $\sigma_2=1.26$; prefectural GDP: $\mu_2=0$, $\sigma_2=0.759$).
After normalization by the median, the distributions of kokudaka and gross prefectural domestic product largely overlap across periods. Their central bodies are close to lognormal, while their tails show power-law-like shapes close to $1/x$. This indicates that the distributional shape of regional economic scale is relatively stable, apart from changes in overall scale caused by macroeconomic growth.
}
\label{a_fig_kokudaka_gdp}
\end{figure*}

The upper panels of Fig.~\ref{a_fig_kokudaka_gdp} show kokudaka in the Edo period. The lower panels show modern prefectural GDP. Fig.~\ref{a_fig_kokudaka_gdp}(a) shows the upper cumulative distributions of kokudaka by domain. Both omotedaka and jitsudaka show power-law-like tails with an exponent close to $\alpha=1$. Figure~\ref{a_fig_kokudaka_gdp}(e) shows the distributions of nominal prefectural GDP from 1960 to 2020. The prefectural GDP distribution shifts to the right over time. At the same time, its overall shape changes little.\par

Fig.s~\ref{a_fig_kokudaka_gdp}(c) and (d) show the kokudaka distributions normalized by their medians. Similarly, Figs.~\ref{a_fig_kokudaka_gdp}(g) and (h) show the prefectural GDP distributions normalized by their medians. After median normalization, both kokudaka and prefectural GDP distributions largely overlap across periods. Roughly speaking, the central body is close to a log-normal distribution, and the tail is close to a power law with $\alpha \simeq 1$. The overall shape can also be approximated by the one-sided dPlN distribution shown by the thick gray line (Eq.~\eqref{app_eq_dPlN}). This feature corresponds to the kofun size distributions shown in the main text. The power-law exponents of the upper cumulative distributions estimated by the method of Clauset et al. were $\alpha=1.38$ for omotedaka and $\alpha=1.03$ for jitsudaka.\par

Fig.~\ref{a_fig_kokudaka_gdp}(b) shows the relation between omotedaka and jitsudaka in the Edo period. They are approximately proportional. Jitsudaka is about 1.3 times omotedaka. This suggests that the economic scale of many domains expanded from the early seventeenth century to around the nineteenth century. At the same time, the relative size relations among domains changed little. This is consistent with the fact that, under the bakuhan system, the framework of territorial rule was maintained for a long period. Large-scale wars and territorial realignments were also limited during the Edo period.\par

\begin{figure*}[t]
    \begin{center}
    \begin{overpic}[
        width=8cm
    ]{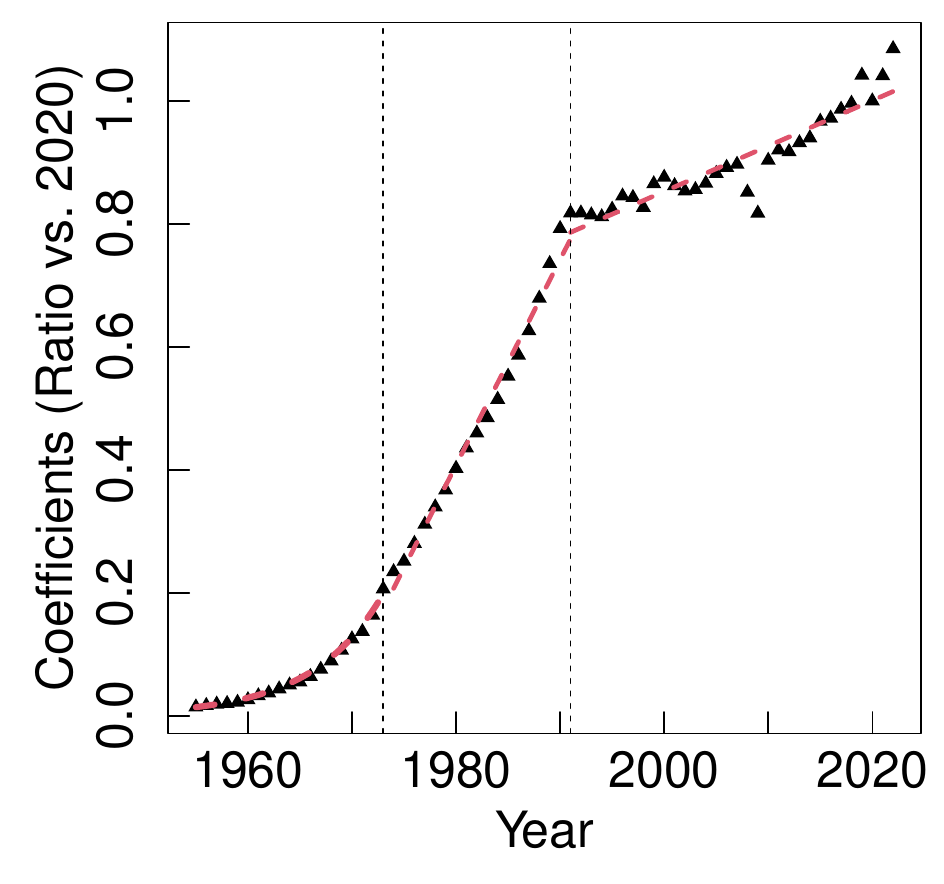}
        \put(50,190){\color{black}\Large\bfseries (a)}
    \end{overpic} 
     \begin{overpic}[
        width=8cm
    ]{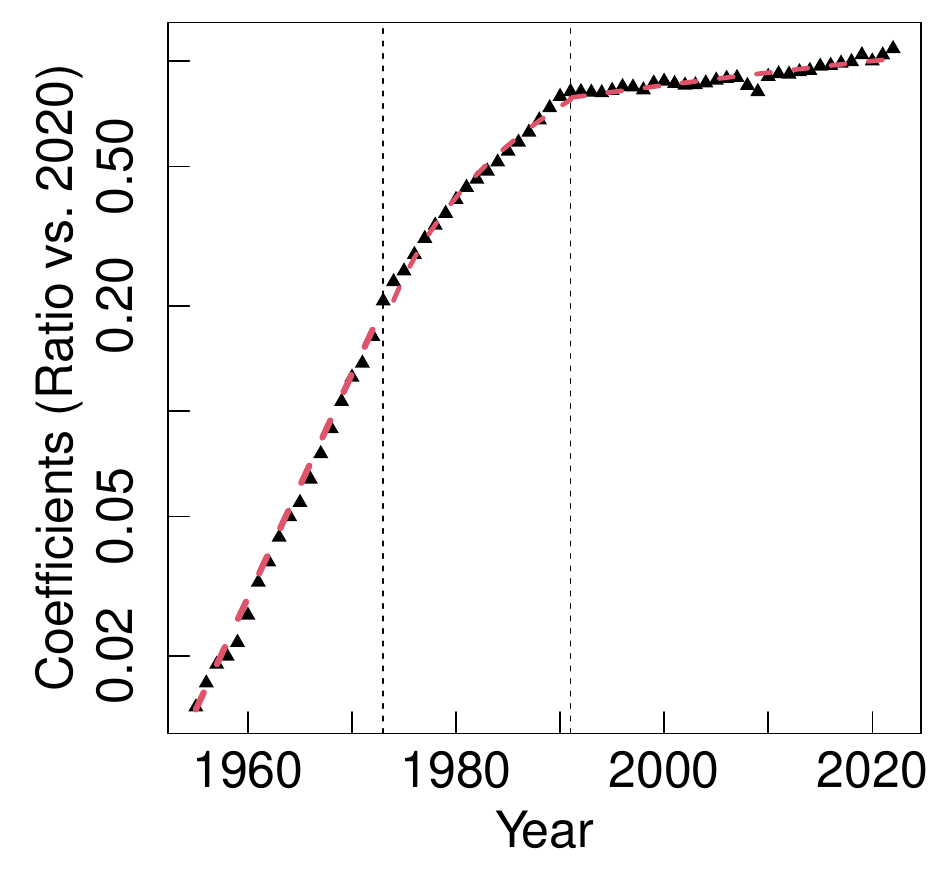}
        \put(50,190){\color{black}\Large\bfseries (b)}
    \end{overpic} 
   \begin{overpic}[
        width=8cm
    ]{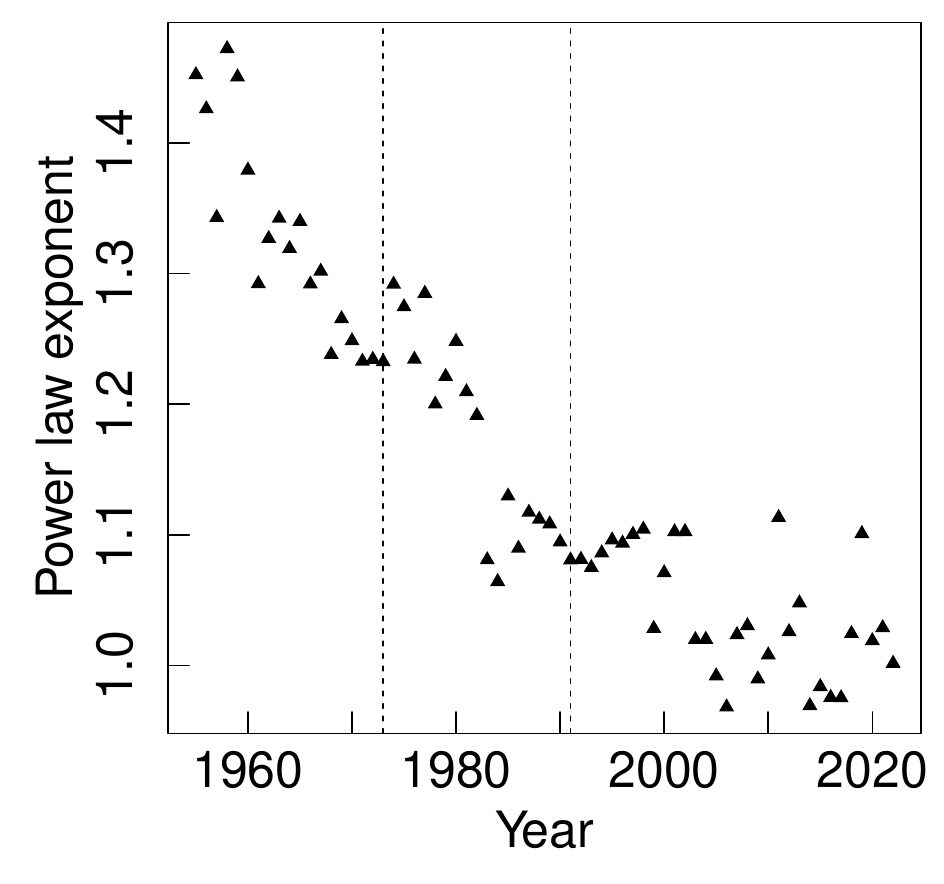}
        \put(195,190){\color{black}\Large\bfseries (c)}
    \end{overpic} 
    \begin{overpic}[
        width=8cm
    ]{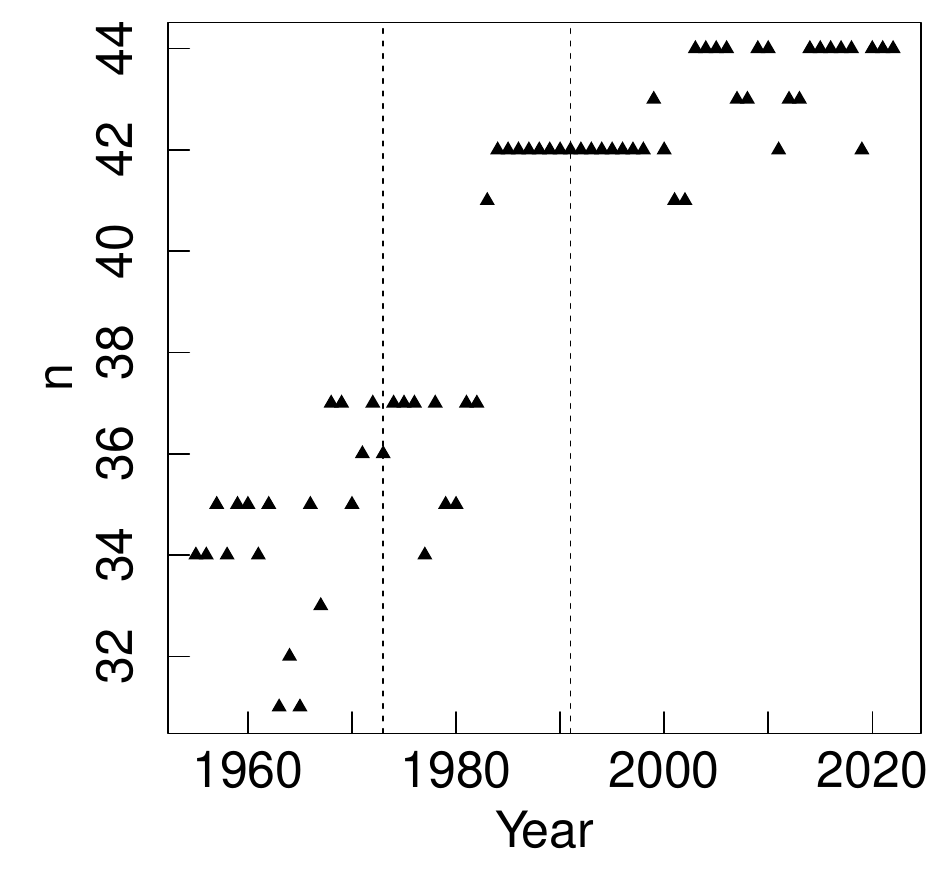}
        \put(50,190){\color{black}\Large\bfseries (d)}
    \end{overpic} 
\end{center}
    \caption{
        Temporal changes in the distribution of nominal gross prefectural domestic product by prefecture.
(a) Proportionality coefficient of gross prefectural domestic product in each year relative to that in 2020. The coefficient $a$ is obtained by regressing the gross prefectural domestic product in each year on that in 2020 using $y=ax$, where $x$ is the gross prefectural domestic product in 2020 and $y$ is that in each year.
(b) Semi-logarithmic representation corresponding to (a).
(c) Power-law exponent estimated for the distribution of gross prefectural domestic product in each year.
(d) Number of prefectures included in the power-law region. The identification of the power-law region and the parameter estimation were performed using the method of Clauset et al.
The vertical dashed lines in (a) and (b) indicate the timing of the first oil shock in 1973 and the collapse of the asset-price bubble in 1991. The red dashed lines are piecewise approximations to the temporal change in the proportionality coefficient. They represent $a=\exp(0.147t-292)$ for 1955--1973, $a=0.0334t-65.8$ for 1974--1991, and $a=0.00740t-13.94$ after 1991, where $t$ denotes the calendar year. Before 1973, the proportionality coefficient increased exponentially, whereas after 1973 it shifted to a more nearly linear increase. After 1991, the rate of increase further declined.
}
 \label{a_fig_ken_gdp}
\end{figure*}

Fig.~\ref{a_fig_kokudaka_gdp}(f) shows the relation between prefectural GDP in 2020 and in earlier years. Prefectural GDP in each year is approximately proportional to prefectural GDP in 2020. This indicates that the Japanese economy grew substantially from 1960 to 2020. At the same time, the relative economic size relations among prefectures remained relatively stable. The slope coefficients of the fitted lines in the figure represent the scale of prefectural GDP in each year relative to the 2020 level. Their temporal changes are plotted in Fig.~\ref{a_fig_ken_gdp}.
 \par

\subsection{Timing of share formation of kofun}
\label{a_sec_kokudaka}
The kokudaka and prefectural GDP results above show that the relative structure among regions, which determines the distributional shape, and the macro-level growth process, which shifts the entire distribution to the right, do not necessarily change at the same time. 
In the cases of Edo-period kokudaka and modern prefectural GDP, regional economic size increased over time, whereas the shape of the median-normalized distribution changed little. 
This observation is suggestive for interpreting size distributions in the Kofun period. 
For simplicity, the model in this study was formulated under the assumption that competition over shares and kofun construction proceeded during the same period. 
In actual politico-economic distributions, however, the formation of shares and the subsequent expansion of scale based on those shares may be temporally separated. 
That is, relative shares among groups may first be formed through competition or reorganization in one period. 
In later periods, the overall distribution may shift to the right through economic growth or an increase in available resources, while those relative shares remain largely unchanged. \par
For simplicity, the model in this study was formulated under the assumption that competition over shares and kofun construction proceeded during the same period.
In actual politico-economic distributions, however, the formation of shares and the subsequent expansion of scale based on those shares may be temporally separated.
That is, relative shares among groups may first be formed through competition or reorganization in one period.
In later periods, the overall distribution may shift to the right through economic growth or an increase in available resources, while those relative shares remain largely unchanged.\par

One example is provided by later Japanese distributions: the kokudaka distribution among domains in the seventeenth century and prefectural GDP from the postwar period to the present. The kokudaka and prefectural GDP results above provide useful comparisons because they show cases in which the distributional shape remained relatively stable while the overall scale shifted. \par
The kokudaka and prefectural GDP results above provide useful comparisons because they show cases in which distributional shape remained relatively stable while the overall scale shifted. In the kokudaka case, the approximate proportionality between omotedaka and jitsudaka suggests that the assessed or actual productive scale of domains increased, while their relative size relations remained largely stable. A similar pattern is observed in postwar prefectural GDP, where the distribution shifts to the right but the median-normalized shape changes little.\par This pattern can be interpreted in relation to the political transition from the Sengoku period to the Edo period. During the Sengoku period, territorial competition among regional lords had a zero-sum-like aspect over limited land and productive capacity. By contrast, in the Edo period, a nationwide political order was established under the Tokugawa shogunate, and large-scale military conflicts and territorial realignments among daimyo (Japanese feudal lords who ruled regional domains) were suppressed. The kokudaka distribution of the Edo period can therefore be interpreted as a case in which a politico-economic share structure formed through earlier competition and political reorganization was largely preserved, while the overall scale expanded through economic growth.\par

This example is suggestive for interpreting the distributions of the Kofun period. The fact that an observed distribution is close to Zipf's law does not necessarily mean that share competition occurred during the same period as the observed construction activity. The kokudaka distribution of the Edo period may have preserved, under the early-modern order, the outcomes of competition and political reorganization from the Sengoku period to the beginning of the early-modern period. In the same way, kofun volume distributions may have partially inherited a power-share distribution formed before, or at the beginning of, kofun construction. In the Yayoi period, which preceded the Kofun period, wet-rice agriculture spread widely across the Japanese archipelago. Against this background, politico-economic inequalities among settlements and regions developed. Competition and armed conflict among groups also became more visible. Chinese historical texts contain descriptions suggesting large-scale disturbances in the Japanese archipelago near the end of the Yayoi period. These disturbances are often referred to as the Wakoku War. The Kofun period can then be understood as a period in which a broader political order was formed after these developments. Thus, kofun volume distributions may reflect not only competitive processes during the period of kofun construction itself, but also the inheritance of share structures formed from the late Yayoi period to the beginning of the Kofun period.\par

From this perspective, the term $J_T$ in the model, which represents the distributional shape (Eq.~\ref{eq:QA0J}), can be interpreted as including the power-share structure formed from the late Yayoi period to the beginning of the Kofun period. It need not be interpreted only as the outcome of competition newly formed during the period of kofun construction. By contrast, left-right shifts of the distribution across periods can be understood mainly as changes in the scale factor $QA_0$. This factor represents the overall level of production and the fraction of resources allocated to kofun construction.  \par
At the same time, the Kofun-period distributions did not necessarily reflect only a completely fixed share structure. In Sanyo, especially in the area including Kibi, a heavy tail characteristic of the exceptional group is observed in the fourth and fifth centuries. In the sixth century, however, the distribution moves closer to the typical group. This change suggests that kofun volume distributions reflected not only the inheritance of previously formed share structures, but also new political reorganization within the Kofun period itself. Therefore, kofun volume distributions should be interpreted as the combined result of inherited power-share structures, political reorganization within the Kofun period, and changes in the overall scale factor.\par

In relation to the fixation of shares by political power, large regional powers that could threaten the central authority were also objects of caution and restraint in the Edo period. For example, an anecdote about the Maeda family of Kaga, one of the largest domains in kokudaka, states that its lord pretended to be foolish, even by growing his nose hair, in order to avoid suspicion from the shogunate. This anecdote is only illustrative. However, it symbolically shows that large regional powers could be viewed with caution under a central authority. Similarly, the result of this study shows that Sanyo and Kibi, which built large kofun comparable to those of the central region in the fifth century, moved closer to the typical group in the sixth century. This is only an analogy based on distributional shape. Even so, it suggests an interpretation in terms of changing relations between the central power and powerful regional groups.\par

%
\clearpage
\section{Extension of the Stopping Process and the Lognormal Distribution}
\label{app_sec_teisi}
In this section, we extend the stopping process discussed in Sec.~\ref{sec_teisi}. We examine the distribution of kofun volumes when the stopping probability changes over time. In the main model, we used a Kesten process in which growth stops with a constant probability at each time step, and a new local group enters again. In this case, the growth duration follows a geometric distribution. In the continuous-time approximation, this corresponds to an exponential lifetime distribution. As a result, a power-law-like shape appears in the tail of the distribution.\par

However, in the distribution of round mounds in particular, the tail is sometimes thinner than a pure power law. It may also show a rounded shape close to a lognormal distribution. This is not inconsistent with the interpretation based on the Kesten process used in this study. First, when the sample size is limited, the realized sample can show large fluctuations in the tail. Thus, even if the theoretical distribution has a power-law-like tail, the observed sample may appear close to a lognormal distribution. This point is examined in Sec.~\ref{app_sec_teisi_yuragi}. Second, if we introduce a stopping process in which the stopping probability increases over time, the theoretical distribution itself can change continuously from a power-law-like tail to a shape close to a lognormal distribution. This point is examined in Sec.~\ref{app_sec_teisi_dist}. In this section, we provide supplementary checks for these two points.\par

\subsection{Fluctuations of the Distribution in Finite Samples}
\label{app_sec_teisi_yuragi}
\begin{figure}[t]
\begin{center}
    \begin{overpic}[
        width=8cm
    ]{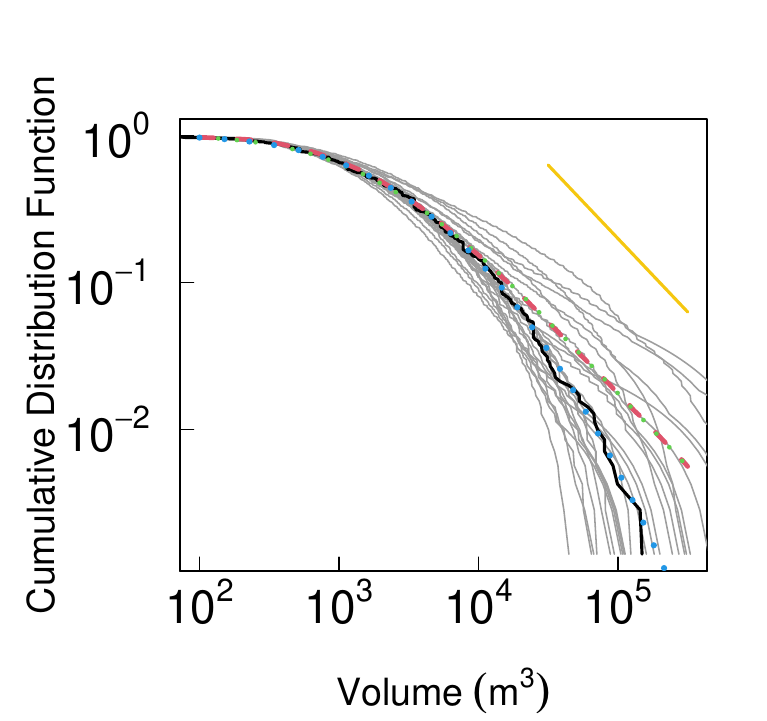}
    \end{overpic} 
\end{center}
 \caption{
Sample fluctuations in Kesten-process simulations (Eq. \ref{eq:kesten}) for the volume distribution of round burial mounds. The thin gray lines show the upper cumulative distributions of 25 simulated samples generated from a Kesten process with an exponential survival-time distribution. The sample size in each simulation was set to $703$, matching the number of empirical round burial mounds. The thick black solid line represents the empirical data for round burial mounds, the red dashed line represents the theoretical distribution derived from the Kesten process (Eq.~\eqref{eq:ccdf}), the green dashed line represents the one-sided dPlN distribution (Eq.~\eqref{app_eq_dPlN}), the blue dashed line represents the lognormal distribution fitted to the empirical data, and the yellow solid line represents the reference line $1/x$.
For the theoretical Kesten distribution, we used $A_0=263$, $b_0=1.16$, and $\alpha=1.00$. For the one-sided dPlN distribution, we used $\mu=6.58$, $\sigma=1.34$, and $\alpha=1.00$. For the lognormal distribution, we used $\mu=7.54$ and $\sigma=1.55$, estimated from the round-mound volume data.
Although the theoretical distribution has a power-law-like tail, sample fluctuations are large in finite samples, and individual simulated samples may exhibit shapes close to a lognormal distribution. This indicates that even when the underlying theoretical distribution has a Zipf-like heavy tail, a finite sample of the size observed for round burial mounds may appear as a lognormal-like distribution. 
}
\label{app_fig_teisi_yuragi}
\end{figure}

Fig.~\ref{app_fig_teisi_yuragi} shows sample fluctuations in the distribution of kofun volumes generated by the Kesten process used in the main text, Eq.~\eqref{eq:kesten}. In each simulation, we generated $703$ data points, which is the same as the sample size of round mounds. The gray lines show individual simulation results. The red dashed line shows the theoretical distribution, the blue dashed line shows the lognormal distribution, and the yellow line shows a reference line proportional to $1/x$.\par

The figure shows that finite-sample fluctuations can substantially affect the empirical shape of the distribution. This effect is especially strong in the tail, where observations are sparse. Thus, a realized sample may appear close to a lognormal distribution, even when the theoretical distribution has a power-law-like tail. Therefore, the fact that the distribution of round mounds is well approximated by a lognormal distribution does not immediately reject a Kesten-type generative process.\par
%
%
\subsection{Generalization of the Stopping Process}
\label{app_sec_teisi_dist}

Next, we consider the case in which the stopping probability changes over time. In the Kesten process used in the main text, Eq.~\eqref{eq:kesten}, the stopping probability is constant at each time step. Therefore, the growth duration corresponds to an exponential-type lifetime distribution. However, for actual local groups or chiefly lineages, the probability of stopping or reorganization may increase as the duration of survival becomes longer. This may be especially relevant for round mounds. Compared with keyhole-shaped mounds, round mounds may include a larger number of smaller construction agents or relatively lower-status groups. Therefore, the duration of these agents, or the social persistence of the buried groups, may have affected the shape of the distribution.\par

Such stopping processes can be described by Weibull and Gompertz distributions. Let $S(k)$ be the survival function at discrete time $k$ (Fig. \ref{app_fig_teisi_main}(d)). Then, the probability that an individual surviving up to time $k$ stops at the next time step, namely the hazard rate, is given by
\begin{equation}
q(k)=1-\frac{S(k+1)}{S(k)} .
\end{equation}
For an exponential-type stopping process, $q(k)$ is constant. In contrast, for Weibull and Gompertz distributions, $q(k)$ can increase over time depending on the parameter values (Fig. \ref{app_fig_teisi_main}(a)). \par

The Weibull-type survival function, shown by the red dashed line in Fig. \ref{app_fig_teisi_main}(d) is given by
\begin{equation}
S_W(k)=\exp\left[-\left(\frac{k}{\lambda}\right)^s\right] .
\label{app_eq_teisi_wei}
\end{equation}
Here, $\lambda$ is the scale parameter, and $s$ is the shape parameter. When $s=1$, this distribution corresponds to the exponential distribution. When $s>1$, the hazard rate increases over time. Examples of the hazard rate are shown by the red dashed line in Fig.~\ref{app_fig_teisi_main}(a).\par

The Gompertz-type survival function, shown by the green dash-dotted line in Fig. \ref{app_fig_teisi_main}(d)  is given by
\begin{equation}
S_G(k)=\exp\left[-\frac{\eta}{\gamma}
\left\{\exp(\gamma k)-1\right\}\right] .
\label{app_eq_teisi_gomp}
\end{equation}
Here, $\eta$ is a parameter corresponding to the initial hazard, and $\gamma$ controls the rate at which the hazard increases. In the limit $\gamma \to 0$, the distribution approaches the exponential distribution. When $\gamma>0$, the hazard rate increases over time. Examples of the hazard rate are shown by the green dash-dotted line in Fig.~\ref{app_fig_teisi_main}(a).\par

We incorporate these stopping processes into the Kesten process used in the main text. As in the main text, the size of an individual that has grown up to age $k$ can be written as
\begin{equation}
x_k=A_0
\left(2U b_0^k+\frac{b_0^k-1}{b_0-1}\right) .
\end{equation}
Here, $U\sim \mathcal{U}(0,1)$ represents variation in the initial size at re-entry. The parameter $A_0$ is the reference scale, and $b_0$ is the growth rate. In this case, the lower and upper bounds of the size at age $k$ are
\begin{equation}
L_k=A_0\frac{b_0^k-1}{b_0-1},
\qquad U_k= L_k+2A_0 b_0^k .
\end{equation}\par

In the stationary state, the age distribution is proportional to the survival function. Thus, we can write
\begin{equation}
\pi_k=\frac{S(k)}{\sum_{\ell=0}^{\infty}S(\ell)} .
\end{equation}
Therefore, the probability that the volume is less than or equal to $v$ is expressed as
\begin{equation}
F(v)=\sum_{k=0}^{\infty}
\pi_k \cdot G\left( \frac{v-L_k}{U_k-L_k}\right) .
\label{app_eq_teisi_kesten}
\end{equation}
Here, $G(z)$ is a linear saturation function defined by
\begin{equation}
G(z)=
\begin{cases}
0 & z \le 0 \\
z &  0<z<1 \\
1 & z\ge 1 .
\end{cases}
\end{equation}
The upper cumulative distribution is then obtained as $\Pr(X>v)=1-F(v)$.\par

In the exponential-type stopping process, the age distribution decreases geometrically. Therefore, a power-law-like tail appears in a sufficiently large-size range. In contrast, in the Weibull-type and Gompertz-type stopping processes, the stopping probability increases with age. This suppresses individuals that continue to grow for a long time and reach extremely large sizes. The corresponding age distributions are shown in Fig.~\ref{app_fig_teisi_main}(d). As a result, the tail becomes thinner than a pure power law. The distribution then changes toward a rounded shape close to a lognormal distribution, as shown in Fig.~\ref{app_fig_teisi_main}(b) and (e). This point is discussed in more detail in Sec.~\ref{app_sec_teisi_enpun}.\par

\subsection{Comparison with the Distribution of Round Mounds}
\label{app_sec_teisi_enpun}

\begin{figure*}[t]
    \begin{center}
    \begin{overpic}[
        width=5.8cm
    ]{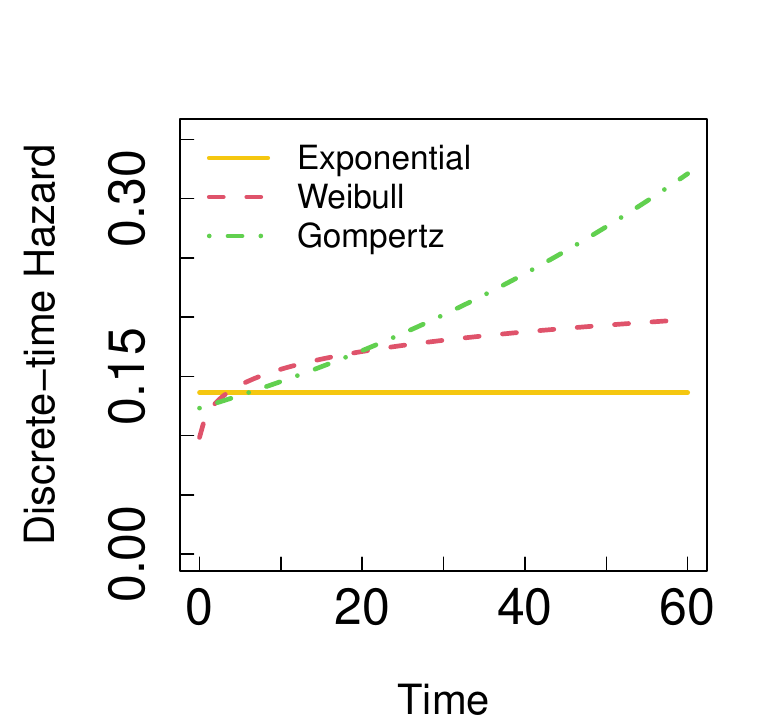}
        \put(50,45){\color{black}\Large\bfseries (a)}
    \end{overpic} 
     \begin{overpic}[
        width=5.8cm
    ]{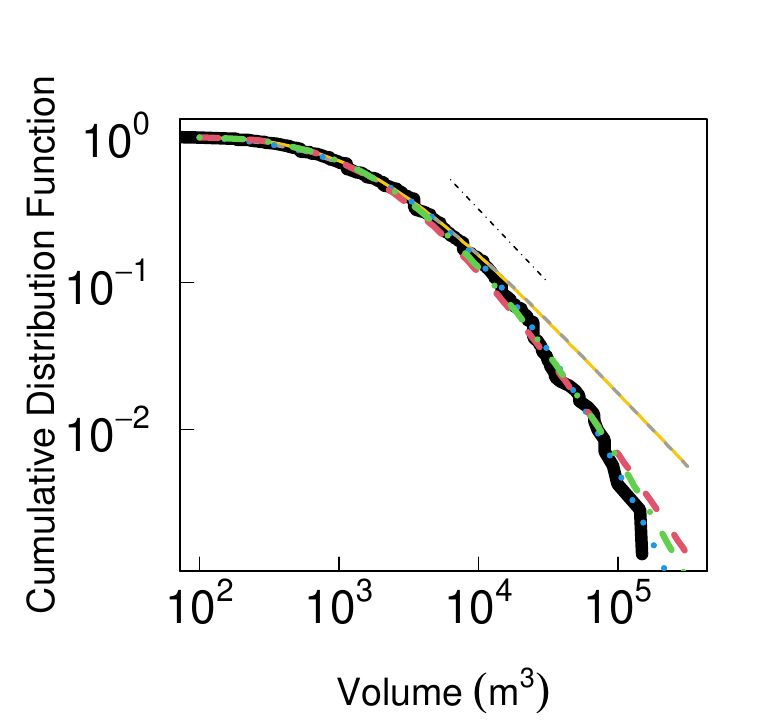}
            \put(45,100){\color{black}\Large\bfseries (b)}
    \end{overpic} 
    \begin{overpic}[
        width=5.8cm
    ]{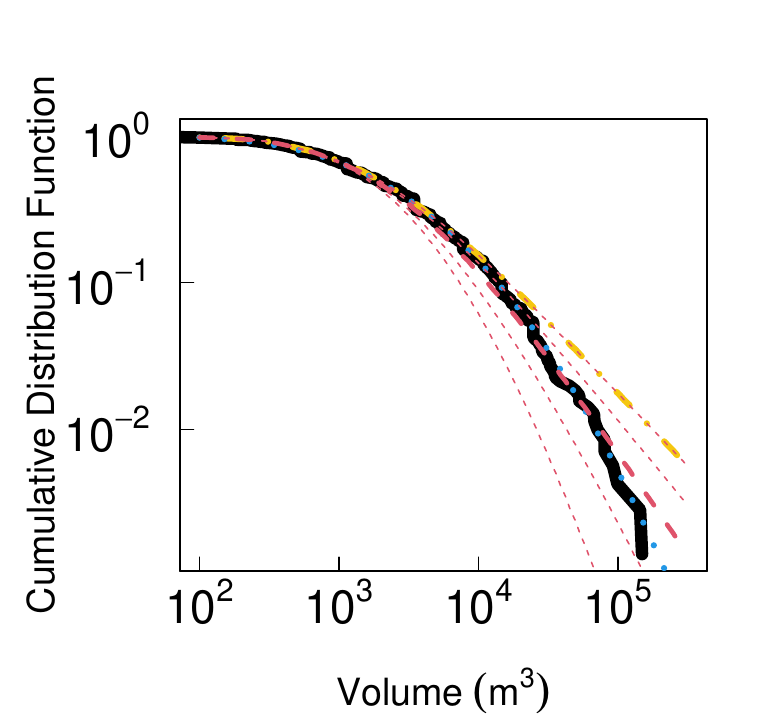}
        \put(45,100){\color{black}\Large\bfseries (c)}
    \end{overpic}
      \begin{overpic}[
        width=5.8cm,height=5.6cm
    ]{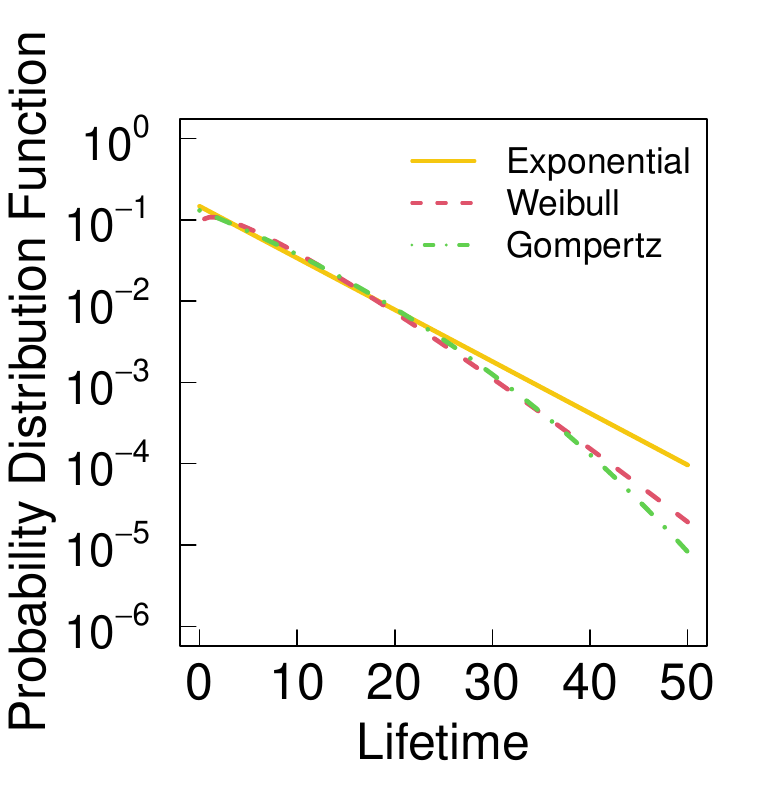}
            \put(45,45){\color{black}\Large\bfseries (d)}
    \end{overpic} 
    \begin{overpic}[
        width=5.8cm
    ]{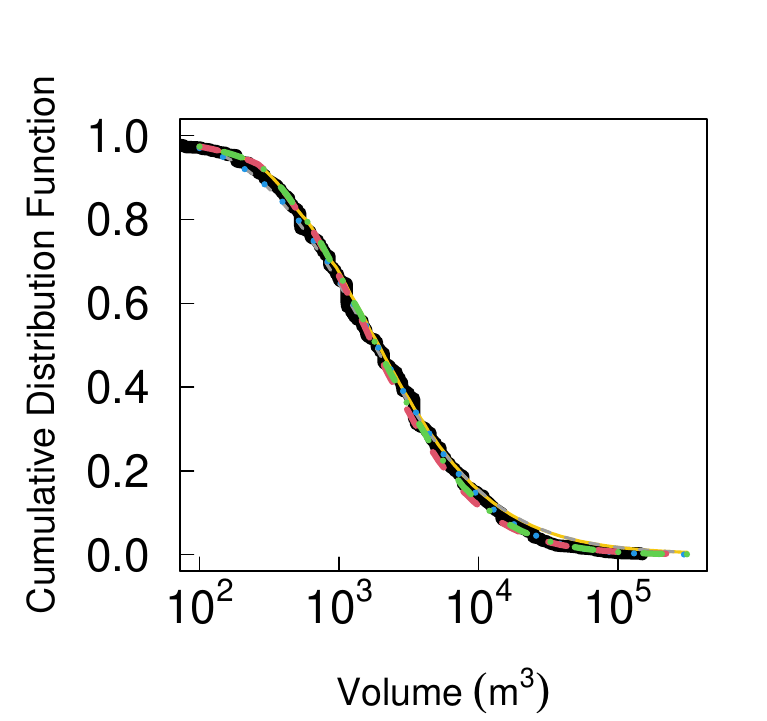}
        \put(40,100){\color{black}\Large\bfseries (e)}
    \end{overpic}  
    \begin{overpic}[
        width=5.8cm
    ]{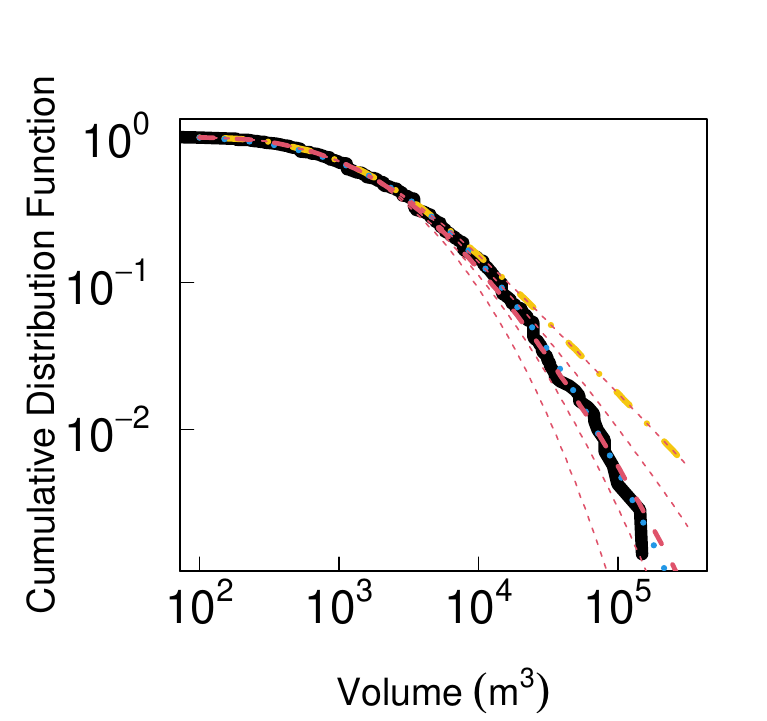}
        \put(45,100){\color{black}\Large\bfseries (f)}
    \end{overpic} 
\end{center}

\caption{
Comparison of kofun volume distributions under different stopping processes.
(a) Stopping probability for each stopping process. This panel shows the conditional probability (discrete-time hazard) that a group surviving until time $t$ stops in the next step, that is, between times $t$ and $t+1$. The yellow solid line represents the exponential distribution, the red dashed line the Weibull distribution, and the green dash-dotted line the Gompertz distribution. In the exponential case, this stopping probability is constant, whereas in the Weibull and Gompertz cases it increases with time.
(d) Probability density functions of the corresponding lifetime distributions. The exponential distribution corresponds to a constant stopping probability, whereas the Weibull and Gompertz distributions correspond to lifetime distributions in which stopping becomes more likely as duration increases.
(b) Log--log plot comparing the volume distributions obtained under different stopping processes. The thick black line represents the empirical data for round burial mounds, the yellow solid line the exponential-type stopping process, the red dashed line the Weibull-type stopping process, and the green dash-dotted line the Gompertz-type stopping process. The gray reference line represents $1/x$, the blue dotted line the lognormal distribution, and the thin gray dashed line the one-sided dPlN distribution (Eq.~\eqref{app_eq_dPlN}). The common parameters of the extended Kesten process (Eq.~\eqref{app_eq_teisi_kesten}) are $A_0=262.65$, $b_0=1.158$, and $\alpha=1$. For the Weibull-type stopping process (Eq.~\eqref{app_eq_teisi_wei}), we use $s=1.15$, and for the Gompertz-type stopping process (Eq.~\eqref{app_eq_teisi_gomp}), we use $\gamma=0.018$. The parameters of the lognormal distribution are $\mu=7.54$ and $\sigma=1.55$, and those of the one-sided dPlN distribution are $\mu=6.58$, $\sigma=1.34$, and $\alpha=1$.
(e) Semi-logarithmic plot corresponding to (b). The distributions generated by the Weibull-type and Gompertz-type stopping processes show shapes very close to a lognormal distribution in the central body.
(c) Log--log plot showing parameter dependence of the stopping processes. The thin red dashed lines represent the volume distributions for the Weibull-type stopping process when the shape parameter is varied as $1$, $1.07$, $1.15$, $1.3$, and $1.5$. The yellow solid line represents the exponential-type stopping process, and the thick black line represents the empirical data for round burial mounds.
(f) Gompertz-type stopping process corresponding to (c). The thin red dashed lines represent the volume distributions for the Gompertz-type stopping process when the shape parameter is varied as $0$, $0.01$, $0.02$, $0.03$, and $0.05$.  As the parameters of the stopping process are varied, the tail of the volume distribution continuously shifts from a lognormal-like shape to a power-law-like shape.
}
\label{app_fig_teisi_main}
\end{figure*}

Fig.~\ref{app_fig_teisi_main}(a) and (d) compare stopping processes corresponding to the exponential, Weibull, and Gompertz distributions. Fig.~\ref{app_fig_teisi_main}(a) shows the conditional probability that a group surviving up to time $t$ stops at the next time step $t+1$. This is the discrete-time hazard. In the exponential distribution, the stopping probability is constant. In contrast, in the Weibull and Gompertz distributions, the stopping probability increases over time. Fig.~\ref{app_fig_teisi_main}(d) shows the corresponding lifetime distributions. These differences in the stopping process are reflected in the tail of the generated volume distribution.\par

Fig.~\ref{app_fig_teisi_main}(b) and (e) compare the theoretical distributions under different stopping processes with the empirical data for round mounds. Fig.~\ref{app_fig_teisi_main}(b) shows the distributions on log--log axes, and Fig.~\ref{app_fig_teisi_main}(e) shows them on semi-log axes. In the exponential-type stopping process, a power-law-like shape remains in the tail. In contrast, in the Weibull-type and Gompertz-type stopping processes, the tail becomes thinner. The resulting distributions are closer to the empirical distribution of round mounds and to a lognormal distribution. This result suggests that the more lognormal-like shape of the round-mound distribution, compared with the keyhole-mound distribution, may be interpreted in terms of a time-dependent stopping probability.\par

Furthermore, Fig.~\ref{app_fig_teisi_main}(c) and (f) show the volume distributions obtained by changing the parameters of the Weibull and Gompertz distributions. Fig.~\ref{app_fig_teisi_main}(c) shows the results on log--log axes, and Fig.~\ref{app_fig_teisi_main}(f) shows them on semi-log axes. For the Weibull distribution, when the shape parameter $s$ is close to $1$, the stopping process is close to the exponential type, and the tail is power-law-like. As $s$ increases, the time dependence of the stopping probability becomes stronger, and the tail becomes rounded toward a lognormal-like shape. Similarly, for the Gompertz distribution, the case $\gamma=0$ corresponds to the exponential type. As $\gamma$ increases, the tail becomes thinner. Thus, by generalizing the stopping process, one can continuously describe distributions ranging from a power-law-like tail to a lognormal-like shape.\par

These results suggest two possible interpretations of the lognormal-like distributional shape of round mounds. First, under finite samples, the power-law-like tail generated by the Kesten process may not be clearly observed because of sample fluctuations. Second, when a process with an increasing stopping probability over time is introduced, the theoretical distribution itself can change toward a shape close to a lognormal distribution. Under the latter interpretation, the distribution of round mounds may reflect a process in which the duration or size growth of construction agents was more easily constrained than in the case of keyhole-shaped mounds. However, this is a mathematical interpretation based on the distributional shape. The social position of round mounds and the nature of their construction agents should therefore be examined carefully in combination with archaeological evidence.\par
%
%
%
%

\clearpage
\section{Distribution of Kofun Sizes Aggregated by Prefecture}
\label{a_sec_kenkohu}

In this section, we examine the distribution of aggregated  keyhole-shaped kofun-size measures at the prefectural level. For each prefecture, we calculated the number of kofun, the maximum volume, the total volume, the mean volume, the median volume, and the mean volume excluding the maximum. Let $N_p$ be the number of kofun in prefecture $p$, $M_p$ be the maximum volume, and $S_p$ be the total volume. Then, when $N_p>1$, the mean volume excluding the maximum is defined as $(S_p-M_p)/(N_p-1)$.\par

\begin{figure*}[t]
    \centering
\begin{overpic}[
        width=5.8cm
    ]{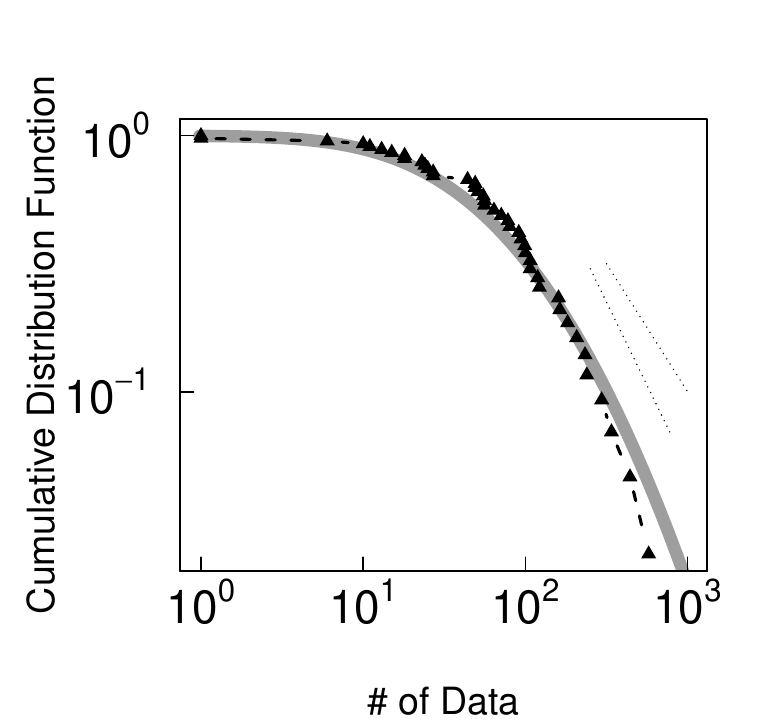}
        \put(120,112){\color{black}\Large\bfseries (a)}
    \end{overpic} 
     \begin{overpic}[
        width=5.8cm
    ]{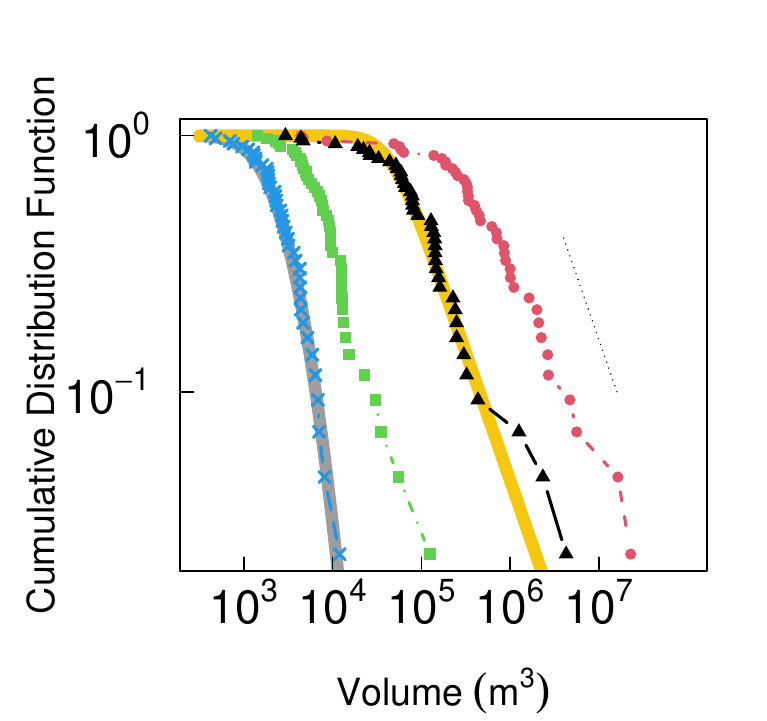}
        \put(120,112){\color{black}\Large\bfseries (b)}
    \end{overpic} 
    \begin{overpic}[
        width=5.8cm
    ]{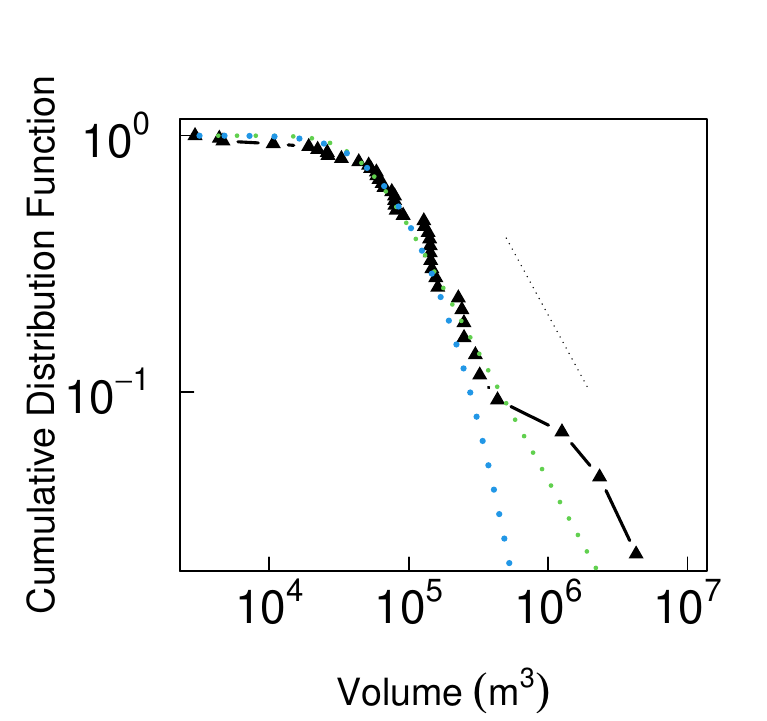}
        \put(120,112){\color{black}\Large\bfseries (c)}
    \end{overpic} 
    \begin{overpic}[
        width=5.8cm
    ]{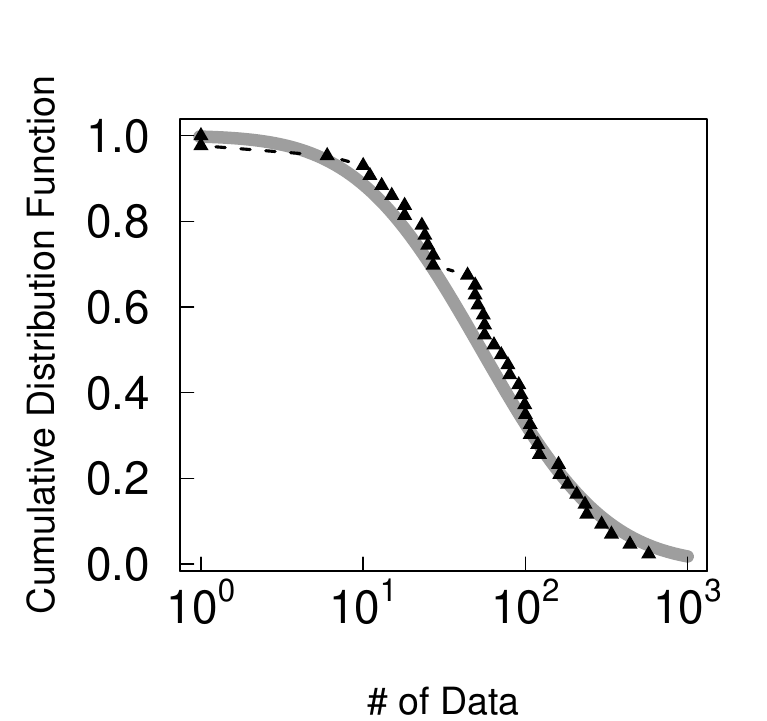}
        \put(120,112){\color{black}\Large\bfseries (d)}
    \end{overpic}
     \begin{overpic}[
        width=5.8cm
    ]{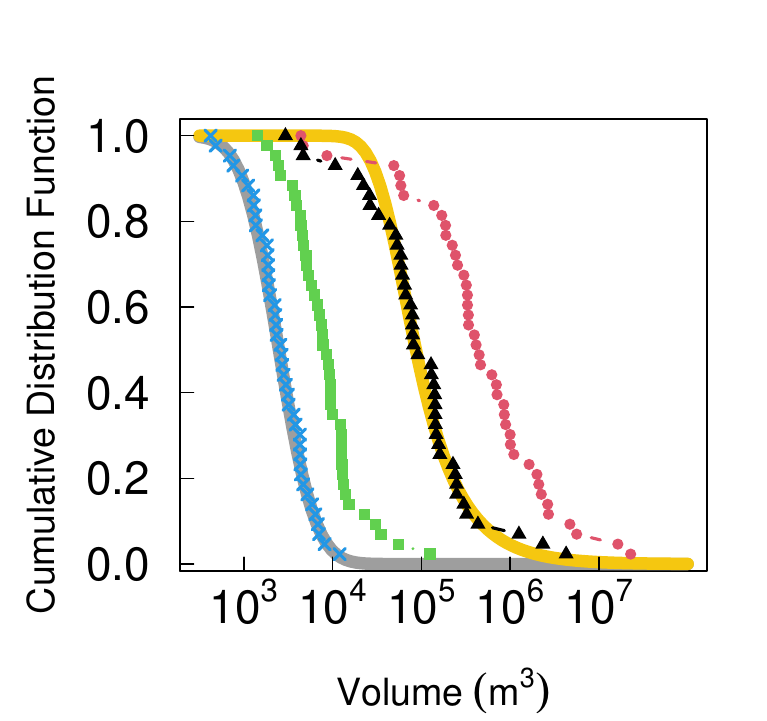}
        \put(120,112){\color{black}\Large\bfseries (e)}
    \end{overpic} 
    \begin{overpic}[
        width=5.8cm
    ]{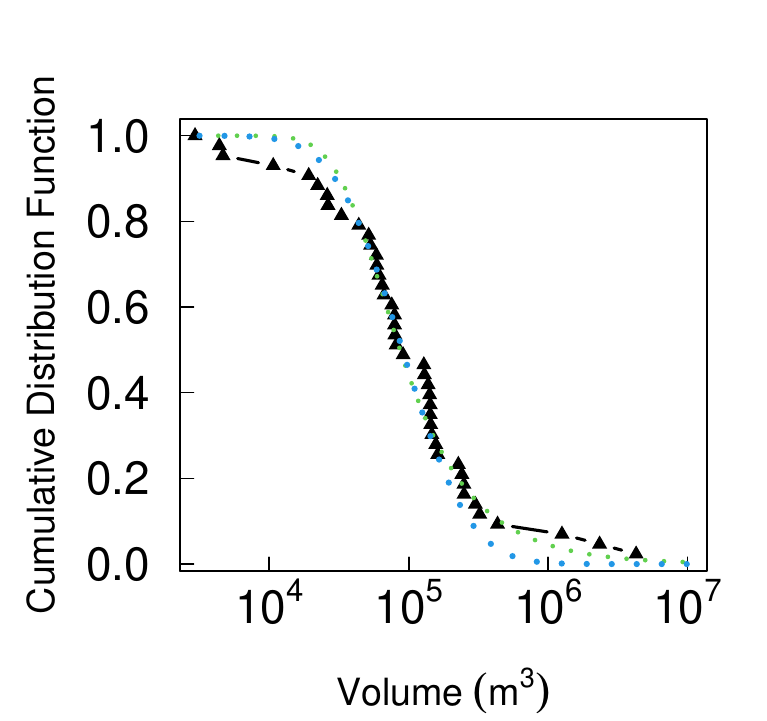}
        \put(120,112){\color{black}\Large\bfseries (f)}
    \end{overpic} 
  \caption{
Distributions of the number of  keyhole-shaped kofun data and kofun volume indicators aggregated by prefecture. The upper panels (a)--(c) use log--log axes, and the lower panels (d)--(f) show the corresponding semi-logarithmic plots.
(a)(d) Upper cumulative distribution of the number of kofun data by prefecture. Black triangles represent the empirical data, and the thick gray line represents a lognormal distribution. The parameters of the lognormal distribution for the number of prefecture-level data points, $N_p$, are $\mu=3.98$ and $\sigma=1.39$. The black dashed lines represent reference lines proportional to $1/x$ and $1/x^{1.3}$. The number of kofun by prefecture is close to a lognormal distribution in the central body and shows a power-law-like shape with a cumulative exponent of approximately $1.3$ in the tail.
(b)(e) Upper cumulative distributions of kofun volume indicators aggregated by prefecture. Black triangles, red circles, green squares, and blue crosses represent the maximum, total, mean, and median values, respectively. The thick yellow line represents a one-sided dPlN distribution with parameters $\mu=10.57$, $\sigma=0.54$, and $\alpha=1.00$. The thick gray line represents a lognormal distribution fitted to the distribution of the median values, with $\mu=7.81$ and $\sigma=0.74$. The black dashed line represents the reference line $1/x$. The maximum and total values show heavy tails close to exponent 1, whereas the median values are close to a lognormal distribution and show a relatively weaker heavy-tail structure.
(c)(f) Distribution of prefecture-level maximum values only. Black triangles represent the empirical data, the green dashed line represents the one-sided dPlN distribution, the blue dotted line represents the lognormal distribution, and the black dashed line represents the reference line $1/x$.
The one-sided dPlN distribution and the log-normal distribution were fitted using the same parameter values as those used in panel (b). 
The distribution of prefecture-level maximum values is close to a lognormal distribution in the central body and shows a power-law-like shape close to $1/x$ in the tail. Thus, the lognormal-like central body and the tail close to exponent 1 observed in the individual kofun volume distribution are largely preserved even after aggregation into prefecture-level maximum values.
}
\label{a_fig_kenkohun_main}
\end{figure*}

Fig.~\ref{a_fig_kenkohun_main}(a) and (d) show the upper cumulative distribution of the number of kofun, $N_p$, in each prefecture. The prefectural number of kofun is close to a lognormal distribution in the central part. In the tail, it shows a power-law-like shape with a cumulative exponent of about $1.3$. The Clauset method gives an upper cumulative exponent of $\alpha=1.29$. This indicates that the number of kofun itself varies greatly across prefectures.\par

Fig.~\ref{a_fig_kenkohun_main}(b) and (e) show the upper cumulative distributions of the kofun-volume measures aggregated by prefecture. The distribution of the maximum volume is approximately close to a $1/x$-type power law in the upper tail. The Clauset method gives an exponent of $\alpha=1.05$ for the upper cumulative distribution. Thus, the maximum kofun size in each prefecture has a heavy tail with an exponent close to one. This is similar to the volume distribution of individual kofun.\par

This result can also be interpreted naturally from the viewpoint of extreme-value statistics. If the volume distribution of individual kofun has a heavy tail, the distribution of maxima taken from each prefecture is also expected to keep a heavy tail. Fig.~\ref{a_fig_kenkohun_main}(c) and (f) show only the distribution of the maximum values. The distribution of maxima is close to a lognormal distribution in the central part. At the same time, it shows a power-law-like shape close to $1/x$ in the upper tail. It can also be roughly approximated by the one-sided dPlN distribution, Eq.~\eqref{app_eq_dPlN}. This suggests that the structure consisting of a lognormal-like central part and a power-law tail close to exponent one is largely preserved even after aggregation to prefectural maxima.\par

In contrast, the distribution of the median volume is very different from that of the maximum volume. As shown in Fig.~\ref{a_fig_kenkohun_main}(b) and (e), the median is close to a lognormal distribution. A heavy power-law-like tail is not clearly observed. The Clauset method gives a large upper cumulative exponent of $\alpha=4.14$ for the upper tail of the median. However, only six observations are included in the tail in this estimate. Therefore, the important point is not the exponent itself. Rather, the distribution of the median has a much thinner tail than that of the maximum. Since the median represents the typical kofun size in each prefecture, it is not expected to reflect regional inequality as strongly as the maximum does.\par

\begin{figure*}[t]
    \centering
    \begin{overpic}[
        width=18cm
    ]{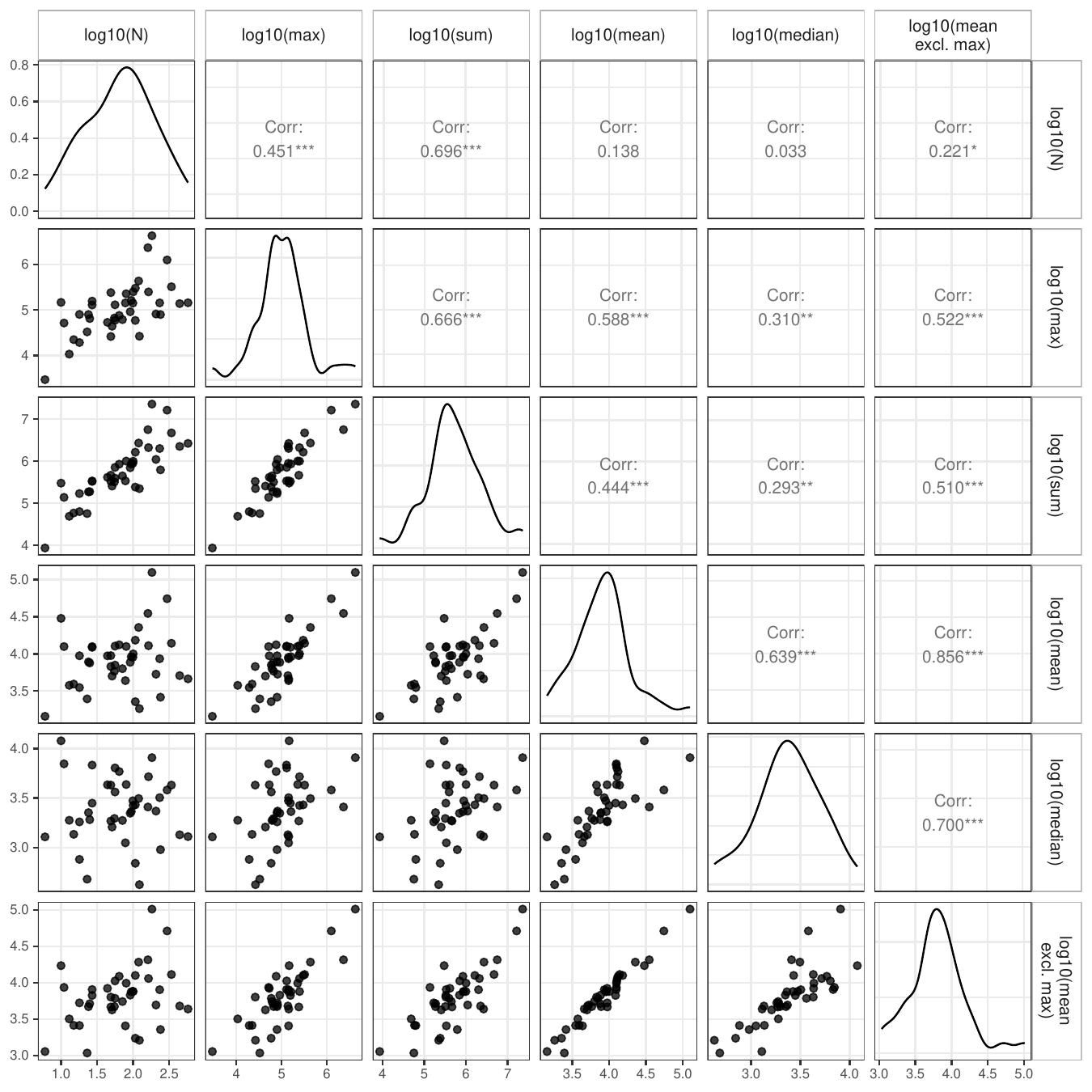}
    \end{overpic}     
    \caption{
      Correlations among keyhole-shaped kofun size indicators aggregated by prefecture. For each prefecture, we calculated the number of kofun, $N$, the maximum volume, $\mathrm{max}$, the total volume, $\mathrm{sum}$, the mean volume, $\mathrm{mean}$, the median volume, $\mathrm{median}$, and the mean volume excluding the maximum value, $\mathrm{mean\ excl.\ max}=(\mathrm{sum}-\mathrm{max})/(N-1)$, and compared their logarithmic values. The diagonal panels show the distribution of each indicator, the lower-left panels show scatter plots, and the upper-right panels show Kendall's rank correlation coefficients. The total volume is strongly positively correlated with both the number of kofun and the maximum volume, whereas the mean volume and median volume show only weak correlations with the number of kofun. Prefectures with larger maximum volumes also tend to have larger mean volumes even after excluding the maximum value.  
}
 \label{a_fig_kenkohun_xy}
\end{figure*}

The distribution of the mean volume also has a thinner tail than that of the maximum. The upper cumulative exponent for the mean volume is $\alpha=1.68$. This indicates that typical kofun size differs across prefectures, but this difference is smaller than the difference in maximum values. Fig.~\ref{a_fig_kenkohun_xy} also shows that the mean volume has a positive relation with the maximum volume. At the same time, it is strongly related to the median. Thus, the mean volume reflects the typical size within each prefecture.\par

The distribution of the total volume has an even heavier tail than that of the maximum. The Clauset method gives an upper cumulative exponent of $\alpha=0.781$ for the total volume. This is because the total volume depends not only on the maximum value, but also strongly on the number of kofun in each prefecture. In other words, the total volume is a composite measure. It includes the size of the largest kofun, the number of kofun, and the size differences among kofun other than the maximum.\par

Fig.~\ref{a_fig_kenkohun_xy} shows the relations among these aggregated measures. The number of kofun is strongly correlated with the total volume. It also shows a positive correlation with the maximum volume. Prefectures with a larger maximum volume also tend to have a larger mean volume after excluding the maximum. To evaluate these relations symmetrically, we performed Deming regression after log transformation. Unlike ordinary least-squares regression, Deming regression takes into account errors or variability in both the explanatory and dependent variables. As a result, the relation between the maximum volume $M_p$ and the number of kofun $N_p$ was approximated as
\begin{equation}
M_p \propto N_p^{1.06} .
\end{equation}
If the upper cumulative distribution of individual kofun volumes is close to $P(X>x)\propto x^{-1}$, the typical scale of the maximum among $N_p$ samples is expected to be roughly proportional to $N_p$. Therefore, this relation is consistent with the extreme-value statistics of a heavy-tailed distribution with an exponent close to one. However, this result does not directly show that regions with many kofun had an additional tendency to construct giant kofun. It can first be interpreted as a consequence of the heavy tail of the individual kofun-size distribution and the effect of sample size.\par

We also performed Deming regression for the total volume $S_p$. The results were
\begin{equation}
S_p \propto N_p^{1.36}
\end{equation}
and
\begin{equation}
S_p \propto M_p^{1.28} .
\end{equation}
Thus, the prefectural total volume tends to increase more strongly than simple proportionality with respect to both the number of kofun and the maximum volume.  \par
Overall, the prefecture-level analysis shows that the heavy tail of individual kofun volumes is largely preserved in prefectural maximum volumes, whereas median and mean volumes have thinner tails. Thus, the Zipf-like tail of the individual distribution is reflected more strongly in the largest kofun in each region than in typical regional size. Since modern prefectures were not political units in the Kofun period, this analysis should be regarded as a supplementary check of aggregated statistics rather than direct evidence for regional generative mechanisms.\par

\clearpage
\section{Details of dataset and methods}
\label{app_sec_method}

\subsection{Dataset}
\label{app_sec_method_data}
This study uses the National Kofun Database published by the Center for Ancient Studies and Sacred Sites, Nara Women's University \cite{NaraWomensUniversityKofunDatabase}. This database is a nationwide database of kofun. It includes keyhole-shaped kofun and square-fronted kofun. It also includes major round mounds, square mounds, and other kofun, including those designated as national or prefectural historic sites. In this study, we used the version last updated on March 31, 2025. This version contains a total of 6779 kofun.\par

The database includes information on the location, mound shape, chronological period, mound dimensions, excavated artifacts, and related attributes of each kofun. In this study, we used mound shape, location, chronological period, and mound dimensions. The analysis focuses on three types of kofun for which sufficient sample sizes are available: keyhole-shaped kofun, round mounds, and square-fronted kofun.\par

Approximately 160,000 kofun and horizontal tombs have been identified across Japan \cite{AgencyCulturalAffairs2021BuriedSites}. Therefore, the National Kofun Database does not cover all kofun in the Japanese archipelago. This limitation is especially important for round mounds. The database mainly includes major round mounds, such as those designated as historic sites. Thus, it cannot be treated as a complete population representing all round mounds in Japan. By contrast, the database has high coverage for keyhole-shaped kofun. The total number of known keyhole-shaped kofun in Japan is estimated to be about 4800 \cite{Izuta2016KeyholeDistribution}, while the database contains 4809 examples. The same is also true for square-fronted kofun. Their total number in Japan is estimated to be about 500 \cite{Wada2009OtherWorldKofun}, while the database contains 469 examples. Thus, for keyhole-shaped and square-fronted kofun, the coverage of known examples is considered to be high.\par
Some mound dimensions are missing in the database. For kofun with missing values in some of the dimensions needed for volume estimation, we applied PCA-based imputation, as described below. By contrast, kofun for which all major dimensions needed for volume estimation were missing were excluded from the analysis of volume distributions. In analyses by chronological period, kofun with unknown periods were also excluded. Therefore, the number of kofun recorded in the database does not necessarily match the number of kofun used in each analysis. The sample sizes used for each classification are shown in Table~\ref{app_tab:mound_counts} ,\ref{tab_table1}, \ref{app_tab_table1} and ~\ref{app_tab_table2}.\par

\begin{table}[hbt]
\centering
\caption{Number of mounds for which volume could be estimated, by century and mound type.}
\label{app_tab:mound_counts}
\begin{tabular}{lccc}
\hline
Century & Keyhole-shaped & Circular & Square-fronted \\
\hline
2nd     & 2    & -- & -- \\
3rd     & 1    & 1  & 4 \\
4th     & 597  & 73 & 229 \\
5th     & 486  & 204 & 12 \\
6th     & 1005 & 224 & 21 \\
7th     & 2    & 95 & 2 \\
8th     & --   & 5 & -- \\
Missing & 2452 & 103 & 187 \\
Total   & 4545 & 705 & 455 \\
\hline
\end{tabular}
\end{table}

\subsection{Missing-value imputation and estimation of mound volume}
\label{app_sec_method_volume}

In this study, mound volume is used as an indicator of the politico-economic scale of kofun (see section \ref{app_sec_model_volume}). Mound volume was estimated from the main dimensions of each kofun using simple geometric approximations. All lengths are measured in meters. The estimated volume is expressed in $\mathrm{m}^3$. \par

Because the dimensional data include missing values, missing-value imputation was performed before calculating volumes. For this purpose, we used the \texttt{imputePCA} function in the R package \texttt{missMDA} \cite{Josse2016MissMDA}. For each type of kofun, we constructed a data matrix consisting of the main dimensions used for volume calculation. PCA-based imputation was then applied with the number of components set to $ncp=1$. Kofun with at least one observed main dimension were included in the imputation. Kofun for which all main dimensions were missing were excluded from volume estimation.\par

The first principal component was used for imputation because the main dimensions of kofun strongly depend on a common size scale. In other words, kofun with larger total mound length or diameter tend to have larger lengths, widths, and heights in their component parts. Therefore, the first principal component can be interpreted as an axis representing the overall scale of each kofun. In the observed values before imputation, the dimensions of each component also show approximately proportional relationships with total mound length or diameter. This relation is examined in Appendix~\ref{app_sec_shape}.\par

For keyhole-shaped kofun, the volume was approximated by separating the front square part and the rear round part. Let $W_f$, $L_f$, and $H_f$ denote the width, length, and height of the front part. Let $D_r$ and $H_r$ denote the diameter and height of the rear round part. Let $L$ denote the total mound length. PCA-based imputation was applied using the following six variables:
\begin{equation}
W_f,\quad L_f,\quad H_f,\quad D_r,\quad H_r,\quad L .
\end{equation}
Using the imputed values, the volume of a keyhole-shaped kofun, $V_{\mathrm{key}}$, was approximated as
\begin{equation}
V_{\mathrm{key}}=W_f L_f H_f+
\pi\left(\frac{D_r}{2}\right)^2 H_r .
\end{equation}
The first term approximates the volume of the front part as a rectangular prism. The second term approximates the rear round part as a cylinder.\par

For square-fronted kofun, the front part and the rear square part were each approximated as rectangular prisms. Let $W_f$, $L_f$, and $H_f$ denote the width, length, and height of the front part. Let $W_b$, $L_b$, and $H_b$ denote the width, length, and height of the rear square part. Let $L$ denote the total mound length. PCA-based imputation was applied using the following seven variables:
\begin{equation}
W_f,\quad L_f,\quad H_f,\quad W_b,\quad L_b,\quad H_b,\quad L .
\end{equation}
Using the imputed values, the volume of a square-fronted kofun, $V_{\mathrm{sq}}$, was approximated as
\begin{equation}
V_{\mathrm{sq}}=W_f L_f H_f+W_b L_b H_b .
\end{equation}
Here, the first term represents the volume of the front part. The second term represents the volume of the rear square part.\par

For round mounds, volume was approximated from diameter and height. When both the east-west diameter $D_{\mathrm{EW}}$ and the north-south diameter $D_{\mathrm{NS}}$ were recorded, the representative diameter $D$ was defined as
\begin{equation}
D=\sqrt{D_{\mathrm{EW}}D_{\mathrm{NS}}}.
\end{equation}
This corresponds to the diameter of a circle with the same area as an ellipse-like plan shape. When the east-west and north-south diameters were not available, the recorded diameter of the round mound was used.\par

For round mounds, PCA-based imputation was applied using two variables: the representative diameter $D$ and the height $H$. Using the imputed values, the volume of a round mound, $V_{\mathrm{round}}$, was approximated as
\begin{equation}
V_{\mathrm{round}}=\pi\left(\frac{D}{2}\right)^2 H .
\end{equation}
\par

These volume approximations do not reproduce the exact three-dimensional shapes of the mounds. However, this study aims to analyze relative mound size and distributional shape, not to reconstruct precise volumes. For this purpose, an approximate volume measure is sufficient for comparing distributions and estimating power-law exponents.
 \par
 
\subsection{Estimation of power-law exponents}
\label{app_sec_method_powerlaw}

We estimated power-law exponents for several size distributions, including kofun volume distributions. The purpose was to examine whether their tails follow power-law distributions. For the estimation, we used the R package \texttt{poweRlaw} \cite{Gillespie2015Powerlaw}. This package provides methods based on the standard procedure proposed by Clauset et al. \cite{Clauset2009PowerLaw}. We used a continuous power-law model. The lower cutoff $x_{\min}$ was estimated as the value that minimizes the Kolmogorov--Smirnov distance. The power-law exponent was then estimated by maximum likelihood for the data satisfying $x\ge x_{\min}$.\par

The exponent estimated by \texttt{poweRlaw} corresponds to the exponent $\zeta$ of the probability density function,
\begin{equation}
p(x)\propto x^{-\zeta}.
\end{equation}
In this study, however, we mainly use the upper cumulative distribution,
\begin{equation}
P(X>x)\propto x^{-\alpha}.
\end{equation}
Therefore, we report the exponent of the upper cumulative distribution as
\begin{equation}
\alpha=\zeta-1.
\end{equation}
\par

The uncertainty of the estimates was evaluated by bootstrap resampling. For each distribution, we performed 1000 bootstrap resamplings. For each bootstrap sample, we re-estimated $x_{\min}$ and the PDF exponent $\zeta$. The resulting bootstrap distribution of $\zeta$ was smoothed by kernel density estimation. The value at which this density was maximized was used as the representative PDF exponent (i.e., the mode of the bootstrap samples was taken as the representative value). We then subtracted one from this value. This gave the representative exponent $\alpha$ of the upper cumulative distribution. The 95\% bootstrap confidence interval for $\alpha$ was obtained in the same way. Specifically, we subtracted one from the 2.5\% and 97.5\% quantiles of the bootstrap distribution of $\zeta$ (Tables~\ref{tab_table1}, \ref{app_tab_table1}, and \ref{app_tab_table2}).\par

\subsection{Clustering of Distributions by Region and Period}
\label{app_sec_distribution_clustering}

In this section, we describe the clustering method used in Fig.~\ref{fig_map}(b). This analysis was conducted as a supplementary visualization of differences in distributional shape across regions and periods.\par

Let the kofun-size data for each region--period group be
\begin{equation}
X^{(g)}={x^{(g)}_1,x^{(g)}_2,\ldots,x^{(g)}_n}.
\end{equation}
Here, $g$ denotes a combination of region and period. When comparing distributions, we normalized the data in each group by its median. This reduces the effect of absolute scale differences across regions and periods. Specifically, we used
\begin{equation}
\tilde{x}^{(g)}_i=\frac{x^{(g)}_i}{\mathrm{Median}(X^{(g)})}.
\end{equation}\par

For any two groups $g$ and $h$, we generated bootstrap samples from the normalized data $\tilde{X}_g$ and $\tilde{X}_h$. We then calculated the one-dimensional Wasserstein distance between their empirical distributions. The one-dimensional Wasserstein distance used in this study is written in terms of quantile functions as
\begin{equation}
W_1(F_g,F_h)=\int_0^1\left|F_g^{-1}(u)-F_h^{-1}(u)\right|du .
\end{equation}
Here, $F_g$ and $F_h$ are the empirical distribution functions of groups $g$ and $h$, respectively. Thus, the one-dimensional Wasserstein distance represents the average distance between corresponding quantiles of the two distributions.\par

Let the distance for bootstrap sample $b$ be
\begin{equation}
d_b(g,h)=W_1\left(\hat{F}_{g}^{(b)},\hat{F}_{h}^{(b)}\right).
\end{equation}
Here, $\hat{F}_{g}^{(b)}$ and $\hat{F}_{h}^{(b)}$ are the empirical distributions obtained from the bootstrap samples. We repeated this procedure $B=1000$ times. This gave a bootstrap distribution of distances between the two groups.\par

We used the mode of the bootstrap distance distribution as the representative distance between the two groups. In the implementation, the bootstrap distance distribution was smoothed by kernel density estimation. The representative distance was then defined as the value at which this density was maximized:
\begin{equation}
d(g,h)=\arg\max_r {\hat{f}_{g,h}(r)}.
\end{equation}
Here, $\hat{f}_{g,h}(r)$ is the kernel density estimate of the distances $d_b(g,h)$ obtained from the $B=1000$ bootstrap samples. This procedure produced the distance matrix $D=(d(g,h))$ between the region--period distributions.\par

In practice, small numerical differences may arise between $d(g,h)$ and $d(h,g)$ because of the bootstrap procedure. We therefore symmetrized the distance matrix as
\begin{equation}
D \leftarrow \frac{D+D^\mathsf{T}}{2}
\end{equation}
and set the diagonal elements to zero. Finally, we applied hierarchical clustering to this distance matrix.
The average linkage method (\texttt{average}) was used for clustering. 
\clearpage
\section{Details of the Kesten-process model}
\label{app_sec_model}
This section describes in detail the Kesten-type model introduced in Sec.~\ref{sec_model}. 
We explain how the model generates a distribution with a log-normal-like body and a power-law tail, and how its parameters can be interpreted in relation to kofun volume distributions.
\subsection{Kofun volume as a proxy for mobilizing capacity}
\label{app_sec_model_volume}
Because kofun are three-dimensional earthen mounds, volume is a natural quantity for representing their physical scale. 
Moreover, the earthwork component of construction cost can be expected to increase roughly with the amount of earth piled up \cite{Obayashi1985Nintoku}.  
In this sense, volume is closer than mound length to the amount of resources and labor invested in kofun construction.  \par
For example, consider Daisenry\=o Kofun, the largest kofun, with a mound length of approximately 500~m and a volume of approximately 1.4 million~$\mathrm{m}^3$. According to an estimate by Obayashi Corporation, a Japanese construction company, building this mound using ancient construction techniques would require approximately 16 years and a total of approximately 6.8 million person-days (an average of approximately 1{,}200 people per day, or 4.9~person-days$/\mathrm{m}^3$) \cite{Obayashi1985Nintoku}. Nishida Kazuhiko and colleagues similarly estimate that Minegazuka Kofun, with a total length of 96.0~m (fill volume of 55{,}000~$\mathrm{m}^3$), would have required approximately 1{,}070 days and a total of 260{,}000 person-days (an average of approximately 240 people per day, or 4.8~person-days$/\mathrm{m}^3$), assuming a maximum daily workforce of 500 people (\textit{Journal of the Japan Society of Civil Engineering History}, Vol.~13, pp.~281--288, 1993). In addition, according to the Kamitsukenosato Museum, Hodota Hachimanzuka Kofun, with a total length of approximately 96~m (fill volume of 14{,}500~$\mathrm{m}^3$), is estimated to have required a total of 62{,}000 person-days (4.3~person-days$/\mathrm{m}^3$). 
These estimates converge on a labor requirement of roughly 4 to 5 person-days per~$\mathrm{m}^3$, suggesting that mound volume can serve as a reasonably consistent proxy for construction labor across different kofun. \par
In this study, we further use kofun volume as a proxy for the resources and mobilizing capacity of the local group associated with its construction. 
This assumption can be understood as follows. 
Kofun construction involved substantial real costs. 
If a group built a mound far beyond its actual capacity, the construction would likely have placed a heavy burden on its resource base and made it difficult to maintain its power. 
At the same time, by observing earlier and neighboring kofun, local groups may have been able to judge the approximate mound size appropriate to their status and mobilizing capacity. 
Through such cost constraints and comparison with surrounding examples, a broad correspondence between kofun volume and mobilizing capacity may have been formed. Alternatively, the correspondence may have arisen through the practical process of construction itself. Builders responsible for kofun construction may have estimated the feasible mound size from the labor force and resources that the local group could mobilize. In that case as well, larger local groups would naturally have constructed larger kofun.

\subsection{Model formulation}
Following the principle of parsimony, or Occam's razor, this study aims to describe the distribution of kofun volumes with a minimal number of parameters.
Specifically, we adopt the following discrete-time model based on a multiplicative random growth process, namely the Kesten process \cite{Kesten1973RandomDifference,SornetteCont1997Multiplicative,TakayasuEtAl1997RandomAmplification}.\par

Let the state variable $x(t)$ denote the politico-economic scale of a chief or local group at time $t$.
At each time step, the scale $x(t)$ grows multiplicatively and receives an additive resource $A_0$ with probability $p_0$,
whereas extinction or reorganization of the group occurs with probability $1-p_0$.
When extinction or reorganization occurs, a new group enters with a size proportional to the characteristic economic scale $A_0$.
This process is written as follows:
\begin{equation}
x(t+1) =
\begin{cases}
b_0 \, x(t) + A_0 & \text{with probability } p_0 \\
A_0 \cdot x_{\mathrm{new}} & \text{with probability } 1 - p_0
\end{cases}
\label{app_eq:kesten}
\end{equation}
Here, $b_0>1$ is the multiplicative growth factor, and $A_0$ is the characteristic economic scale of the system.
The variable $x_{\mathrm{new}}$ is a random variable representing the relative scale of a newly entering group.
It is sampled from a uniform distribution normalized to have expectation 1, that is, a uniform distribution on the interval from 0 to 2:
\begin{equation}
x_{\mathrm{new}} \sim \mathcal{U}(0,\,2).
\label{eq:uniform}
\end{equation}
The kofun volume $y(t)$ is expressed using the ratio $Q$ of kofun construction cost to economic scale as
\begin{equation}
y(t) = Q \cdot x(t).
\label{eq:volume}
\end{equation}

\subsection{Relation to the dPlN model and the one-sided Pareto tail}
\label{app_sec_model_dpln}
The process in Eq.~\eqref{eq:kesten} generates both a log-normal-like central part, or body, and a Pareto power-law heavy tail through the same mechanism as the double Pareto-lognormal (dPlN) model of Reed and Jorgensen, namely the combination of multiplicative growth and a random stopping time \cite{ReedJorgensen2004dPlN} (Appendix~\ref{app_sec_dPlN}).
However, the distribution generated by the model used in this study does not correspond to the two-sided dPlN distribution, but rather to its one-sided counterpart: only the tail corresponding to large-volume kofun follows a power law. This model shows good approximate agreement with the one-sided dPlN distribution; however, an (almost) exact mathematical correspondence holds only for a special case of the present model, as discussed in Appendix~\ref{app_sec_dPlN_kesten}. \par
A similar mathematical structure, especially the emergence of heavy tails through exponential growth and random stopping as in the present model, is also found in Jones and Kim's top-income inequality model (the simple model in \cite{Jones2018}).\par

\subsection{Size distribution of the model}
\label{subsec:kesten_analysis}

\subsubsection{Determination of the power-law exponent}
\label{app_sec_model_power}
Eq.~\eqref{eq:kesten} is a special case of the Kesten process $x(t+1)=b(t)\,x(t)+f(t)$.
According to the general theory of Kesten processes, when the multiplier $b(t)$ satisfies
\begin{equation}
\langle b(t)^{\alpha} \rangle = 1,
\label{eq:kesten_cond}
\end{equation}
the tail of the upper cumulative distribution of the random variable $X$ follows a power law,
\begin{equation}
\Pr(X > x) \propto x^{-\alpha}
\label{eq:pareto}
\end{equation}
\cite{SornetteCont1997Multiplicative,TakayasuEtAl1997RandomAmplification}.

In the present model, the multiplier $b(t)$ takes two values:
$b(t)=b_0$ with probability $p_0$, and $b(t)=0$ with probability $1-p_0$.
In this case, condition~\eqref{eq:kesten_cond} becomes
\begin{equation}
p_0\, b_0^{\alpha} = 1,
\label{eq:kesten_binary}
\end{equation}
which gives the relation between the growth factor and the survival probability,
\begin{equation}
b_0 = p_0^{-1/\alpha}.
\label{app_eq:b0_alpha}
\end{equation}
In particular, for $\alpha=1$, corresponding to Zipf's law, we obtain $b_0=1/p_0$.
\subsubsection{Zero-sum-like interpretation of the Zipf condition}

The condition for Zipf's law, $\alpha=1$, can be written as $p_0 b_0=1$.
This condition has an intuitive interpretation as a zero-sum-like balance in relative shares.
For example, when $p_0=1/2$, half of the groups are reset on average, and the surviving groups double their relative shares by absorbing the shares of the reset groups.
This interpretation applies to the multiplicative part of the model.
Because the additive term $f(t)=A_0$ is also present, the full process is not a strictly zero-sum system.
Rather, it can be viewed as a quasi-zero-sum competitive process in which externally supplied resources are added and then distributed through competition.
As one possible intuition, the multiplicative term may be viewed as capturing share competition through conflict, political reorganization, or differential access to scarce resources and exchange networks, such as iron materials and technologies. By contrast, the additive term may be viewed as capturing gradual resource accumulation, such as annual agricultural income or surplus.

\subsubsection{Intuitive origin of the log-normal-like body}
We next give an intuitive explanation for why the central part of the model distribution can be approximated by a log-normal distribution.

In the present model, the origin of this log-normal-like body can be understood intuitively from the fact that, for groups that stop after a short growth duration, the additive term $A_0$ has a relatively large effect.\par

In particular, when the scale $x(t)$ is still small and the parameter corresponding to the empirical data satisfies $b_0 \simeq 1$,
the contribution of the additive term dominates over multiplicative amplification.
In this case,
\begin{equation}
x(t+1) \simeq x(t)+A_0
\end{equation}
and therefore the scale after $T$ growth steps can be roughly approximated as
\begin{equation}
x_T \simeq A_0 T + A_0 x_{\mathrm{new}} .
\end{equation}
Thus, for short-lived groups, the scale increases approximately in proportion to the growth duration $T$.\par

The growth duration $T$ is the number of consecutive growth steps before reset, and follows a geometric distribution with stopping probability $1-p_0$.
Therefore, under the above approximation, $x_T$ is approximately a linear transformation of $T$ and has a and has a density close to a monotonically decreasing exponential form in real space. 
By contrast, when this distribution is viewed in terms of $\log x_T$, it becomes unimodal, and the central part of this distribution can be approximated by a normal distribution.
This provides an intuitive explanation for why a log-normal-like body appears in real space.
A more detailed discussion of this point is given in Appendix~\ref{app_sec_kesten_lnorm}.\par
\subsubsection{Full-range distribution and scale--shape decomposition}
\label{app_sec_model_decomposition}

We next write the distribution over the full range, including the tail, while keeping the above intuition in mind.
The economic scale of a group that has grown for $T$ steps can be written as the sum of a geometric series:
\begin{equation}
x_T = A_0 \left( 2U \cdot b_0^{T} + \frac{b_0^{T}-1}{b_0-1} \right).
\label{eq:xt}
\end{equation}
Here, $U \sim \mathcal{U}(0,1)$ is a uniform random variable derived from the initial size at entry.
The growth duration $T$ is the number of consecutive growth steps before reset and follows a geometric distribution,
\begin{equation}
\Pr(T=k)=(1-p_0)p_0^k, \qquad k=0,1,2,\dots .
\end{equation}
\par

If we define the dimensionless random variable inside the parentheses as
\begin{equation}
J_T =2U b_0^T + \frac{b_0^T-1}{b_0-1},
\label{eq:J}
\end{equation}
then the scale can be written as
\begin{equation}
x_T = A_0 \cdot J_T .
\label{eq:xJt}
\end{equation}
This decomposition allows us to interpret $A_0$ as a scale factor that shifts the entire distribution to the left or right.
By contrast, $J_T$ can be interpreted as a dimensionless competitive outcome determined by the history of growth, stopping, and re-entry, and hence as the factor that determines the distributional shape after removing scale.
\par

We can make this scale--shape decomposition explicit by deriving the probability density of $J_T$.
For a fixed growth duration $T=k$, define
\begin{equation}
\ell_k=\frac{b_0^k-1}{b_0-1},
\qquad
u_k=\ell_k+2b_0^k .
\label{eq:ell_u}
\end{equation}
Then Eq.~\eqref{eq:J} can be written as
\begin{equation}
J_T=\ell_k+(u_k-\ell_k)U
\qquad (T=k).
\end{equation}
Because $U$ is uniformly distributed on $[0,1]$, the conditional distribution of $J_T$ given $T=k$ is uniform on the interval $[\ell_k,u_k]$:
\begin{equation}
f_{J\mid T=k}(j)=\frac{1}{u_k-\ell_k}
\mathbf{1}_{\{\ell_k \le j \le u_k\}} .
\label{eq:fJ_cond}
\end{equation}
Here, $\mathbf{1}_{\{A\}}$ equals 1 if the condition $A$ holds and 0 otherwise.
Since $T$ follows a geometric distribution, the unconditional density of $J_T$ is the following mixture of uniform distributions:
\begin{equation}
f_J(j)=\sum_{k=0}^{\infty} (1-p_0) p_0^k \frac{1}{u_k-\ell_k} \mathbf{1}_{\{\ell_k \le j \le u_k\}} .
\label{eq:fJ}
\end{equation}
Thus, the shape of the stationary distribution is determined by the dimensionless density $f_J$, while the scale is determined by $A_0$.
Indeed, since $X=A_0J$, the density of $X$ is
\begin{equation}
f_X(x)=\frac{1}{A_0} f_J\left(\frac{x}{A_0}\right).
\label{eq:fX}
\end{equation}
If an additional proportionality factor $Q$ is used to convert the model scale into kofun volume, so that $Y=QA_0J$, then
\begin{equation}
f_Y(y)=\frac{1}{QA_0} f_J\left(\frac{y}{QA_0}\right).
\label{eq:fY}
\end{equation}
This shows explicitly that $A_0$, or $QA_0$ for volume, controls the horizontal scale of the distribution, whereas $f_J$ controls its scale-free shape.
\par

\subsubsection{Stationary upper cumulative distribution}
\label{app_sec_model_prob}

The upper cumulative distribution can be obtained by integrating the density in Eq.~\eqref{eq:fJ}.
Here, the upper cumulative distribution denotes the probability of taking a value greater than or equal to a given value.
We define the linearly saturated function on the interval $[0,1]$ as
\begin{equation}
G(z) =
\begin{cases}
0  & z \le 0, \\
z  & 0 < z < 1, \\
1  & z \ge 1 .
\end{cases}
\label{eq:G}
\end{equation}
For a fixed growth duration $T=k$, the conditional probability that $J_T$ is greater than or equal to $j$ is
\begin{equation}
\Pr(J_T\ge j\mid T=k)=
1-G\left(\frac{j-\ell_k}{u_k-\ell_k}\right).
\end{equation}
By averaging over the geometric distribution of $T$, we obtain
\begin{equation}
\bar{F}_J(j)=\Pr(J\ge j)=1-\sum_{k=0}^{\infty}
(1-p_0) p_0^k \cdot G\left(\frac{j-\ell_k}{u_k-\ell_k}\right).
\label{eq:ccdf_J}
\end{equation}
Since $X=A_0J$, this gives
\begin{equation}
\bar{F}_X(x)=\Pr(X\ge x)=1-\sum_{k=0}^{\infty}(1-p_0) p_0^k \cdot G\left(\frac{x/A_0-\ell_k}{u_k-\ell_k}\right).
\label{eq:ccdf}
\end{equation}
Equivalently, by defining the dimensional lower and upper bounds
\begin{equation}
L_k=A_0\ell_k,
\qquad
U_k=A_0u_k ,
\end{equation}
Eq.~\eqref{eq:ccdf} can be written as
\begin{equation}
\bar{F}_X(x)=1-\sum_{k=0}^{\infty}(1-p_0) p_0^k \cdot G\left(\frac{x-L_k}{U_k-L_k}\right).
\label{eq:ccdf_dimensional}
\end{equation}
If the model scale is converted into volume as $Y=QA_0J$, then $x/A_0$ in Eq.~\eqref{eq:ccdf} should be replaced by $y/(QA_0)$.
\par

The distribution given by Eq.~\eqref{eq:ccdf} shows a central part whose shape is close to a log-normal distribution over a finite size range, and asymptotically approaches the power law in Eq.~\eqref{eq:pareto} in the tail.
Thus, the present model explains both the log-normal-like body and the Pareto power-law tail observed in the kofun volume distribution through a single process of growth, stopping, and re-entry.
The red dashed line in Fig.~\ref{fig_volume}(a) shows this theoretical distribution.
\par

\subsection{The Additive Term in the Kesten Process and the Lognormality of the Body}
\label{app_sec_kesten_lnorm}

The intuition for the body of the proposed Kesten-type model, Eq.~\ref{eq:kesten}, can be understood from the accumulation of the additive term during short growth periods. The size after $T$ growth steps following a reset can be written as
\begin{equation}
x_T=A_0\left(
2U b_0^T+\frac{b_0^T-1}{b_0-1}
\right) .
\end{equation}
When $T$ is small and $b_0$ is close to one, we can approximate
\begin{equation}
\frac{b_0^T-1}{b_0-1}
\simeq
\frac{\log b_0}{b_0-1}T .
\end{equation}
Thus, the size of a short-lived group can be regarded approximately as
\begin{equation}
x_T\simeq aT ,
\end{equation}
where
\begin{equation}
a=A_0\frac{\log b_0}{b_0-1} . \label{app_eq_a}
\end{equation}
This approximation can be understood intuitively as follows. When $b_0$ is close to one, the multiplicative amplification during the early stage is weak. The growth dynamics are therefore close to
\begin{equation}
x(t+1)=b_0 x(t)+A_0 \simeq x(t)+A_0 .
\end{equation}
In this regime, the additive term is accumulated almost linearly, and the size of a short-lived group is approximately
\begin{equation}
x(t)\simeq A_0 t .
\end{equation}
This agrees with Eq.~\eqref{app_eq_a}, since $\log b_0/(b_0-1)\simeq 1$ for $b_0\simeq 1$, giving $a\simeq A_0$. \par
\begin{figure}[t]
    \centering
    \begin{overpic}[
        width=8cm
    ]{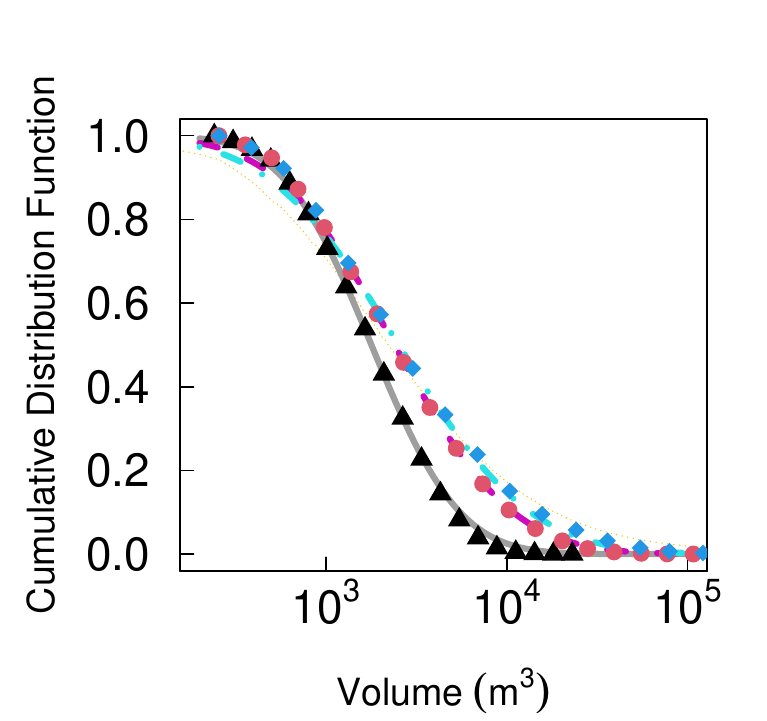}
    \end{overpic}     
  \caption{
Accumulation of additive terms and the formation of a lognormal-like body in the Kesten process (Eq.~\eqref{eq:kesten}). The yellow dotted line shows the simulation result of the Kesten process with $A_0=255.81$, $b_0=1.128$, $\alpha=1$, and $p_0=b_0^{-\alpha}$, using $10^5$ samples. The black triangles, red circles, and blue diamonds represent the distributions of $y_1$, $y_2$, and $y_3$, respectively, obtained by retaining terms up to first, second, and third order in $T$ in the expansion of Eq.~\eqref{app_eq_lnorm}. Specifically, we set
$y_1=aT+2A_0U$,
$y_2=aT+cT^2+2A_0U$, and
$y_3=aT+cT^2+dT^3+2A_0U$,
where $T$ is the growth duration and $U$ is a uniform random variable on $[0,1]$.
These quantities represent the growth effect in the short-lived regime where the additive term is dominant. For example, when $b_0\simeq 1$, $y_1\sim A_0T$, corresponding approximately to growth driven only by the additive term, $y(t+1)=y(t)+A_0$.
The smooth curves are lognormal distributions fitted to $y_1$, $y_2$, and $y_3$ using $\mu=\mathrm{median}(\log y)$ and $\sigma=\mathrm{MAD}(\log y)$. The corresponding parameter values are $(\mu,\sigma)=(7.49,0.881)$, $(7.78,1.18)$, and $(7.83,1.31)$, respectively.
As the higher-order terms $T^2$ and $T^3$ are included, the distributions approach the simulation result of the Kesten process and are also well approximated by lognormal distributions. This result shows that the lognormal-like body of the distribution can arise approximately from the accumulation of additive terms in short-lived groups.
%
}
\label{a_fig_lnorm}
\end{figure}

The growth duration $T$ follows a geometric distribution. In the continuous approximation, this corresponds to an exponential-type lifetime distribution. Therefore, under the first-order approximation $x_T\simeq aT$, the size distribution of short-lived groups is also close to an exponential distribution. The logarithm of an exponentially distributed variable is not exactly normally distributed. 
However, on the logarithmic scale, it forms a unimodal distribution and has a shape close to a normal distribution around the center as shown by the black triangles in Fig. \ref{a_fig_lnorm}. For this reason, many groups with short growth durations form a body that is close to a lognormal distribution on the logarithmic scale.

In this respect, the present model produces a distributional shape similar to that of the dPlN distribution. In the dPlN distribution, a lognormal-like body is introduced by explicitly including a normal component in log size. 
In contrast, the Kesten-type model used in this study can produce a lognormal-like body, even without explicitly assuming either a lognormal entry-size distribution or normally distributed noise in the growth rate.
It emerges from the accumulation of the additive term and the mixture of short growth durations. Thus, the two models differ in the mechanism that generates the body, but they produce very similar shapes in the central part of the distribution.

More precisely, we can expand
\begin{equation}
\frac{b_0^T-1}{b_0-1}=
\frac{\log b_0}{b_0-1}T
+
\frac{(\log b_0)^2}{2(b_0-1)}T^2
+
\frac{(\log b_0)^3}{6(b_0-1)}T^3
+\cdots . \label{app_eq_lnorm}
\end{equation}
By including the $T^2$ and $T^3$ terms, the approximation improves from the short-lived region to the intermediate region. 
Numerically, this gives a central body that is closer to a lognormal distribution (Fig. \ref{a_fig_lnorm}). However, this lognormal-like shape is not derived from an exact normal component. Instead, it arises approximately from the additive term. For short growth periods, the additive contribution grows almost linearly with $T$, with higher-order corrections. Because the growth duration follows a geometric distribution, which is approximated by an exponential distribution in continuous time, most groups are relatively short-lived. These short-lived groups therefore make up the central lognormal-like body of the distribution. \par
On the other hand, in the region where $T$ is large, the multiplicative term becomes dominant, and
\begin{equation}
x_T\sim A_0 b_0^T .
\end{equation}
The combination of this exponential growth and the geometrically distributed growth duration generates a Pareto-type power law in the tail. Thus, the present model simultaneously generates a lognormal-like body through the additive accumulation of short-lived groups and a Pareto-type tail through the multiplicative growth of long-lived groups.

%
%
%
%
%
\subsection{Dependence of the Kesten Model on the Entry-Size Distribution}
\label{app_sec_additive}

In this section, we examine how the stationary distribution changes when the distribution of initial sizes at new entry is varied in the Kesten process used in the main text, Eq.~\eqref{eq:kesten}. In the main model, the relative size at new entry was assumed to follow a uniform distribution. That is, when reorganization or reset occurs, the initial size was assumed to be drawn from a uniform distribution with mean one.\par

To check the robustness of this assumption, we introduce a random variable $v$ that represents the relative size at new entry. We compare the cases of a uniform distribution, a truncated normal distribution, a lognormal distribution, a gamma distribution, a beta distribution, and constant entry. In all cases, the scale was adjusted so that the mean entry size was approximately the same.\par


The exponent of the tail in the Kesten process is mainly determined by the multiplicative process. As discussed in the main text, in the present model $b(t)=b_0$ with probability $p_0$, and $b(t)=0$ with probability $1-p_0$. Therefore, the cumulative exponent $\alpha$ of the tail is determined by
\begin{equation}
p_0 b_0^\alpha = 1 .
\end{equation}
Thus, as long as the entry-size distribution $V$ does not have an extremely heavy tail, changing its detailed form is not expected to greatly change the exponent of the tail of the stationary distribution. On the other hand, the entry-size distribution may affect the lower and central parts of the distribution. We therefore examine this effect numerically.\par

\begin{figure*}[t]
    \begin{center}
    \begin{overpic}[
        width=5.8cm
    ]{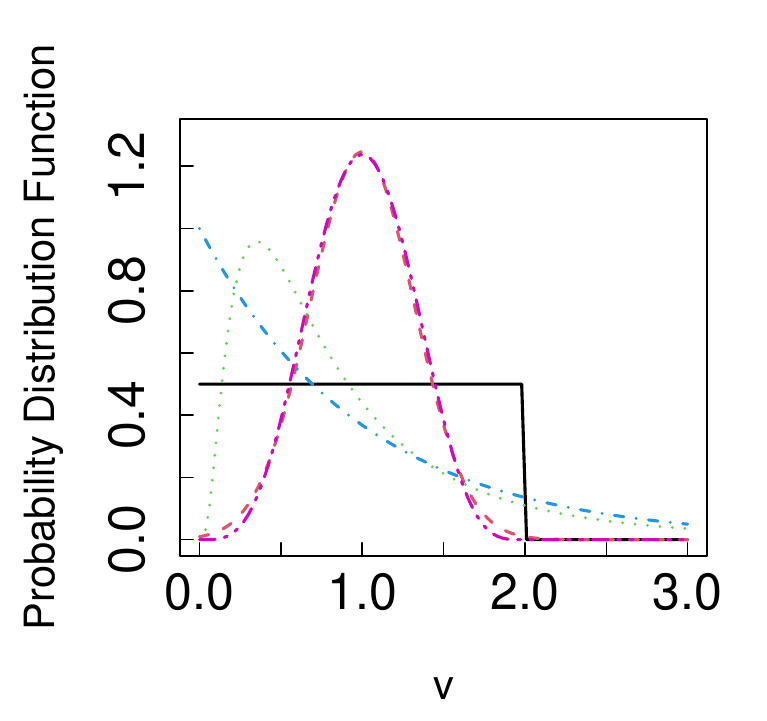}
        \put(120,113){\color{black}\Large\bfseries (a)}
    \end{overpic} 
     \begin{overpic}[
        width=5.8cm
    ]{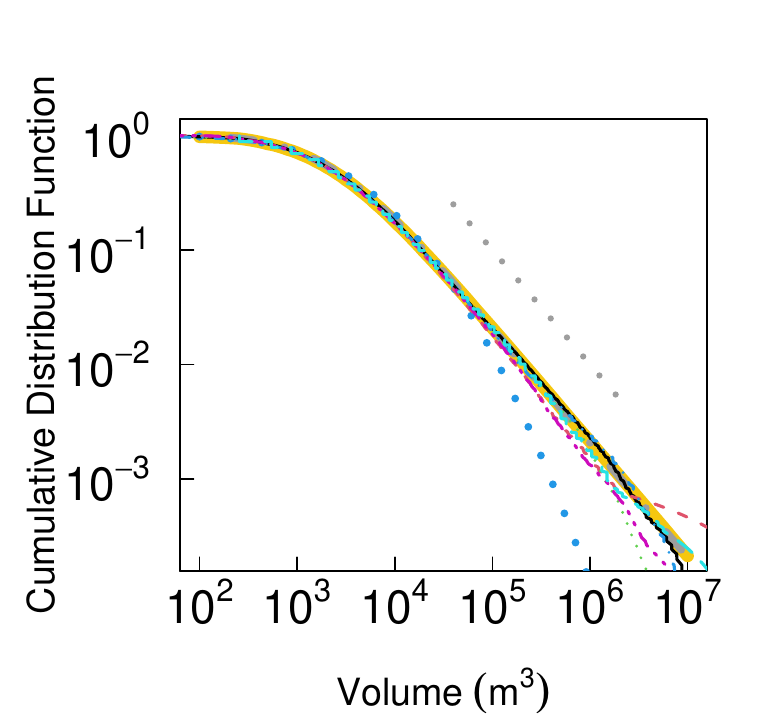}
        \put(120,113){\color{black}\Large\bfseries (b)}
    \end{overpic} 
   \begin{overpic}[
        width=5.8cm
    ]{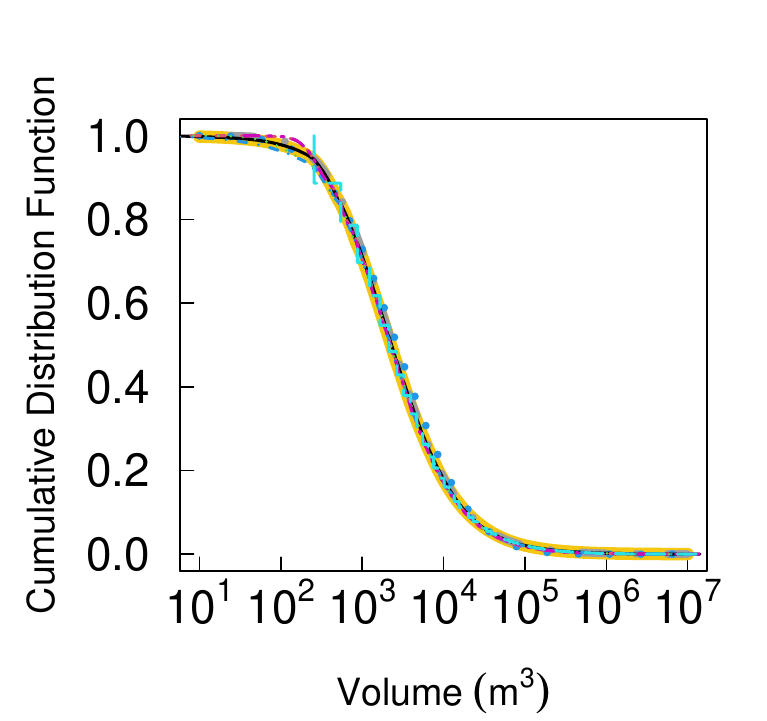}
        \put(120,113){\color{black}\Large\bfseries (c)}
    \end{overpic}
\end{center}
 \caption{
Dependence of the volume distribution on the entry distribution in the Kesten process (Eq.~\eqref{eq:kesten}).
(a) Probability density distributions of the entry variable $v$. The black solid line represents the uniform distribution $U(0,2)$, the red dashed line the normal distribution $N(\mu=1,\sigma=0.32)$, the green dashed line the lognormal distribution $\log N(\mu=-0.35,\sigma=0.83)$, the blue dash-dotted line the exponential distribution $\mathrm{Exp}(1)$, and the pink dash-dot-dotted line the beta distribution extended to the interval $[0,2]$, $2\times \mathrm{Beta}(s_1=5.06,s_2=5.06)$. All of these distributions are chosen so that their means are approximately 1.
(b) Upper cumulative distributions of volume obtained under the respective entry distributions (log--log plot). The colors and line styles of the entry distributions are the same as in (a), and the cyan long-dashed line represents the constant-entry case $v=1$. The thick yellow line represents the theoretical distribution of the Kesten process for the uniform-entry case, with parameters $A_0=255.81$, $b_0=1.128$, and $\alpha=1.00$. The thick gray dash-dotted line represents the one-sided dPlN distribution (Eq.~\eqref{app_eq_dPlN}), with parameters $\mu=6.88$, $\sigma=1.23$, and $\alpha=1.00$. The thick blue dashed line represents the lognormal distribution, with parameters $\mu=7.89$ and $\sigma=1.62$. The gray reference line represents $1/x$.
(c) Semi-logarithmic plot corresponding to (b). The volume distributions obtained from all entry distributions almost overlap in the central body, indicating that they depend little on the details of the entry distribution and share a lognormal-like central part. By contrast, in the tail, the theoretically expected power-law-like behavior of the $1/x$ type is preserved. In the constant-entry case, shown by the cyan long-dashed line, the distribution appears step-like because the entry value is fixed and the growth process is discrete in time.
}
\label{app_fig_f_main}
\end{figure*}

Fig.~\ref{app_fig_f_main}(a) shows the entry-size distributions compared in this section. Fig.~\ref{app_fig_f_main}(b) and (c) show the volume distributions generated by the Kesten process with these entry-size distributions. Even when the entry-size distribution is changed to a uniform, truncated normal, lognormal, gamma, or beta distribution, the resulting volume distributions almost overlap. In all cases, the central part of the distribution has a shape close to a lognormal distribution. The tail shows a power-law-like shape close to $1/x$.\par

In the case of constant entry, there is no fluctuation in the initial size at entry. The distribution therefore shows a slightly step-like shape. This occurs because individuals with the same growth duration are concentrated around almost the same size. Even in this case, however, the overall shape from the central part to the tail is not very different from that of the uniform-entry model used in the main text.\par

These results show that the qualitative conclusions of the Kesten process used in this study do not strongly depend on the detailed form of the initial-size distribution at new entry. In other words, the structure consisting of a lognormal-like central part and a power-law-like tail with an exponent close to one is not specific to the uniform entry-size distribution assumed in the main text. It also appears under a wider class of entry-size distributions. This suggests that the detailed form of the entry-size distribution is not the essential element of the model. In Sec.~\ref{app_sec_kesten_lnorm}, we discuss the relation between the additive term and the lognormal-like central part of the distribution.\par
%
%
%
%
%
%
%

\clearpage

\section{Double Pareto-lognormal distribution}
\label{app_sec_dPlN}
\subsection{Random-variable representation and derivation of the distribution}
In this section, we describe the double Pareto-lognormal distribution, or dPlN distribution, using a random-variable representation on the logarithmic scale \cite{ReedJorgensen2004dPlN}.
The one-sided dPlN distribution used in this study is treated as a special case in Sec.~\ref{app_sec_dPlN_kesten}.
For a positive random variable $X$, we define
\begin{equation}
Y=\log X .
\end{equation}
The dPlN distribution is obtained when $Y$ is expressed as the sum of a normal random variable and a two-sided exponential random variable.
More specifically, it can be written as
\begin{equation}
Y=\nu+\tau Z+U-V .
\end{equation}
Here,
\begin{equation}
Z\sim N(0,1),\qquad
U\sim {\rm Exp}(\alpha),\qquad
V\sim {\rm Exp}(\beta),
\end{equation}
and these random variables are assumed to be mutually independent.
The parameter $\alpha>0$ determines the exponent of the tail.
The parameter $\beta>0$ determines the exponent of the left tail.
In this representation, $U-V$ follows an asymmetric two-sided exponential distribution.
Its density function is given by
\begin{equation}
g(w)=
\frac{\alpha\beta}{\alpha+\beta}
\begin{cases}
e^{\beta w} & w<0,\\
e^{-\alpha w} & w\geq 0 .
\end{cases}
\end{equation}
Thus, on the logarithmic scale, the dPlN distribution can be interpreted as the sum of a normal distribution and a two-sided exponential distribution.

\subsubsection{Cumulative distribution of $X$}

It is convenient to standardize $Y$ by introducing
\begin{eqnarray}
\zeta \equiv Z+T,
\qquad T\equiv\frac{U-V}{\tau}, \nonumber \\
\qquad\text{so that}\qquad
Y=\nu+\tau\zeta,\qquad X=e^{Y}=e^{\nu+\tau\zeta}.
\end{eqnarray}

{\bf Step 1: cumulative distribution of $T=(U-V)/\tau$.}\\
The density of $W\equiv U-V$ is $g(w)$ given above, so $T=W/\tau$ has density $g_T(t)=\tau\,g(\tau t)$. Integrating this piecewise exponential density directly (no integration by parts needed) gives
\begin{equation}
H_T(t)\equiv P(T\le t)=
\begin{cases}
\dfrac{\alpha}{\alpha+\beta}\,e^{\beta\tau t}, & t<0,\\[6pt]
1-\dfrac{\beta}{\alpha+\beta}\,e^{-\alpha\tau t}, & t\ge 0 .
\end{cases}
\end{equation}
For example, for $t<0$,
\begin{equation}
H_T(t)=\int_{-\infty}^{t}\tau\frac{\alpha\beta}{\alpha+\beta}e^{\beta\tau s}\,ds
=\frac{\alpha}{\alpha+\beta}e^{\beta\tau t},
\end{equation}
and the branch for $t\ge0$ follows in the same way, starting from $H_T(0)=\alpha/(\alpha+\beta)$.

{\bf Step 2: a Gaussian identity.} \\
Here, $\phi(u)$ and $\Phi(u)$ denote the probability density function and
the cumulative distribution function of the standard normal distribution,
respectively, and $\Phi^{c}(u) \equiv 1-\Phi(u)$ denotes its complementary
cumulative distribution function. 
The following identity is known: for a constant $c$,
\begin{eqnarray}
\int_{-\infty}^{z}\phi(u)\,e^{cu}\,du=e^{c^2/2}\,\Phi(z-c), \nonumber \\
\int_{z}^{\infty}\phi(u)\,e^{-cu}\,du=e^{c^2/2}\,\Phi^c(z+c).
\end{eqnarray}

This identity can be derived as follows. Completing the square in the exponent of $\phi(u)e^{cu}$,
\begin{align}
\phi(u)e^{cu}
&=\frac{1}{\sqrt{2\pi}}
\exp\left\{-\frac{u^2}{2}+cu\right\}\\
&=\frac{1}{\sqrt{2\pi}}
\exp\left\{-\frac{1}{2}(u-c)^2+\frac{c^2}{2}\right\}\\
&=e^{c^2/2}\,\phi(u-c),
\end{align}
where we used $-\dfrac{u^2}{2}+cu=-\dfrac{1}{2}\big[(u-c)^2-c^2\big]$.
Replacing $c\to -c$ gives, in the same way,
\begin{equation}
\phi(u)e^{-cu}=e^{c^2/2}\,\phi(u+c).
\end{equation}
Integrating the first identity over $u\in(-\infty,z)$ and
substituting $v=u-c$ (so that $u=z\Rightarrow v=z-c$),
\begin{align}
\int_{-\infty}^{z}\phi(u)\,e^{cu}\,du
&=e^{c^2/2}\int_{-\infty}^{z}\phi(u-c)\,du\\
&=e^{c^2/2}\int_{-\infty}^{z-c}\phi(v)\,dv\\
&=e^{c^2/2}\,\Phi(z-c).
\end{align}
Similarly, integrating the second identity over $u\in(z,\infty)$ and
substituting $v=u+c$ (so that $u=z\Rightarrow v=z+c$),
\begin{align}
\int_{z}^{\infty}\phi(u)\,e^{-cu}\,du
&=e^{c^2/2}\int_{z}^{\infty}\phi(u+c)\,du\\
&=e^{c^2/2}\int_{z+c}^{\infty}\phi(v)\,dv\\
&=e^{c^2/2}\,\Phi^c(z+c).
\end{align}

{\bf Step 3: convolving $H_T$ with the normal density.} \\
Since $Z$ and $T$ are independent, the cumulative distribution function of $\zeta=Z+T$ is the convolution of the standard normal density $\phi$ with the cumulative distribution $H_T$:
\begin{equation}
F(z)\equiv P(\zeta\le z)=P(Z\le z-T)
=\int_{-\infty}^{\infty}\phi(u)\,H_T(z-u)\,du .
\end{equation}
Since $z-u<0$ for $u>z$ and $z-u\ge0$ for $u\le z$, we split the integral according to the two branches of $H_T$:
\begin{eqnarray}
&&F(z)=\int_{-\infty}^{z}\phi(u)\left[1-\frac{\beta}{\alpha+\beta}e^{-\alpha\tau(z-u)}\right]du \nonumber \\
&&+\int_{z}^{\infty}\phi(u)\,\frac{\alpha}{\alpha+\beta}e^{\beta\tau(z-u)}\,du .
\end{eqnarray}
Pulling out the $u$-independent factors,
\begin{eqnarray}
&&F(z)=\Phi(z) 
-\frac{\beta}{\alpha+\beta}e^{-\alpha\tau z}\int_{-\infty}^{z}\phi(u)e^{\alpha\tau u}\,du \nonumber \\
&&+\frac{\alpha}{\alpha+\beta}e^{\beta\tau z}\int_{z}^{\infty}\phi(u)e^{-\beta\tau u}\,du .
\end{eqnarray}
Applying the identity from Step 2 with $c=\alpha\tau$ and $c=\beta\tau$, respectively,
\begin{eqnarray}
&&F(z)=
\Phi(z)
-\frac{\beta}{\alpha+\beta}\,e^{-\alpha\tau z+\alpha^2\tau^2/2}\,\Phi(z-\alpha\tau) \nonumber \\
&&+\frac{\alpha}{\alpha+\beta}\,e^{\beta\tau z+\beta^2\tau^2/2}\,\Phi^c(z+\beta\tau).
\end{eqnarray}

{\bf Step5: Change of variables to $x$.} \\
Since $X=e^{\nu+\tau\zeta}$ is a strictly increasing function of $\zeta$,
\begin{equation}
F(x)\equiv P(X\le x)=P\!\left(\zeta\le \frac{\log x-\nu}{\tau}\right)=F(z),
\qquad
z=\frac{\log x-\nu}{\tau} .
\end{equation}
Using $\alpha\tau z=\alpha(\log x-\nu)$ and $\beta\tau z=\beta(\log x-\nu)$, this reproduces the cumulative distribution function of $X$:
\begin{equation}
\begin{aligned}
F(x)
=&\ \Phi(z)\\
&-\frac{\beta}{\alpha+\beta}
\exp\left\{
-\alpha(\log x-\nu)
+\frac{\alpha^2\tau^2}{2}
\right\}
\Phi(z-\alpha\tau)\\
&+
\frac{\alpha}{\alpha+\beta}
\exp\left\{
\beta(\log x-\nu)
+\frac{\beta^2\tau^2}{2}
\right\}
\Phi^c(z+\beta\tau).
\end{aligned}
\end{equation}

\subsubsection{Density of $X$} 
Differentiating $F(x)$ with respect to $x$ (using $dz/dx=1/(\tau x)$; the terms proportional to $\phi(z)$ cancel between the three pieces) gives the density function
\begin{equation}
\begin{aligned}
f(x)
=\frac{\alpha\beta}{\alpha+\beta}
\Bigg[
&
x^{-\alpha-1}
\exp\left\{
\alpha\nu+\frac{\alpha^2\tau^2}{2}
\right\}
\Phi(z-\alpha\tau)
\\
&+
x^{\beta-1}
\exp\left\{
-\beta\nu+\frac{\beta^2\tau^2}{2}
\right\}
\Phi^c(z+\beta\tau)
\Bigg].
\end{aligned}
\end{equation}

From this expression, the tail satisfies
\begin{equation}
P(X>x)\propto x^{-\alpha}
\qquad (x\to\infty).
\end{equation}
The left tail satisfies
\begin{equation}
P(X<x)\propto x^{\beta}
\qquad (x\to 0).
\end{equation}
Therefore, the dPlN distribution has a shape close to a lognormal distribution in its central part.
At the same time, it has Pareto-type power-law tails on both sides.
\subsection{One-sided dPlN distribution}
\label{app_sec_dPln_right}

From the representation above, the one-sided dPlN distribution can be understood as a limiting case of the dPlN distribution.
For example, consider the limit
\begin{equation}
\beta\to\infty .
\end{equation}
Then $V\sim{\rm Exp}(\beta)$ concentrates at 0.
In this limit, we obtain
\begin{equation}
Y=\nu+\tau Z+U .
\end{equation}
This gives a one-sided dPlN distribution with an exponential tail only on the right side.

Its cumulative distribution function is
\begin{equation}
F_{+}(x)
=\Phi(z)-
\exp\left\{
-\alpha(\log x-\nu)
+\frac{\alpha^2\tau^2}{2}
\right\}
\Phi(z-\alpha\tau).
\end{equation}
The density function is
\begin{equation}
f_{+}(x)=\alpha
x^{-\alpha-1}
\exp\left\{
\alpha\nu+\frac{\alpha^2\tau^2}{2}
\right\}
\Phi(z-\alpha\tau) . \label{app_eq_dPlN_pdf}
\end{equation}
This distribution has a tail of the form
\begin{equation}
P(X>x)\propto x^{-\alpha}.
\end{equation}
On the other hand, it does not have a Pareto-type tail on the left side.


Thus, the one-sided dPlN distribution can be regarded as the limiting case in which one of the exponential components is removed from the dPlN distribution.\par

In the main text, we denote the one-sided dPlN distribution by
\begin{equation}
X\sim {\rm dPlN}_{+}(\mu,\sigma,\alpha) . \label{app_eq_dPlN}
\end{equation}
This notation is used to make the correspondence with the lognormal distribution clear.
Here, $\mu$ and $\sigma$ are used as the location and scale parameters of the lognormal component.
The corresponding probability density function is given by Eq.~\eqref{app_eq_dPlN_pdf}.
The parameters correspond as $\mu=\nu$ and $\sigma=\tau$.

\subsection{One-sided dPlN distribution and the Kesten process}
\label{app_sec_dPlN_kesten}

The one-sided dPlN distribution can also be understood as a special limiting case of a Kesten-type framework.
Consider the case in which the additive term during growth is removed.
We also assume that the entry size after resetting follows a lognormal distribution.

Specifically, let
\begin{equation}
x_{t+1}=b_0 x_t
\end{equation}
and assume that the entry size after resetting is given by
\begin{equation}
x_{\rm new}\sim {\rm Lognormal}(\mu,\sigma^2).
\end{equation}
Then, after $T$ growth steps following a reset, the size is
\begin{equation}
x_T=x_{\rm new} b_0^T .
\end{equation}
Taking the logarithm gives
\begin{equation}
\log x_T
=
\log x_{\rm new}
+
T\log b_0 .
\end{equation}

Here, $\log x_{\rm new}$ follows a normal distribution.
The growth duration $T$ follows a geometric distribution.
Therefore, under a continuous approximation, $T\log b_0$ becomes an exponentially distributed component.
Thus, we can write
\begin{equation}
\log x_T
\simeq
\mu+\sigma Z+U,
\qquad
U\sim{\rm Exp}(\alpha),
\end{equation}
where
\begin{equation}
\alpha
=
-\frac{\log p_0}{\log b_0}.
\end{equation}
This is the random-variable representation of the one-sided dPlN distribution.

Thus, the one-sided dPlN distribution can be regarded as a continuous approximation of a Kesten-type process without an additive term.
In this limiting case, the entry size follows a lognormal distribution.
In contrast, the model used in this study includes the additive term $A_0$.
It also uses a simple uniform distribution for the entry size.

Therefore, in the present model, the lognormal-like central part is not produced by assuming a lognormal entry distribution.
Instead, it arises from the accumulation of additive terms and the mixture of short-lived groups.
In this respect, the Kesten-type model used in this study shares the mechanism that generates the tail with the one-sided dPlN distribution.
However, it differs in the mechanism that generates the central part of the distribution.

\subsection{Derivation of the dPlN distribution from a random walk and a stopping process}
We next show that the dPlN distribution arises naturally from a random walk and a stopping process.
Suppose that the log-size $Y(t)$ follows a continuous-time random walk.
That is, we write
\begin{equation}
Y(t)=Y_0+mt+sB_t .
\end{equation}
Here, $B_t$ is standard Brownian motion.
The parameter $m$ is the drift, and $s$ represents the magnitude of fluctuations.
We assume that the initial value follows
\begin{equation}
Y_0\sim N(\nu,\tau^2).
\end{equation}
We also assume that the growth process stops at a random time $T$, where
\begin{equation}
T\sim {\rm Exp}(\lambda).
\end{equation}
The observed log-size is then
\begin{equation}
Y=Y(T)=Y_0+mT+sB_T .
\end{equation}
The key point is that
\begin{equation}
Q=mT+sB_T
\end{equation}
follows a two-sided exponential distribution.
Indeed, conditional on $T=t$, we have
\begin{equation}
Q\mid T=t \sim N(mt,s^2t).
\end{equation}

{\bf Known facts used below.} \\
We use the following two standard facts about characteristic functions.
\begin{itemize}
\item[(i)] If $X\sim N(\mu,\sigma^2)$, then
\begin{equation}
E\left[e^{iuX}\right]=\exp\left\{iu\mu-\frac{1}{2}\sigma^2u^2\right\},
\qquad u\in\mathbb{R}.
\end{equation}
\item[(ii)] If $T\sim {\rm Exp}(\lambda)$, then
\begin{equation}
E\left[e^{\theta T}\right]=\frac{\lambda}{\lambda-\theta},
\qquad \mathrm{Re}(\theta)<\lambda .
\end{equation}
In particular, $E[e^{iuT}]=\lambda/(\lambda-iu)$.
\end{itemize}

{\bf Characteristic function of $Q=mT+sB_T$.} \\
By fact (i) applied to $Q\mid T=t\sim N(mt,s^2t)$,
\begin{align}
E\left[e^{iuQ}\mid T=t\right]
&=\exp\left\{ium\,t-\frac{1}{2}s^2u^2t\right\}\\
&=\exp\left\{t\left(ium-\frac{1}{2}s^2u^2\right)\right\}.
\end{align}
Since $\mathrm{Re}\!\left(ium-\frac12 s^2u^2\right)=-\frac12 s^2u^2<\lambda$,
we may apply fact (ii) with $\theta=ium-\frac12 s^2u^2$, averaging over $T$:
\begin{align}
E\left[e^{iuQ}\right]
&=E\left[E\left[e^{iuQ}\mid T\right]\right]\\
&=E\left[\exp\left\{T\left(ium-\frac{1}{2}s^2u^2\right)\right\}\right]\\
&=\frac{\lambda}{\lambda-\left(ium-\frac{1}{2}s^2u^2\right)}\\
&=\frac{\lambda}{\lambda-ium+\frac{1}{2}s^2u^2}.
\end{align}

{\bf Characteristic function of $U-V$.}\\
Let $U\sim{\rm Exp}(\alpha)$ and $V\sim{\rm Exp}(\beta)$ be independent.
Using fact (ii) twice (with $\theta=iu$ for $U$, and $\theta=-iu$ for $V$),
\begin{align}
E\left[e^{iu(U-V)}\right]
&=E\left[e^{iuU}\right]E\left[e^{-iuV}\right]\\
&=\frac{\alpha}{\alpha-iu}\cdot\frac{\beta}{\beta+iu}\\
&=\frac{\alpha\beta}{(\alpha-iu)(\beta+iu)}\\
&=\frac{\alpha\beta}{\alpha\beta+u^2+iu(\alpha-\beta)}.
\end{align}

{\bf Matching the two characteristic functions.} \\
Multiplying the numerator and denominator of $E[e^{iuQ}]$ by $2/s^2$,
\begin{equation}
E\left[e^{iuQ}\right]
=\frac{2\lambda/s^2}
{u^2-iu\left(2m/s^2\right)+2\lambda/s^2}.
\end{equation}
Both $E[e^{iuQ}]$ and $E[e^{iu(U-V)}]$ are ratios of a constant to a monic
quadratic in $u$, with the constant equal to the quadratic's value at $u=0$
(both characteristic functions equal $1$ at $u=0$).
Hence $E[e^{iuQ}]=E[e^{iu(U-V)}]$ for all $u$ if and only if the two
denominators agree coefficient by coefficient:
\begin{align}
\alpha\beta&=\frac{2\lambda}{s^2}, \label{eq_ab_prod}\\
\beta-\alpha&=\frac{2m}{s^2}. \label{eq_ab_diff}
\end{align}

{\bf Solving for $\alpha,\beta$.}\\
From \eqref{eq_ab_diff}, $\beta=\alpha+2m/s^2$. Substituting into \eqref{eq_ab_prod},
\begin{equation}
\alpha\left(\alpha+\frac{2m}{s^2}\right)=\frac{2\lambda}{s^2}
\;\Longrightarrow\;
\alpha^2+\frac{2m}{s^2}\,\alpha-\frac{2\lambda}{s^2}=0 .
\end{equation}
Since $\alpha>0$, the quadratic formula selects the positive root:
\begin{align}
\alpha
&=\frac{\sqrt{m^2+2\lambda s^2}-m}{s^2}.
\end{align}
Correspondingly,
\begin{align}
\beta&=\alpha+\frac{2m}{s^2}\\
&=\frac{\sqrt{m^2+2\lambda s^2}+m}{s^2}.
\end{align}
Thus, with $\alpha,\beta$ given by
\begin{equation}
\alpha
=
\frac{\sqrt{m^2+2\lambda s^2}-m}{s^2},
\qquad
\beta
=
\frac{\sqrt{m^2+2\lambda s^2}+m}{s^2},
\end{equation}
the characteristic function of $Q=mT+sB_T$ coincides with that of $U-V$, so
\begin{equation}
mT+sB_T
\overset{d}{=}
U-V .
\end{equation}
Here, $\overset{d}{=}$ denotes equality in distribution.
It follows that
\begin{equation}
\begin{aligned}
Y
&=
Y_0+mT+sB_T \\
&\overset{d}{=}
\nu+\tau Z+U-V .
\end{aligned}
\end{equation}
Therefore,
\begin{equation}
X=e^Y
\end{equation}
follows a dPlN distribution.
This derivation shows that the dPlN distribution can be generated by three elements.
These are lognormal variation in the initial size, random-walk-like growth, and a stopping time that follows an exponential distribution.
In particular, the exponential stopping time produces large variation in the growth duration $T$.
As a result, a two-sided exponential component appears on the logarithmic scale.
On the original scale, this component generates Pareto-type heavy tails.

\clearpage
\section{Pareto-type distributions in premodern data}
\label{app_sec_old}
This section reviews studies on distributions of wealth and related socioeconomic proxies in premodern societies.
Here, we focus not on urban scale, which is often estimated from variables such as city-wall length, but on wealth-related proxies such as property holdings, landholdings, residential size, and grave size.
Studies of wealth distributions in this sense remain limited for premodern societies. \par

Existing cases include house sizes in Akhetaten, Egypt, in the 14th century BCE \cite{AbulMagd2002}, dwelling sizes in Pompeii in the first century CE \cite{Danon2022PompeiiSenatorialWealth}, landholding records from fourth-century Hermopolis in Egypt \cite{Danon2025ReconstructingWealthDistributions}, property declarations from 15th-century Florence \cite{Danon2025ReconstructingWealthDistributions}, grave sizes in Neolithic Chinese settlements \cite{YuEtAl2019GraveSizesInequality}, dolmen volumes in Gochang, Korea \cite{NohKim2026SouthKoreaInequality}, and residential sizes at ancient Maya sites \cite{StrawinskaZankoEtAl2018MayaInequality,BrownEtAl2012PoorMayapan}.
However, most of these studies analyze distributions at the individual, household, or single-site level.
Premodern datasets available at the collective-unit or country-wide scale are extremely scarce.

One of the few examples at the collective-unit level is the number of serf families owned by medieval Hungarian nobles \cite{HegyiEtAl2007}.
In that study, the distribution of serf-family holdings was reported to follow a power-law distribution with a cumulative exponent of approximately 1.
In this study, we also analyze, for comparison, the kokudaka of daimyo, or regional lords, during Japan's Edo period from the 17th to 19th centuries.
Kokudaka is a rice-yield-based measure of economic scale (Appendix \ref{a_sec_kokudaka}).
The kokudaka distribution can also be characterized as having a cumulative exponent close to 1.\par

However, both medieval Hungary and Edo-period Japan were societies in which institutional infrastructures, including written administration and monetary exchange, were already present.
Therefore, these cases do not directly answer whether Zipf's law can arise in collective-unit resource distributions before such infrastructures were fully established.\par

Table \ref{app_table_tab_old} summarizes the power-law exponents reported in these studies or used for comparison in the present study.
However, because the objects of analysis, sample sizes, variable definitions, and estimation methods differ across studies, these values should be treated as reference values rather than as strict comparisons.

\begin{table*}[t]
\centering
\scriptsize
\setlength{\tabcolsep}{1.7pt}
\renewcommand{\arraystretch}{1.03}
\caption{
Examples of Pareto-like or power-law-like distributions in premodern
wealth-related proxies.
Exponents $\alpha$ are expressed as complementary cumulative distribution function
(CCDF) exponents and are rounded to the displayed precision.
``Upper-tail MLE'' denotes a Clauset-type maximum-likelihood fit to the
upper tail.
Here, $N$ denotes the total sample size; tail sample sizes are given in
the notes where available.
}
\label{app_table_tab_old}
\begin{tabular}{
@{}
p{1.85cm}
p{1.65cm}
p{1.85cm}
p{1.85cm}
p{1.85cm}
p{0.85cm}
p{0.85cm}
p{4.30cm}
@{}
}
\hline
Reference
& Region
& Period
& Unit of analysis
& Scale measure
& $N$
& $\alpha$
& Notes \\
\hline

\multicolumn{8}{l}{\textbf{Broad-scale collective or political units}} \\
\hline

Hegyi et al. (2007) \cite{HegyiEtAl2007}
& Hungary
& c.~1550
& Noble families and religious institutions
& Serf households controlled
& 1,399
& 0.92
& Entities controlling fewer than ten serf households excluded;
1,283 noble families and 116 religious institutions. \\

This study
& Japan
& Mid-third to seventh century CE
& Local groups represented by keyhole-shaped kofun
& Kofun volume
& 4,545
& 1.00
& Proxy for collective mobilization rather than individual wealth. \\

This study
& Japan
& Seventeenth-century Edo period
& Daimyo domains
& Omotedaka
& 270
& 1.38
& Official rice-yield-based assessment of domain economic scale. \\

This study
& Japan
& Nineteenth-century Edo period
& Daimyo domains
& Jitsudaka
& 270
& 1.03
& Estimate of actual rice-yield-based domain economic scale. \\

\hline
\multicolumn{8}{l}{
\textbf{Local or regional individual-, household-, residence-,
landholding-, and burial-level proxies}
} \\
\hline
Abul-Magd (2002) \cite{AbulMagd2002}
& Akhetaten, Egypt
& Fourteenth century BCE
& Households or residences
& House area
& 498
& 3.76
& Whole-distribution generalized Lotka--Volterra fit;
an alternative cutoff specification gives $\alpha=1.59$. \\

Danon (2025) \cite{Danon2025ReconstructingWealthDistributions}
& Hermopolis, Egypt
& Fourth century CE
& Landholders
& Landholding area
& 233
& 0.70
& Upper-tail MLE; records from one of four urban districts. \\

Danon (2022) \cite{Danon2022PompeiiSenatorialWealth}
& Pompeii, Italy
& 79 CE
& Households or residences
& Residential ground-floor area
& 366
& 1.82
& Upper-tail MLE with 10,000 bootstraps;
95\% HPD interval: 1.52--2.15. \\

Danon (2025) \cite{Danon2025ReconstructingWealthDistributions}
& Florence, Italy
& 1427
& Households
& Declared household wealth
& 8,349
& 1.45
& Upper-tail MLE based on fiscal declarations. \\

Strawinska-Zanko et al. (2018)
\cite{StrawinskaZankoEtAl2018MayaInequality}
& Komch\'en, Mexico
& c.~350--150 BCE
& Residential structures
& House area
& 324
& 2.38
& Upper-tail MLE, tail $n=66$; whole distribution closer to exponential. \\

Strawinska-Zanko et al. (2018)
\cite{StrawinskaZankoEtAl2018MayaInequality}
& Komch\'en, Mexico
& c.~350--150 BCE
& Residential structures
& Construction volume
& 324
& 1.89
& Same method, tail $n=40$; limited reliability. \\

Strawinska-Zanko et al. (2018)
\cite{StrawinskaZankoEtAl2018MayaInequality}
& Palenque, Mexico
& c.~600--800 CE
& Residential structures
& House area
& 1,135
& 1.87
& Upper-tail MLE, tail $n=253$. \\

Strawinska-Zanko et al. (2018)
\cite{StrawinskaZankoEtAl2018MayaInequality}
& Palenque, Mexico
& c.~600--800 CE
& Residential structures
& Construction volume
& 1,135
& 1.47
& Same method, tail $n=63$; volume proxies labor investment. \\

Strawinska-Zanko et al. (2018)
\cite{StrawinskaZankoEtAl2018MayaInequality}
& Sayil, Mexico
& c.~800--1000 CE
& Residential structures
& House area
& 767
& 1.56
& Upper-tail MLE, tail $n=125$. \\

Strawinska-Zanko et al. (2018)
\cite{StrawinskaZankoEtAl2018MayaInequality}
& Mayap\'an, Mexico
& c.~1200--1450 CE
& Residential structures
& House area
& 1,214
& 2.60
& Upper-tail MLE, tail $n=418$; reanalysis of Brown et al. (2012). \\

Yu et al. (2019) \cite{YuEtAl2019GraveSizesInequality}
& Dawenkou, China
& 6000--4500 cal yr BP
& Graves or buried individuals
& Grave area
& 133
& 1.95
& Whole-distribution asymmetric double Pareto fit;
Gini coefficient 0.38. \\

Yu et al. (2019) \cite{YuEtAl2019GraveSizesInequality}
& Liangzhu, China
& 5500--4000 cal yr BP
& Graves or buried individuals
& Grave area
& 80
& 3.19
& Same method; three cemeteries combined;
Gini coefficient 0.28. \\

Noh and Kim (2026) \cite{NohKim2026SouthKoreaInequality}
& Southeastern Gochang, South Korea
& c.~300 BCE--1 CE
& Dolmens
& Capstone volume
& 415
& 2.26
& Upper-tail MLE, tail $n=53$; 95\% CI: 1.17--3.23;
power law and lognormal not distinguishable;
Gini coefficient 0.537. \\

\hline
\end{tabular}
\end{table*}


\end{document}